\documentclass[]{jfm}

\usepackage{graphicx}
\usepackage{newtxtext}
\usepackage{newtxmath}
\usepackage{natbib}
\usepackage{hyperref}
\hypersetup{
    colorlinks = true,
    urlcolor   = blue,
    citecolor  = black,
}

\newcommand{\RomanNumeralCaps}[1]

\title{Vortex dynamics of accelerated flow past a mounted wedge}

\author{Jiten C. Kalita\aff{1}	
	\and Pankaj Kumar\aff{2}
	\corresp{\email{jiten@iitg.ac.in}}}

\affiliation{\aff{1}Department of Mathematics, Indian Institute of Technology Guwahati, Guwahati 781039, Assam, India
	\aff{2}Department of Mechanical Engineering, Indian Institute of Technology Guwahati, Guwahati 781039, Assam, India}

\begin{document}
	\maketitle
	
	\begin{abstract}
		This study is concerned with the simulation of a complex fluid flow problem involving flow past a wedge mounted on a wall for channel Reynolds numbers $Re_c=1560$, $6621$ and $6873$ in uniform and accelerated flow medium. The transient Navier-Stokes (N-S) equations governing the flow has been discretized using a recently developed second order spatially and temporally accurate compact finite difference method on a nonuniform Cartesian grid by the authors. All the flow characteristics of a well-known laboratory experiment of \cite{pullin1980} have been remarkably well captured by our numerical simulation, and we provide a qualitative and quantitative assessment of the same. Furthermore, the influence of the parameter $m$, controlling the intensity of acceleration, has been discussed in detail along with the intriguing consequence of non-dimensionalization of the N-S equations pertaining to such flows. The simulation of the flow across a time span significantly greater than the aforesaid lab experiment is the current study's most noteworthy accomplishment. For the accelerated flow, the onset of shear layer instability leading to a more complicated flow towards transition to turbulence have also been aptly resolved. The existence of coherent structures in the flow validates the quality of our simulation, as does the remarkable similarity of our simulation to the high Reynolds number experimental results of \cite{lian1989} for the accelerated flow across a typical flat plate. All three steps of vortex shedding, including the exceedingly intricate three-fold structure, have been captured quite efficiently.
	\end{abstract}

	\noindent

	\vspace*{0.3cm}
	\noindent 
	\textbf{Keywords: }  \\
	\textbf{Mathematics Subject Classifications:} 
	\section{Introduction}
	Owing to the intrinsic and complicated characteristics of the flow phenomena, especially the separation of flows, research on flow past sharp edges of bluff bodies has attracted a great deal of interest for many years. Here, the boundary layer gets constricted near the edge, intensifies and eventually forms a vortex. The boundary layer separates and forms a coherent spiral shear-layer as the vortex evolves in intensity and size, which is actually a roll-up characteristic of vortex formation.
	These flows can be used to investigate the free shear layer and vortex dynamics. The current study aims to investigate a classic example of fluid flow past a bluff body with a sharp edge, namely flow past a wedge.
	
	The flow past a thin plate normal to the free stream is a classic example of bluff body flow, which corresponds to wedges with zero angle of incidence. \citet{prandtl1904} was the pioneer in such studies, who experimented with the flow past a normal flat plate in 1904 in the first study of this kind. Later, \citet{anton1956} and \citet{wedemeyer1961} presented their experimental studies for the same. \citet{pierce1961} used the spark shadowgraph technique to show photographic evidence of separation at sharp angles. \citet{taneda1971} employed the aluminum dust method to visualize the flow past impulsively started and uniformly accelerated flat plates. \citet{lian1989} used the hydrogen bubble technique to visualize the flow past accelerating plates. \citet{dennis1993} did an experimental investigation for the impulsively started flat plate at moderate Reynolds numbers. In a similar fashion, \citet{pullin1980} used dye in water to study the development of the separating shear layer past thin and thick wedges. They presented the trajectory of the vortex center through detailed analysis for different acceleration parameters $m$ and wedge angles. They also compared their observations via inviscid similarity theory \citet{kaden1931,pullin1978}.  Recent experiments are seen to focus their attention on studying the leading-edge flow past a plate and three-dimensional tip vortex effects \citet{kriegseis2013}; \citet{ringuette2007}) associated with it. Another study worth mentioning in this direction is that of \citet{koumoutsakos1996}, who  numerically simulated the flow past an impulsively started ($m=0$)  and uniformly accelerated ($m=1$) flat plates.
	
	In contrast to other bluff body flows, the wedge's separation point is marked by a sharp edge. Although it appears to be easy, obtaining the numerical solutions are difficult owing to edge singularity. Many scholars have published analytical results for this sort of flow \citet{villat1930}; \citet{wedemeyer1961}; \citet{pullin1978}; \citet{krasny1991}) assuming inviscid evolution of the vorticity field by properly assessing circulation growth at the wedge tip and distant from it.
	Furthermore, numerical simulation of these types of flows is difficult due to the singularities at the tips. As a result, the majority of the computational techniques devised to explore this flow were initially limited to steady-state flow for low and moderate Reynolds values. To avoid singularity, one way is to construct an extremely fine grid around the edge.
	
	\citet{davies1995} (in 1995) numerically studied the interactions of coastal currents with topographic indentations that mimics the flow past a sharp wedge. Recently, \citet{xu2015} numerically studied viscous flow past a flat plate in accelerating flow. Their results were in good agreement with the experimental study of \citet{pullin1980} for thin wedges with an angle of $5^0$. \citet{xu2016} then numerically studied vortex formation in the impulsively started viscous flow past an infinite wedge for wedge angles ranging from $60^0$ to $150^0$ in uniform flow, where potential flow theory was employed to solve this problem. All the previous research, both experimental and numerical \citet{kiya1988,perry1987,kiya1980,chein1988,tamaddon1994}, including those by \citet{pullin1980} and \citet{xu2015,xu2016} on the flat plate or wedge, were conducted only for the early part in the near wake region.
	
	The current research is mainly inspired by the experimental visualization of \citet{pullin1980} for the flow past a mounted wedge for an wedge angle $60^\circ$. Our study attempts to build on other previous works of research ( \citet{pullin1978,koumoutsakos1996,xu2015,xu2016}) by addressing flow simulation in the following two areas: uniform flow in laminar region for channel Reynolds number $Re_c=1560$ and accelerated flow till transition to turbulence from laminar state for $Re_c=6621$ $Re_c=6873$. A recently developed second order spatially and temporally accurate compact finite difference method on a nonuniform Cartesian grid by the authors (\cite{kumar2020}) has been used to discretize the transient Navier-Stokes equations governing the flow. Apart from replicating the flow visualization of \citet{pullin1980} for the earliest part of the flow, where the flow was stopped at a non-dimensional time much less than unity, we continued our simulations up to a non-dimensional time $t=3.0$. This allowed us to observe hitherto unreported vortex structures, which were observed earlier only in the experimental visualization of accelerated flow past a flat plate (\cite{lian1989}). We have also provided a detailed account of the flow separation, vortex formation and the movement of the vortex centers for  $Re_c=6873$, which has not been reported earlier.
	
	The paper has been arranged in six sections. Section \ref{flow_des} deals with the flow description and the governing equations, section \ref{discr} with a brief discussion on discretization procedure and numerical method, section \ref{result}, and finally, section \ref{concl} summarizes the whole work. 
	
	\section{The flow description and the governing equations}\label{flow_des}
	\begin{figure}
		\begin{center}
			\includegraphics[width=5.5in]{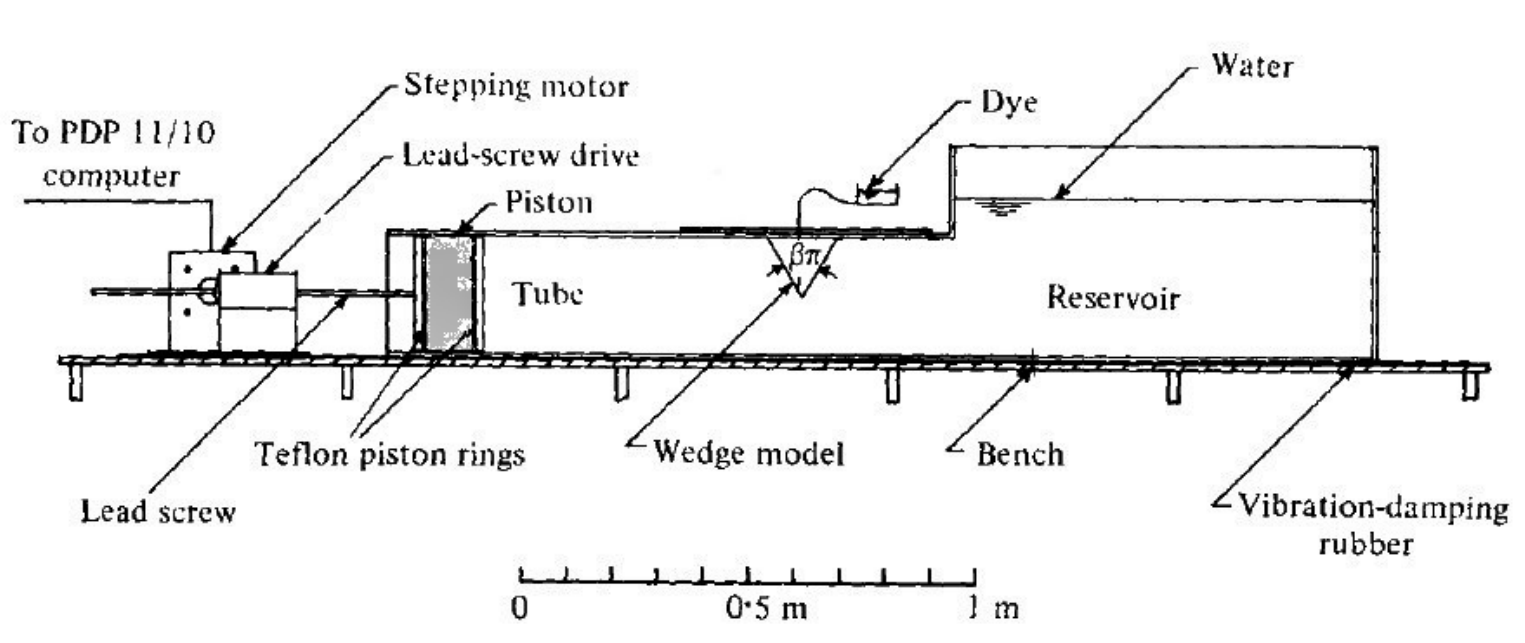}
			\caption{Experimental set-up for starting vortex flow visualization by (\citet{pullin1980}). The movement of the piston, shown in shade, generates the acceleration in the flow.}
			\label{fig:pulin}
		\end{center}
	\end{figure}
	
	The problem considered here is the accelerated flow past a wedge mounted on a wall, which was experimentally carried out by \citet{pullin1980}. The experimental set-up used by them for the starting vortex flow visualization is reproduced  in figure \ref{fig:pulin}. The wedge of height $h$ is kept normal to the accelerated flow in a channel of width $H$  and is fixed to the wall with an angle $60^0$, defined as the wedge angle $\theta=\beta\pi$, so that the case undertaken here corresponds to  $\beta=\frac{1}{3}$. The flow was driven by the rectangular piston as shown in the figure, which was also responsible for imparting acceleration to the flow. Note that flow past a flat plate is a limiting example of flow past a wedge of zero angle. This sharp edge fixes the point of separation which leads to generation of spiral shear layer and vortex formation. For other details of the set-up, readers may refer to the same work by \citet{pullin1980}.

	The flow is governed by the Navier-Stokes equations for incompressible viscous flows, the streamfunction-velocity ($\psi$-$v$) formulation (\cite{gupta2005,kalita2010,sen2013_2,sen2015}) of which, in dimensional form, can be written as
	\begin{equation}\label{sv}
		\frac{\partial}{\partial \tilde{t}}(\nabla^2\tilde{\psi})=\nu\nabla^4\tilde{\psi}+\tilde{u}\nabla^2\tilde{v}-\tilde{v}\nabla^2\tilde{u},
	\end{equation}
	where the quantities $\tilde{t}$, $\tilde{\psi}$, $\tilde{u}$, $\tilde{v}$  are respectively time, streamfunction, and velocities along the horizontal $\tilde{x}$- and vertical $\tilde{y}$-directions and $\nu$ is the kinematic viscosity. Their units are $\tilde{t}\; (s)$, $\tilde{\psi} \; ({cm}^2 s^{-1})$, $\tilde{u} \; ({cm}s^{-1})$, $\tilde{v} \; ({cm}s^{-1})$ and $\nu \; ({cm}^2 s^{-1})$.
	
	The $\psi$-$v$ formulation can easily be written in pure Streamfunction form as
	\begin{equation}\label{pure}
		\frac{\partial}{\partial \tilde{t}}(\nabla^2\tilde{\psi})=\nu\nabla^4\tilde{\psi}-\frac{\partial \tilde{\psi}}{\partial \tilde{y}}\frac{\partial}{\partial \tilde{x}}\nabla^2\tilde{\psi}+\frac{\partial \tilde{\psi}}{\partial \tilde{x}}\frac{\partial}{\partial \tilde{y}}\nabla^2\tilde{\psi}
	\end{equation}
	
	As mentioned earlier, the flow is driven by the piston moving with a velocity $\tilde{v}_p=\tilde{u}_0+A\tilde{t}^m$ where $\tilde{u}_0$ (set as zero in the actual experiment) is the 	velocity at the inlet, $A$ and $m$ are pre-chosen constants. $\tilde{v}_p$ enables acceleration of the viscous fluid. For example, $m=0$ renders a uniform velocity at the inlet indicating an impulsive start, while $m=1$ and $2$ stand for uniform and linear acceleration respectively \citet{xu2015}. Equations \eqref{sv}-\eqref{pure} are non-dimensionalized using the channel width $H$ and, the constants $A$ and $m$. Defining the non-dimensional variables
	
	\begin{equation}\label{nondim}
		(x,y)=\frac{(\tilde{x},\tilde{y})}{H},\; (u,v)=\frac{(\tilde{u},\tilde{v})}{H\left(\frac{A}{H}\right)^{\frac{1}{1+m}}}, \; \psi=\frac{\tilde{\psi}}{H^2\left(\frac{A}{H}\right)^{\frac{1}{1+m}}}\; {\rm and } \; t=\frac{\tilde{t}}{\left(\frac{H}{A}\right)^{\frac{1}{1+m}}},
	\end{equation}
	equation \eqref{sv} in non-dimensional form can be written as 
	\begin{equation}\label{non_sv}
		\frac{\partial}{\partial t}(\nabla^2\psi)=\frac{1}{Re_c}\nabla^4\psi+u\nabla^2v-v\nabla^2u,
	\end{equation}
	where $\displaystyle Re_c$ is the channel Reynolds number defined as $\displaystyle Re_c=\frac{H^2\left(\frac{A}{H}\right)^{\frac{1}{1+m}}}{\nu}$ (\citet{pullin1980}). For the values of $m$
	used in this study, the dimensional variables can be retrieved by using the following table \ref{t1} where apart from tabulating the pre-chosen constants $A$, $m$ and the corresponding channel Reynolds numbers $Re_c$, we have also provided the instants of time at which the experiments were stopped and the corresponding non-dimensional times following \eqref{nondim}.
	
	\begin{table}
		\caption{Flow parameters used in the current study.}
		\begin{center}
			\begin{tabular}{ccccc}\hline
				m &$A \; (cm \; s^{-(1+m)})$ &$Re_c$ &$\tilde{t}_f\;(s)$ &$t_f$\\ \hline
				0 &0.63 &1560 &12.52 &0.310535\\
				0.45 &0.86 &6621 &5.96 &0.577061 \\
				0.88 &0.38 &6873 &7.08 &0.757289\\
				\hline
			\end{tabular}
		\end{center}
		\label{t1}
	\end{table}
	
	\citet{pullin1980} also introduced a scale Reynolds number defined by 
	\begin{equation}
		Re_s(\tilde{t})=\alpha_0^{\frac{2}{2-n}}\left[\left(\frac{A}{H}\right)^{\frac{1}{1+m}}\tilde{t}\right]^{2M-1}Re_c,
		\label{scale}
	\end{equation}
	with $\displaystyle n=\frac{1}{2-\beta}$, $\displaystyle M=\frac{1+m}{2-n}$ and $\alpha_0$ is a dimensionless constant whose value depends on the wedge angle $\beta\pi$ and can be found from figure 14 of \citet{pullin1980}. Note that the expression within the parenthesis on the right hand side of \eqref{scale} is the non-dimensional time and therefore \eqref{scale} may be recast as 
	\begin{equation}
		Re_s({t})=\alpha_0^{\frac{2}{2-n}}{t}^{2M-1}Re_c,
		\label{scale_t}
	\end{equation}
	which will be used to show the trajectory of the scaled vortex center later on in section \ref{inv_scl}.
	
	\section{Discretization and numerical method}\label{discr}
	A second order time and space accurate compact finite difference scheme has recently been devised (\cite{kumar2020}) for the $\psi$-$v$ form of the transient-state Navier-Stokes (N-S) equations on nonuniform Cartesian grids without transformation. On a nine point stencil, equation \eqref{non_sv} in discretized form has the following form:

	\begin{equation}\label{dis1}
		\begin{aligned}
			&\dfrac{2}{x_f +x_b}\left[\dfrac{\psi_{i+1,j}^{n+1}}{x_f} + \dfrac{\psi_{i-1,j}^{n+1}}{x_b}-\left(\dfrac{1}{x_f}+\dfrac{1}{x_b}\right)\psi_{i,j}^{n+1}\right] + \dfrac{2}{y_f +y_b}\left[\dfrac{\psi_{i,j+1}^{n+1}}{y_f}+\dfrac{\psi_{i,j-1}^{n+1}}{y_b} -\left(\dfrac{1}{y_f}+\dfrac{1}{y_b}\right)\psi_{i,j}^{n+1}\right]\\ &=\dfrac{0.5 \Delta t}{Re_c}\left\{\left( A\psi _{i+1,j+1}^{n}+B\psi _{i,j+1}^{n}+C\psi _{i-1,j+1}^{n}+D\psi _{i+1,j}^{n}+E\psi _{i,j}^{n}+F\psi _{i-1,j}^{n}+G\psi _{i+1,j-1}^{n}+H\psi _{i,j-1}^{n}\right. \right.\\& \left. \left.+I\psi _{i-1,j-1}^{n}\right)-\phi_{i,j}^{n}\right\} +\dfrac{0.5\Delta t}{Re_c}\left\{\left( A\psi _{i+1,j+1}^{n+1}+B\psi _{i,j+1}^{n+1}+C\psi _{i-1,j+1}^{n+1}+D\psi _{i+1,j}^{n+1}+E\psi _{i,j}^{n+1}+F\psi _{i-1,j}^{n+1}\right. \right.\\& \left. \left.+G\psi _{i+1,j-1}^{n+1}+H\psi _{i,j-1}^{n+1}+I\psi _{i-1,j-1}^{n+1}\right)-\phi_{i,j}^{n+1}\right\} \\& +\dfrac{2}{x_f +x_b}\left[\dfrac{\psi_{i+1,j}^{n}}{x_f} + \dfrac{\psi_{i-1,j}^{n}}{x_b}-\left(\dfrac{1}{x_f}+\dfrac{1}{x_b}\right)\psi_{i,j}^{n}\right] + \dfrac{2}{y_f +y_b}\left[\dfrac{\psi_{i,j+1}^{n}}{y_f}+\dfrac{\psi_{i,j-1}^{n}}{y_b} -\left(\dfrac{1}{y_f}+\dfrac{1}{y_b}\right)\psi_{i,j}^{n}\right]\\
			&+\dfrac{x_f-x_b}{x_f +x_b}\left[\dfrac{v_{i+1,j}^{n}}{x_f} + \dfrac{v_{i-1,j}^{n}}{x_b}-\left(\dfrac{1}{x_f}+\dfrac{1}{x_b}\right)v_{i,j}^{n}\right] - \dfrac{y_f-y_b}{y_f +y_b}\left[\dfrac{u_{i,j+1}^{n}}{y_f}+\dfrac{u_{i,j-1}^{n}}{y_b} -\left(\dfrac{1}{y_f}+\dfrac{1}{y_b}\right)u_{i,j}^{n}\right]\\
			&+\dfrac{x_f-x_b}{x_f +x_b}\left[\dfrac{v_{i+1,j}^{n+1}}{x_f} + \dfrac{v_{i-1,j}^{n+1}}{x_b}-\left(\dfrac{1}{x_f}+\dfrac{1}{x_b}\right)v_{i,j}^{n+1}\right] - \dfrac{y_f-y_b}{y_f +y_b}\left[\dfrac{u_{i,j+1}^{n+1}}{y_f}+\dfrac{u_{i,j-1}^{n+1}}{y_b} -\left(\dfrac{1}{y_f}+\dfrac{1}{y_b}\right)u_{i,j}^{n+1}\right]\\
		\end{aligned}
	\end{equation}
	Where $\Delta t$ is the uniform time step and $x_f$, $x_b$, $y_f$, $y_b$ are the nonuniform forward and backward step-lengths in the $x$ and $y$ directions, respectively. The horizontal and vertical velocities, respectively $u$ and $v$, can be retained using $\displaystyle u=\frac{\partial \psi}{\partial y}\;{\rm and}\; v=-\frac{\partial \psi}{\partial x}$. Using this relations, the following implicit relation can be used to approximate the horizontal velocity up to third order accuracy.
	
	\begin{equation}\label{tr_u}
		\begin{aligned}
			\dfrac{y_{b}}{2k}u_{i,j+1} -2u_{i,j} + \dfrac{y_{f}}{2k}u_{i,j-1}  =  \dfrac{3\left(\psi_{i,j+1}-\psi_{i.j-1}\right)}{2k} &-\dfrac{3(y_{f}-y_{b})}{2k}\left[\dfrac{\psi_{i,j+1}}{y_{f}} + \dfrac{\psi_{i,j-1}}{y_{b}} - (\dfrac{1}{y_f}+\dfrac{1}{y_b})\psi_{i,j}\right]\\ & +O\left(y_fy_b(y_f-y_b)\right).
		\end{aligned}
	\end{equation}
	The renowned tridiagonal solver Thomas algorithm \cite{ander95,pletcher} is used to solve the algebraic system produced by \eqref{tr_u}. Similar calculations can be used to approximate the vertical velocity, $v$. In addition to the derivation of the equations \eqref{dis1} through \eqref{tr_u}, the specifics of the coefficients $A,\;B,\;C,\;D,E\;F,\;G,\;H,\;I$ and the function $\phi$ can be found in \cite{kumar2020}. The vorticity $\omega$, when required can be post-processed from the relation $\displaystyle \omega=\frac{\partial v}{\partial x}- \frac{\partial u}{\partial y}$.
	
	\begin{figure}
		\centering
		\includegraphics[width=1\textwidth]{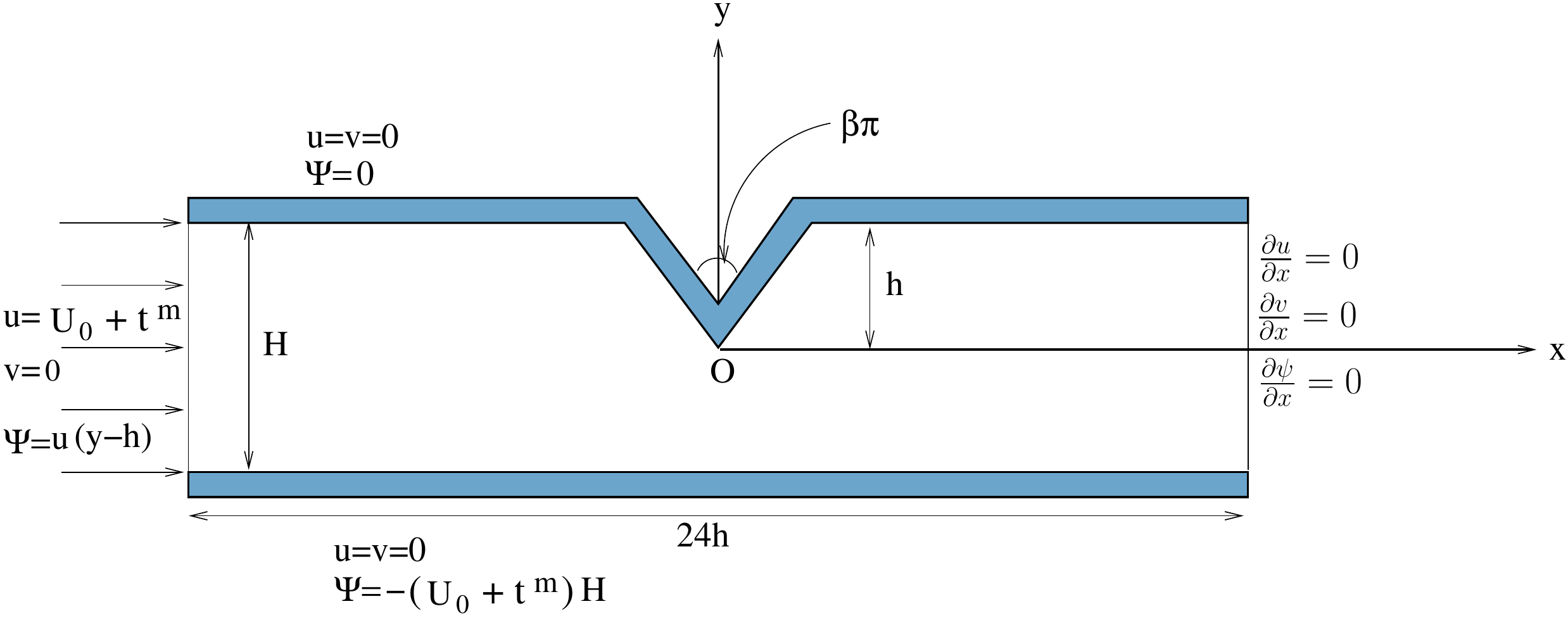} 
		\caption{Problem statement of flow past a wedge in accelerated flow.}
		\label{wedge}
	\end{figure}
	
	The schematic of the flow along with boundary conditions used in our computation is shown in figure \ref{wedge}. The fluid is assumed to be flowing from left to right.  Note that use of equation \eqref{nondim} for non-dimensionalization renders the piston velocity $\tilde{v}_p=\tilde{u}_0+A\tilde{t}^m$ to $u=U_0+t^m$ at non-dimensional time $t$ where $U_0$ is the inlet velocity at $t=0$.  The dimensions of the computational domain was set as $-6.0\leq x \leq 6.0$ and $-0.5\leq y \leq 0.5$ such that the width of the channel is unity and the tip of the wedge is located at the origin.  Following \citet{pullin1980}, the ratio of the channel width and wedge-height is kept at $2:1$ as well. The boundary conditions at the inlet is taken as $u= U_0 + t^m,\  v=0$ and $\psi = u(y-h)$. $\psi$ is scaled this way in order to attain a value zero on the top wall, i.e., $\psi_{top}= 0$ and the same streamline corresponding to $\psi=0$ continues its journey by touching the wedge surface. Consequently  boundary conditions for $\psi$ at the bottom wall are $\psi_{bottom}= -\left(U_0 + t^m\right)H$; also $u =0 ,\; v=0$. At the outlet, we have used the zero gradient boundary condition $\dfrac{\partial \psi}{\partial x}= \dfrac{\partial u}{\partial x}=\dfrac{\partial v}{\partial x}=0 $. 
	
	\begin{figure}
		\begin{tabular}{ccc}
			\hspace{-1.0cm}\includegraphics[width=0.3525\linewidth]{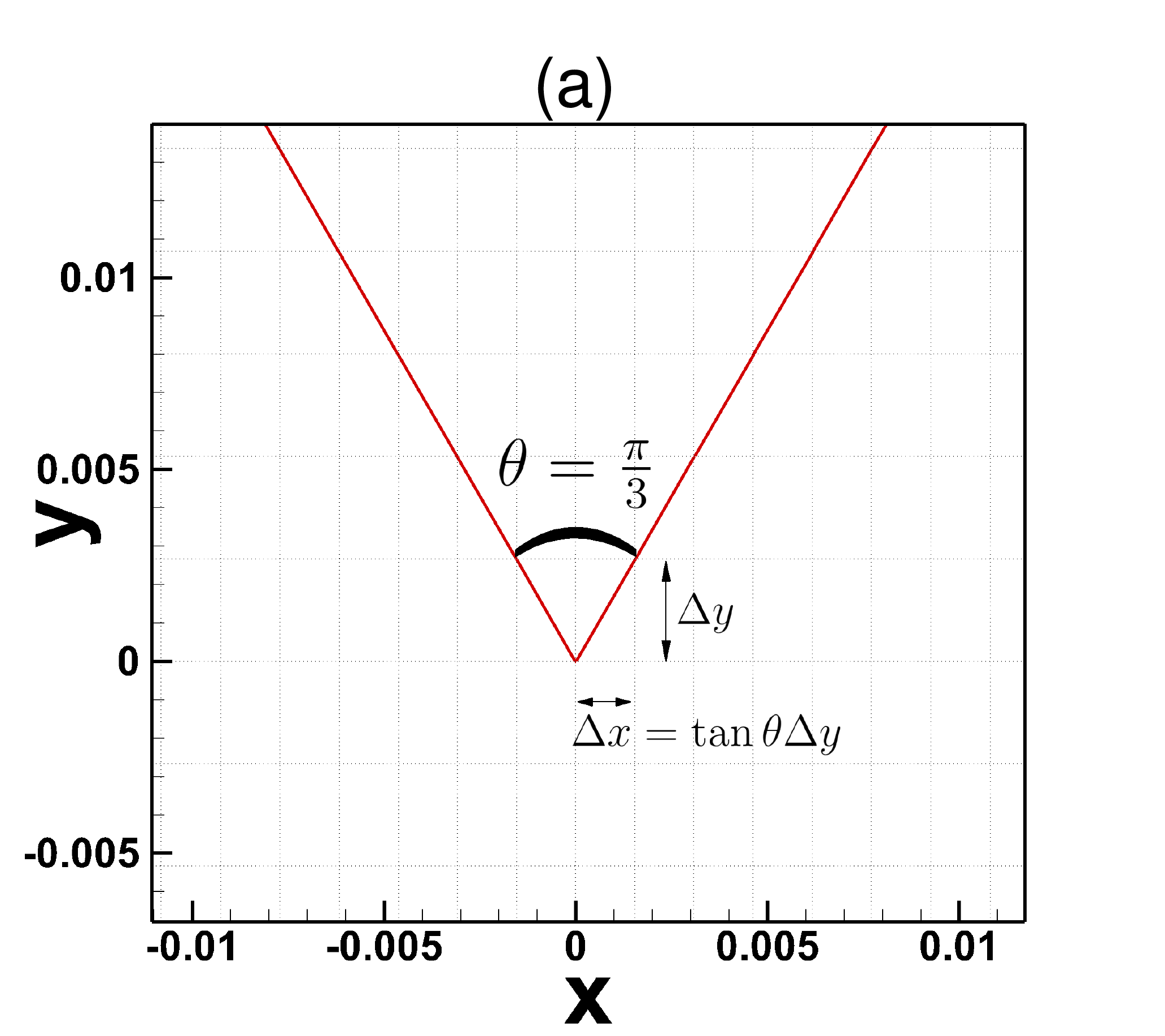}
			&
			\hspace{-0.01cm}\includegraphics[width=0.3\linewidth]{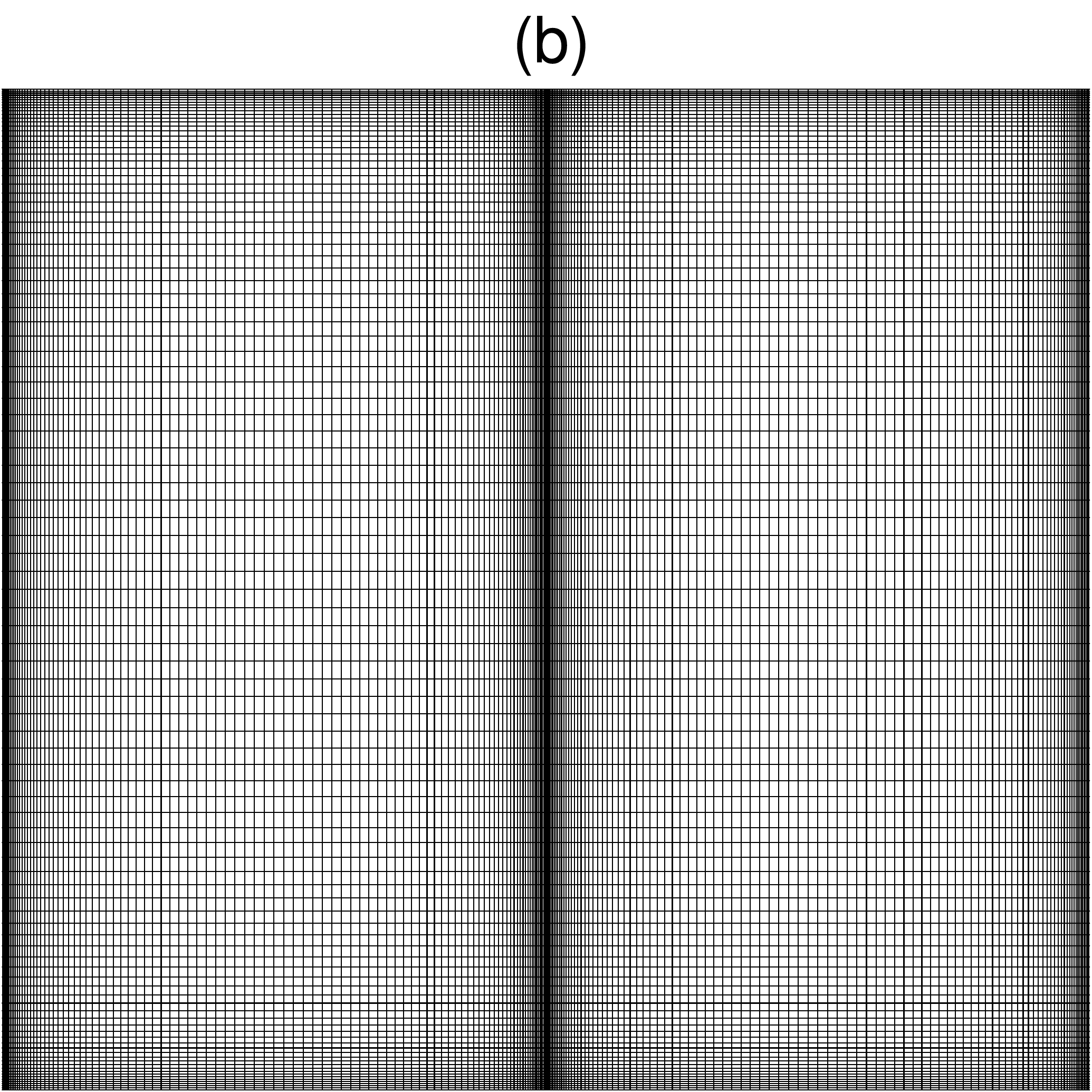}
			&
			\hspace{-0.02cm}\includegraphics[width=0.3\linewidth]{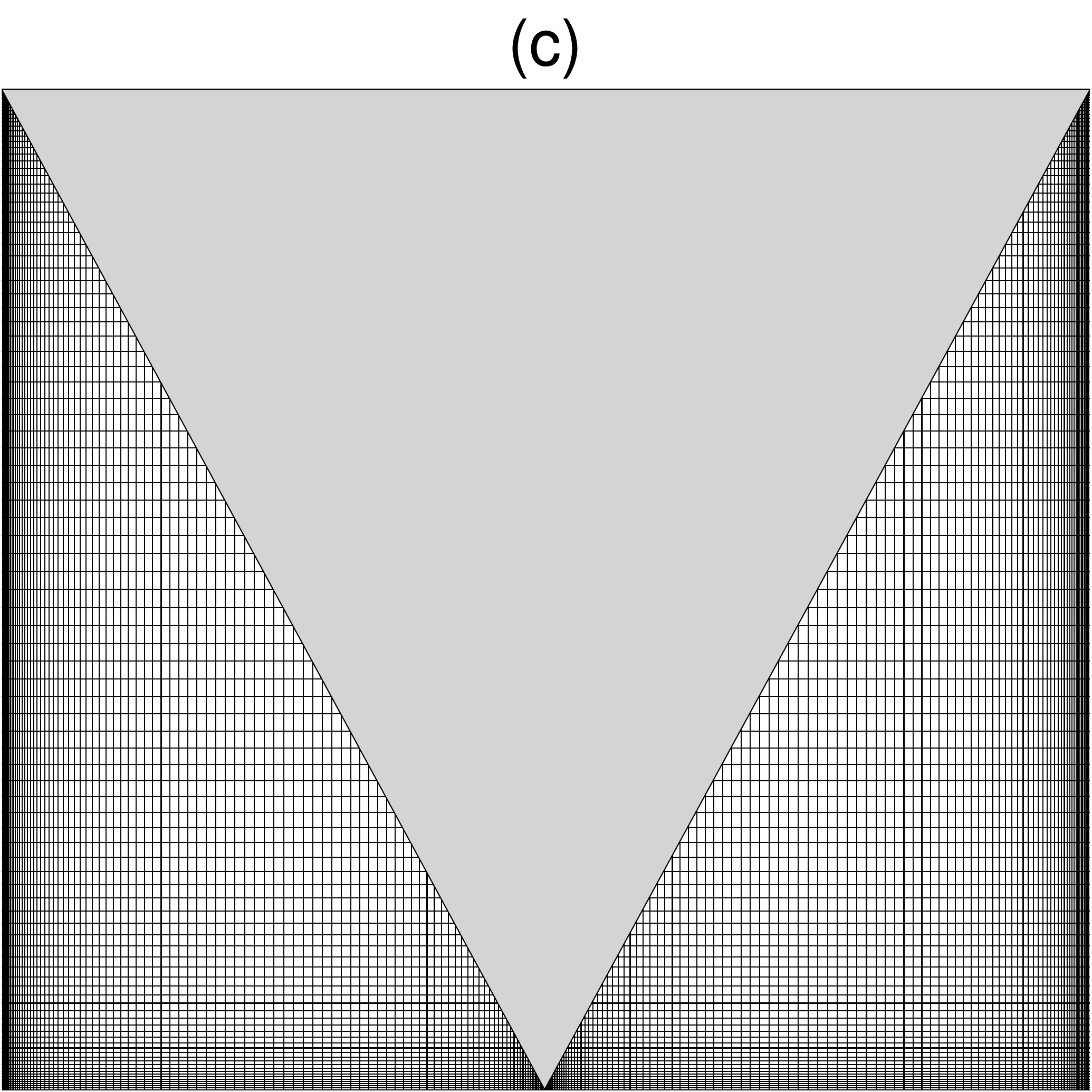}
		\end{tabular}
		\caption{\sl {The grid generating strategy: (a) the transformation used, (b) the fundamental block and (c) the wedge implanted on the block.}}
		\label{gr_wedge}
	\end{figure}
	
\begin{figure}
		\centering\includegraphics[width=0.6\textwidth]{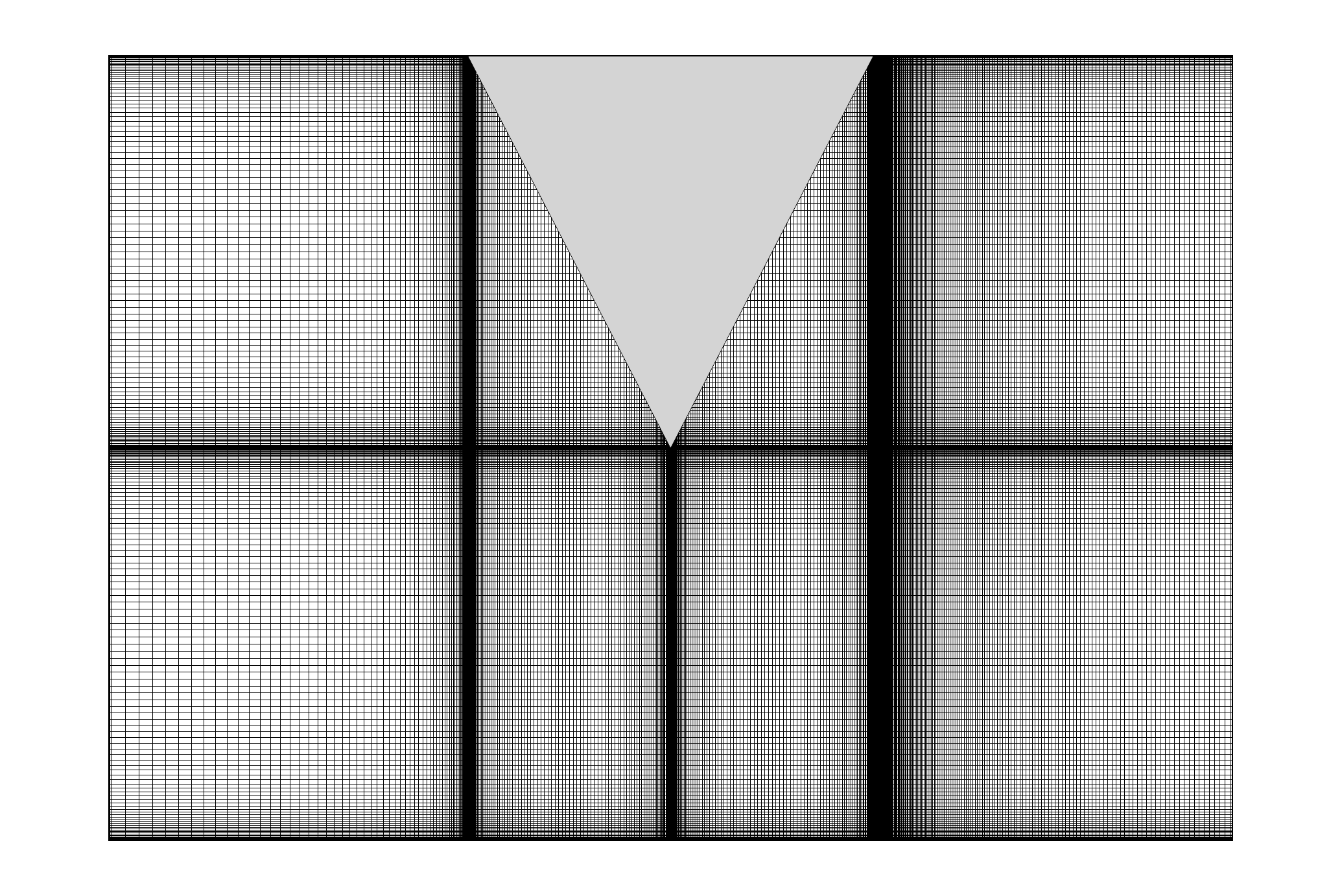} 
		\caption{Schematic of the overall grid used for the accelerated flow past an wedge mounted on wall.}
		\label{wl_wedge}
	\end{figure}

	It is worth mentioning that since we are using a Cartesian grid for discretizing the governing equations, the grid must be generated in such a way that the wedge geometry passes through the nodes. This was accomplished in the following way. Firstly, a square block is created in the domain $-l\leq x \leq l,\; 0\leq y \leq 0.5$, where the value of $l$ depends on the wedge angle. Along $y$-direction, a centro-symmetric grid with clustering at the top and bottom is generated using the stretching function \begin{equation}
		y_j=\frac{j}{j_{max}}-\frac{\lambda}{2\pi}\sin\bigg(\frac{2\pi j}{j_{max}}\bigg),\; 0<\lambda\leq 1
		\label{cl} 
	\end{equation}
	with $\lambda$ determining the intensity of clustering (we used $\lambda=0.4$ here). Then in the  sub-domain $0\leq x \leq l,\; 0\leq y \leq 0.5$, the grid along the $x$-direction is generated using the transformation $\displaystyle x=(\tan\theta)y$, where $\theta$ is the wedge-angle (figure \ref{gr_wedge}(a)). The grid in the sub-domain $-l\leq x \leq 0,\; 0\leq y \leq 0.5$ is obtained by creating a mirror image of the grid in the sub-domain $0\leq x \leq l,\; 0\leq y \leq 0.5$ about the vertical line $x=0$. This completes the generation of grid in the fundamental block (figure \ref{gr_wedge}(b)) which accommodates the wedge geometry to pass through the grid points (figure \ref{gr_wedge}(c)). Following this, the two neigbouring blocks on the left and right of it (see figure \ref{wl_wedge}) are generated using nonuniform grid spacing in the $x$-direction where points are clustered in the neighbourhoods of the vertical boundaries of the fundamental block. This completes the grid in the domain $-6.0 \leq x \leq 6.0,\; 0\leq y \leq 0.5$. The grid in the domain $-6.0 \leq x \leq 6.0,\; -0.5\leq y \leq 0$ is an exact replica of the previous grid just above it as shown in figure \ref{wl_wedge}. Thus the grid generation in the whole computational domain is complete. During the entire process, care must be taken such that the continuity of the grid lines in either direction of the whole computational domain  is maintained. Note that grid points around the neighbourhoods of the three vertices of the triangle defining the wedge are extremely clustered.
	\begin{figure}
		\minipage{0.3\textwidth}
		\centering 
		\includegraphics[width=\linewidth]{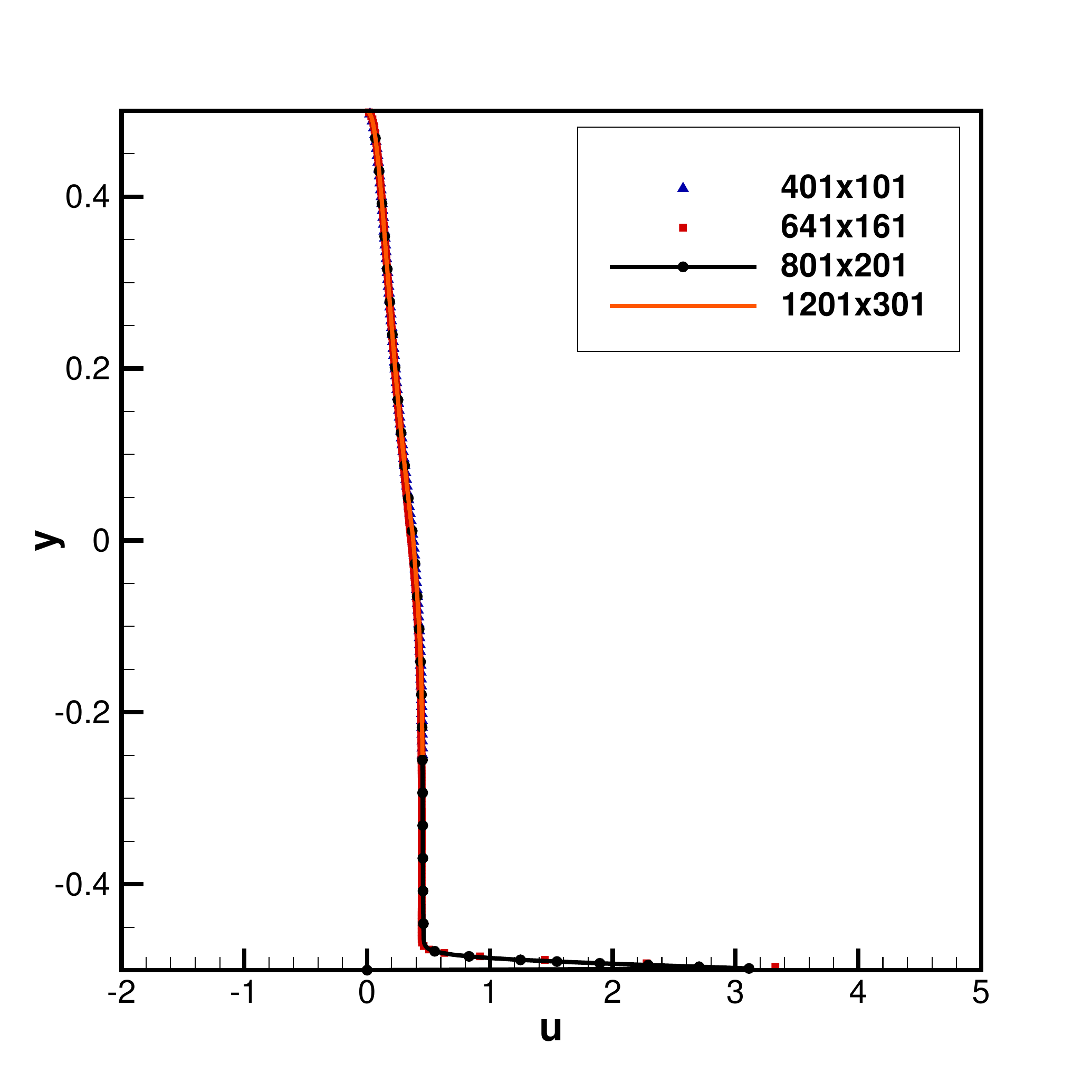}\\
		(a1)
		\endminipage\hfill
		\minipage{0.3\textwidth}
		\centering
		\includegraphics[width=\linewidth]{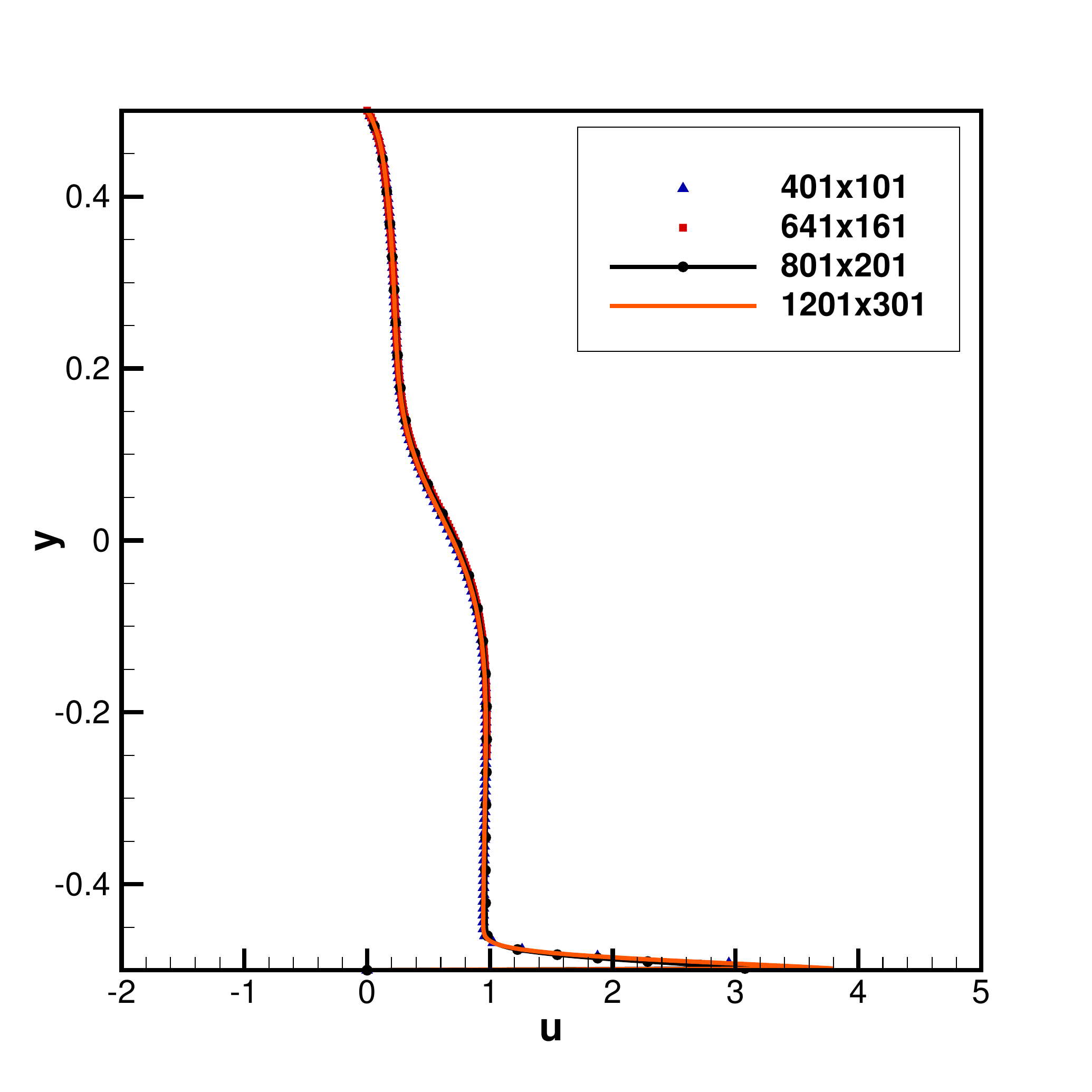}\\(b1)
		\endminipage\hfill
		\minipage{0.3\textwidth}%
		\centering
		\includegraphics[width=\linewidth]{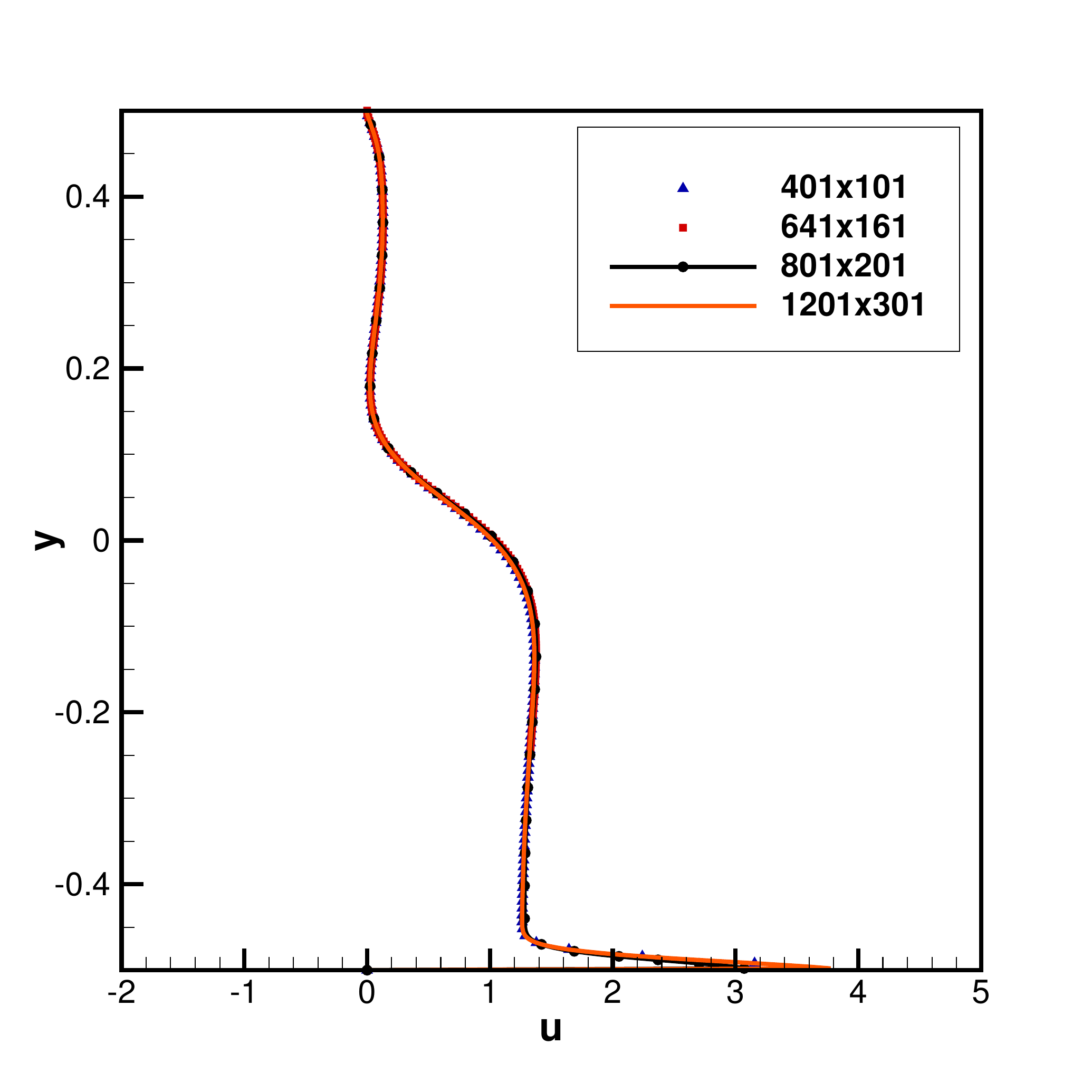}\\
		(c1)
		\endminipage\hfill
		\minipage{0.3\textwidth}
		\centering 
		\includegraphics[width=\linewidth]{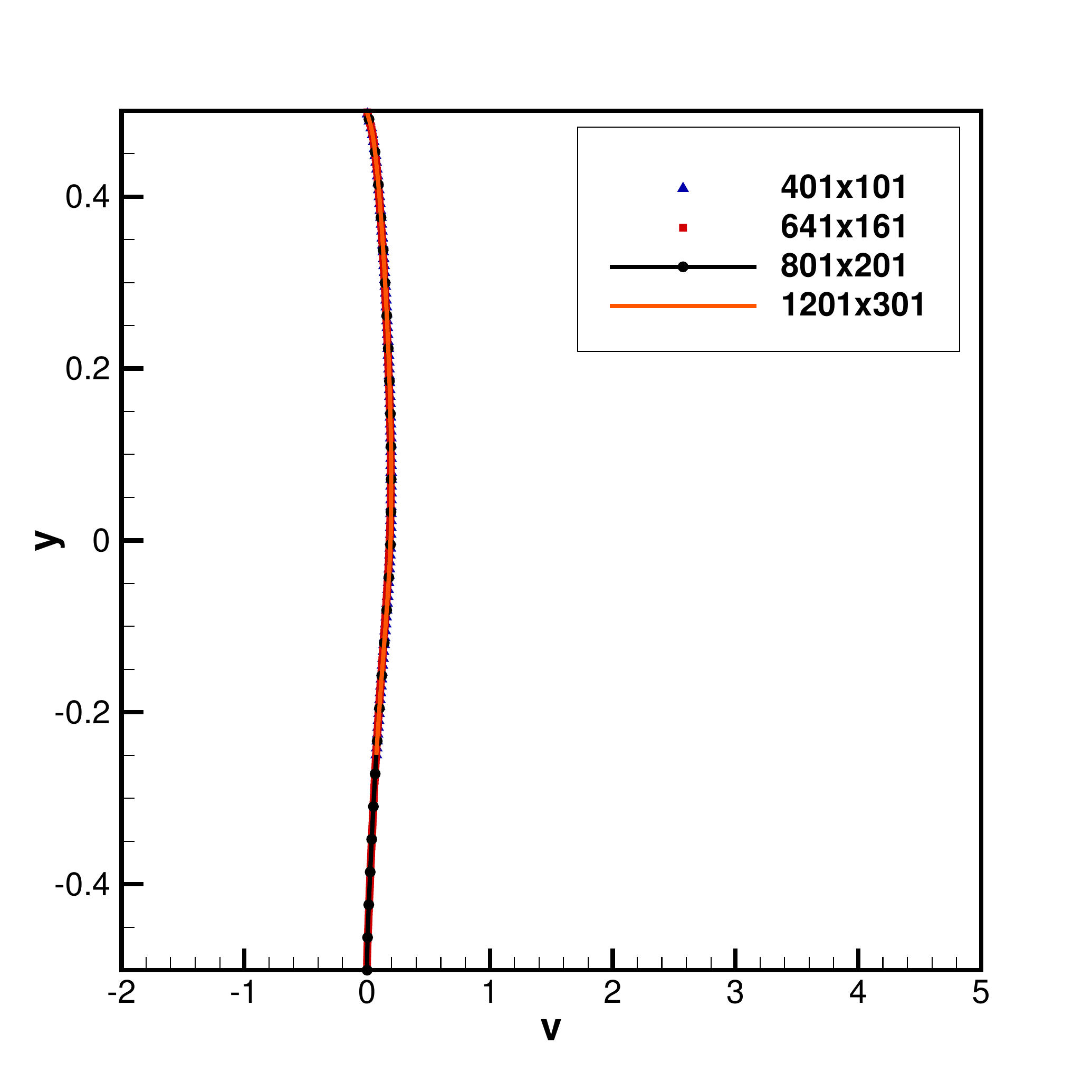}\\
		(a2)
		\endminipage\hfill
		\minipage{0.3\textwidth}
		\centering
		\includegraphics[width=\linewidth]{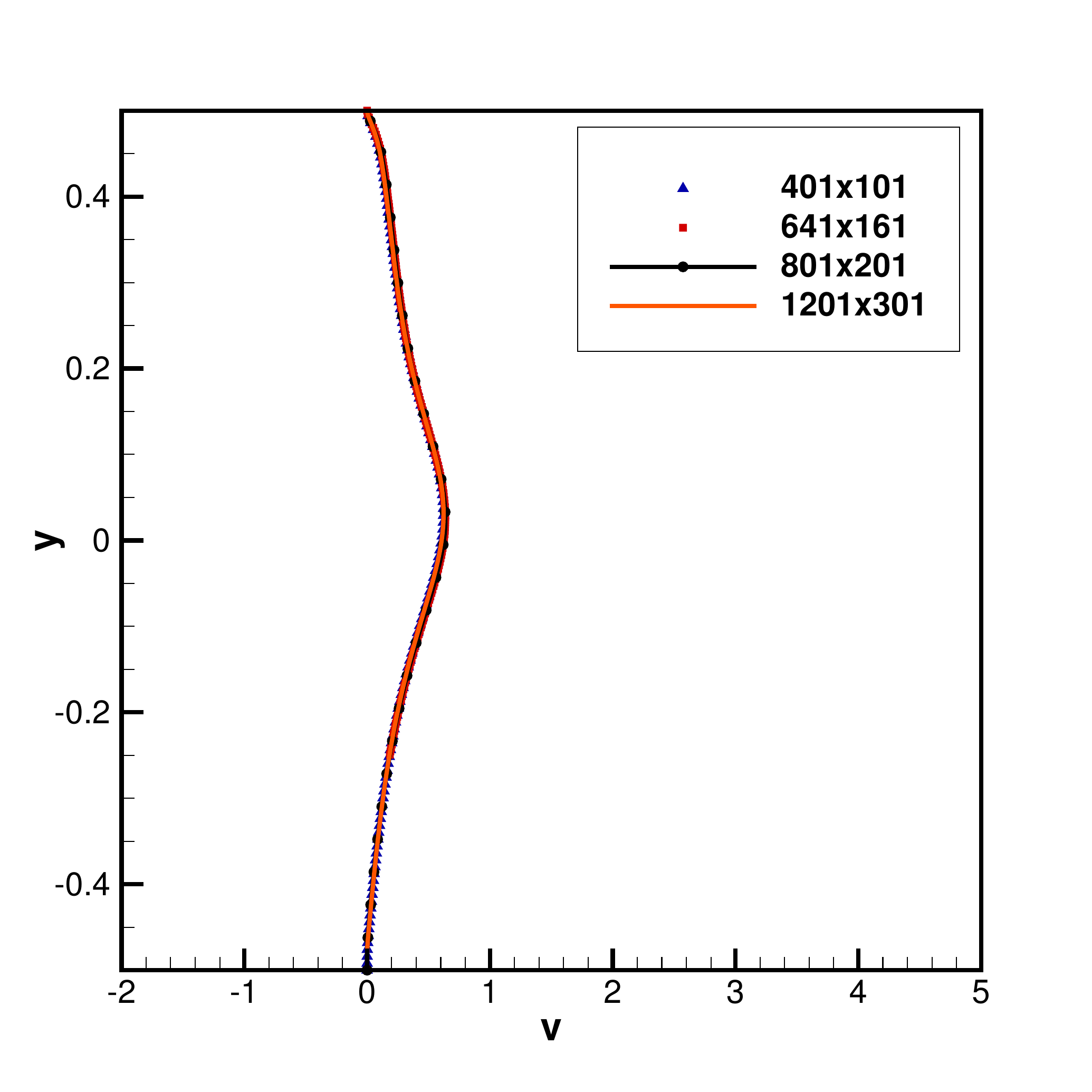}\\(b2)
		\endminipage\hfill
		\minipage{0.3\textwidth}%
		\centering
		\includegraphics[width=\linewidth]{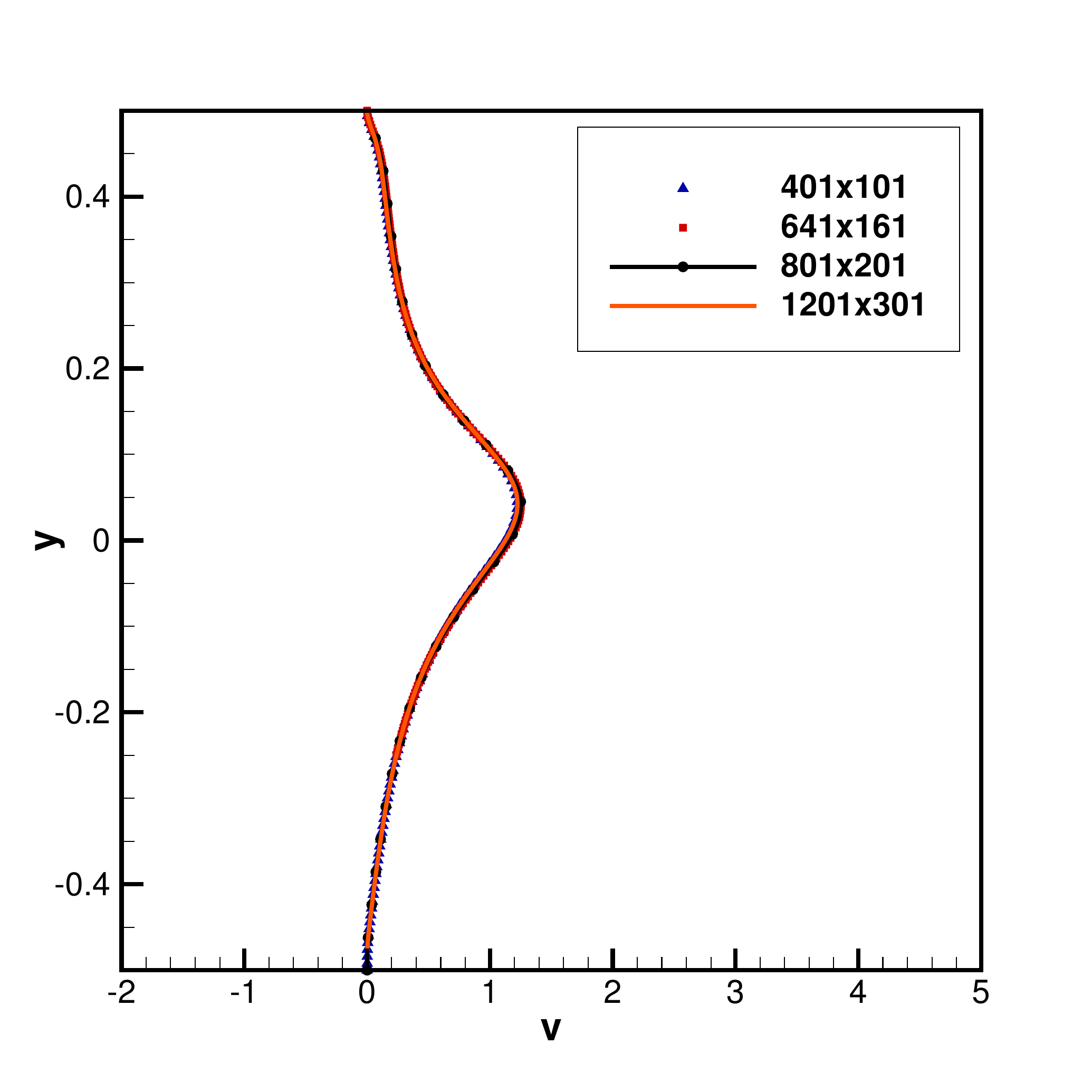}\\
		(c2)
		\endminipage\hfill	
		\caption{{Grid independence of the computed solutions for four different grids at three different time step: $u$ (top) and $v$ (bottom) along the vertical line at $x=0.3$ at times (a1,a2) $t=0.3$, (b1,b2) $t=0.6$ and (c1,c2) $t=0.75$ for $Re_c=6873$.} }
		\label{gr_ind_uv}
	\end{figure}
	We performed computations on grids of sizes $401 \times 101$, $641 \times 161$, $801 \times 201$ and $1201 \times 301$ for channel Reynolds numbers $1560$, $6621$, and $6873$, which correspond to $m=0$, $m=0.45$, and $0.88$ respectively. In the process, we also carry out a grid-convergence study of our computed results. For this, we  present the horizontal and vertical velocities ($u$ and $v$) along the vertical line at $x=0.3$ on all the four different grids used in computation, viz., $401 \times 101$, $641 \times 161$, $801 \times 201$ and $1201 \times 301$ for $Re_c=6873$ at three different instants of non-dimensional time $t=0.3$, $0.6$ and $0.75$ in figure \ref{gr_ind_uv}. For the first three grid, a time step $\Delta t=10^{-5}$ was used, while for the last $\Delta t=5 \times 10^{-6}$ was chosen.  These figures clearly demonstrate the grid-independence of our computed solutions. Note that for all the numerical results that follow, a grid of size $801 \times 201$ with $\Delta t=10^{-5}$ has been used for $Re_c=1560$ and $6621$. For $Re_c=6873$, the results presented here were computed on a grid of size $1201 \times 301$ with $\Delta t=5 \times 10^{-6}$.
	
	\section{Results and Discussion}\label{result}
	As mentioned earlier, the current work is inspired by the experimental visualizations reported by \citet{pullin1980} in their start-stop lab experiments, which we would endeavour to replicate through numerical simulation. Note that, in their experiments, the flow was stopped at actual times as shown in table \ref{t1}, whose corresponding non-dimensional time  $t<1$ \citet{pullin1980}. However, we continued our simulations up to a non-dimensional time $t=3.0$ to delve into the flow beyond the experimental visualization of \citet{pullin1980} and have a better understanding of the flow characteristics.
	
	\subsection{Flow development at the earliest stage}\label{fl_early}
	\begin{figure}
		\begin{tabular}{cccc}
			\hspace{-1.0cm}\includegraphics[width=0.3\textwidth]{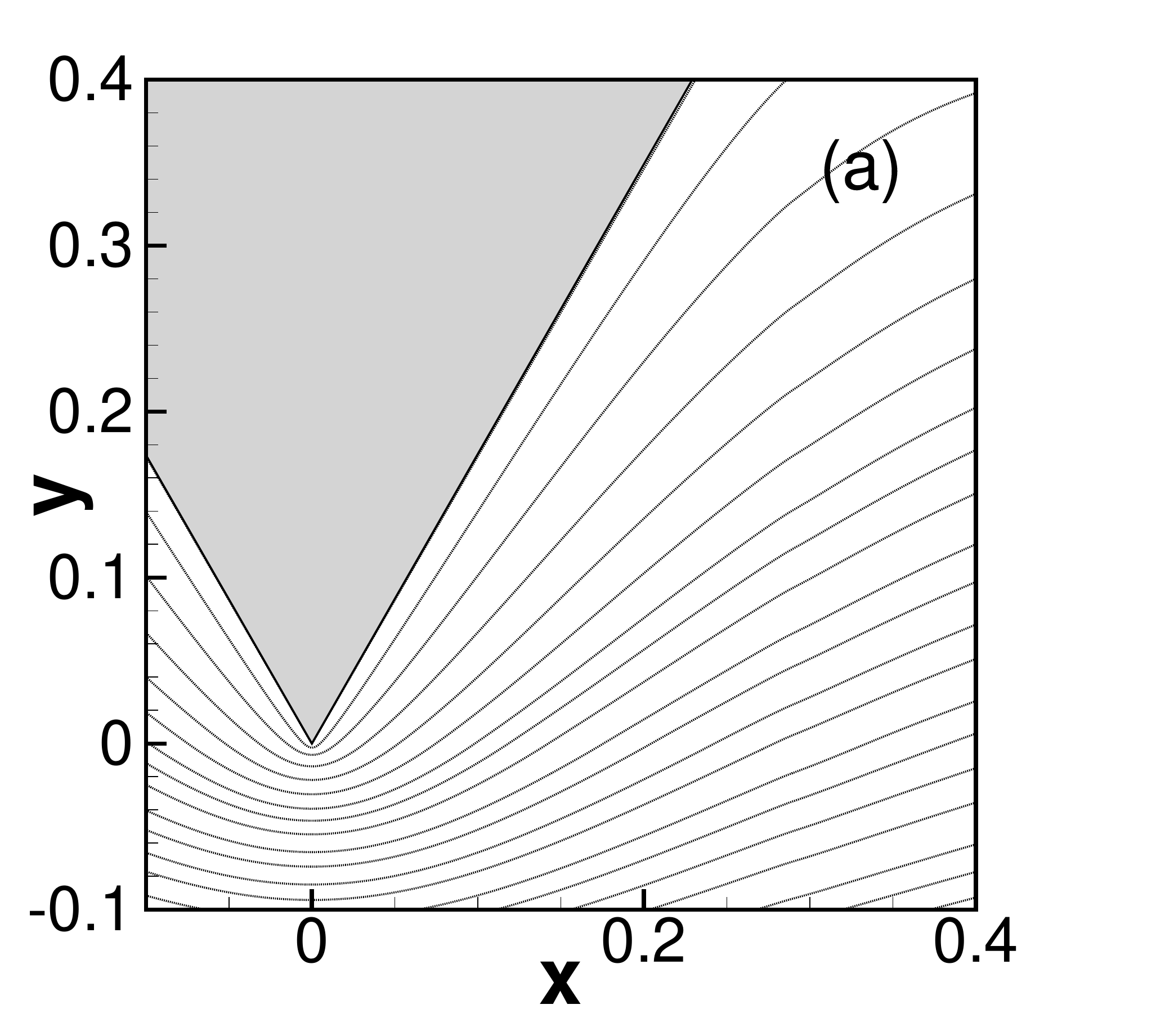}
			&
			\hspace{-0.5cm}\includegraphics[width=0.3\textwidth]{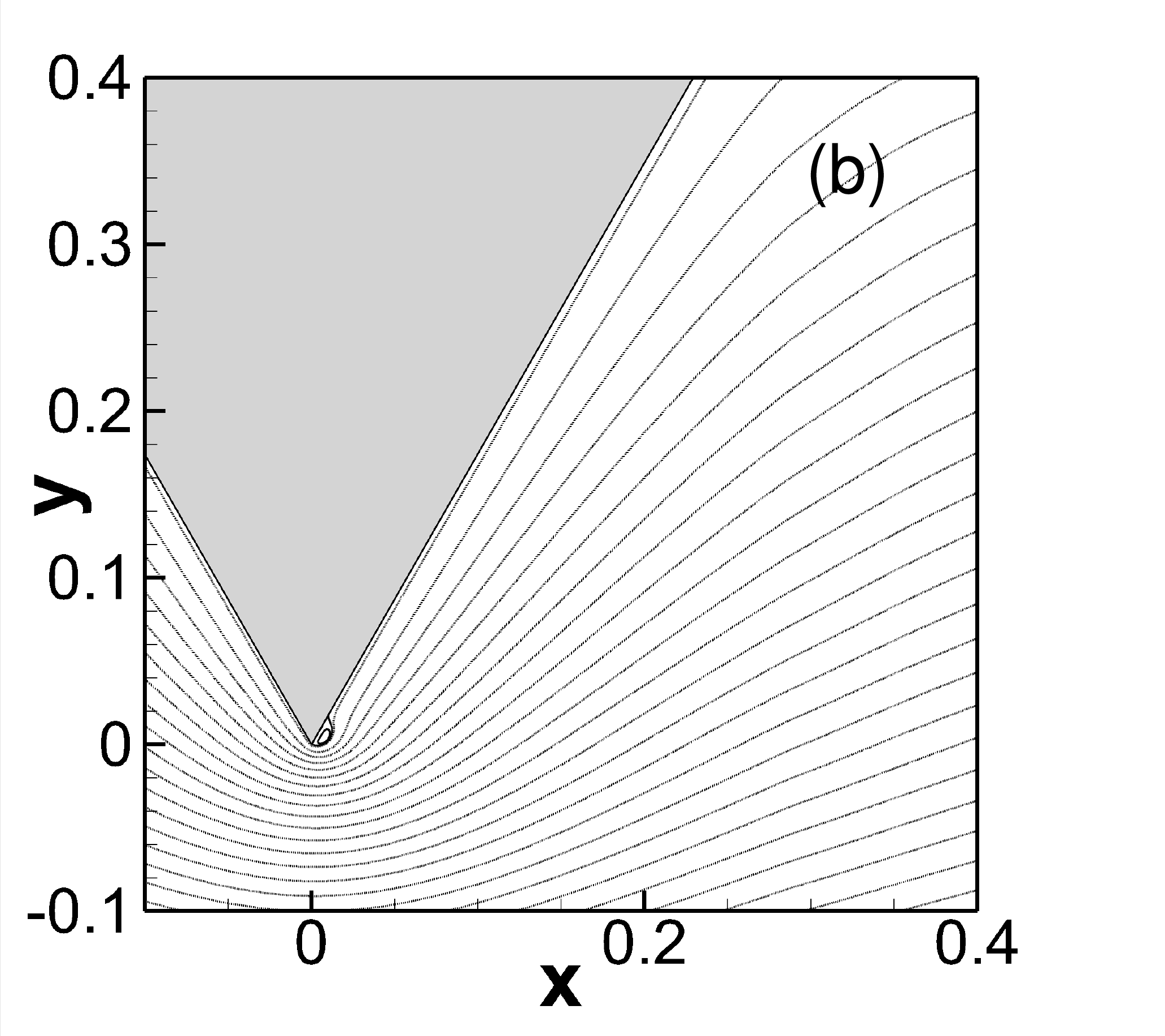}
			\&
			\hspace{-0.5cm}\includegraphics[width=0.3\textwidth]{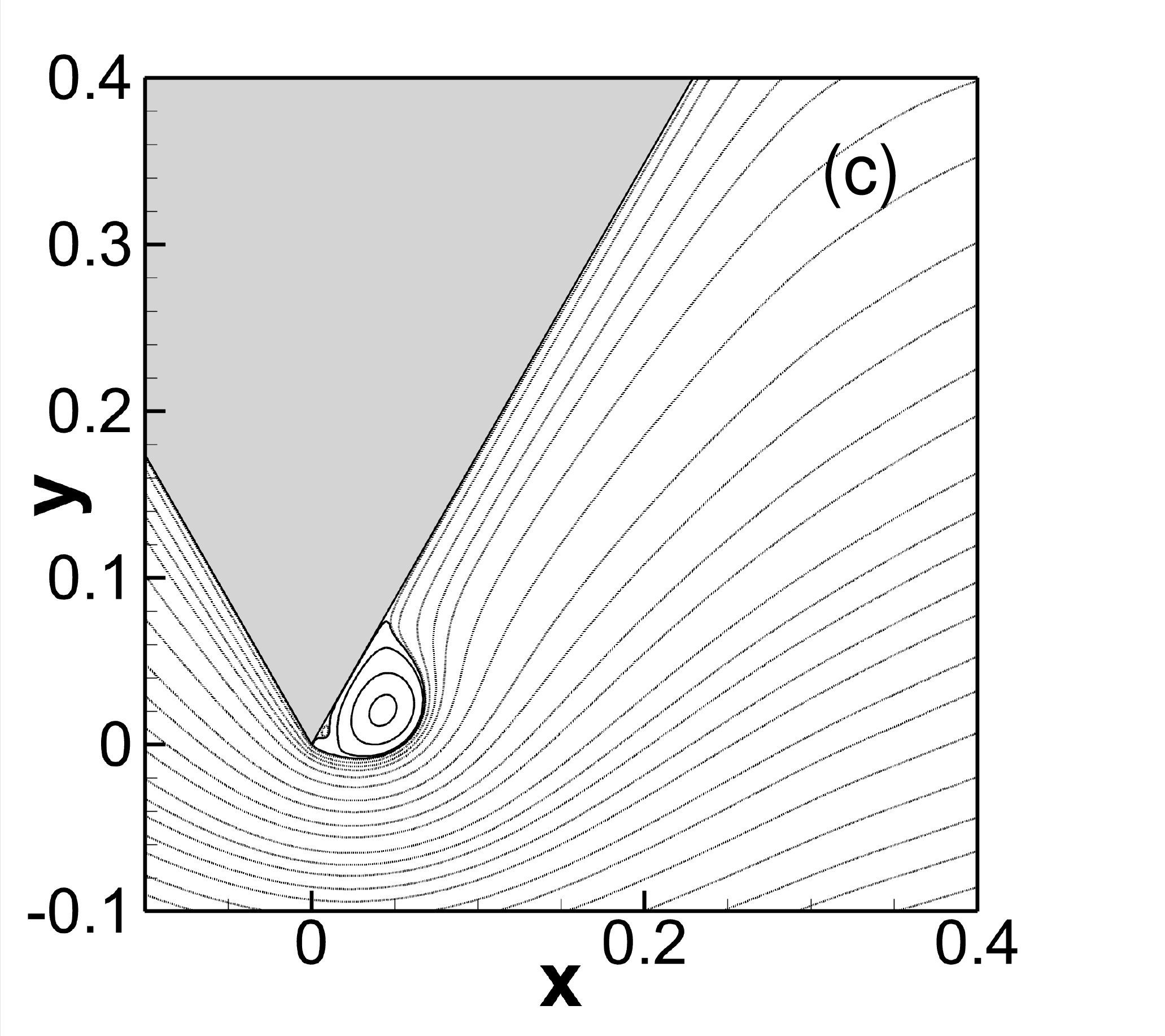}
			&
			\hspace{-0.5cm}\includegraphics[width=0.3\textwidth]{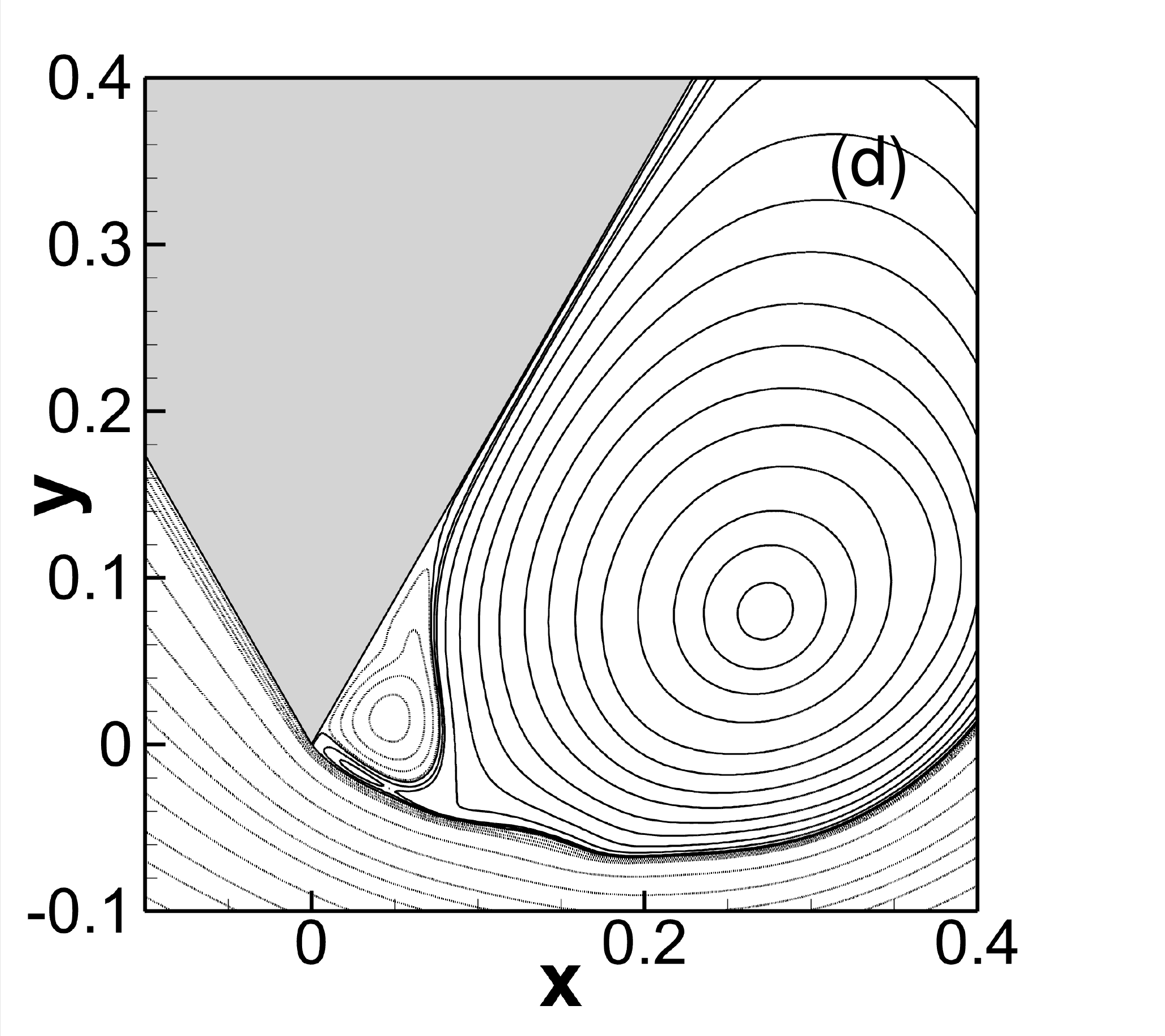}
		\end{tabular}
		\begin{tabular}{cccc}
			\hspace{-1.0cm}\includegraphics[width=0.3\textwidth]{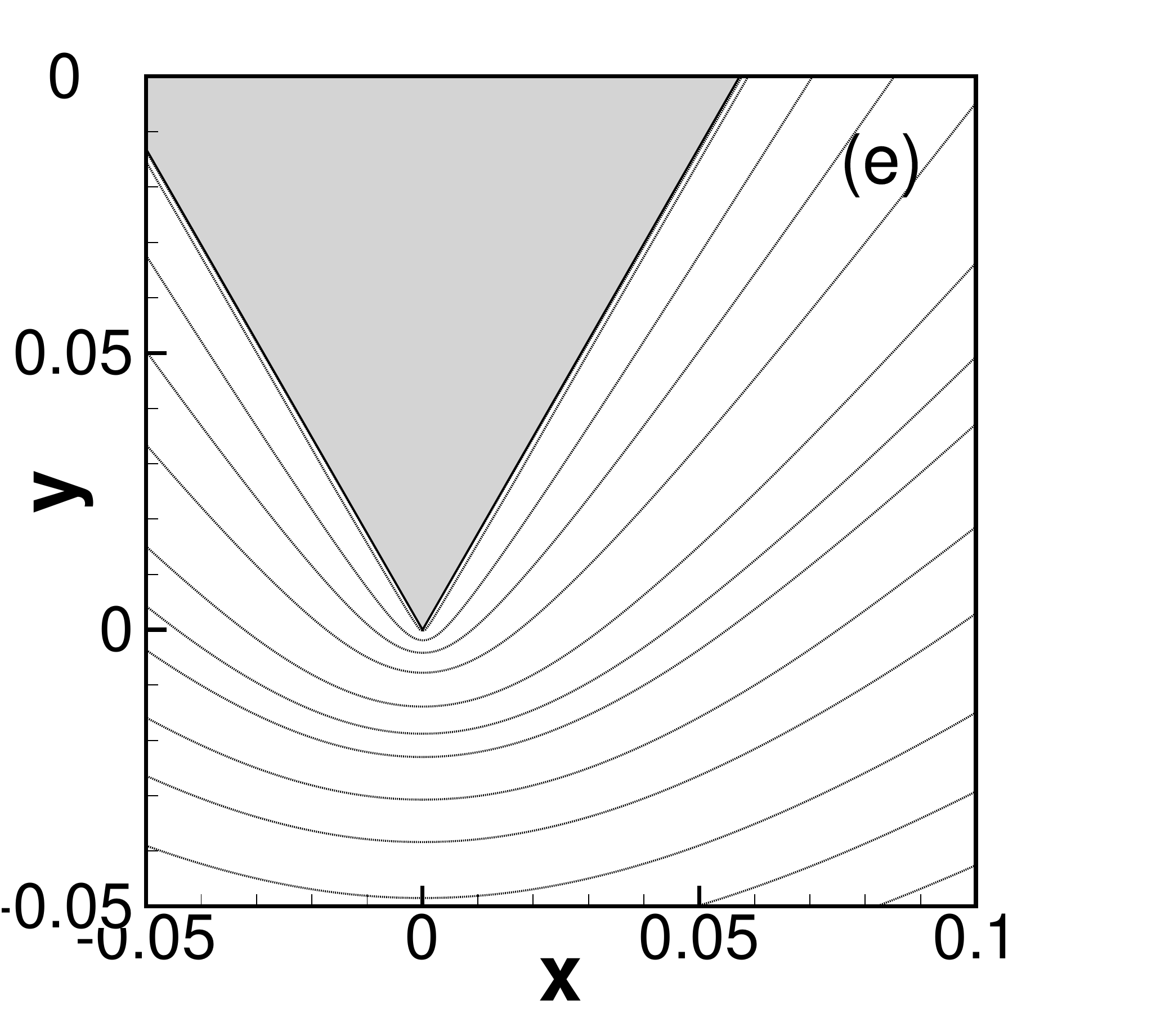}
			&
			\hspace{-0.5cm}\includegraphics[width=0.3\textwidth]{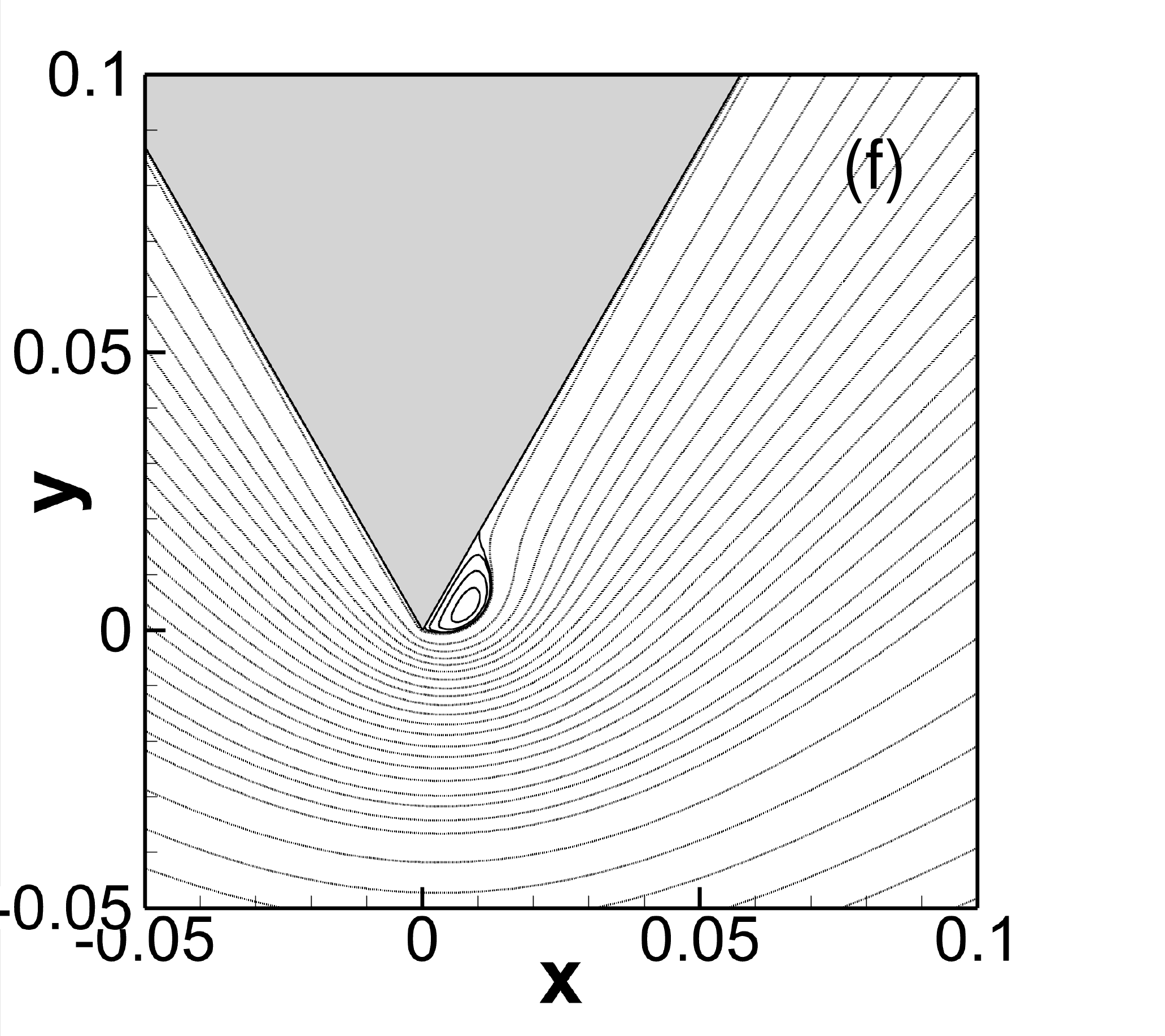}
			\&
			\hspace{-0.5cm}\includegraphics[width=0.3\textwidth]{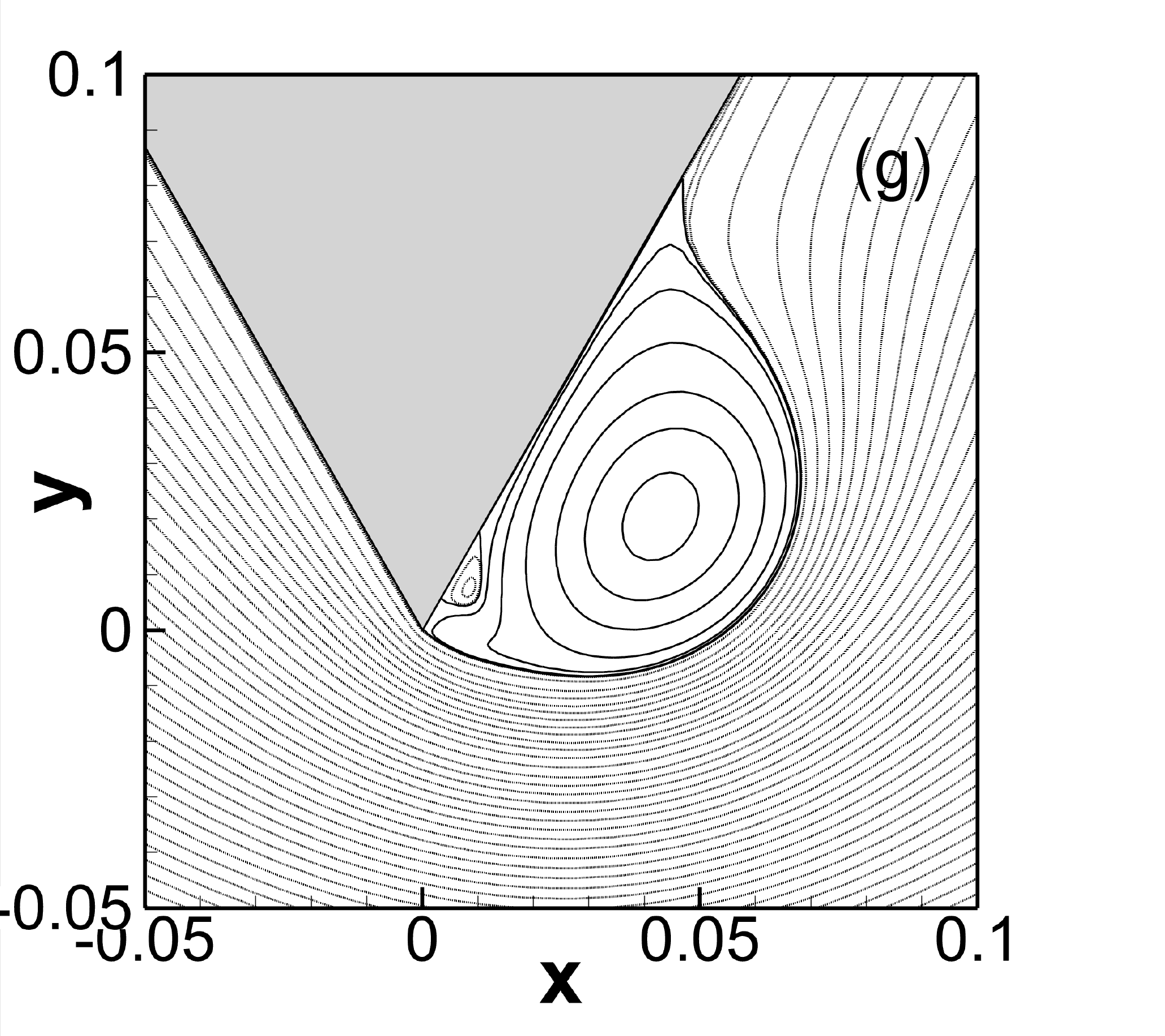}
			&
			\hspace{-0.5cm}\includegraphics[width=0.3\textwidth]{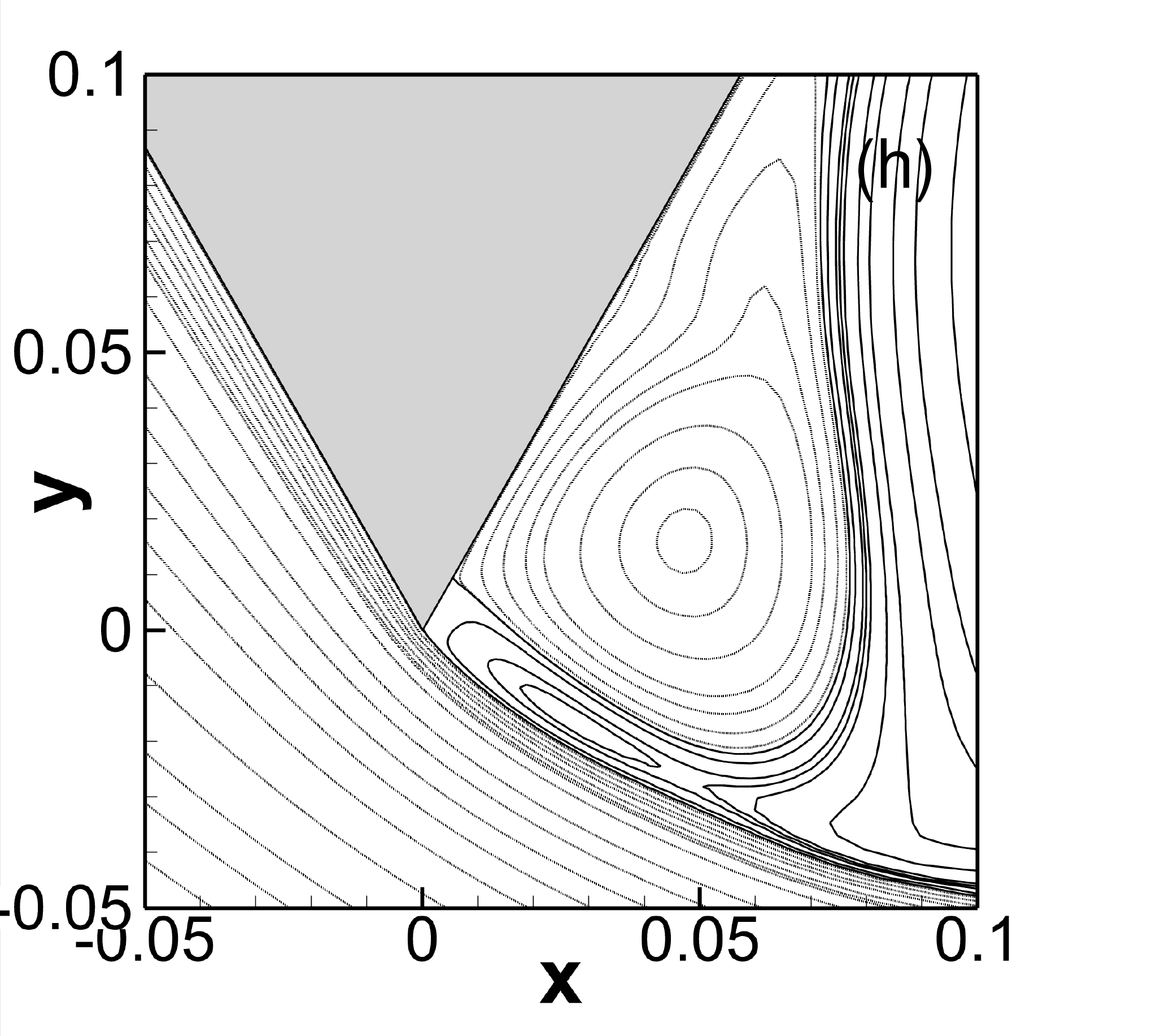}
		\end{tabular}
		\caption{Four stages of streamfunction evolution for $Re_c=6873$ at non-dimensional time (a) $t=0.01$, (b) $t=0.10$, (c) $t=0.30$ and  (d) $t=1.0$, and the corresponding close-up views in (e-h). The positive contours are in solid lines and negative contours are in dotted lines.}
		\label{rayleigh_sf}
	\end{figure}
	\begin{figure}
		\begin{tabular}{cccc}
			\hspace{-1.0cm}\includegraphics[width=0.3\textwidth]{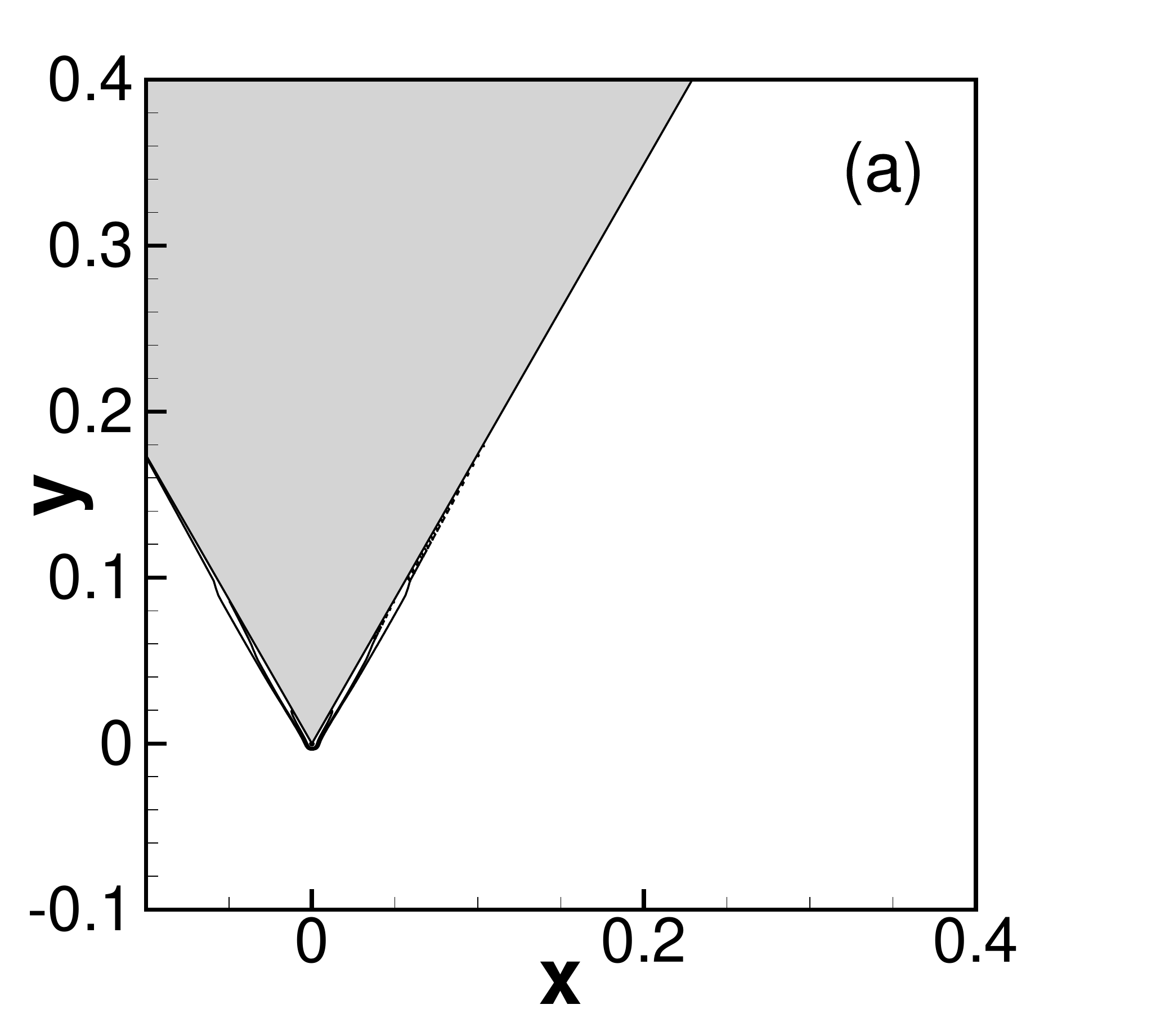}
			&
			\hspace{-0.5cm}\includegraphics[width=0.3\textwidth]{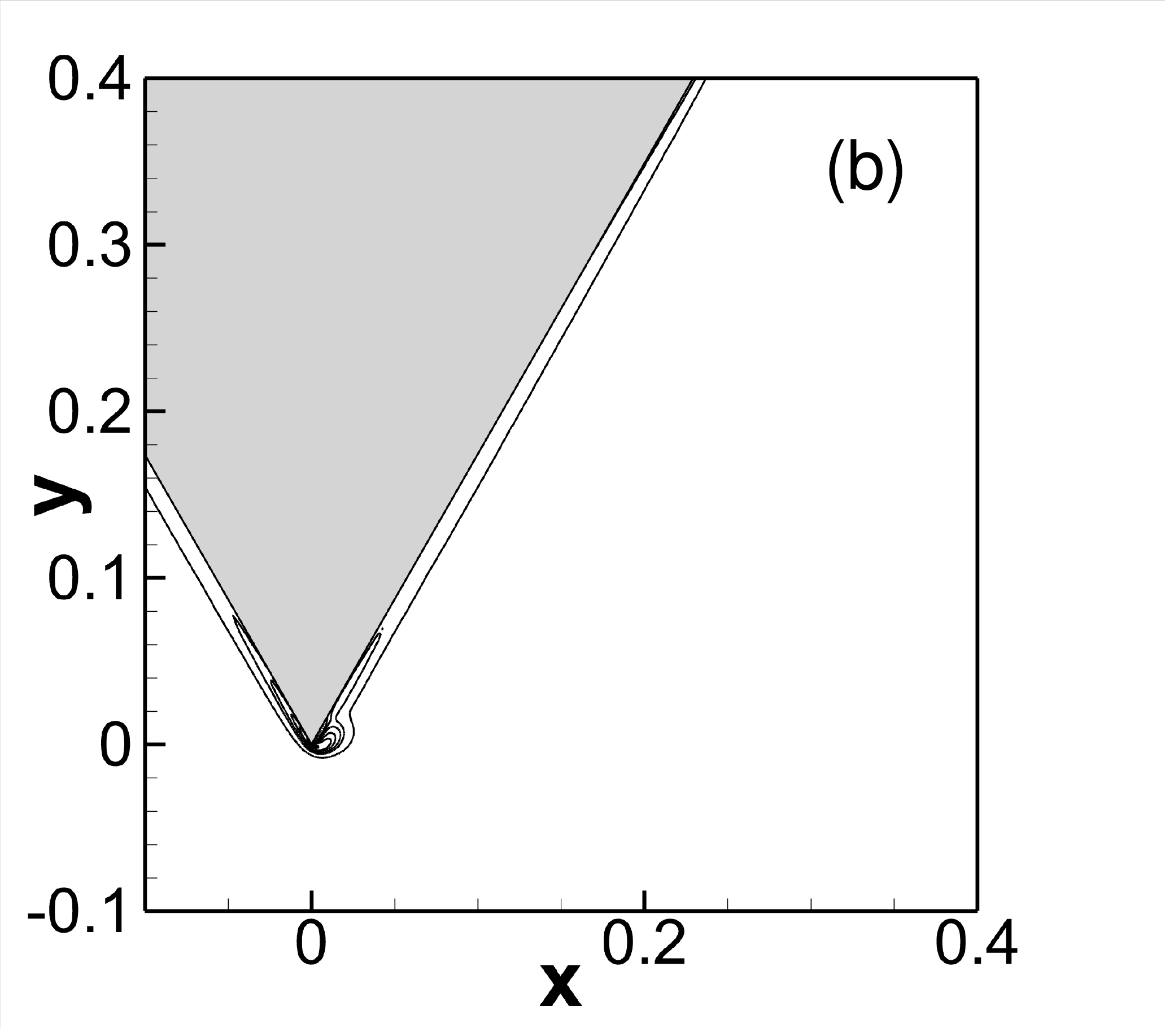}
			\&
			\hspace{-0.5cm}\includegraphics[width=0.3\textwidth]{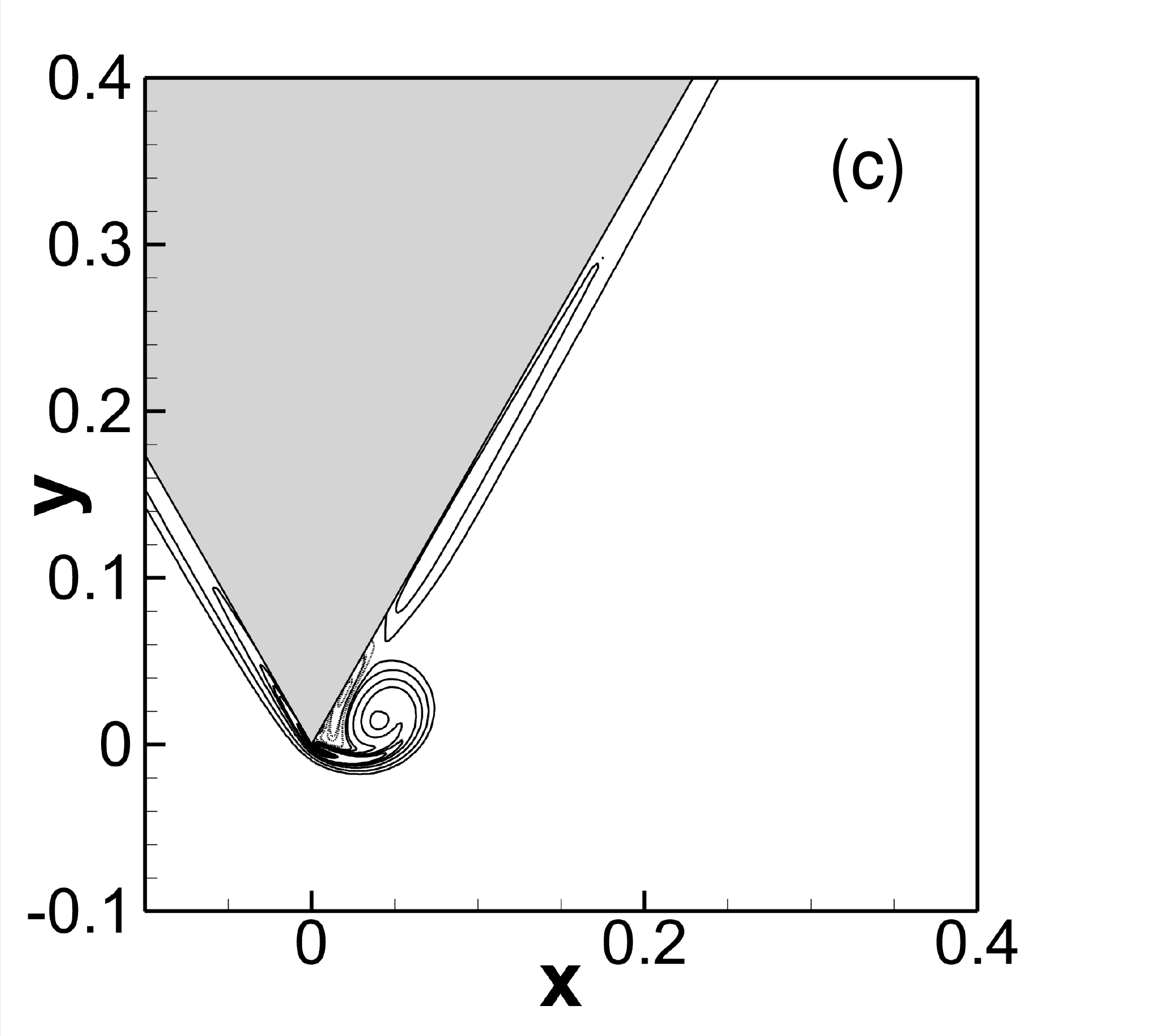}
			&
			\hspace{-0.5cm}\includegraphics[width=0.3\textwidth]{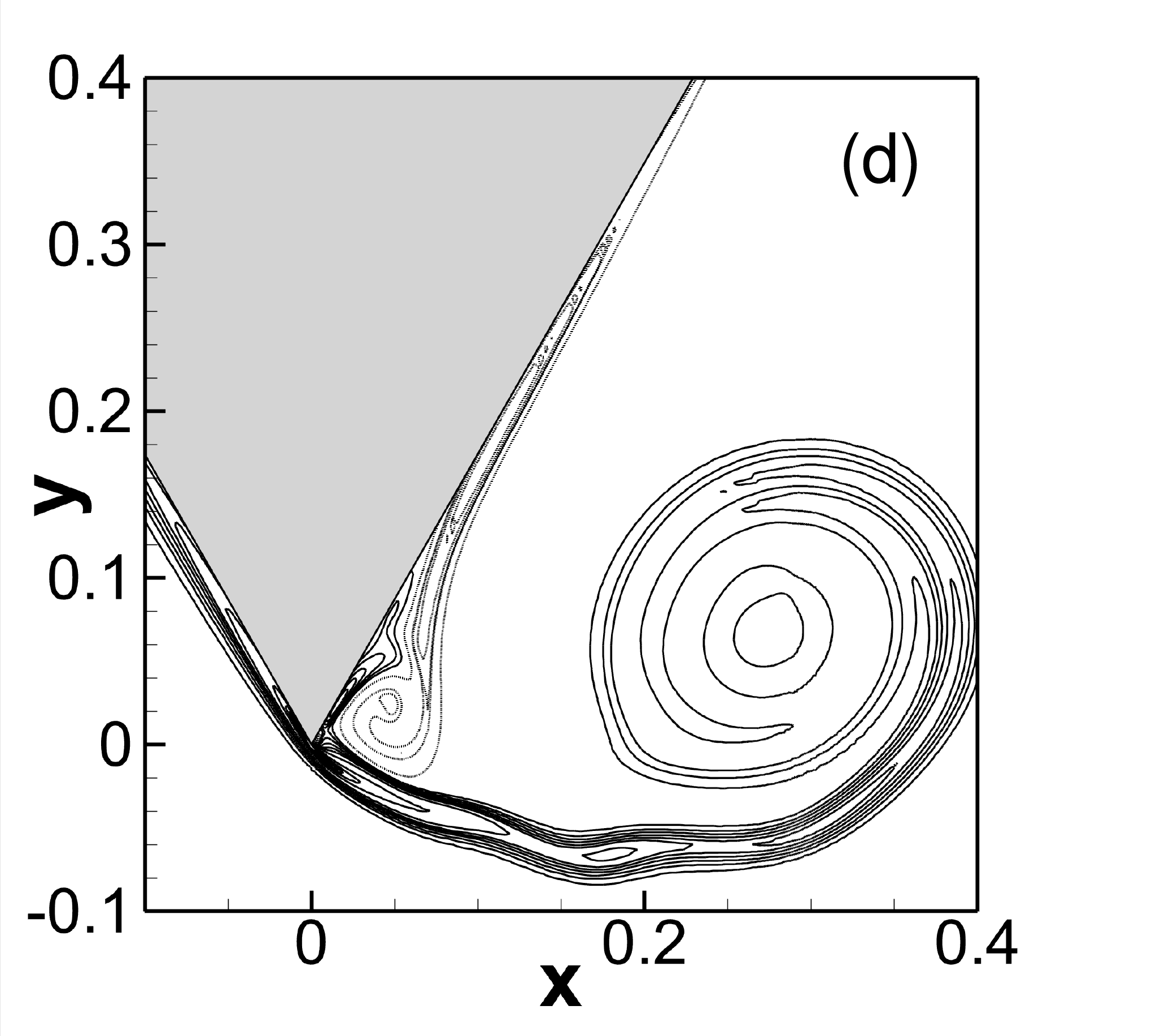}
		\end{tabular}
		\begin{tabular}{cccc}
			\hspace{-1.0cm}\includegraphics[width=0.3\textwidth]{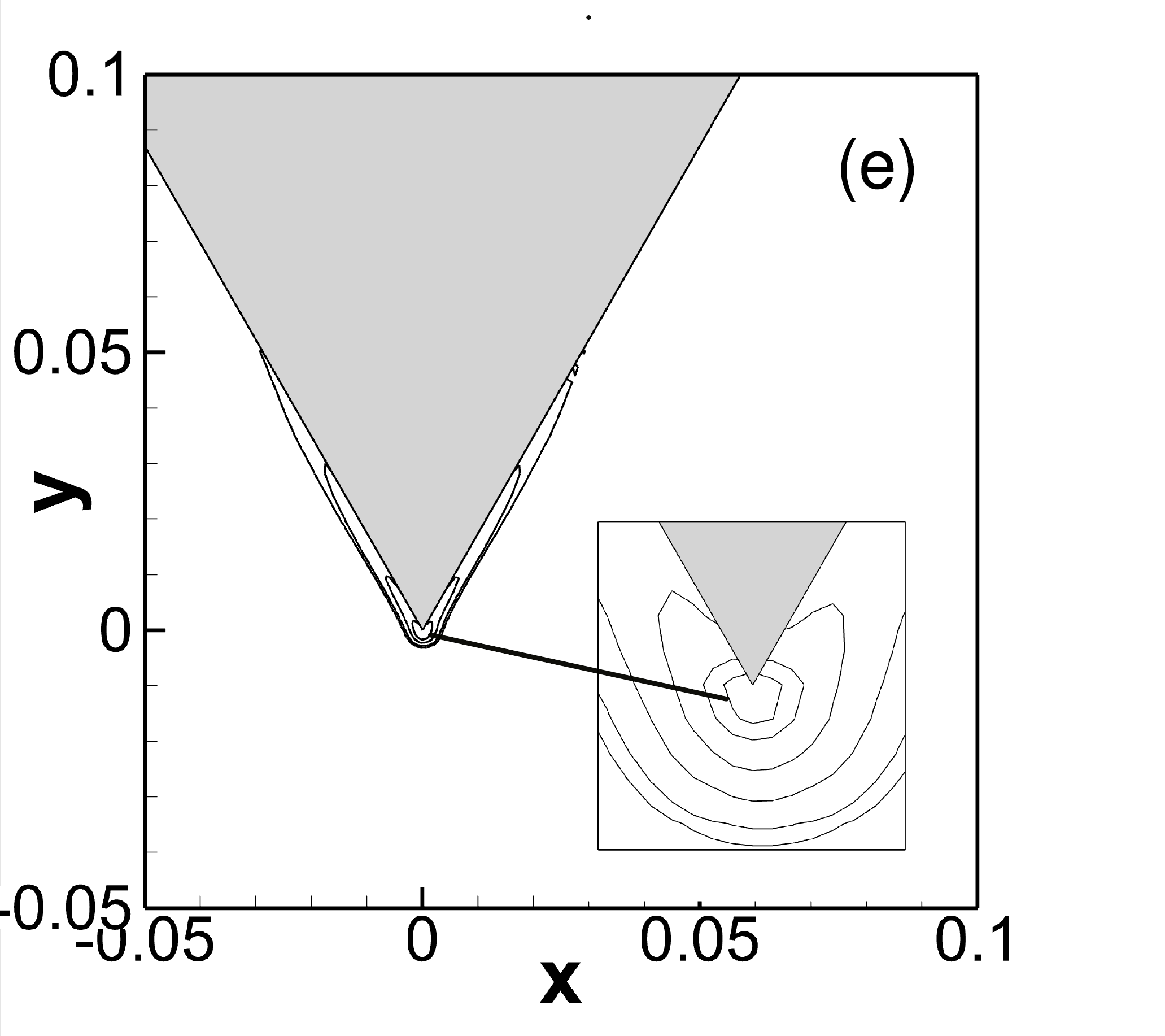}
			&
			\hspace{-0.5cm}\includegraphics[width=0.3\textwidth]{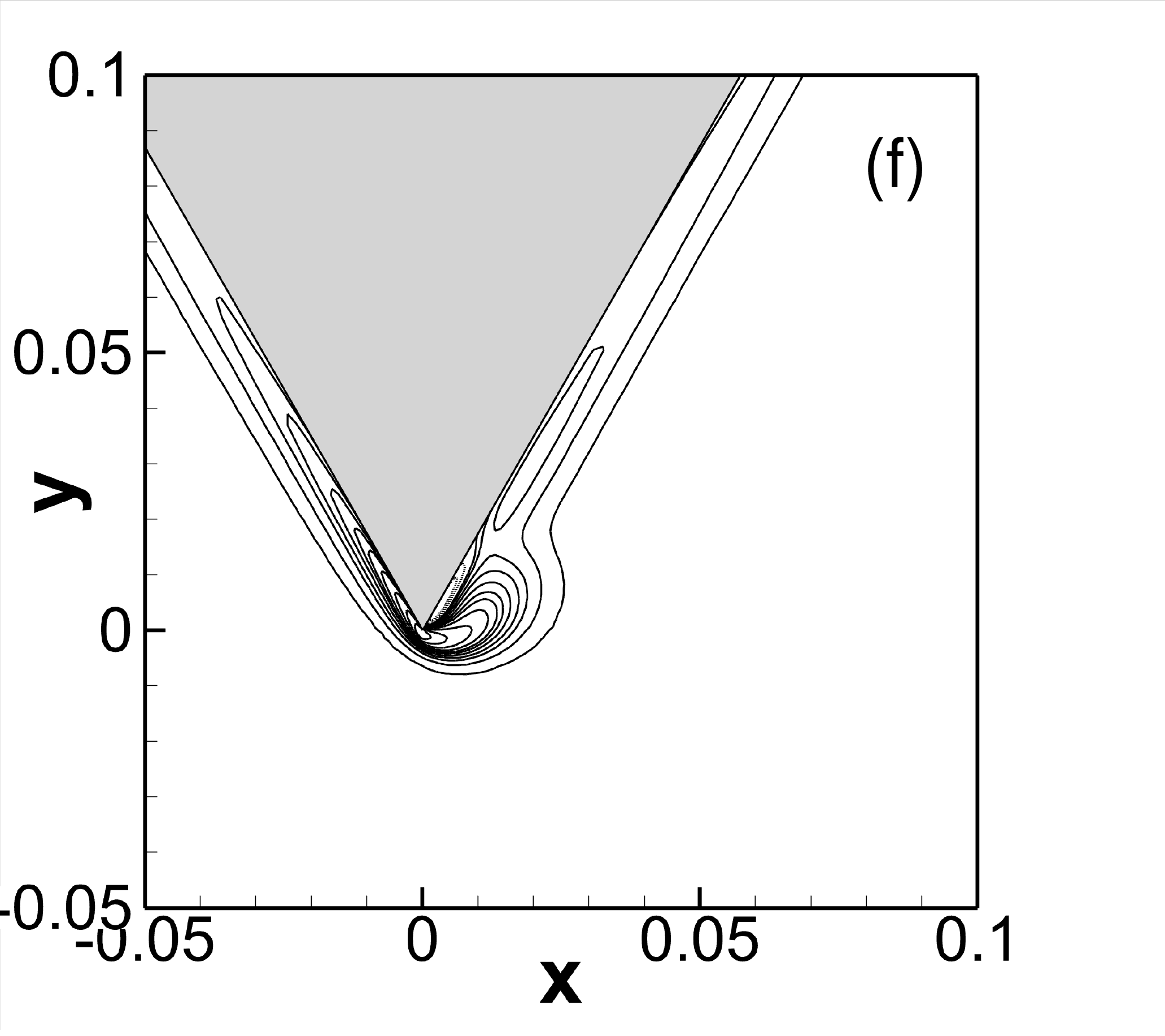}
			\&
			\hspace{-0.5cm}\includegraphics[width=0.3\textwidth]{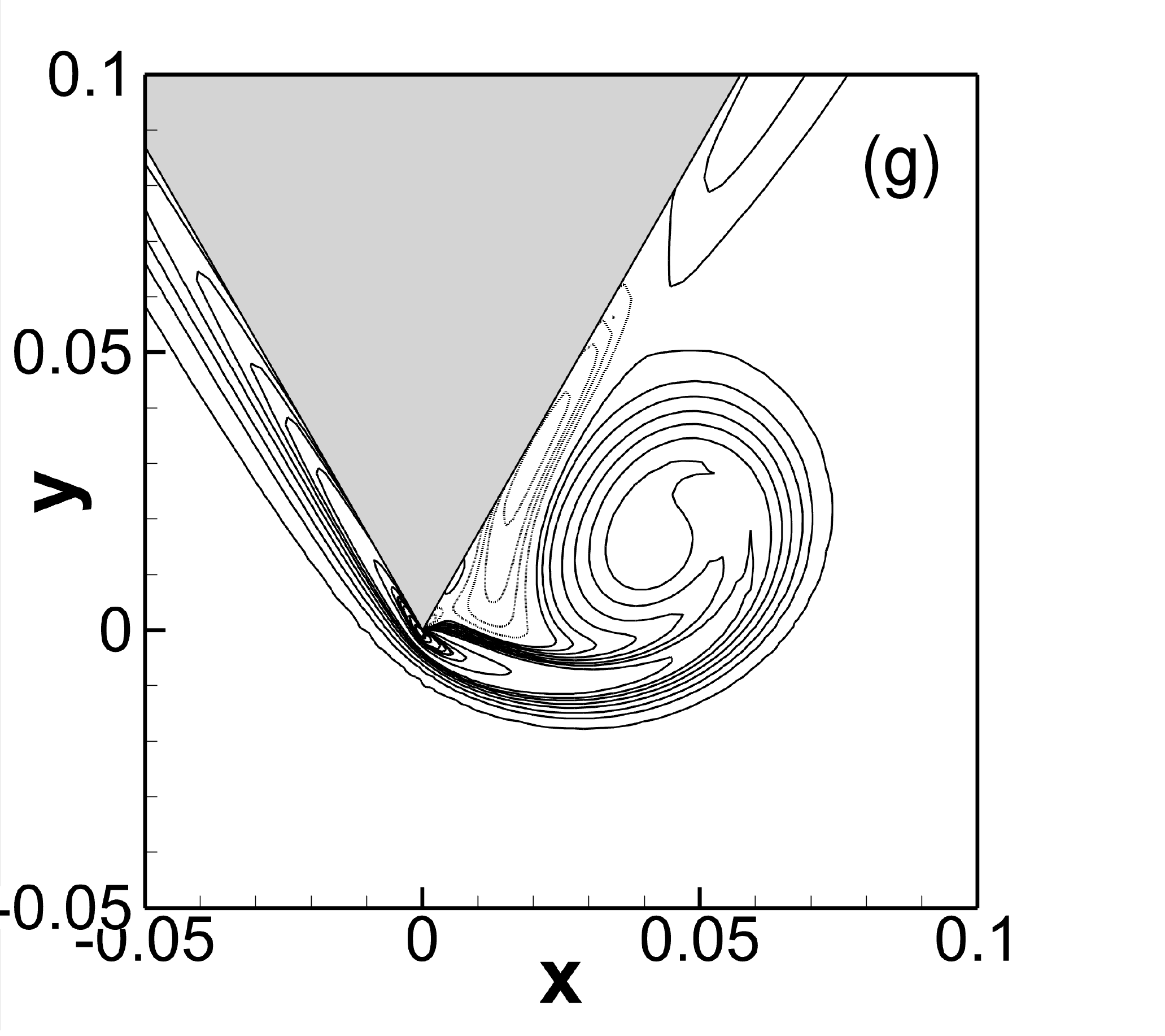}
			&
			\hspace{-0.5cm}\includegraphics[width=0.3\textwidth]{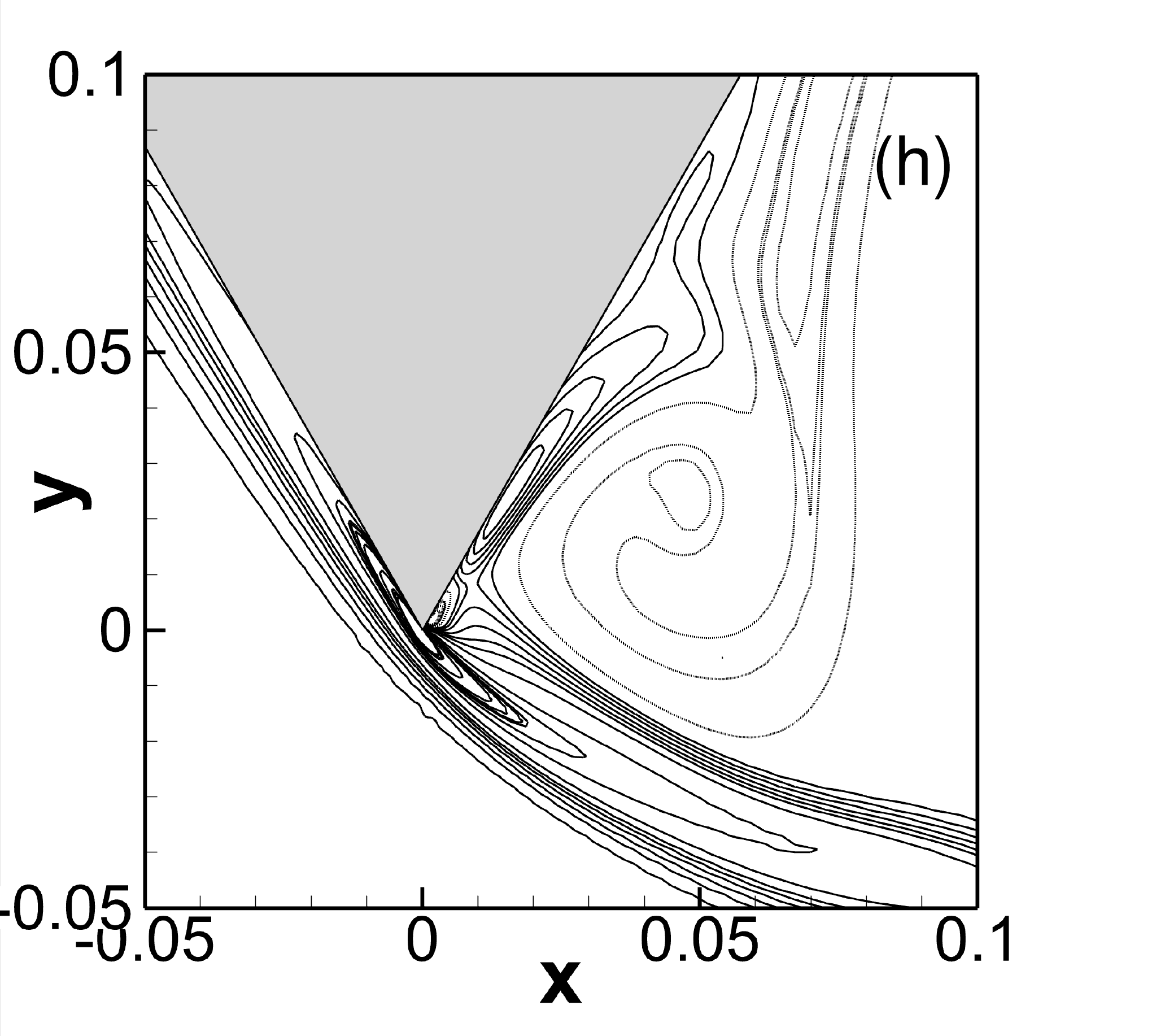}
		\end{tabular}
		\caption{Four stages of vorticity evolution for $Re_c=6873$ at non-dimensional time (a) $t=0.01$, (b) $t=0.10$, (c) $t=0.30$ and  (d) $t=1.0$, and the corresponding close-up views in (e-h). The positive contours are in solid lines and negative contours are in dotted lines.}
		\label{rayleigh_vt}
	\end{figure}
	In figures \ref{rayleigh_sf}(a)-(h) and \ref{rayleigh_vt}(a)-(h), one can see the earliest stages of flow development for $Re_c=6873$. While figure \ref{rayleigh_sf} demonstrates the streamlines and their close-up views, figure \ref{rayleigh_vt} depicts the corresponding vorticity contours. Here, all the four stages of start-up vortex evolution, reported by \cite{xu2015}, \cite{xu2016} and \cite{koumoutsakos1996} can be observed clearly; however, the flow evolution for the wedge subjected to accelerated flow is much more rapid than the one observed in the accelerated flat plate case (\cite{xu2015,koumoutsakos1996}). Initially, the formation of an almost symmetrical vortex is observed at the tip of the wedge (can be seen at $t=0.01$ in figures \ref{rayleigh_sf}(a), \ref{rayleigh_sf}(e), \ref{rayleigh_vt}(a) and \ref{rayleigh_vt}(e)) with a boundary layer of uniform thickness around it. All the vorticity values are positive at this juncture, with the highest value appearing at the wedge tip.  Vorticity contours then start deviating towards the leeward direction, still without any separating streamline (with value $\psi=0$). This is the Rayleigh stage, duration of which ends when a negative vorticity region forms within the starting vortex near the wedge wall with a well-defined center of rotation and bounding streamline as seen in figures \ref{rayleigh_sf}(b), \ref{rayleigh_sf}(f), \ref{rayleigh_vt}(b) and \ref{rayleigh_vt}(f) at time $t=0.10$. This is the Viscous stage where the convective terms of the N-S equations become comparable with the viscous terms and thus the well-defined vortex structures begin to appear. As time passes, the negative vorticity boundary layer thickens, causing the starting vortex to separate from the wedge. After a while, the vorticity acquires a local maximum in the vortex core as seen in figures \ref{rayleigh_sf}(c), \ref{rayleigh_sf}(g), \ref{rayleigh_vt}(c) and \ref{rayleigh_vt}(g) at time $t=0.3$, which was not observed in figures  \ref{rayleigh_sf}(b), \ref{rayleigh_sf}(f), \ref{rayleigh_vt}(b) and \ref{rayleigh_vt}(f). This is the self-similar inviscid stage. At this stage, convective terms become dominant, but the vortex is still small enough to be independent of the geometry except for the local edges. Eventually the outer streamline detaches from the slant wedge wall giving rise to the stage known as vortex expulsion. The initial recirculation bubble opens up, freeing it up from the self-similar growth and the vortex starts lagging behind the body as shown in (figures \ref{rayleigh_sf}(d), \ref{rayleigh_sf}(h), \ref{rayleigh_vt}(d) and \ref{rayleigh_vt}(h) at time $t=1.0$). The starting vortex grows in size and is convected downstream, which in turns induces a secondary vortex at the tip of the wedge as can be seen in figure \ref{rayleigh_sf}(g). The orientation of flow inside this secondary vortex is in the direction opposite to the primary one.
	
	\begin{figure}
		\begin{center}
			\includegraphics[width=0.45\textwidth]{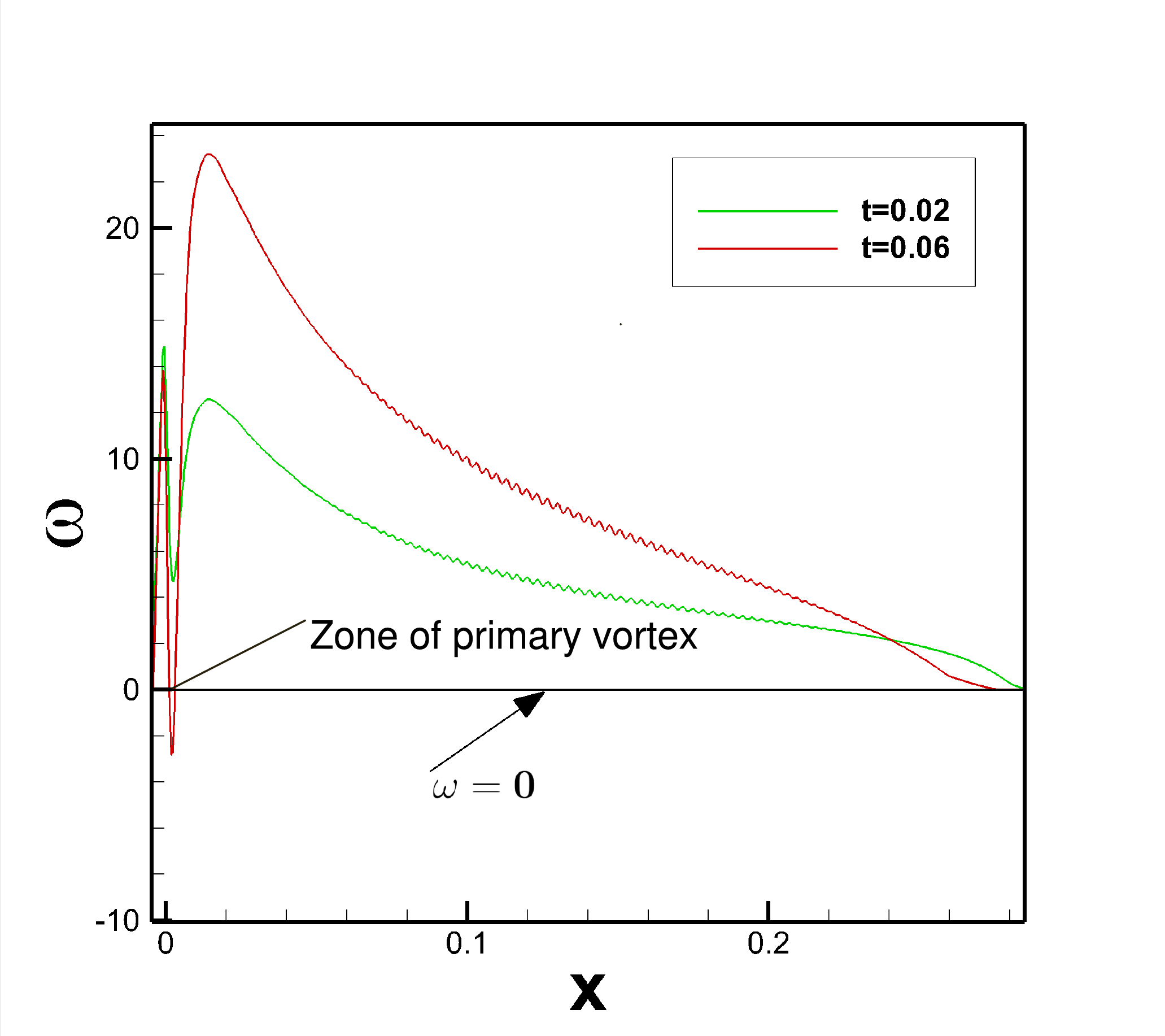}
			{(a)}
			\includegraphics[width=0.45\textwidth]{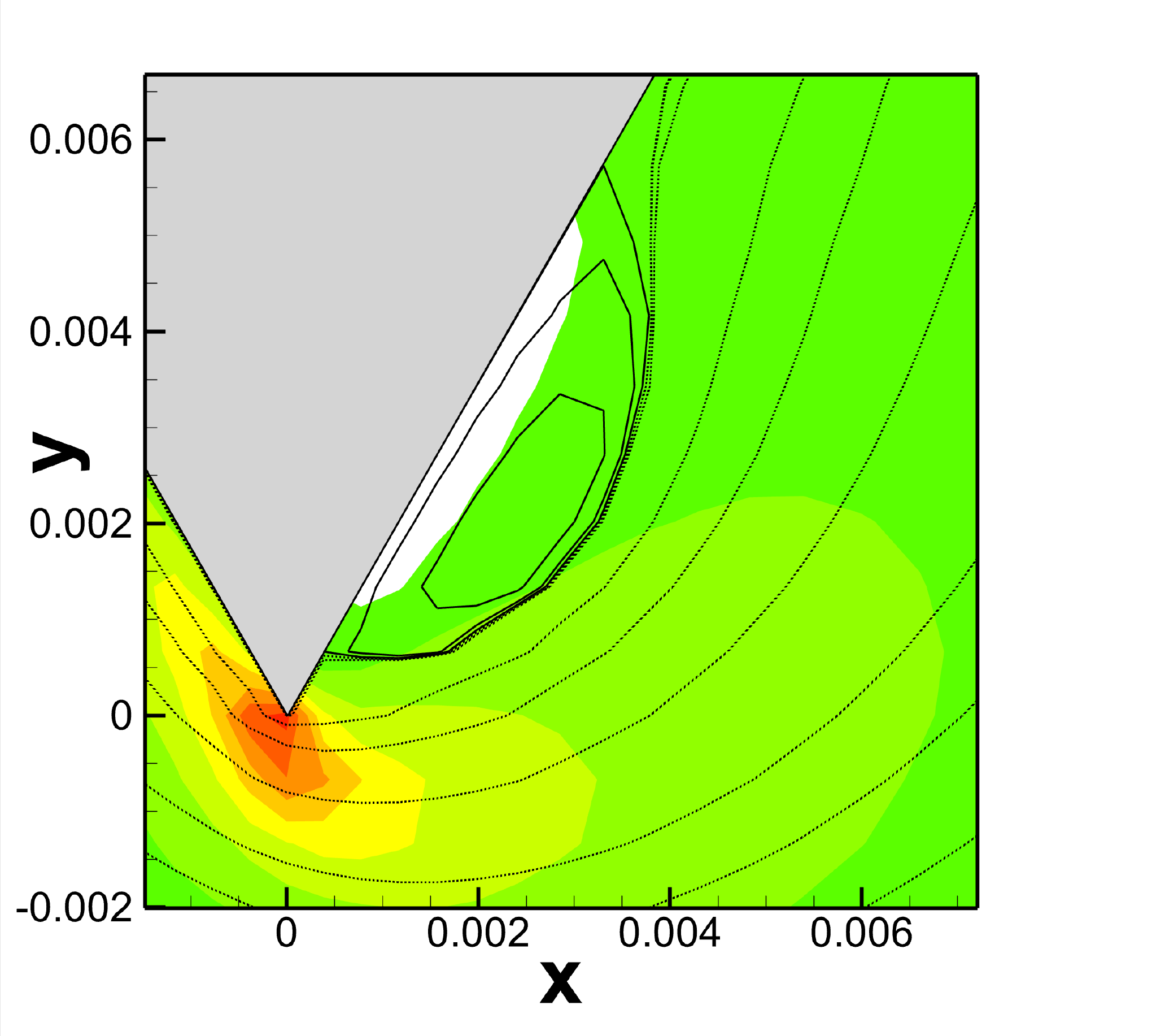}
			{(b)}
			\includegraphics[width=0.45\textwidth]{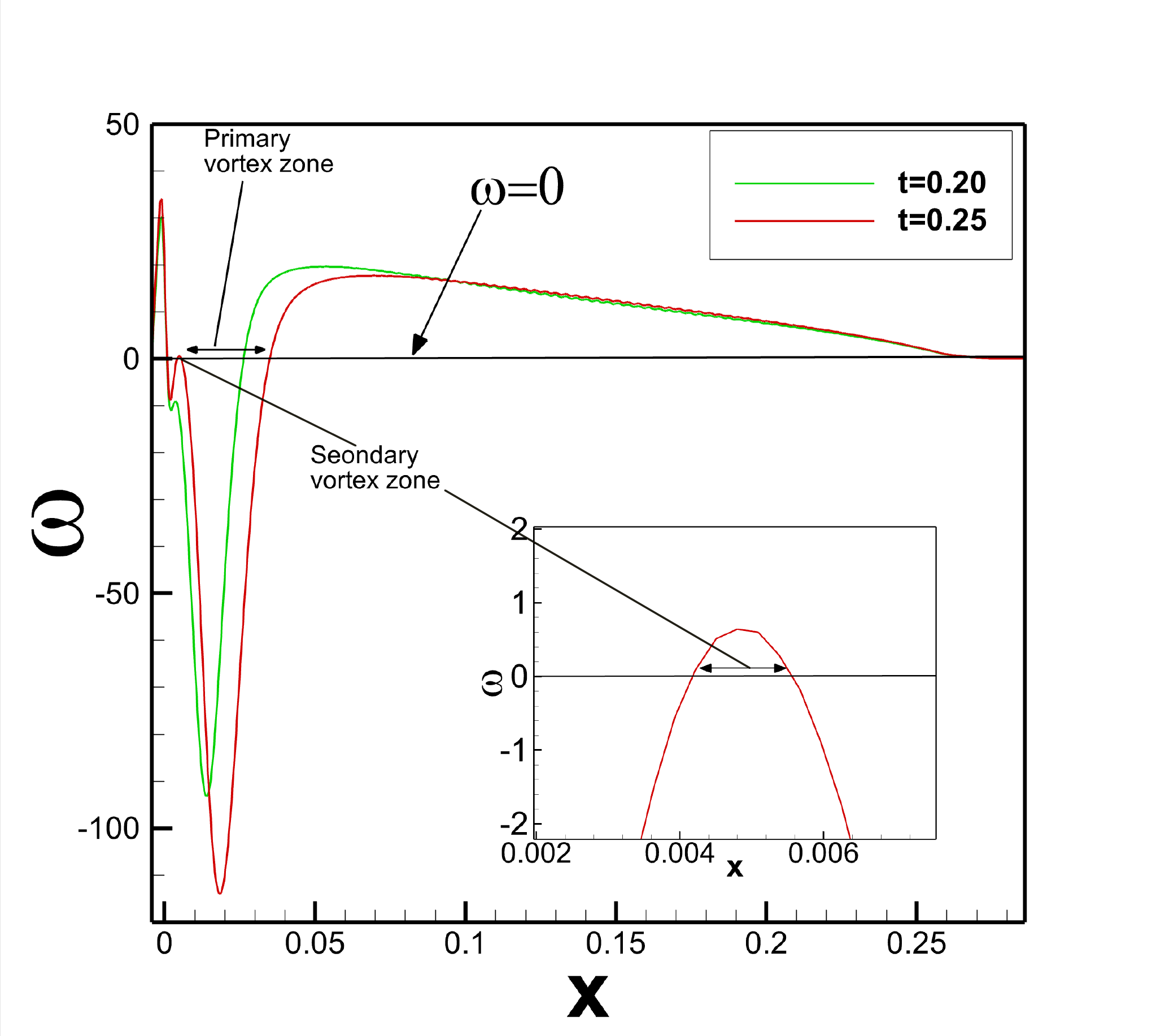}
			{(c)}
			\includegraphics[width=0.45\textwidth]{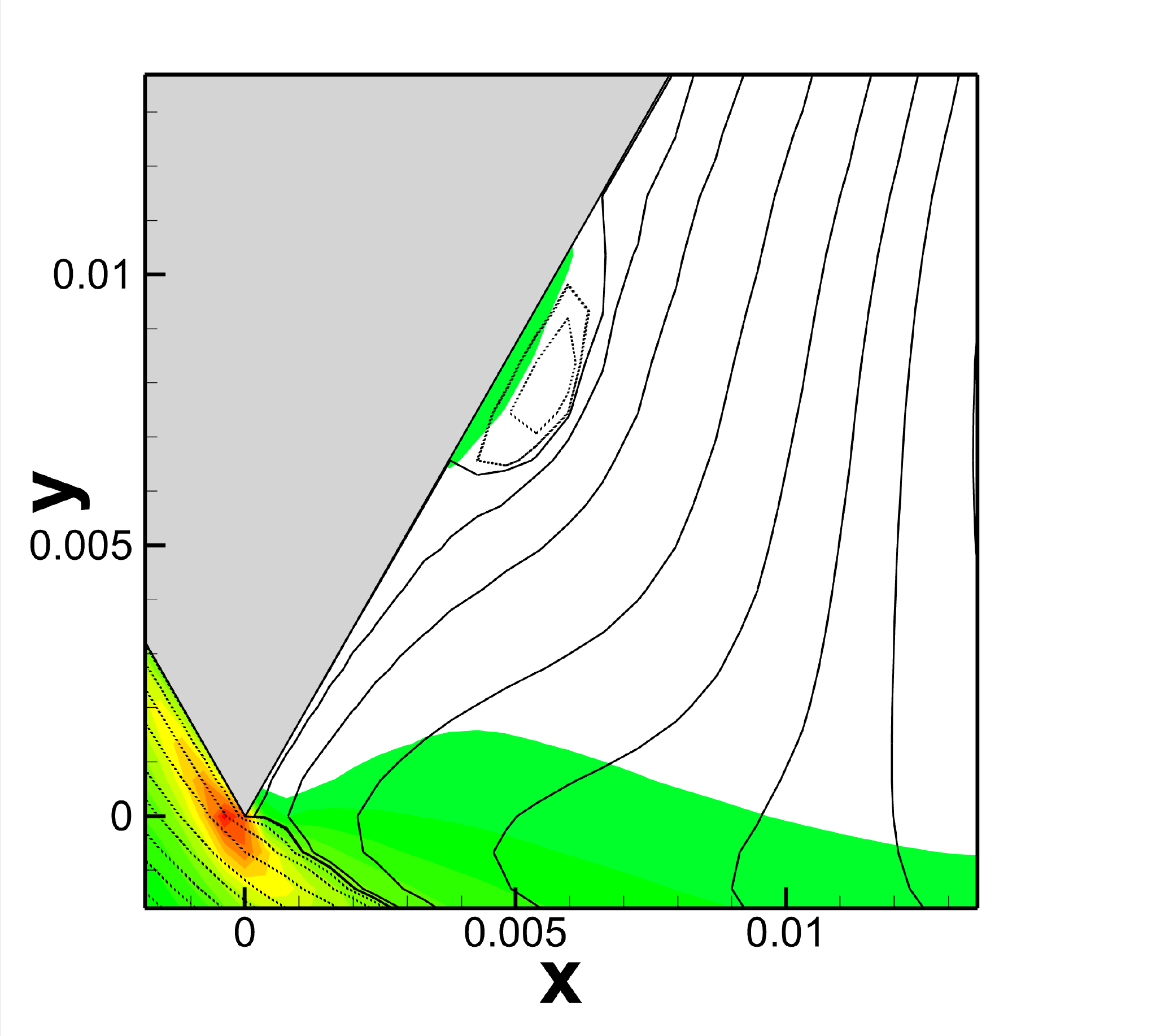}
			{(d)}
			\includegraphics[width=0.45\textwidth]{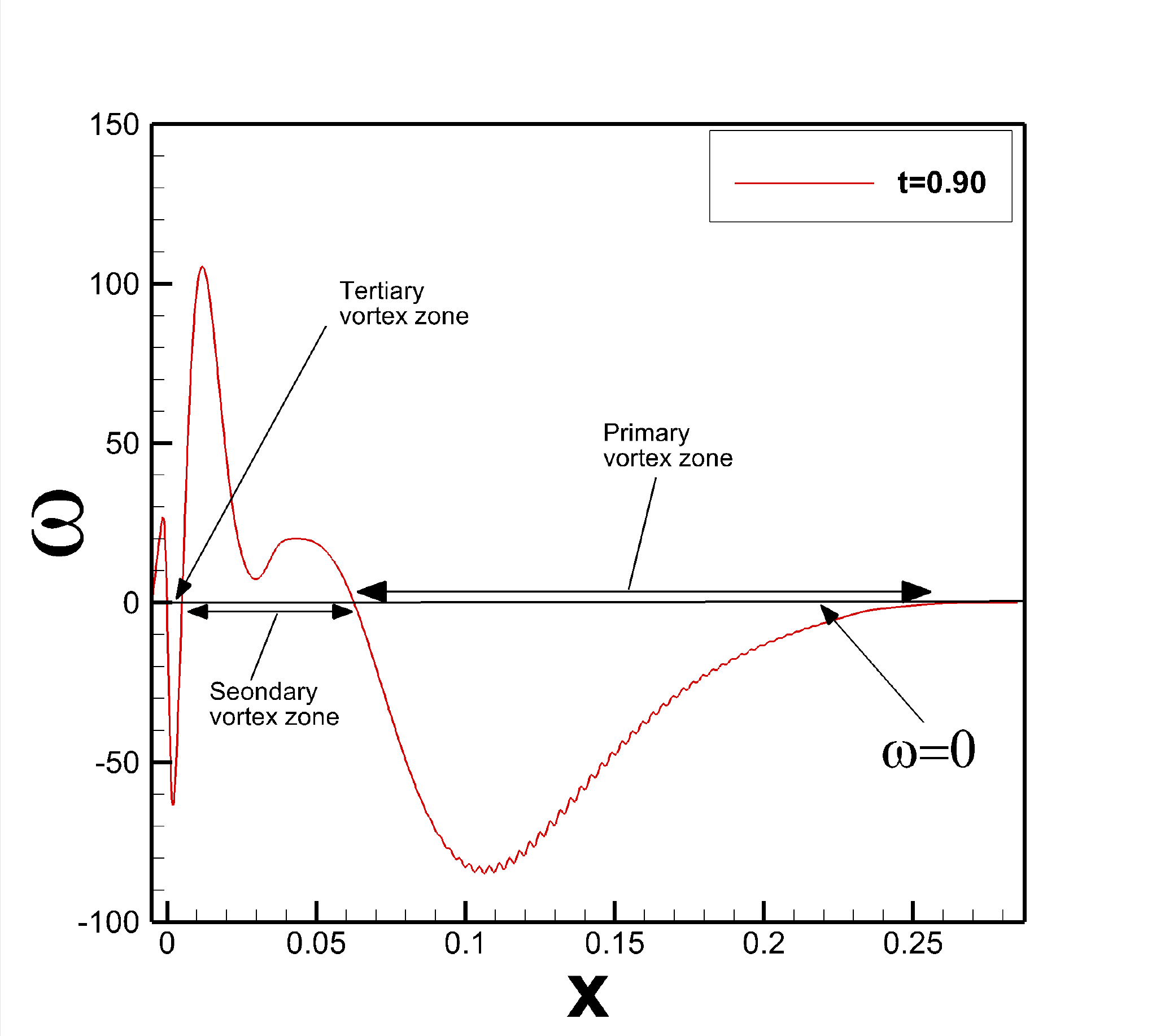}
			{(e)}
			\includegraphics[width=0.45\textwidth]{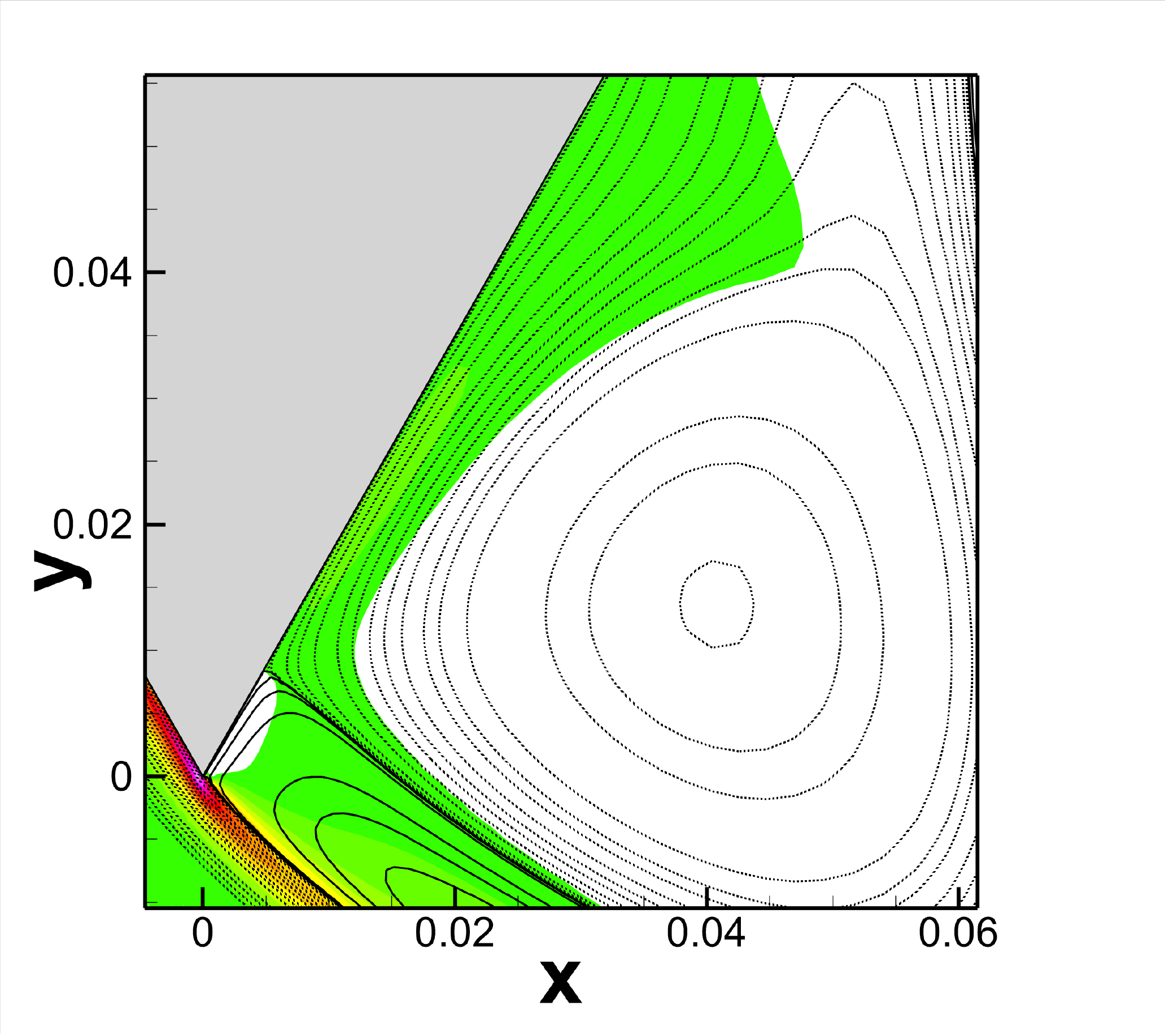}
			{(f)}
			\caption{Flow separation from the wedge wall  for $Re_c=6873$:  Distribution of vorticity along the wedge wall at time (a) $t=0.02$ $\&$ $t=0.06$, c) $t=0.20$ $\&$ $t=0.25$ and (e) $t=0.90$. Combined streamfunction (dotted lines representing  negative values and solid lines representing positive values) and vorticity contours (flooded, white colour representing areas with negative values) at time (b) $t=0.06$, (d) $t=0.25$ and $t=0.90$.}
			\label{separation}
		\end{center}
	\end{figure}
	
	It would be interesting to see approximately at what time flow separates from the wedge wall giving rise to the vortex phenomena mentioned above. It is well known that whenever vorticity changes sign along a solid wall, flow separation takes place paving the way for the creation of a vortex (\cite{ander95,pletcher,kalita2013unsteady}). As such, the number of vortices will depend upon the total number of changes of sign in vorticity values. We illustrate this phenomenon in figures \ref{separation}(a)-(f). In figures \ref{separation}(a), (c) and (e), we have plotted the vorticity distribution against the $x$-coordinate of the wedge wall and different times, while figures \ref{separation}(b), (f) and (f) depicts the corresponding combined streamfunction and vorticity contours. In figure \ref{separation}(a), one can clearly see that at time $t=0.02$, there is no change of sign of vorticity along the wall, similar to the situation depicted in figure \ref{rayleigh_vt}(e) at time $t=0.01$. On the other hand, at a later time $t=0.06$, one can see a very small zone of negative vorticity value which corresponds to the white region in figure \ref{separation}(b), enclosed by the (primary) vortex bounded by the $\psi=0$ streamline. This clearly demonstrates that the first flow separation takes place in the interval $t \in (0.02,0.06]$. Likewise, in figure \ref{separation}(c), one can see that at $t=0.20$, there is only one negative vorticity zone across the wedge wall. However, at $t=0.25$, one can see a very narrow zone of positive vorticity inside the primary vortex zone (see the enlarged view in the inset), giving rise to the second separation. This secondary vortex can be seen more clearly in figure \ref{separation}(d), where the region in green colour is enclosed by the $\psi=0$ streamline and the primary vortex that engulfs the secondary vortex is represented by the white region (see also figures \ref{rayleigh_sf}(g) and \ref{rayleigh_vt}(g) at time $t=0.30$). This indicates a second flow separation in the interval $t \in (0.20,0.25]$. Meanwhile, the secondary vortex grows in size and strength and at around $t=0.6$ divides the primary vortex into two chambers, one of which is the core and the other one gives rise to a tertiary recirculating zone. It becomes more prominent at $t=0.90$ onwards, which can also be observed in figures \ref{rayleigh_sf}(d) and (h). The vorticity distribution along the wall in figure \ref{separation}(c) at $t=0.90$ clearly demonstrates these three primary, secondary and tertiary recirculation zones, with figure \ref{separation}(f) showing the small negative vorticity region in the neighbourhood of the wedge-tip.

	\subsection{Movement of the primary vortex center and the m effect}\label{m_effect}
	\begin{figure}
		\centering\includegraphics[width=0.5\textwidth]{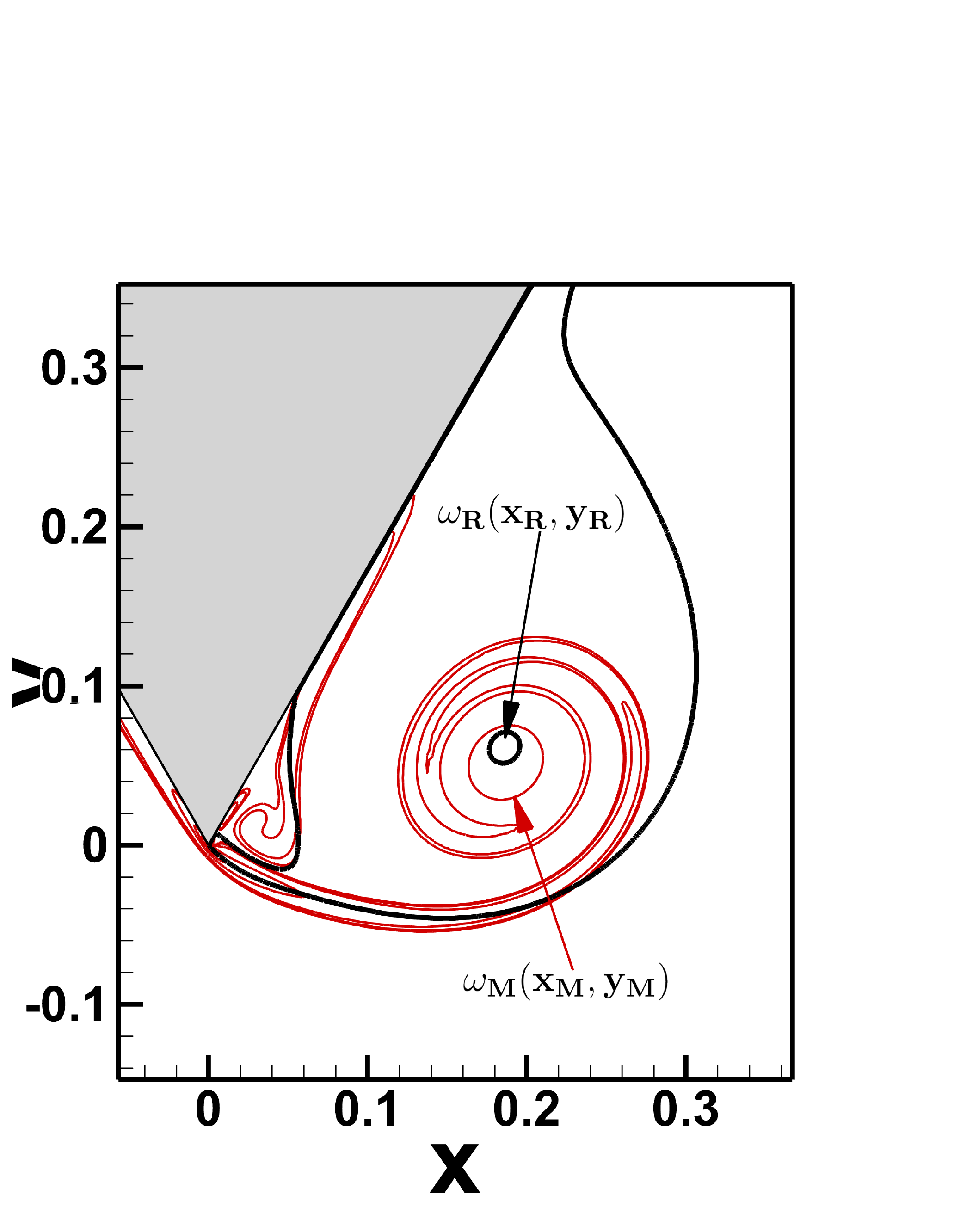}
		\caption{{Schematic of vortex and rotation centers. The vortex center lies inside the small red circular shape and the rotation center, inside the black one, as marked by the arrows.} }
		\label{sc_center}
	\end{figure}
	
\begin{figure}
		\begin{center}
			\includegraphics[width=0.95\textwidth]{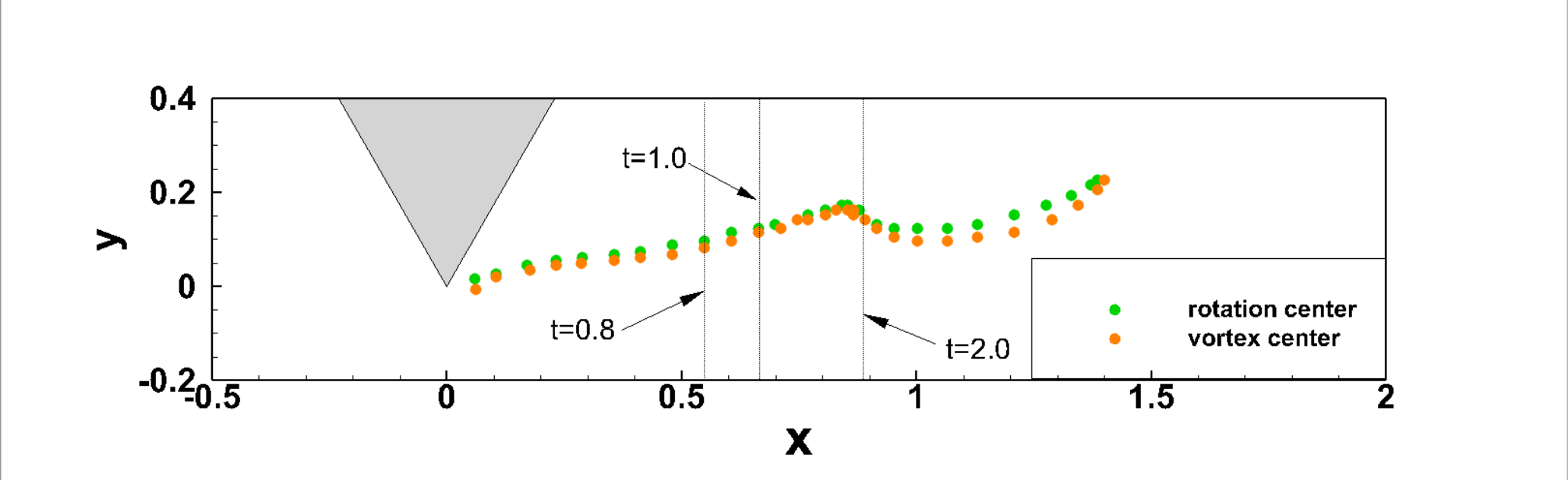}
			{(a)}
			\includegraphics[width=0.95\textwidth]{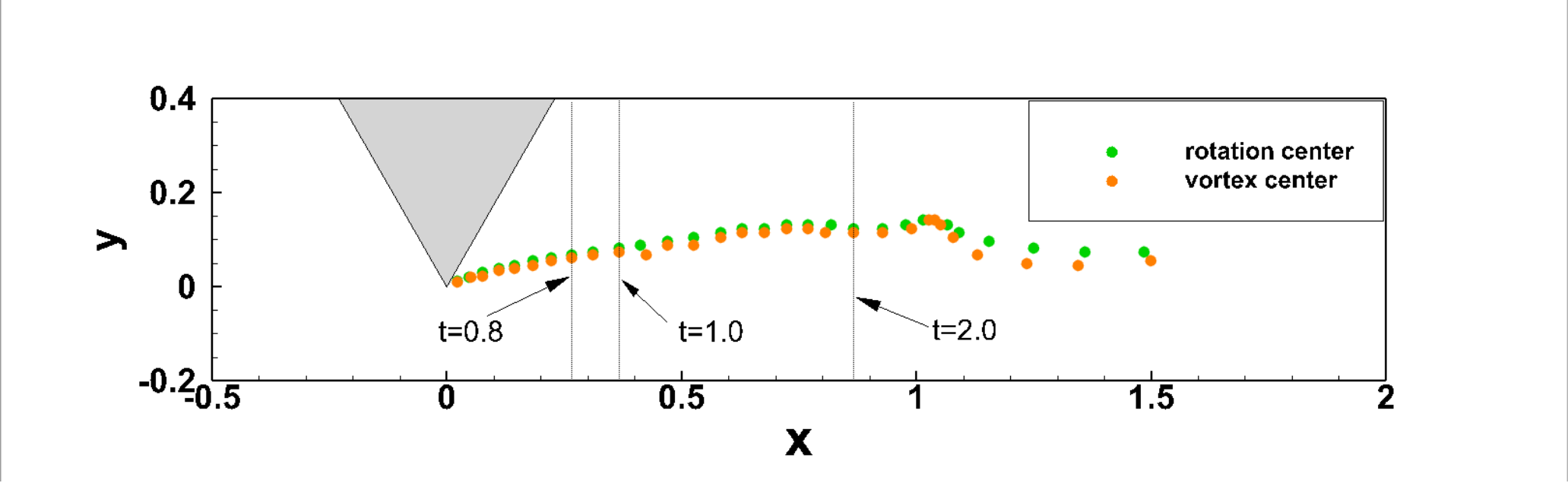}
			{(b)}
			\includegraphics[width=0.95\textwidth]{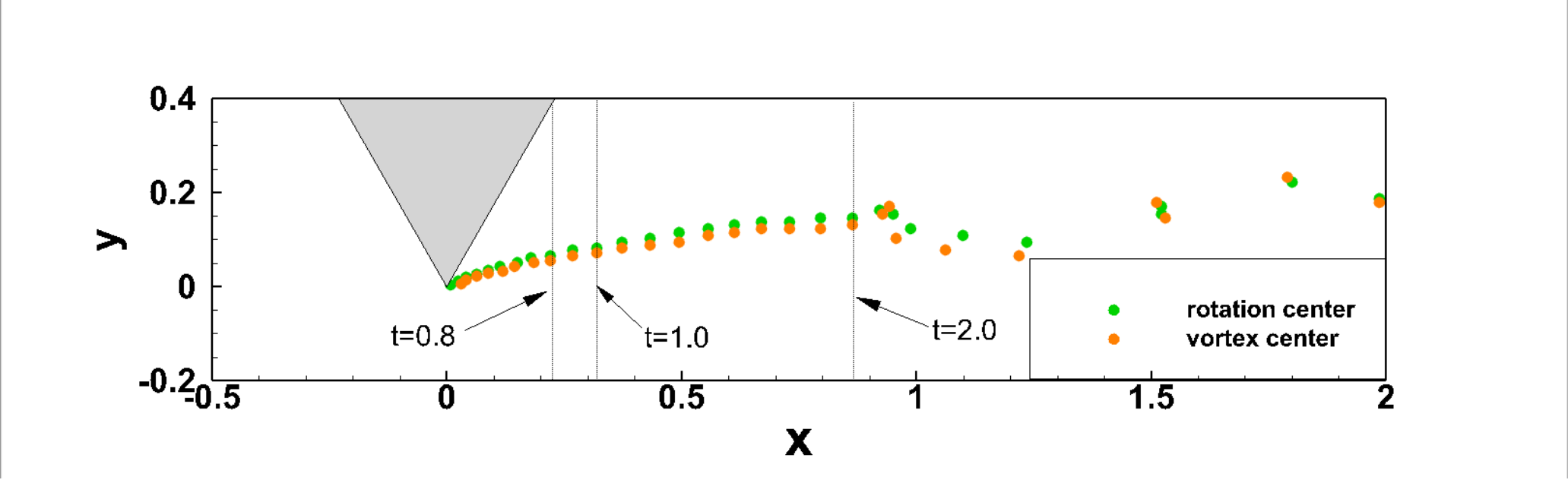}
			{(c)}
		\end{center}
		\caption{{Vortex and rotation center trajectories during $0<t \leq 3.0$ for (a) $Re_c=1560$ ($m=0$), (b) $Re_c=6621$ ($m=0.45$) and (c) $Re_c=6873$ ($m=0.88$).} }
		\label{center}
	\end{figure}

	In their laboratory experiment, \cite{pullin1980} presented the trajectories of the primary vortex center obtained directly from the measured frame by frame projection of their cine film. In the absence of any vorticity measurement in their experiments, they had taken the approximate geometric center, or the center of flow rotation of the vortex streaklines as the vortex center. Following \cite{xu2015,xu2016}, we also define the vortex center $(x_M,y_M)$ as the point in the computational domain having local vorticity maximum $\omega_M$. On the other hand, we define the rotation center $(x_R,y_R)$ as the point having local streamfunction maximum and denote the vorticity thereat as $\omega_R$. The schematic of the vortex and rotation centers are depicted in figure \ref{sc_center}, where the thick curves in black represent the streamlines and the red curves the vorticity contours. In figures \ref{center}(a)-(c), we show the trajectories of these vortex and rotation centers for $Re_c=1560$, $Re_c=6621$ and $Re_c=6873$ respectively. Note that, while a well defined rotation center is observed immediately after the first separation as documented in section \ref{fl_early}, it takes a while for the local maximum vorticity to set in and hence the formation of a well-defined vortex center. Our computation reveals that the first appearance of local vorticity maximums for $Re_c=1560$, $Re_c=6621$, $Re_c=6873$ occurred at instants $t=0.05$, $t=0.08$ and $t=0.2$ respectively. At those respective instants, the locations of the rotation and vortex centers are much closer to the wedge-tip for $m \neq 0$ than for the uniform flow as can be seen from figures \ref{center}(a)-(c).
	
	For the initial part ($t\leq 1$), when the rolled-up vortex is almost self similar, the trajectories almost moves in a straight line and the vortex center is always below the rotation center. However, with the onset of shear layer instability (see section), for $t>1$, their movement becomes more dispersed and rapid.
	
\begin{figure}
		\centering
		\begin{tabular}{cc}
			\includegraphics[width=0.4\textwidth]{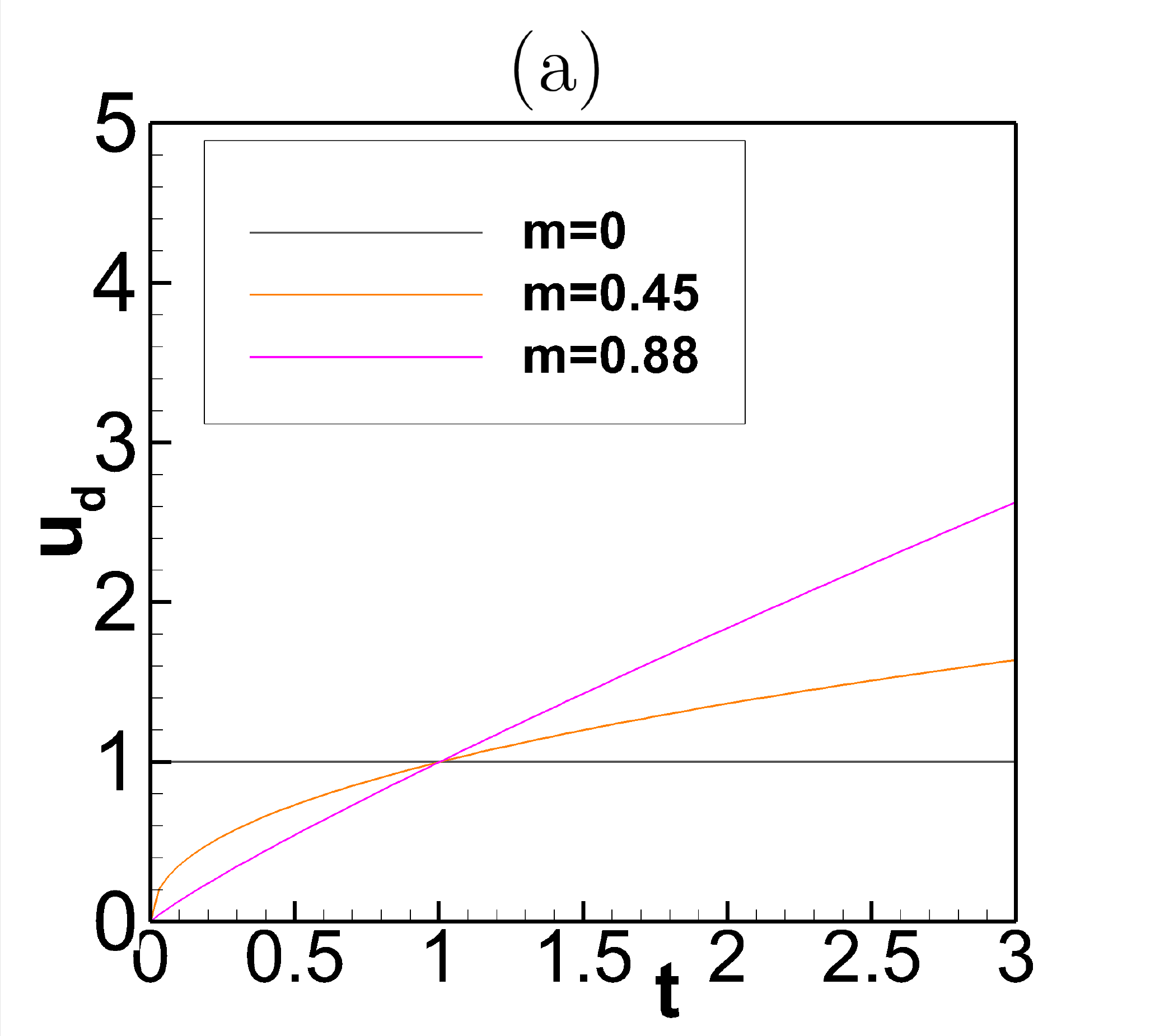}
			&
			\hspace{-0.5cm}\includegraphics[width=0.4\textwidth]{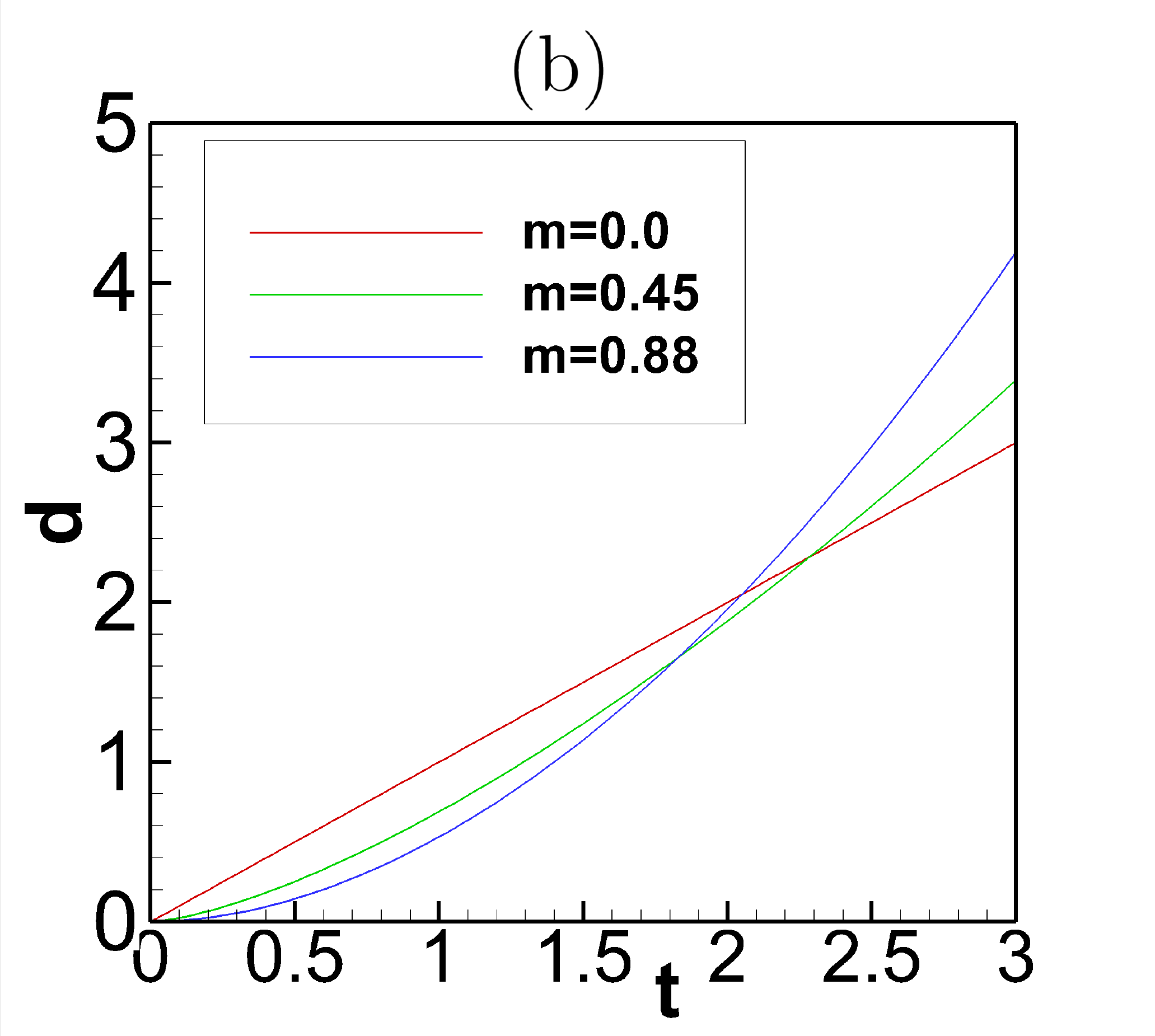}
		\end{tabular}	
		\caption{Effect of $m$ on non-dimensional driving velocity and distance: (a) $u_d$ against $t$ and (b) $d$ against $t$. }\label{eff_m_ut}	
	\end{figure}

	The effect the parameter $m$ on the trajectories of vortex centers may be interpreted in terms of the movement of the wedge driven by the piston depicted in figure \ref{fig:pulin}. If the wedge is moved from rest in the horizontal direction with a non-dimensional driving velocity obeying $u_d=t^m$, it will traverse a distance 
	\begin{equation}
		d=\frac{t^{m+1}}{m+1}.
		\label{nd_dt}
	\end{equation}
	In figures \ref{eff_m_ut}(a)-(b), we plot the graphs of non-dimensional driving velocity $u_d$ and distance $d$ as a function of the non-dimensional time $t$ respectively. The solution at a fixed time depends on $m$, as the wedge would have traversed significantly different distances $d$ for different values of $m$ as the formula suggests. It is obvious from the expressions for $u_d$ and $d$ that for $t\leq 1$, $u_d$ is a decreasing function of $m$, while for $t>1$, it is the opposite scenario. The same can be said about $d$ for approximately $t\leq 2$ and $t>2$. This explains the rapid movement of vortex centers away from the wedge tip for $t>2$ when $m \neq 0$ (figures \ref{center}(b)-(c)), with farthest displacement occurring for $m=0.88$, the largest value of $m$ considered in this study.
	
\begin{figure}
		\centering
		\begin{tabular}{ccc}
			\hspace{-1.0cm}\includegraphics[width=0.4\textwidth]{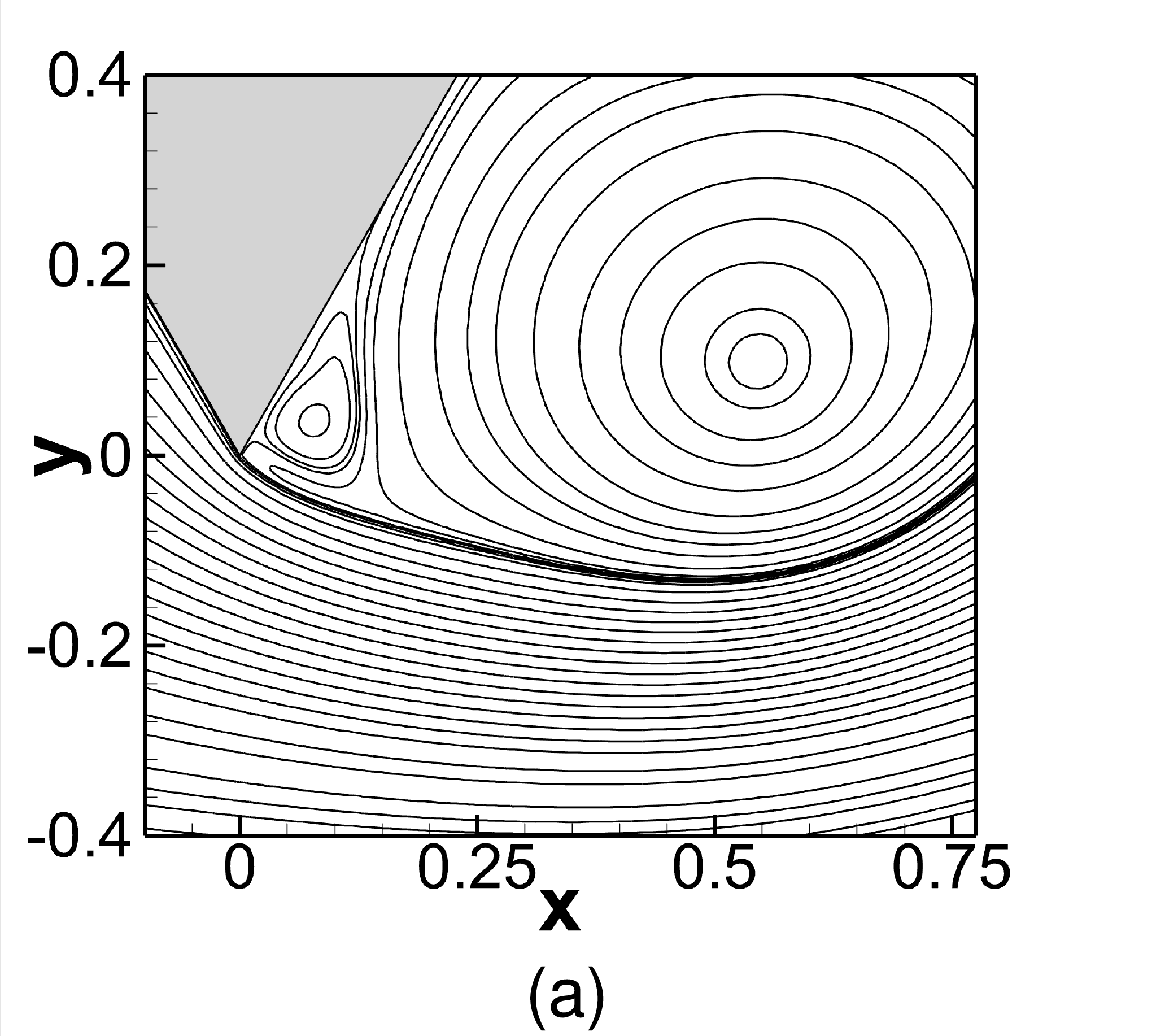}
			&
			\hspace{-1.0cm}\includegraphics[width=0.4\textwidth]{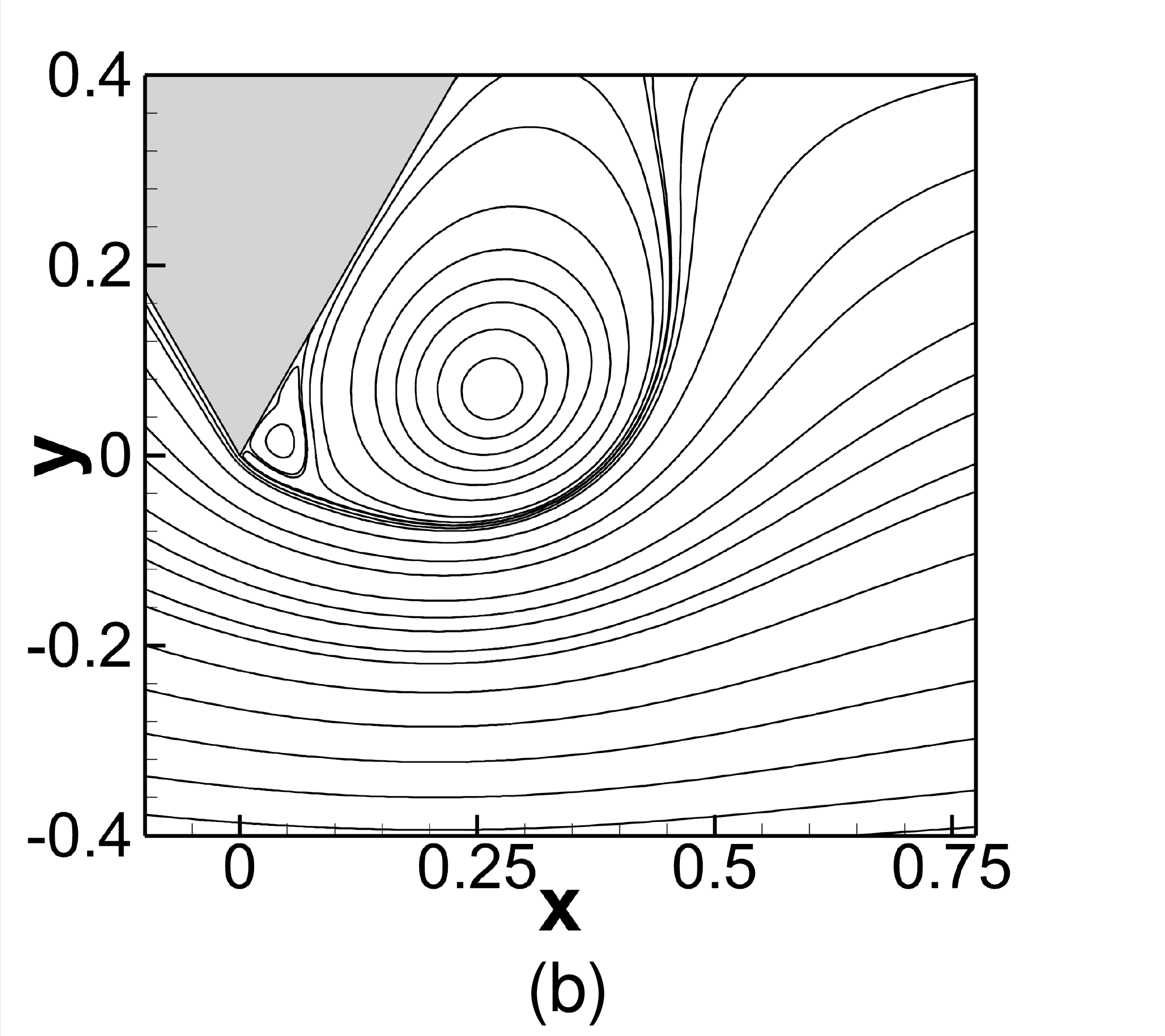}
			&
			\hspace{-1.0cm}\includegraphics[width=0.4\textwidth]{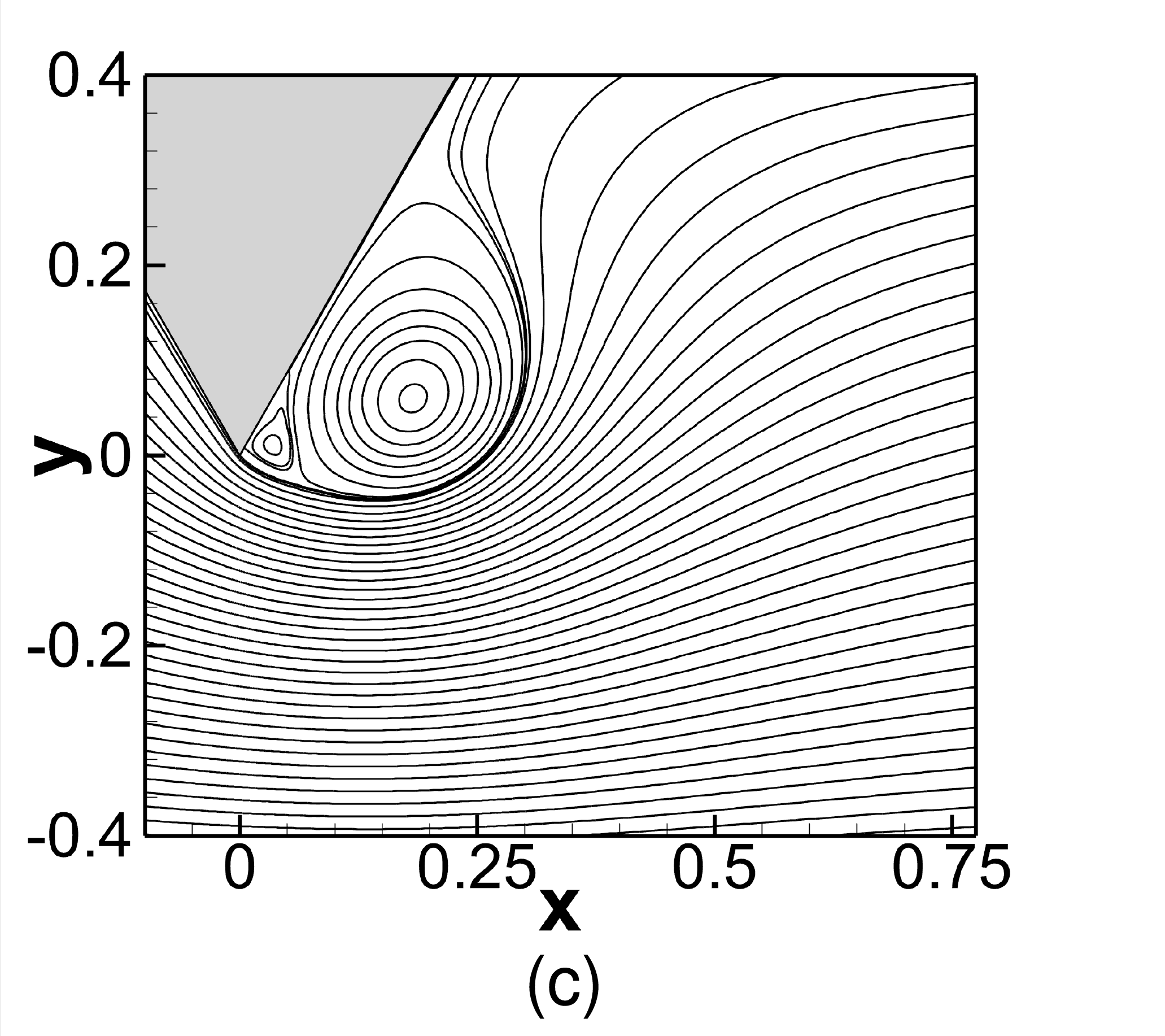}
		\end{tabular}	
		\caption{Streamlines for accelerated flow past a $60^o$ wedge at time  $t=0.8$ for  (a) $Re_c=1560$ ($m=0$), (b) $Re_c=6621$ ($m=0.45$) and  (c) $Re_c=6873$ ($m=0.88$). }\label{eff_m}	
	\end{figure}

\begin{figure}
		\centering\includegraphics[width=0.65\textwidth]{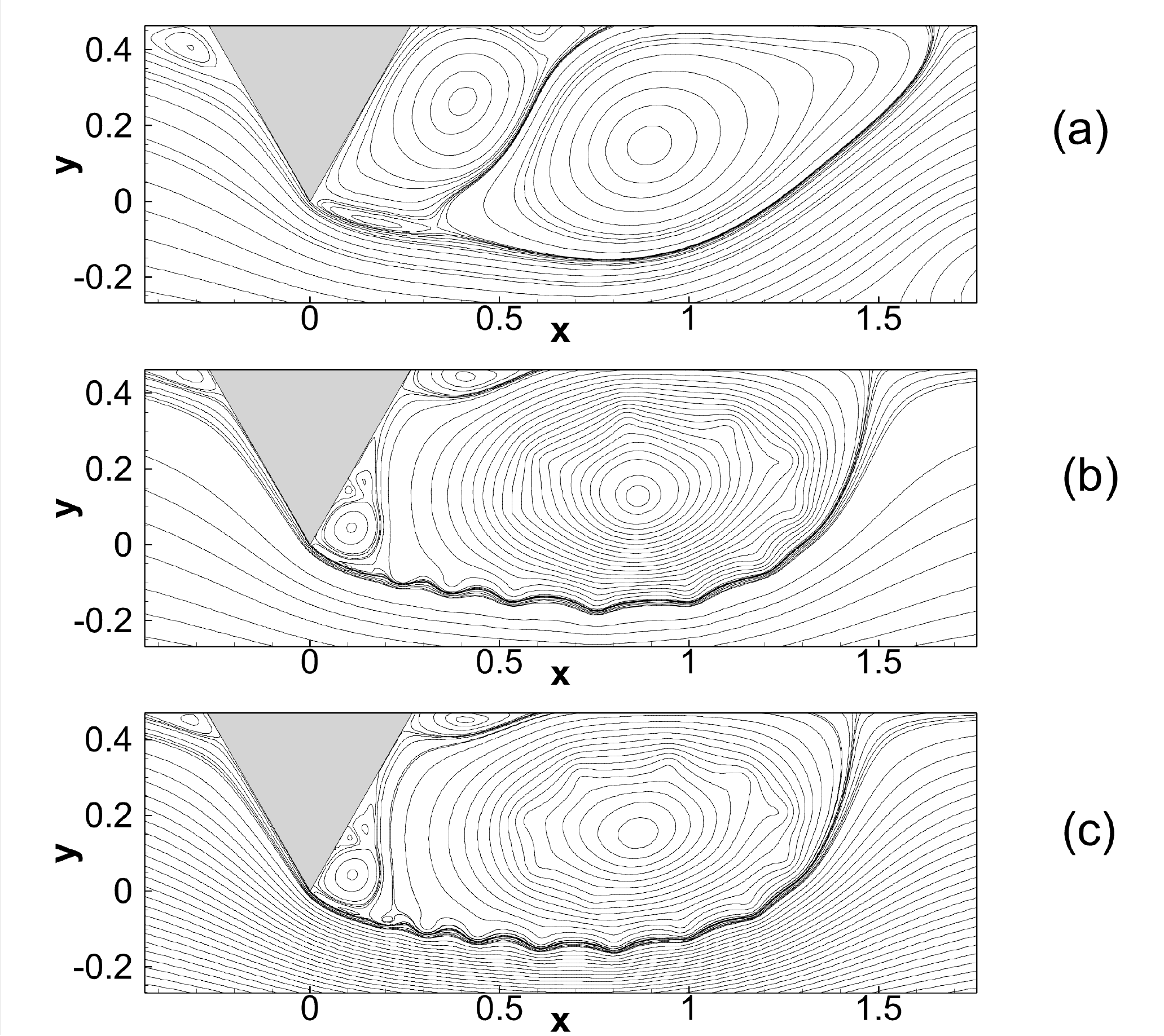}
		\caption{{Streamlines for accelerated flow past a $60^o$ wedge at time  $t=2.0$ for  (a) $Re_c=1560$ ($m=0$), (b) $Re_c=6621$ ($m=0.45$) and  (c) $Re_c=6873$ ($m=0.88$).} }
		\label{eff_m1}
	\end{figure}

	The effect of $m$ is further displayed in figures \ref{eff_m} and  \ref{eff_m1} depicting the streamlines for the three values of $m$ considered at times $t=0.8$ and $t=2.0$ respectively. Note that the displacement of the primary vortex center away from the wedge-tip is related to its growth as well. In figure \ref{eff_m} one can observe that with increase in $m$, the size of primary vortex is reducing. This is because of the fact that both $u_d$ and $d$ are decreasing functions of $m$ for $t\leq 1$. As such, till time $t=1.0$, a stronger velocity field is indicated for $m=0$ which is also obvious from figure \ref{eff_m_ut}(a). On the other hand, as seen from figure \ref{eff_m_ut}(b), from around $t=2.0$, the distances corresponding to $m\neq 0$ overtakes the ones for $m=0$. This is the reason one can see the displacement of vortex centers and the size of the primary vortices almost catching up with each other as in figures \ref{eff_m1}(a)-(c) and  \ref{center}(a)-(c), where the location of the vortex centers are marked at times $t=0.8$ and $t=2.0$ as well. This exemplifies the role of the parameter $m$ above. 

\begin{figure}
		\minipage{0.33\textwidth}
		\centering 
		\includegraphics[width=\linewidth]{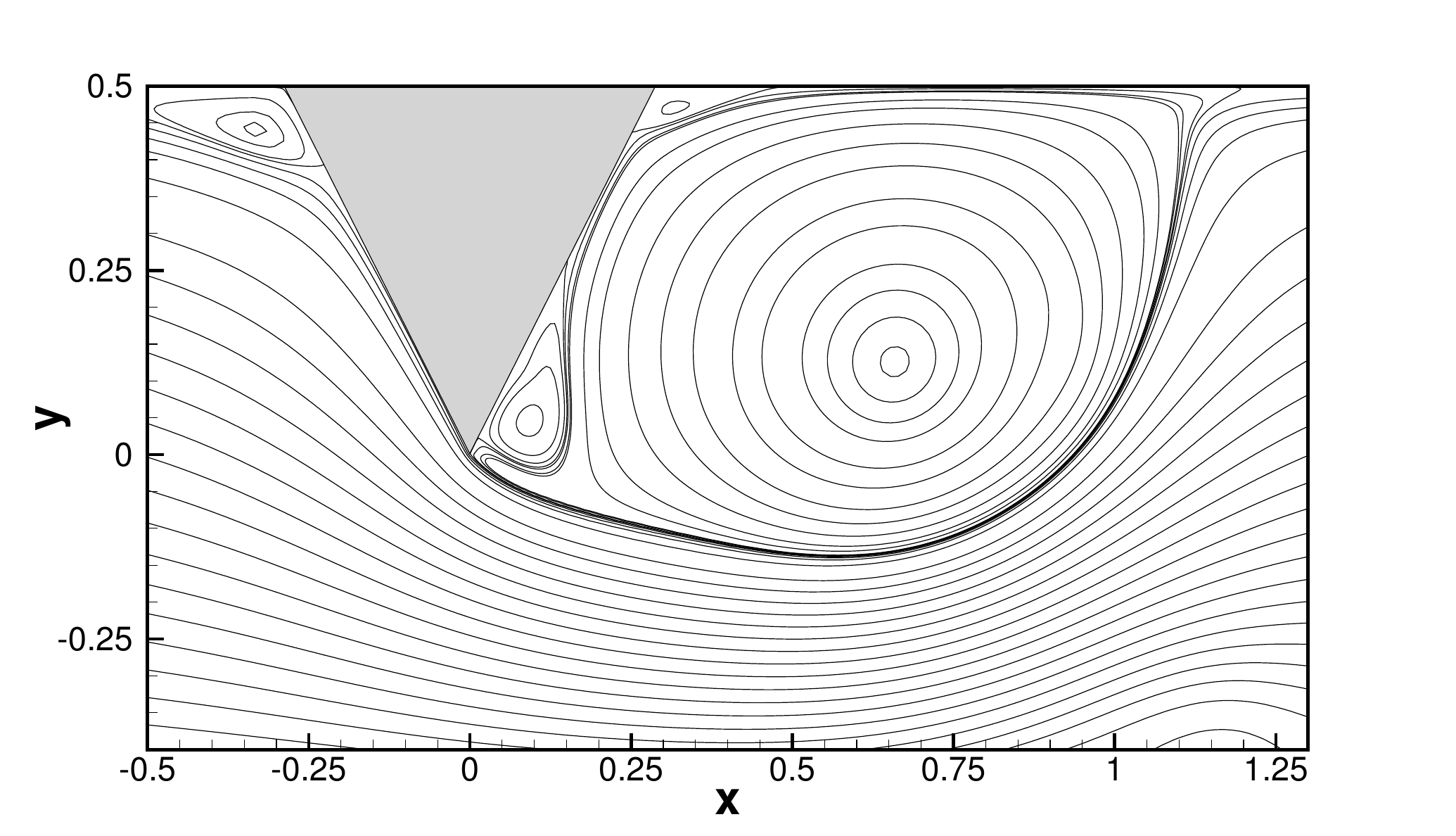}\\
		(a1)
		\endminipage\hfill
		\minipage{0.33\textwidth}
		\centering
		\includegraphics[width=\linewidth]{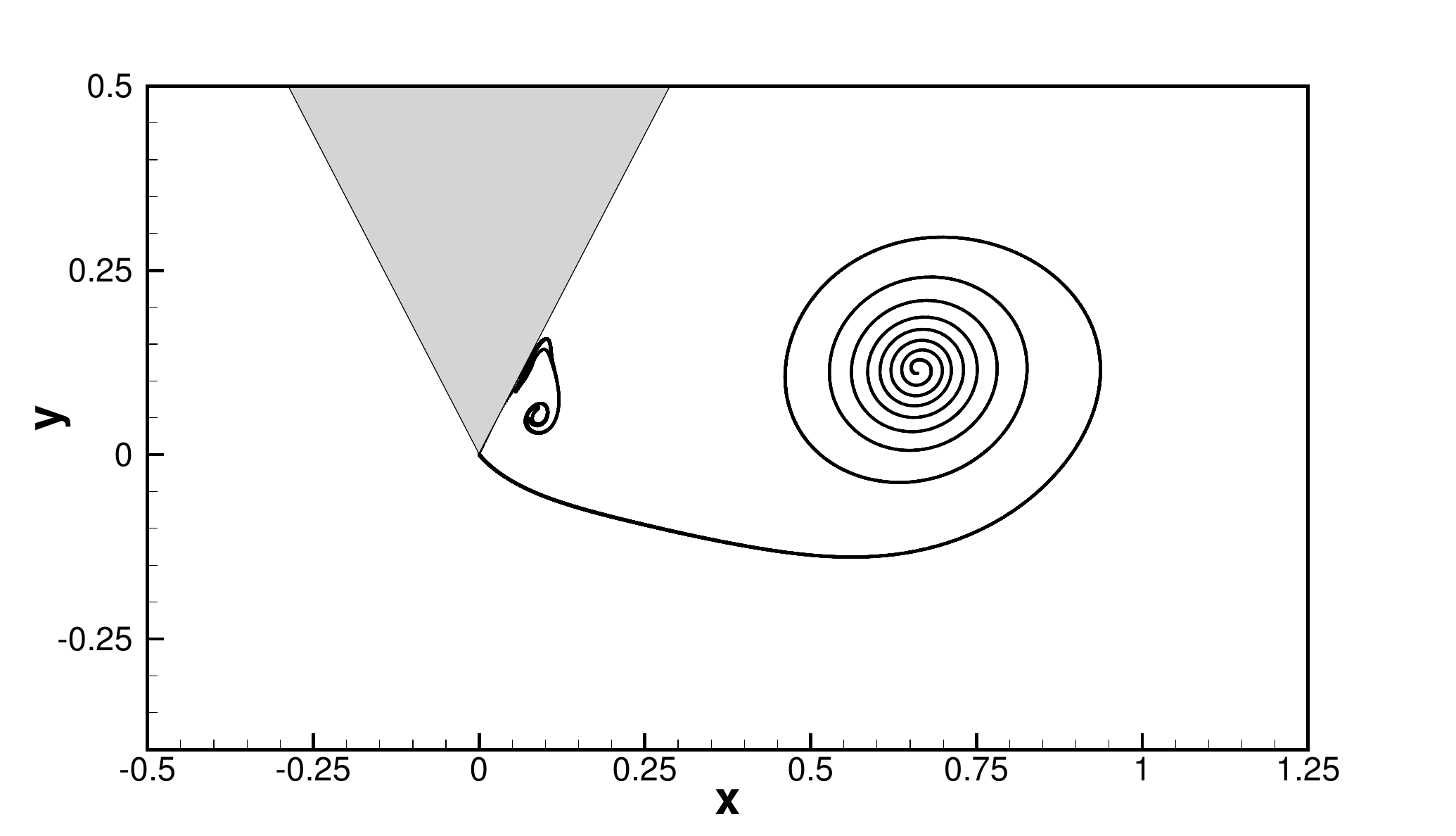}\\(b1)
		\endminipage\hfill
		\minipage{0.33\textwidth}%
		\centering
		\includegraphics[width=\linewidth]{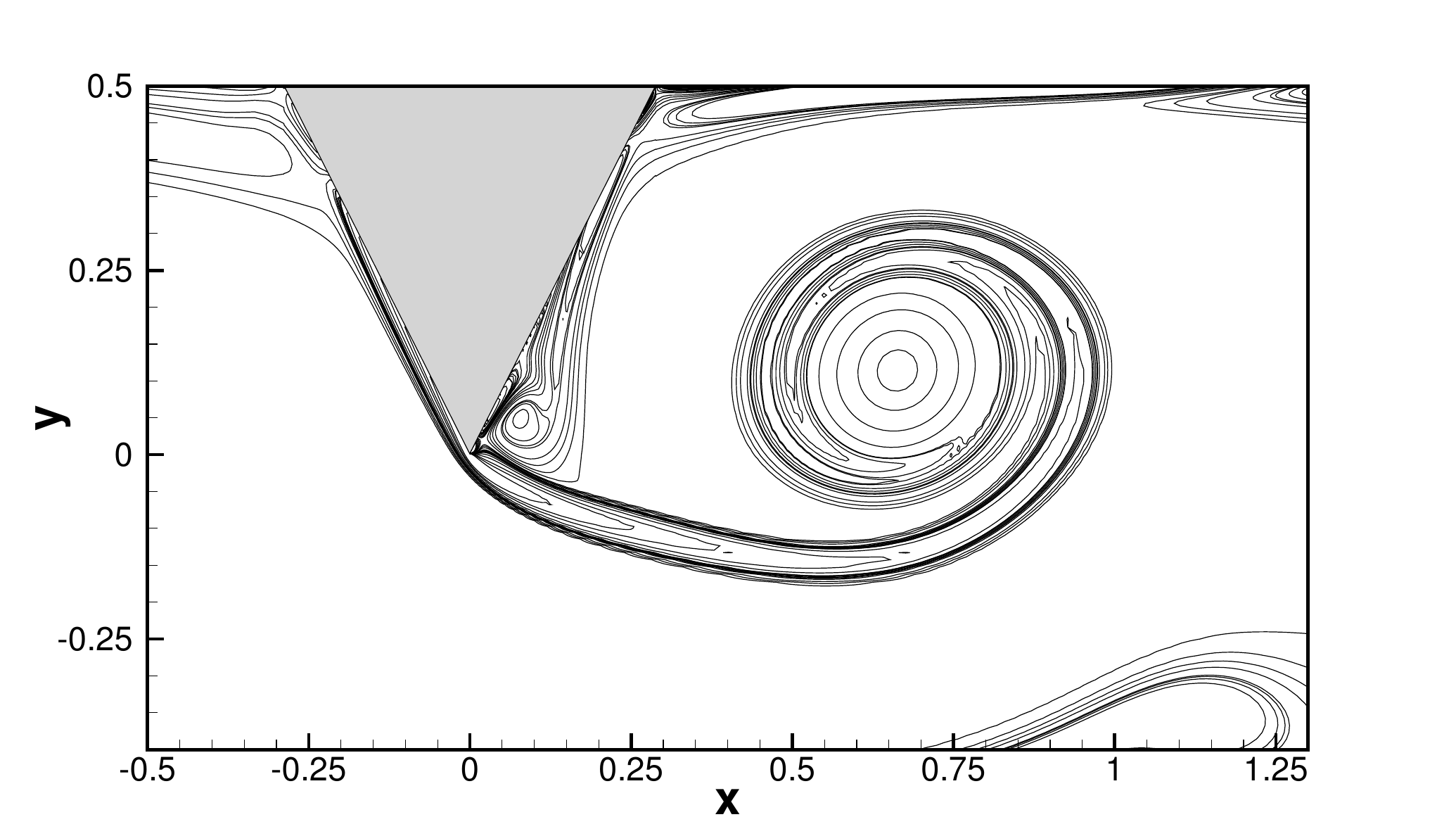}\\
		(c1)
		\endminipage\hfill
		\minipage{0.33\textwidth}
		\centering 
		\includegraphics[width=\linewidth]{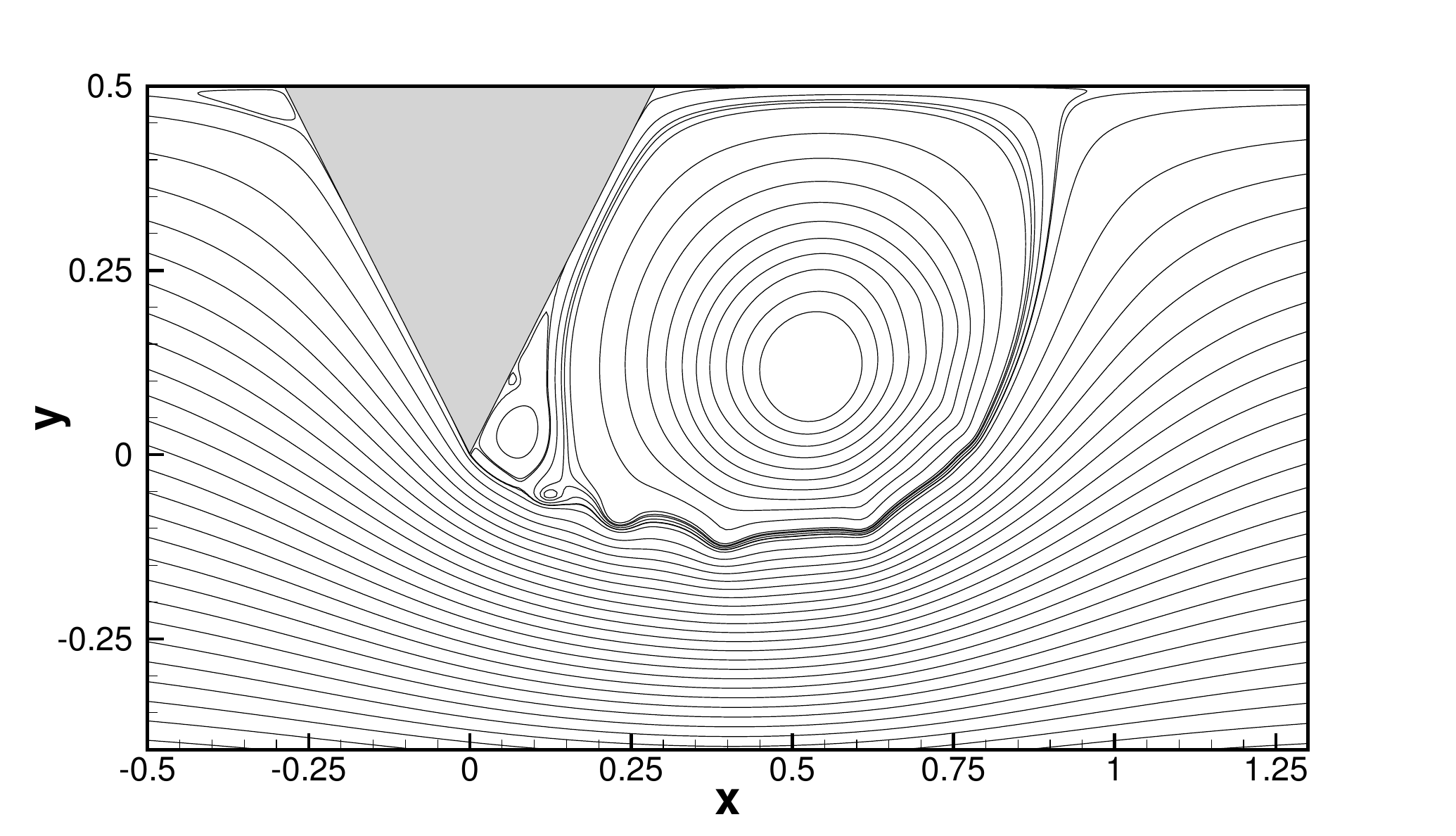}\\
		(a2)
		\endminipage\hfill
		\minipage{0.33\textwidth}
		\centering
		\includegraphics[width=\linewidth]{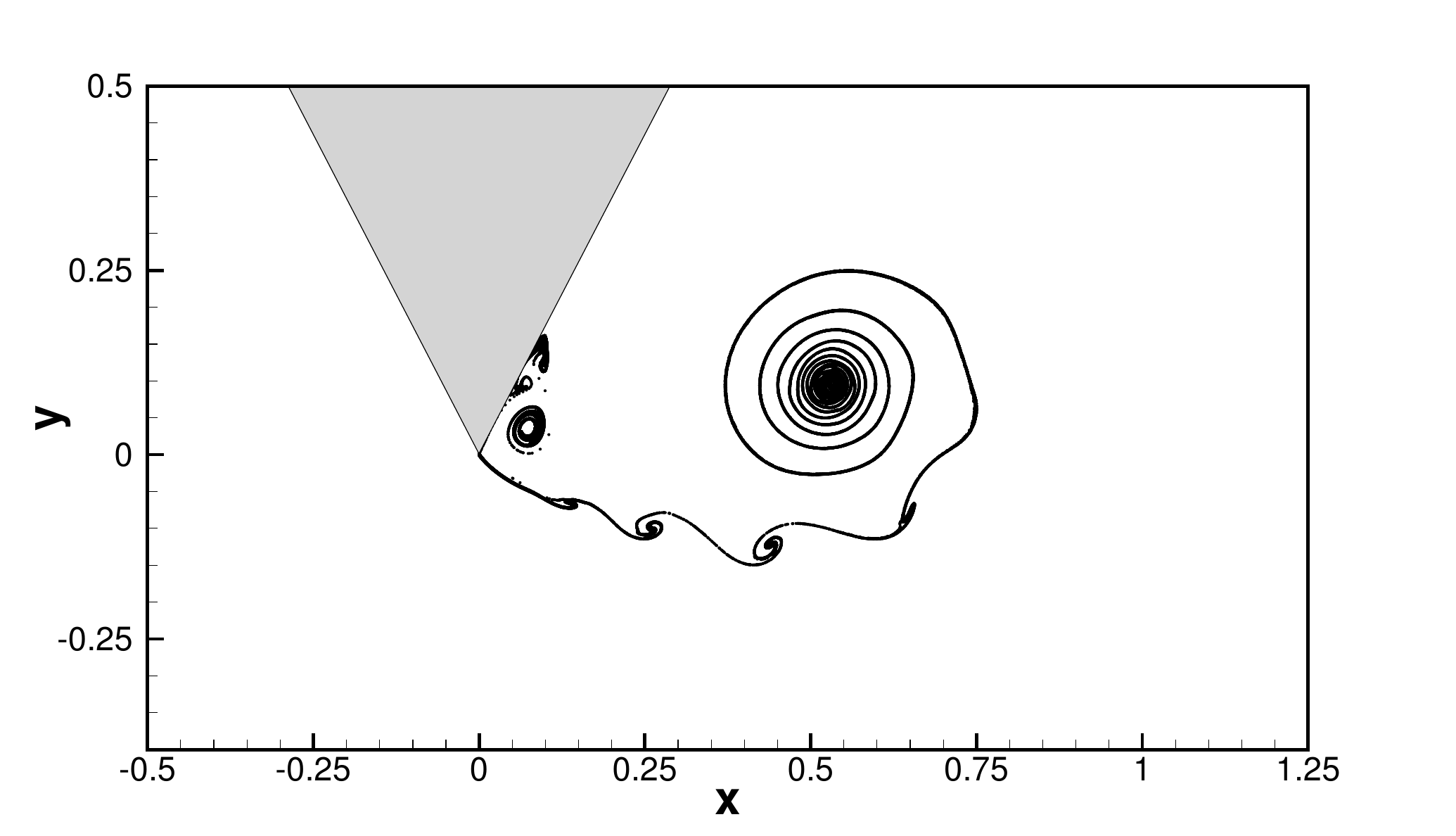}\\(b2)
		\endminipage\hfill
		\minipage{0.33\textwidth}%
		\centering
		\includegraphics[width=\linewidth]{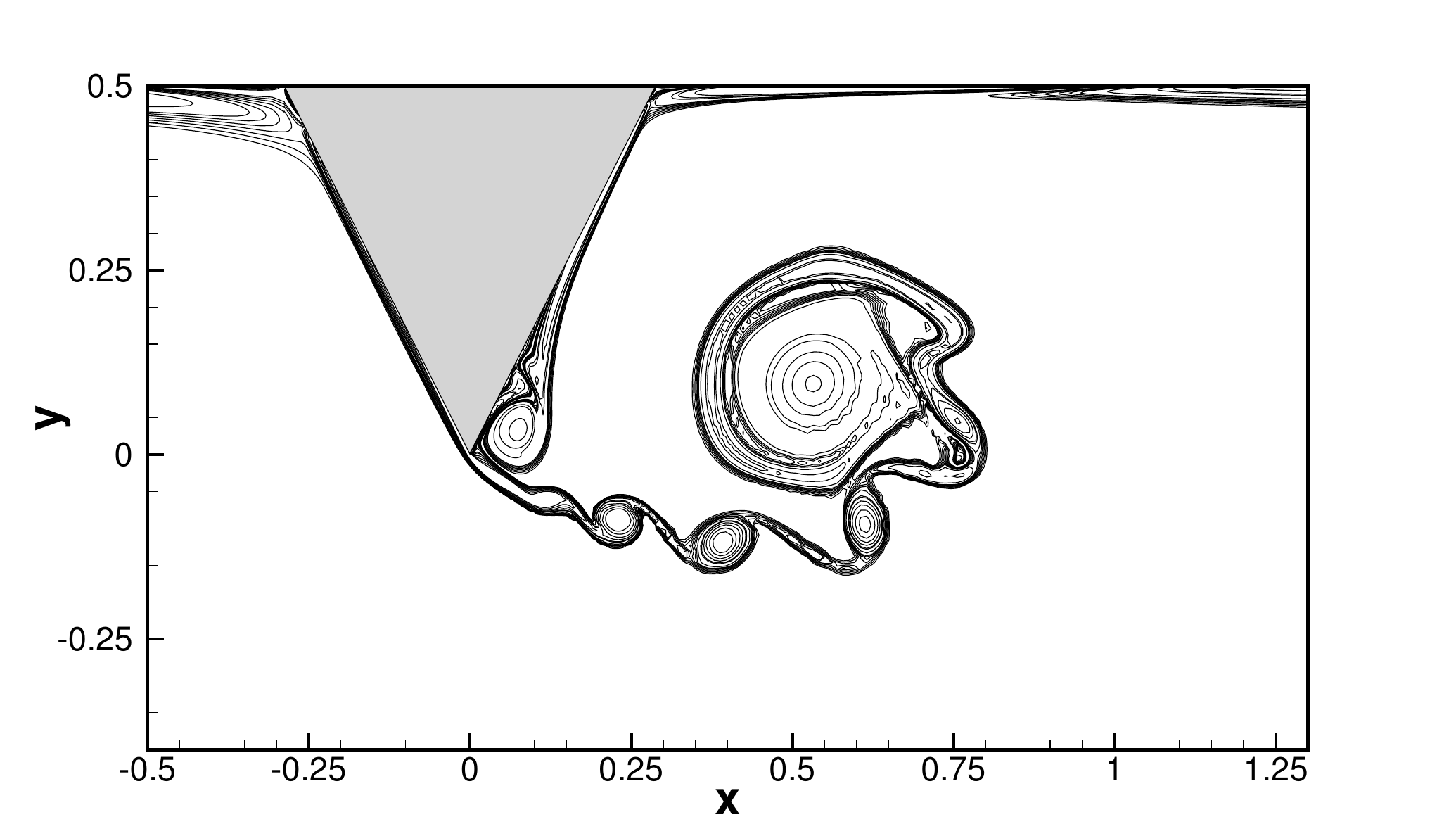}\\
		(c2)
		\endminipage\hfill
		\minipage{0.33\textwidth}
		\centering 
		\includegraphics[width=\linewidth]{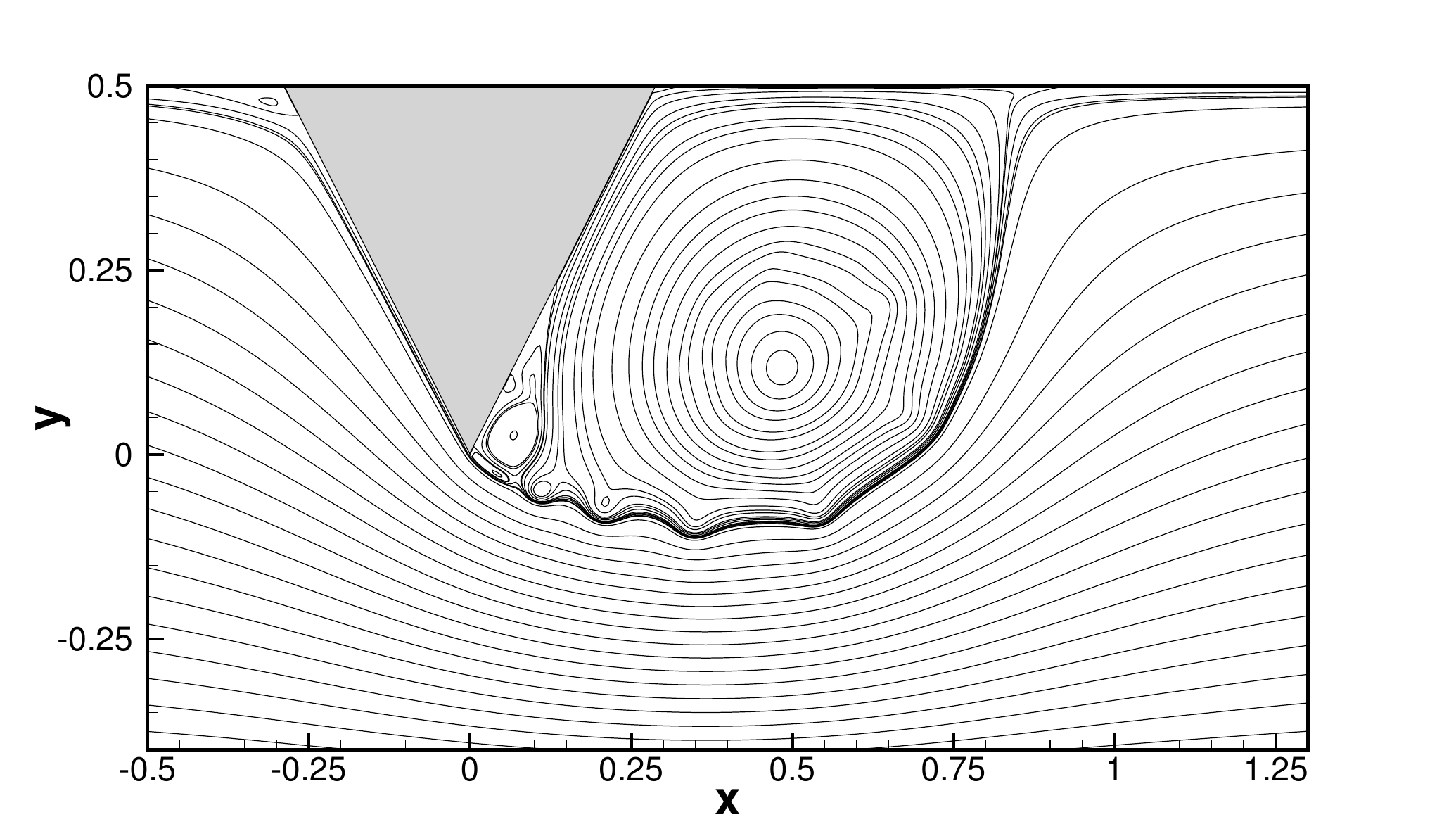}\\
		(a3)
		\endminipage\hfill
		\minipage{0.33\textwidth}
		\centering
		\includegraphics[width=\linewidth]{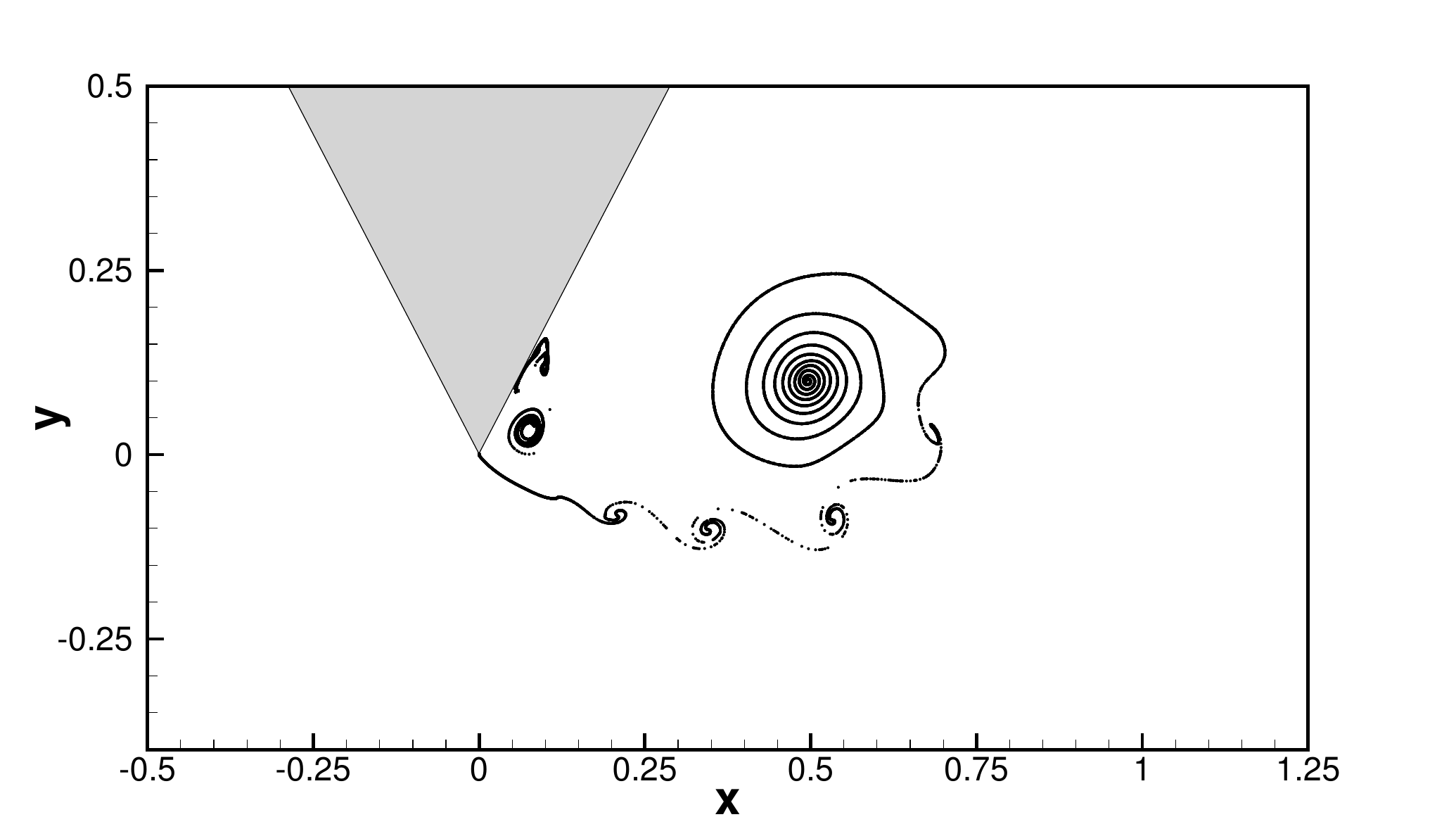}\\(b3)
		\endminipage\hfill
		\minipage{0.33\textwidth}%
		\centering
		\includegraphics[width=\linewidth]{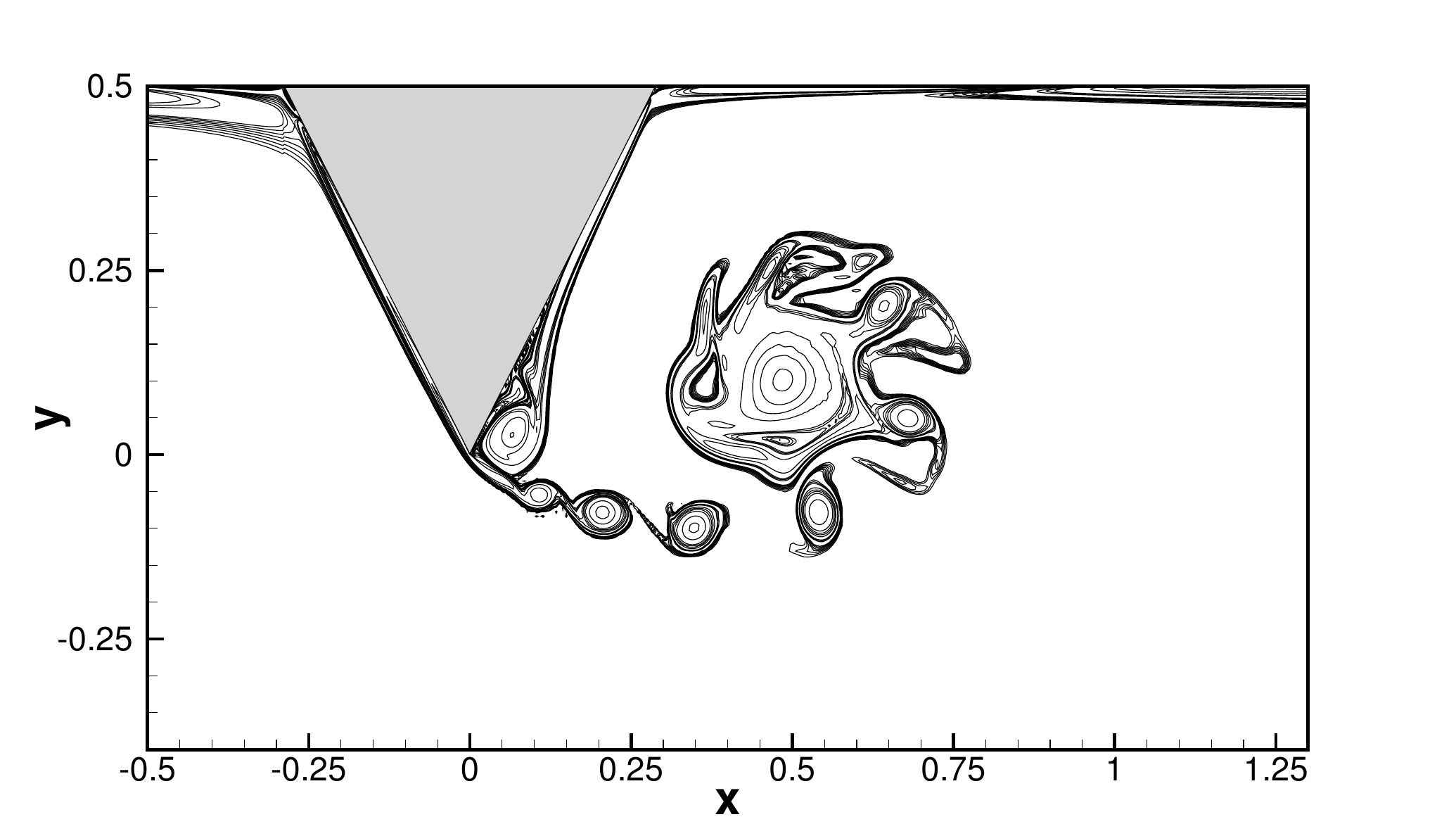}\\
		(c3)
		\endminipage\hfill
		\caption{{ Effect of $m$ on the flow field for a fixed displacement $d=1$: The rows from the top to bottom corresponds to $m=0$, $0.45$ and $0.88$ and the columns represent  (a1-a3) Streamfunction, (b1-b3) streaklines and (c1-c3) vorticity for $Re_c=1560$ (top row), $Re_c=6621$ (middle row) and $Re_c=6873$ (bottom row).  } }
		\label{wd_d1}
	\end{figure}

\begin{figure}
		\centering
		\begin{tabular}{ccc}
			\hspace{-1.2cm}\includegraphics[width=0.375\linewidth]{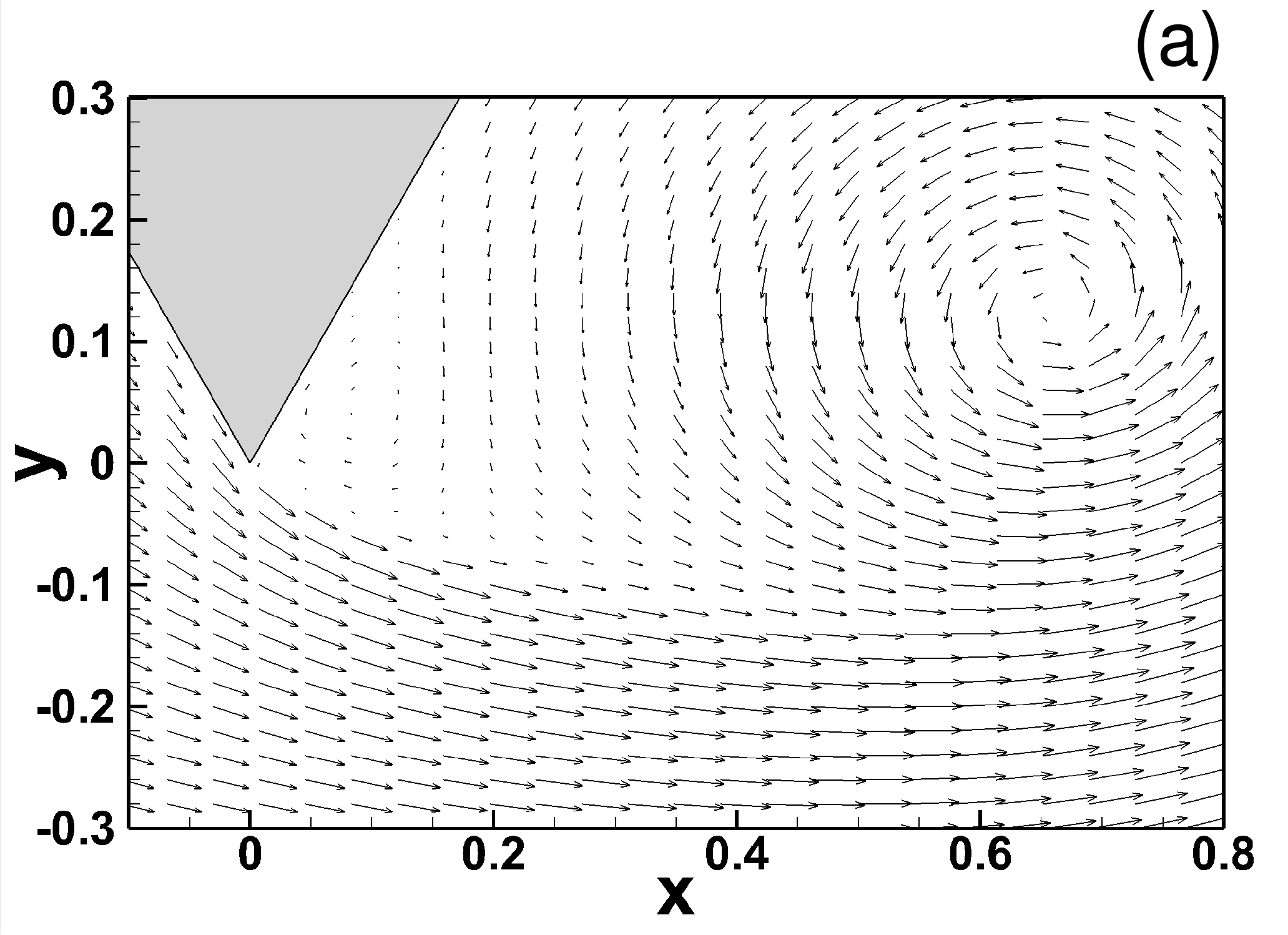}
			&
			\hspace{-0.4cm}\includegraphics[width=0.375\linewidth]{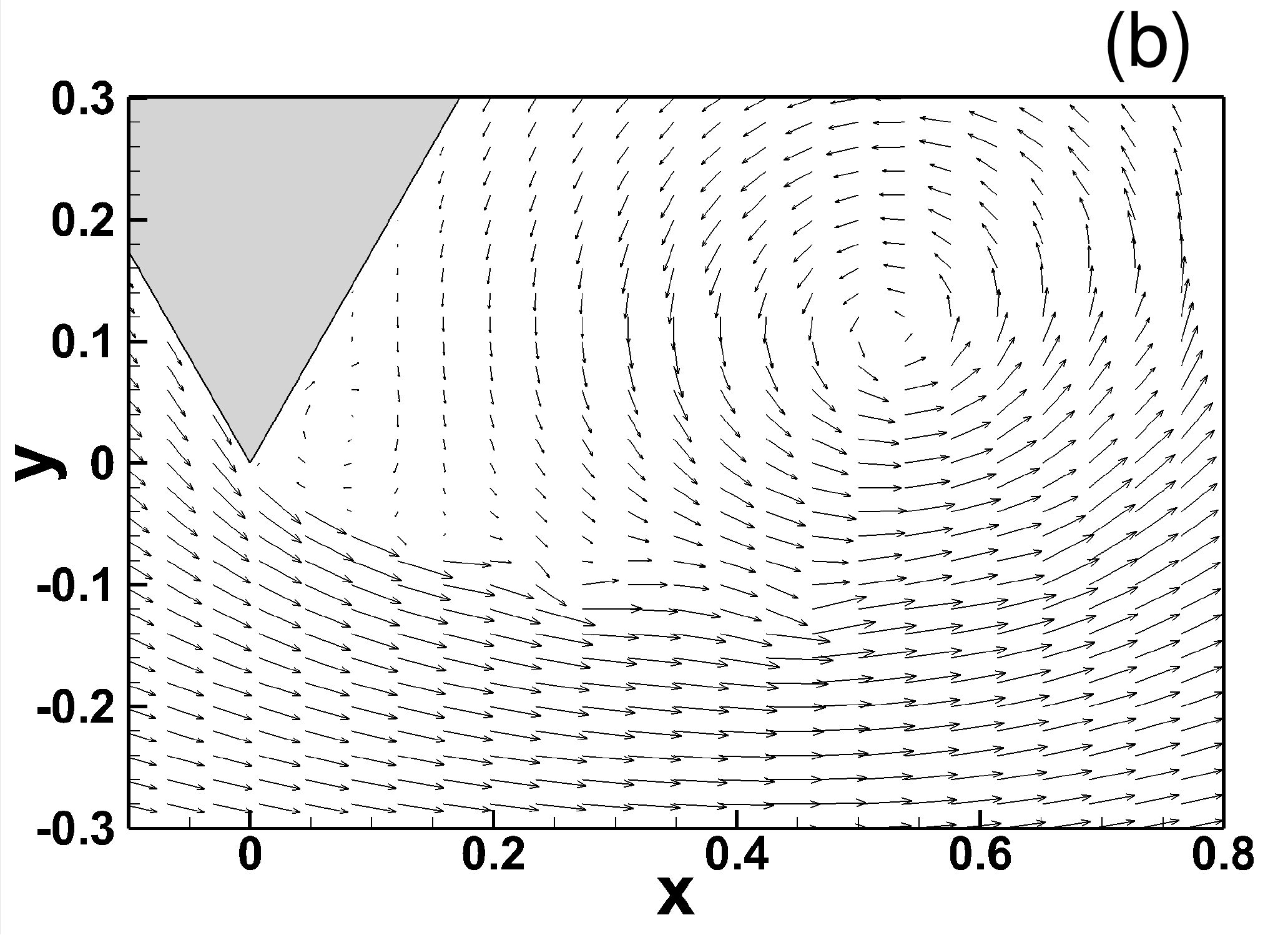}
			&
			\hspace{-0.4cm}\includegraphics[width=0.375\linewidth]{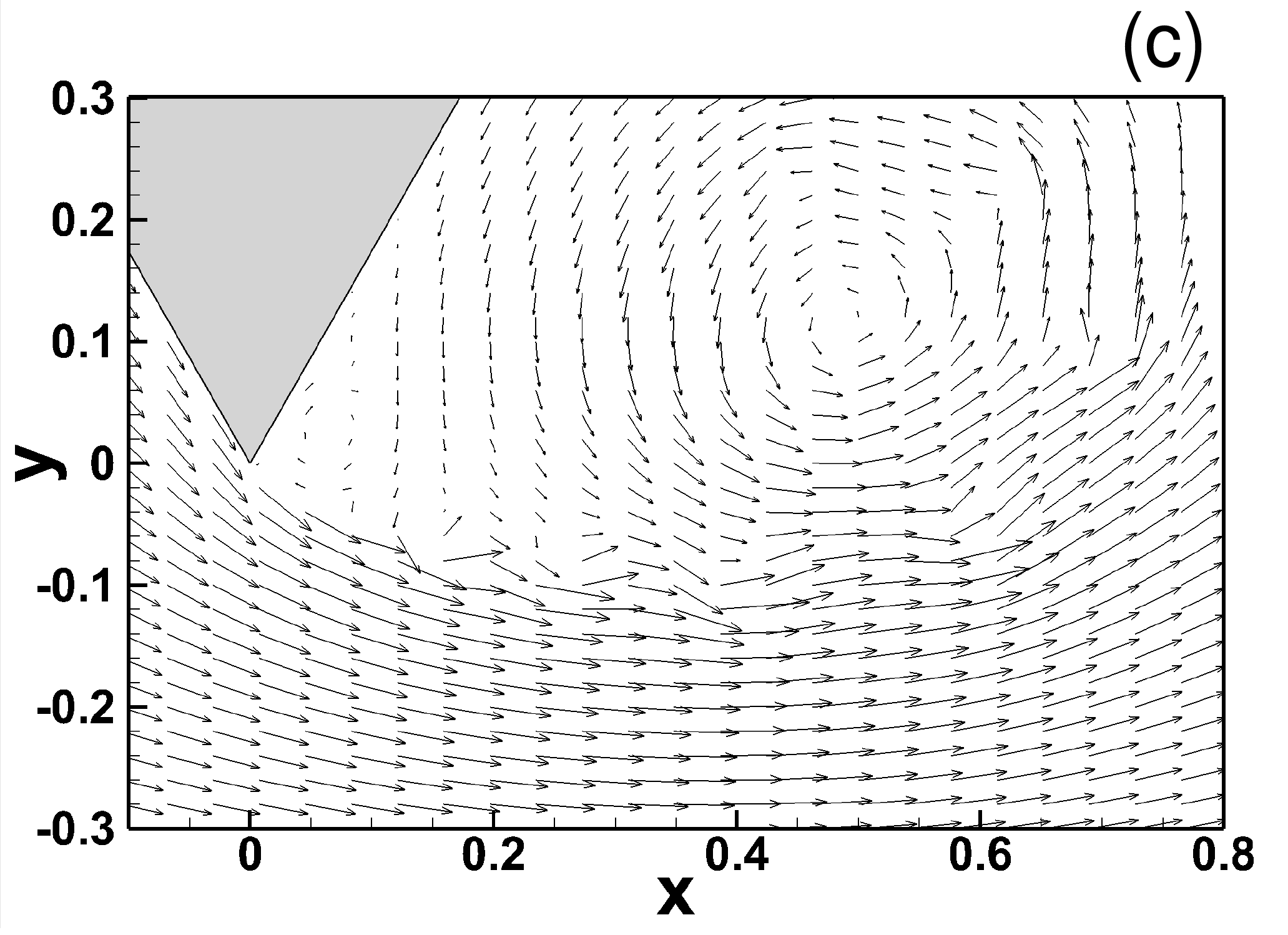}
		\end{tabular}	
		\caption{The vector field close to the wedge tip for $d=1$ and (a) $Re_c=1560$, (b) $Re_c=6621$ and (c) $Re_c=6873$. }\label{d1_vector}
	\end{figure}

\begin{figure}
		\begin{center}
			\includegraphics[width=4.0in]{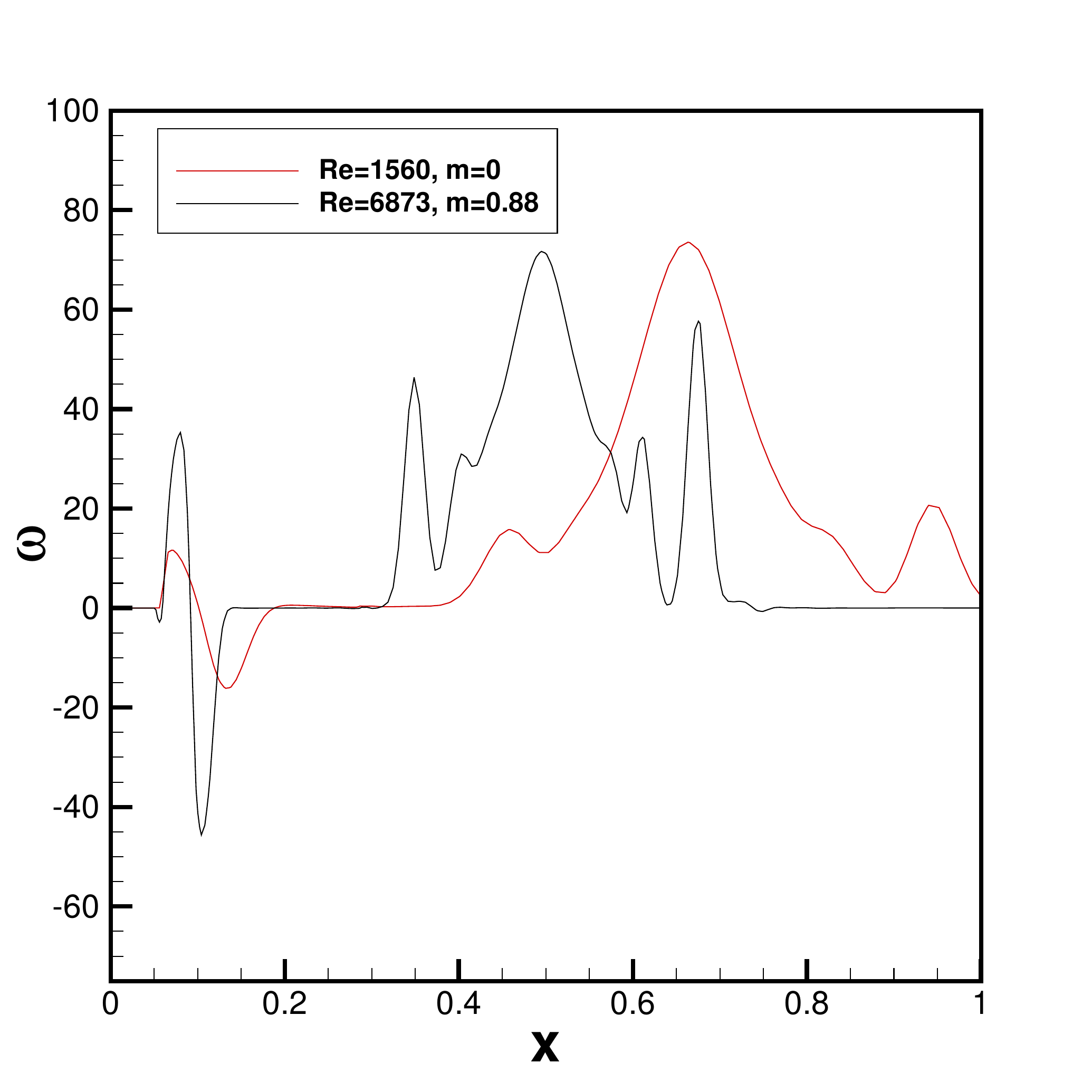}
			\caption{Vorticity distribution across the horizontal line through the core vortex center for $d=1$: (a) $Re_c=1560$ and (b) $Re_c=6873$.}
			\label{vort_dist}
		\end{center}
	\end{figure}

	Next, we look at the flow for a fixed displacement $d$ of the wedge for varying $m$'s, which provides a better understanding of the whole phenomena (\citet{xu2015}). In figure \ref{wd_d1}, we compare the vortcity contours, streaklines and the streamlines for a fixed displacement $d=1$ of the wedge for $m=0$, $0.45$ and $0.88$ (see table \ref{t2} also). The corresponding vector field in the vicinity of the tip of the wedge is shown in figure \ref{d1_vector}. One can again observe gradual decrease in the size of the vortices for the early part of the flow from figures \ref{wd_d1}(a1)-(a3) with  increase in $m$. The development of shear layer is more prominent as $m$ increases, which is also obvious from the velocity vector plots in figures  \ref{d1_vector}(a)-(c). The density of the streamlines below the bounding streamline of the core vortex indicates a stronger velocity field with increase in $m$, which is also reflected by the length of the vectors. The streaklines and vorticity contours in figures \ref{wd_d1}(b2)-(b3) and \ref{wd_d1}(c2)-(c3) clearly demonstrate that for the accelerated flow, shear layer instability has set in, which is indicated by the waviness of the outermost vortex layer of the starting vortex (see figures \ref{eff_m1}(b)-(c) and \ref{d1_vector}(b)-(c) also). This will be discussed in more details in the next section. In figure \ref{vort_dist}, we show the vorticity profile as a function of zero and non-zero values of $m$ which corresponds to the wedge experiencing a uniform and accelerated flow respectively. Here we plot the vorticity values corresponding to $m=0$ and $0.88$, and $d=1$ along the horizontal line passing through the core vortex center. One can clearly see larger maximal vorticity occurring for smaller value of $x$ for non-zero $m$, reflecting the smaller shape of the vortices for the accelerated flow. The shear layer instability for the accelerated case is clearly indicated by the highly oscillating nature of the graph against a much well-behaved one for the uniform flow case.
	
	\begin{table}
		\caption{Effect of non-dimensionalization on the flow field.}
		\begin{center}
			\begin{tabular}{|c|c|ccc|}\hline
				&m &$0$ &$0.45$ &$0.88$ \\ \hline
				t=0.8 &$x_R$ &$0.549$ &$0.266$ &$0.180$ \\
				&$\tilde{t}$ (in sec) &$32.25$ &$8.26$ &$7.48$  \\ \hline
				t=2.0 &$x_R$ &$0.890$ &$0.866$ &$0.865$ \\
				&$\tilde{t}$ (in sec) &$80.64$ &$20.66$ &$18.70$  \\ \hline
				d=1.0 &t &$1.0$ &$1.29$ &$1.40$ \\
				&$\tilde{t}$ (in sec) &$40.32$ &$13.34$ &$13.08$  \\ \hline
			\end{tabular}
		\end{center}
		\label{t2}
	\end{table}

	It must be mentioned that the flow evolution in terms of the non-dimensional variables may portray a slightly different scenario than the real time flow field evolution. This is because of the fact that except displacement, for each of the other flow variables, a non-linear scaling (see equation \eqref{nondim}) has been used.  Therefore, though figure \ref{eff_m1} reveals the proximity of the horizontal distances traversed by the vortex centers for a fixed non-dimensional time $t=2.0$, in reality, when the flow is devoid of any acceleration, the time taken for the same is approximately four times than when it is subjected to acceleration. This fact is also revealed by table \ref{t2}, where we have tabulated the horizontal coordinate $x_R$ of the rotation center at fixed non-dimensional times $t=0.8$ and $t=2.0$ along with their corresponding physical times. Also tabulated here are the dimensional and non-dimensional times against a fixed distance $d=1$ (or a physical distance $\tilde{d}=25.4cm$, see figure \ref{wd_d1} also), which reveals a similar trend. It further reasserts that acceleration in the flow triggers shear layer instability early in the flow (around real time $13\; secs$), while no trace of such instability is seen for uniform flow even after $80\; secs$.
	
\begin{figure}
		\centering
		\begin{tabular}{ccc}
			\hspace{-1.2cm}\includegraphics[width=0.4\linewidth]{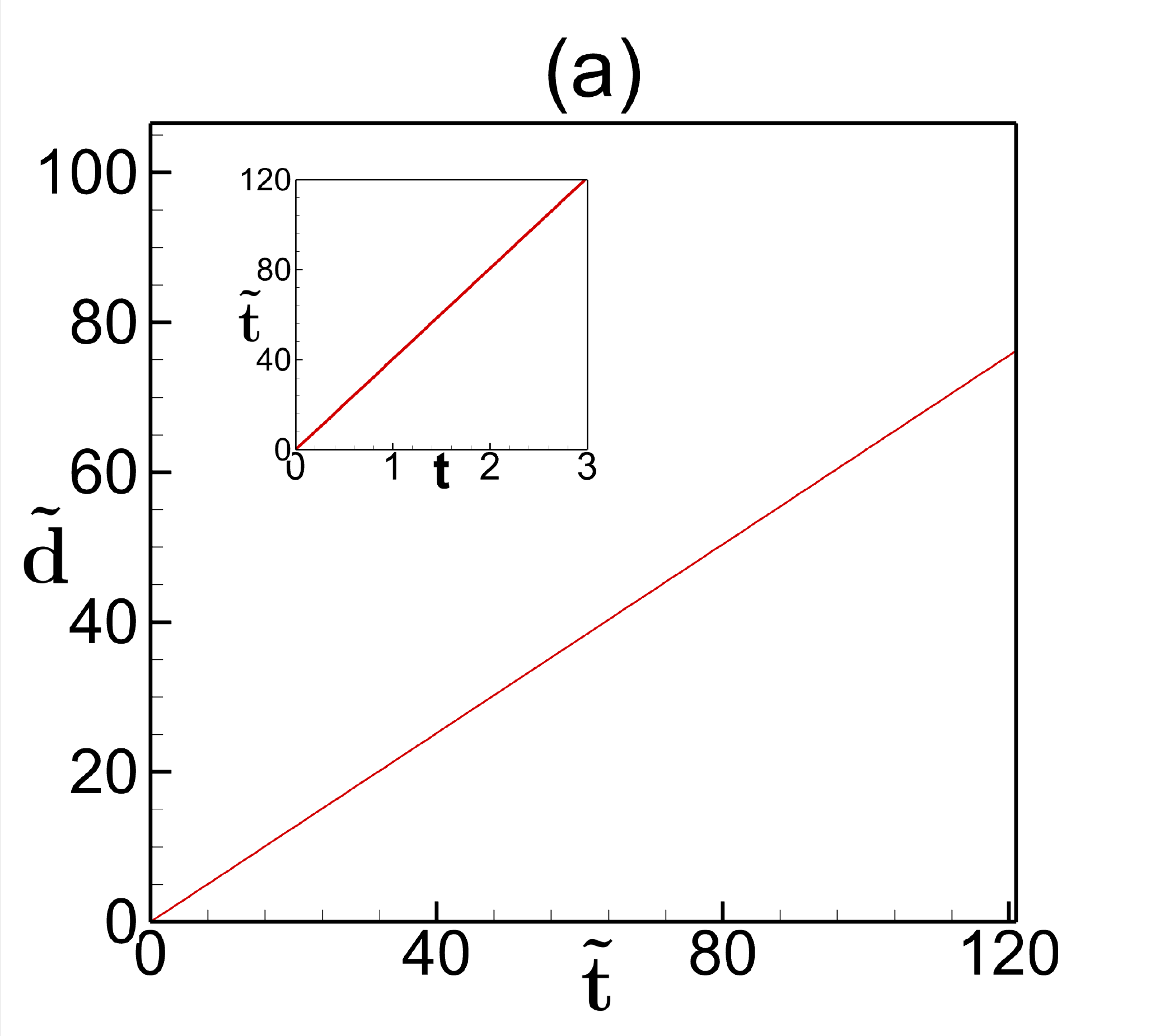}
			\&
			\hspace{-0.8cm}\includegraphics[width=0.4\linewidth]{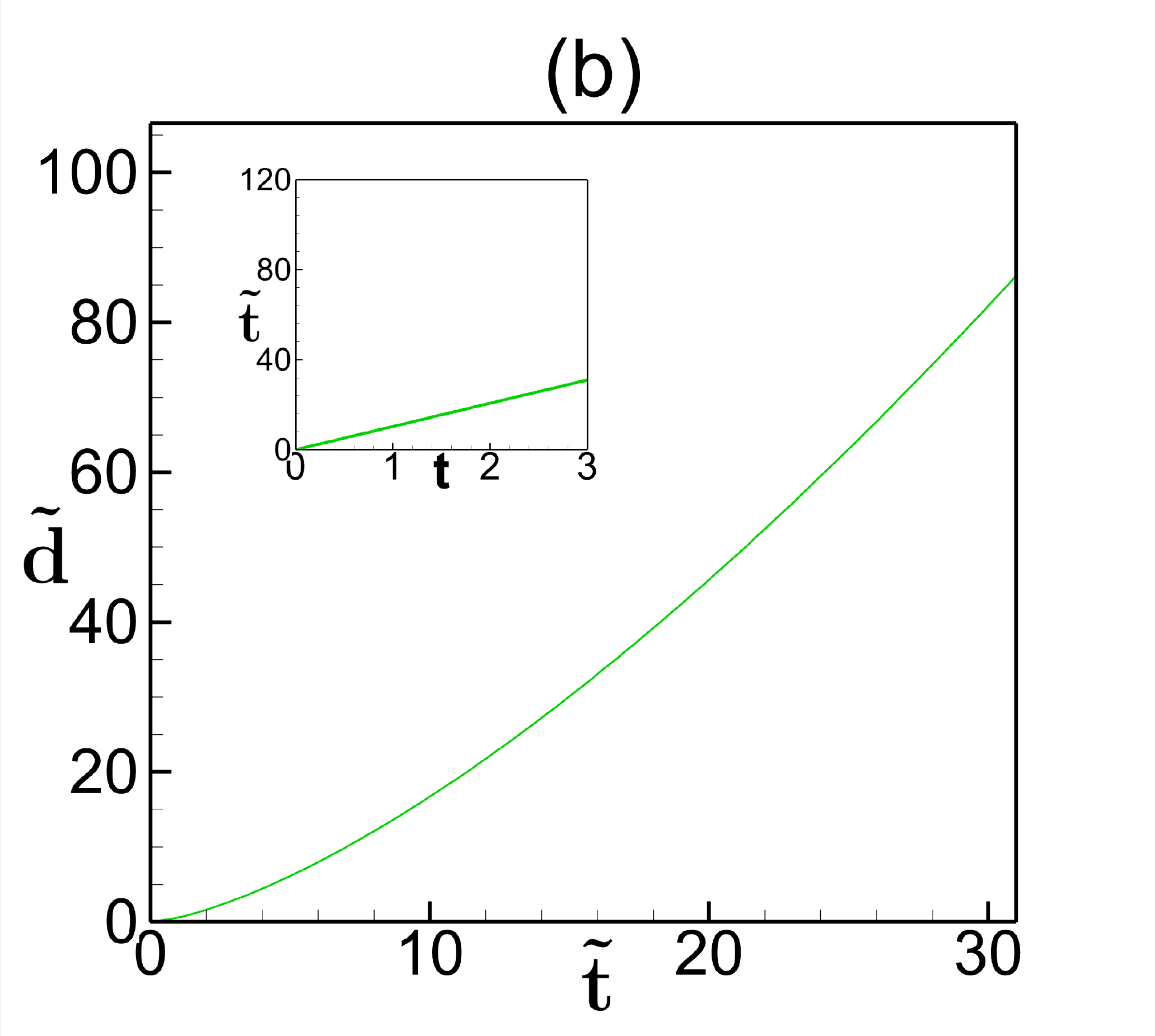}
			&
			\hspace{-0.8cm}\includegraphics[width=0.4\linewidth]{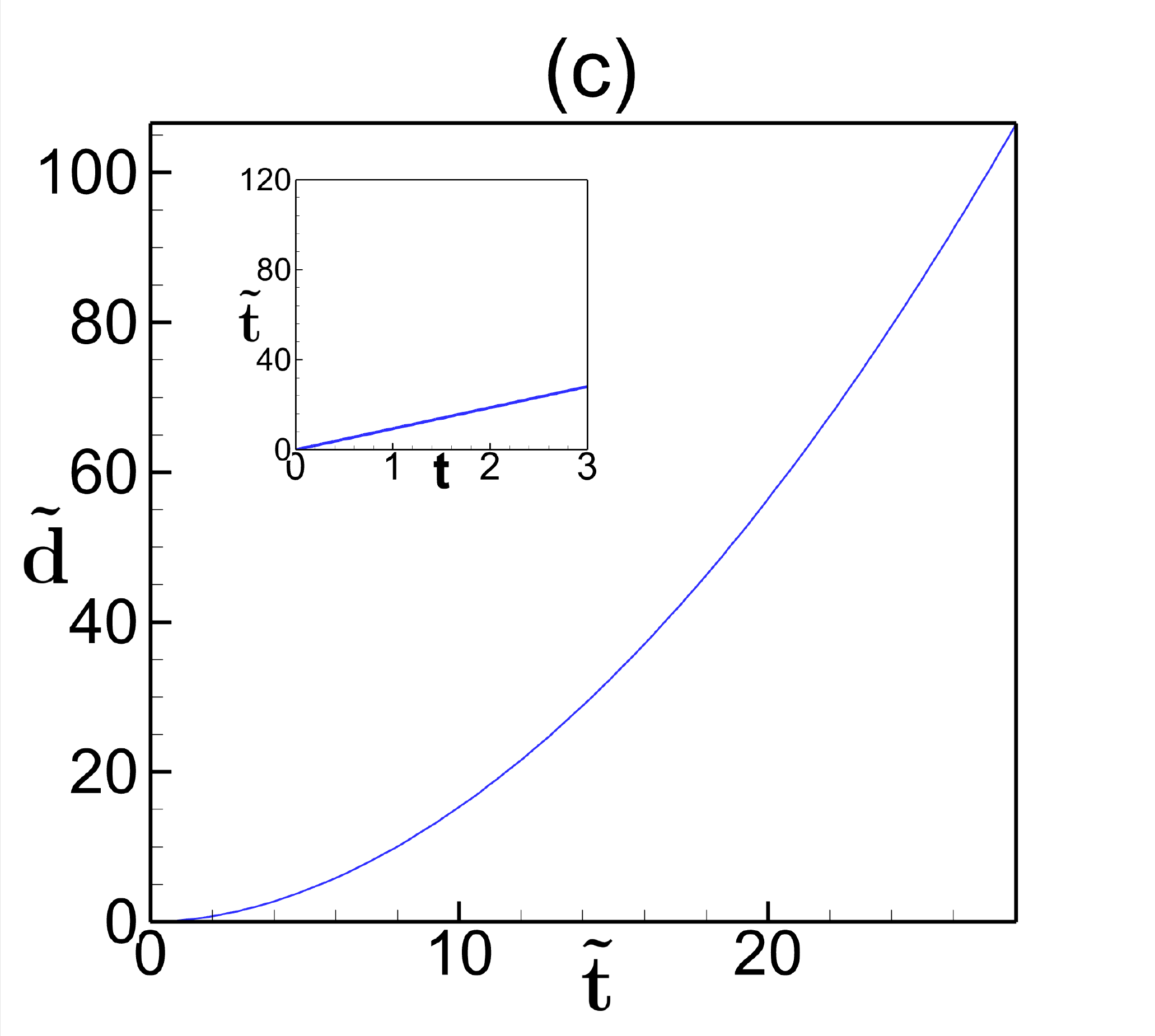}
		\end{tabular}
		\caption{Effect of non-dimensionalization on flow visualization, actual distance travelled in real physical time, with non-dimensional time versus the real time at the inset for: (a) $Re_c=1560\;(m=0)$, (b) $Re_c=6621\;(m=0.45)$ and  (c) $Re_c=6873\;(m=0.88)$. The units of $\tilde{t}$ and $\tilde{d}$ are in $sec$ and $cm$ respectively.}
		\label{nd_effect}
	\end{figure}

	The effect of non-dimensionalization on the flow could also be understood from the formula for distance $d$ traversed by the wedge calculated in terms of the non-dimensional time $t$ as in equation \eqref{nd_dt} above, which will translate into $\displaystyle \tilde{d}=A\frac{\tilde{t}^{m+1}}{m+1} cm$ in real physical time. This is actually the distance traversed by the piston in the laboratory experiment of \cite{pullin1980}. For the benefit of the readers, we also plot these actual distances (in cm) the piston would have travelled in real physical time (in sec), with non-dimensional time versus the real time at the inset corresponding to these $m$ values in figures \ref{nd_effect}(a)-(c). These figures, along with the backdrop of the non-dimensional distances plotted in figure \ref{eff_m_ut}(b) clearly reveal the contrasts between the actual distance travelled against the non-dimensional ones. 

	\subsection{Numerical simulations vis a vis Experimental visualization: the earliest stage of flow evolution}
	\begin{figure}
		\begin{tabular}{cccc}
			\hspace{-1.0cm}\includegraphics[width=0.3\linewidth]{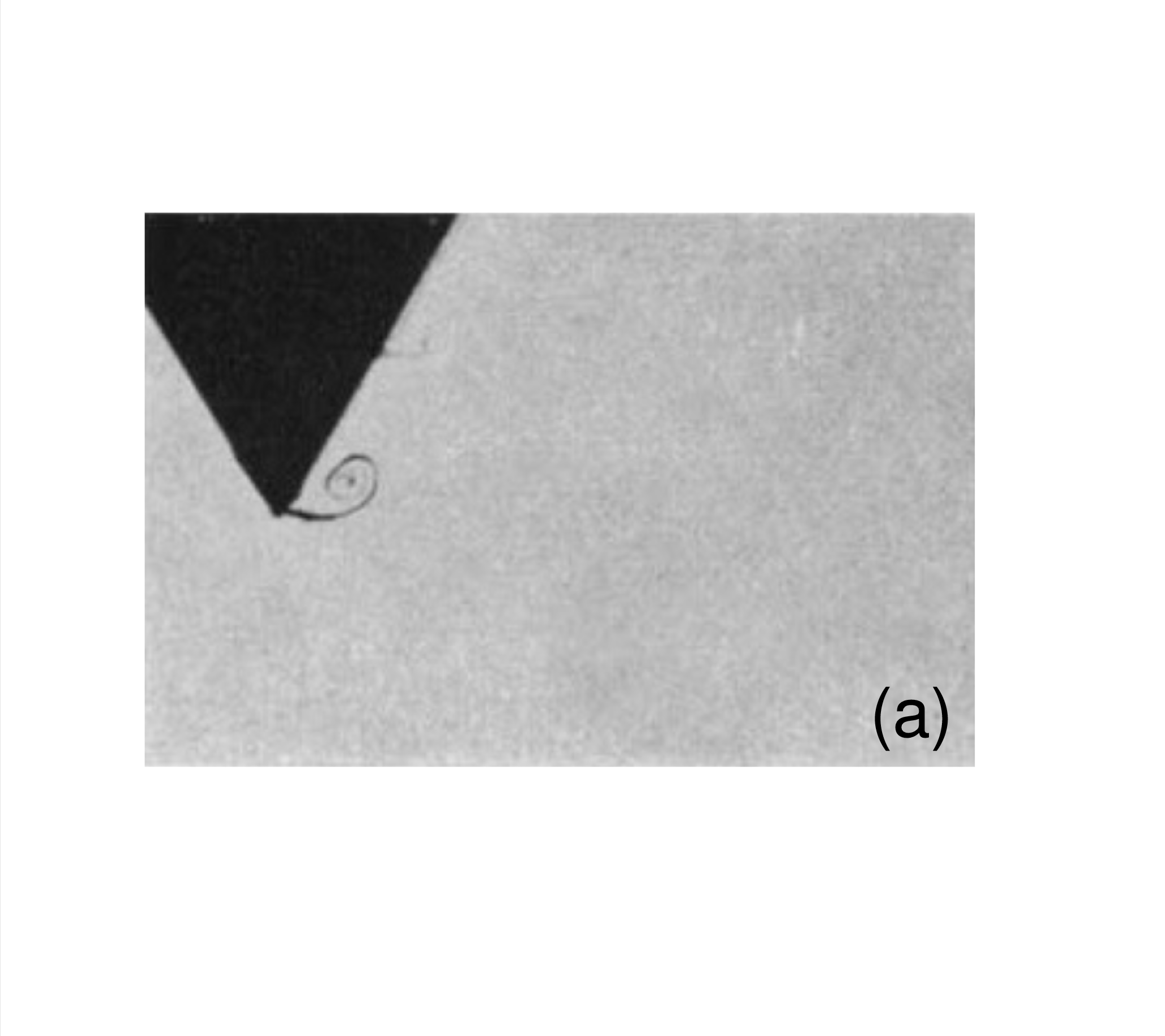}
			&
			\hspace{-0.5cm}\includegraphics[width=0.3\linewidth]{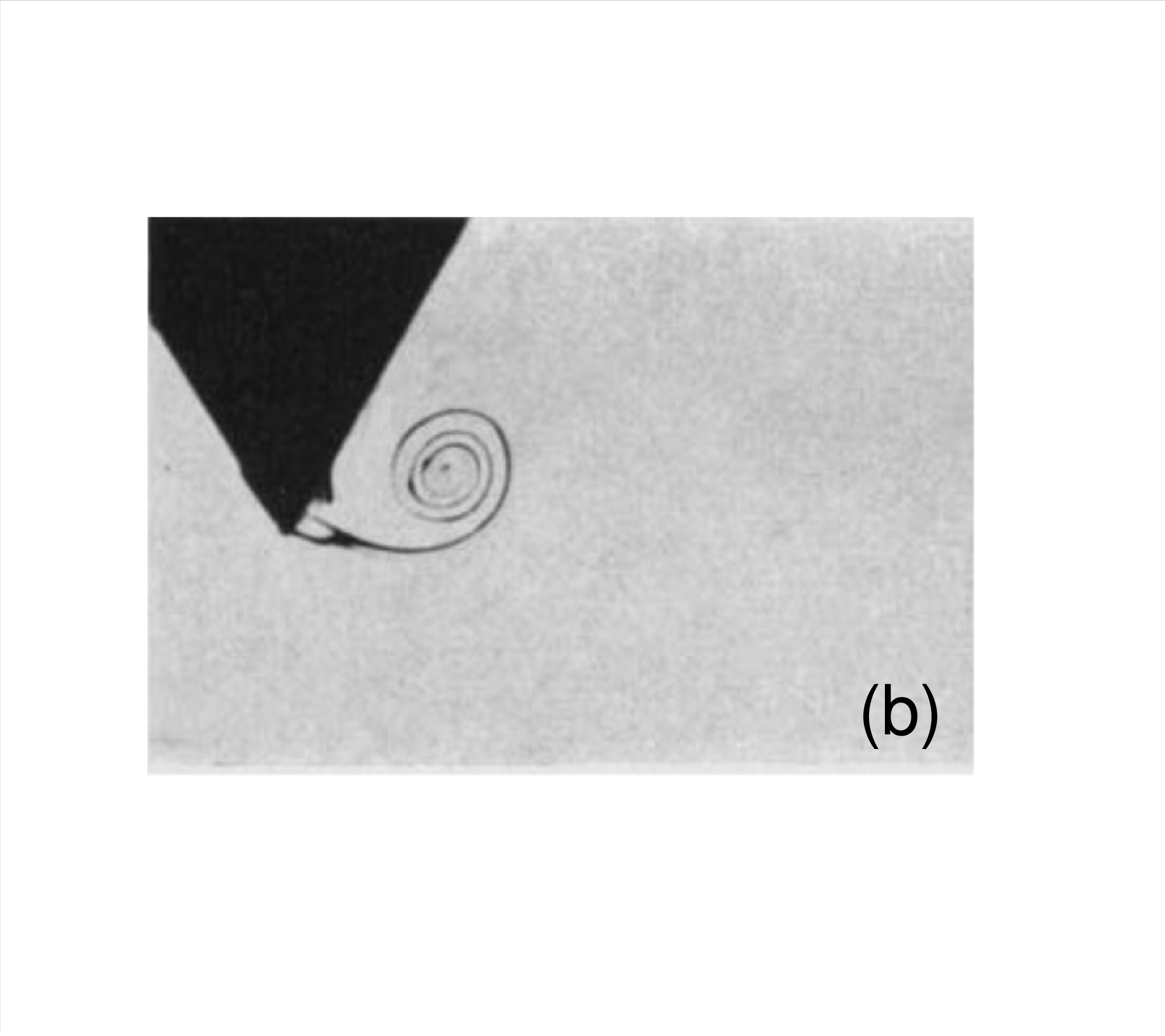}
			\&
			\hspace{-0.5cm}\includegraphics[width=0.3\linewidth]{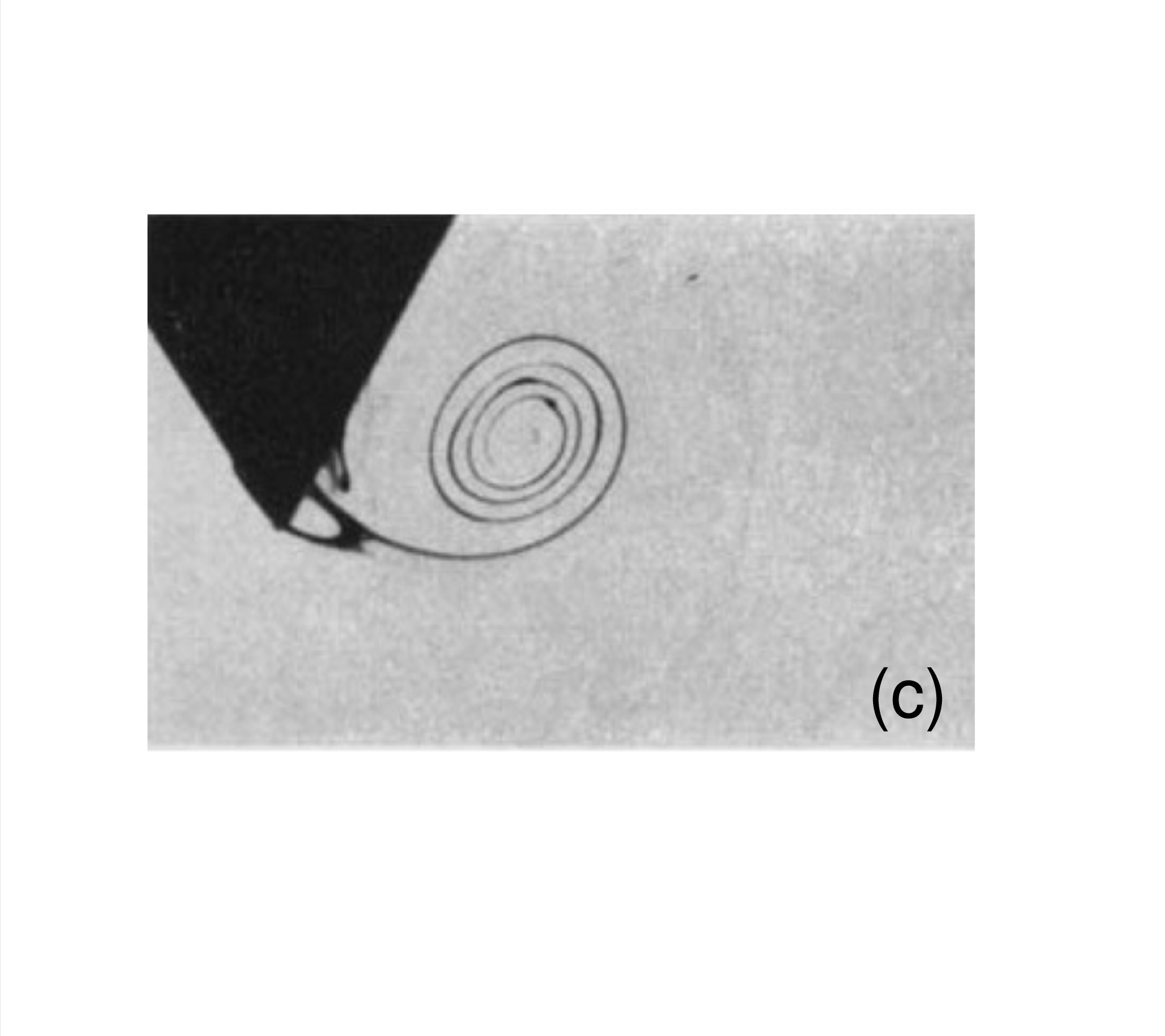}
			&
			\hspace{-0.5cm}\includegraphics[width=0.3\linewidth]{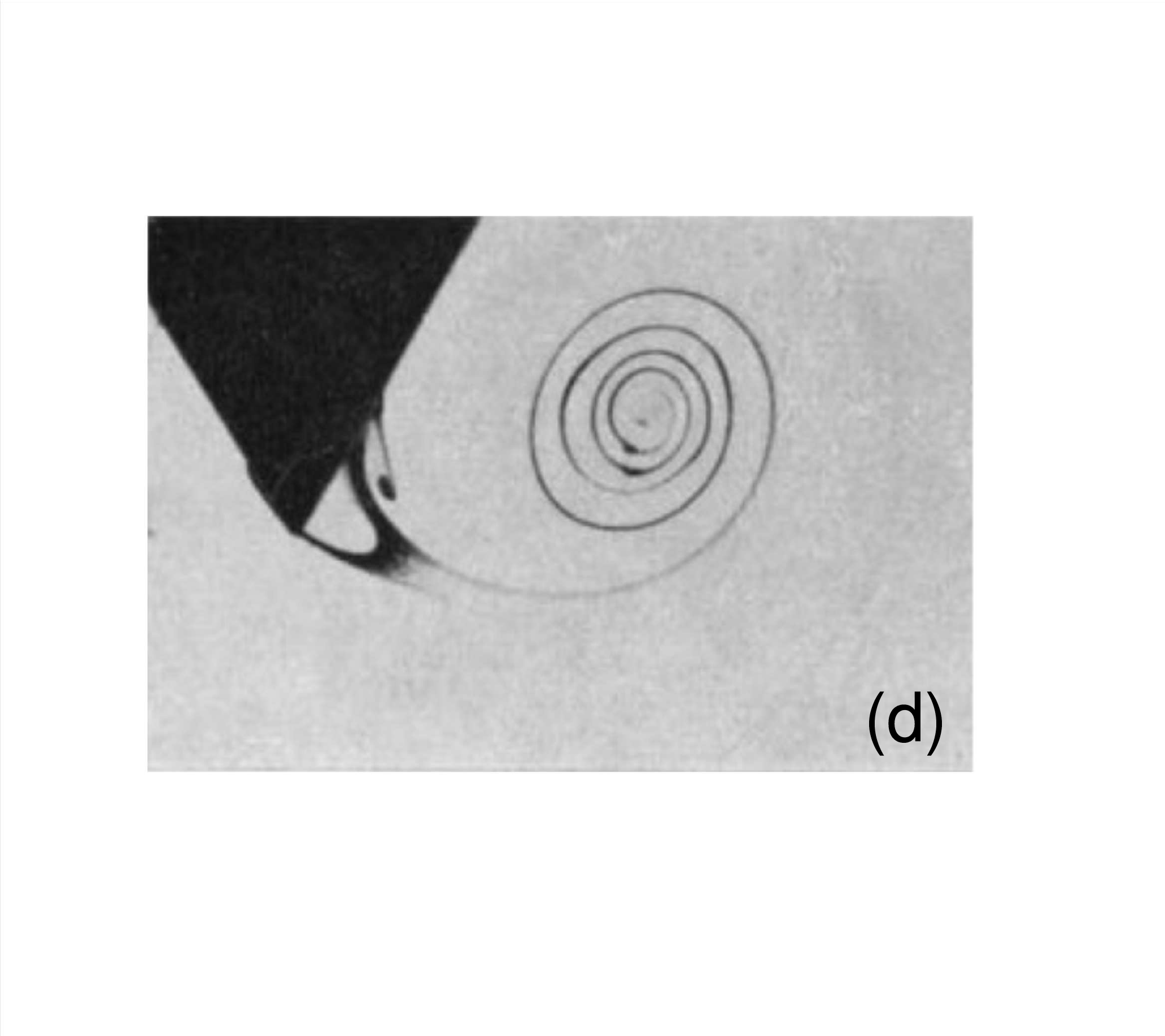}
		\end{tabular}
		\centering
		\begin{tabular}{cccc}
			\hspace{-1.0cm}\includegraphics[width=0.3\linewidth]{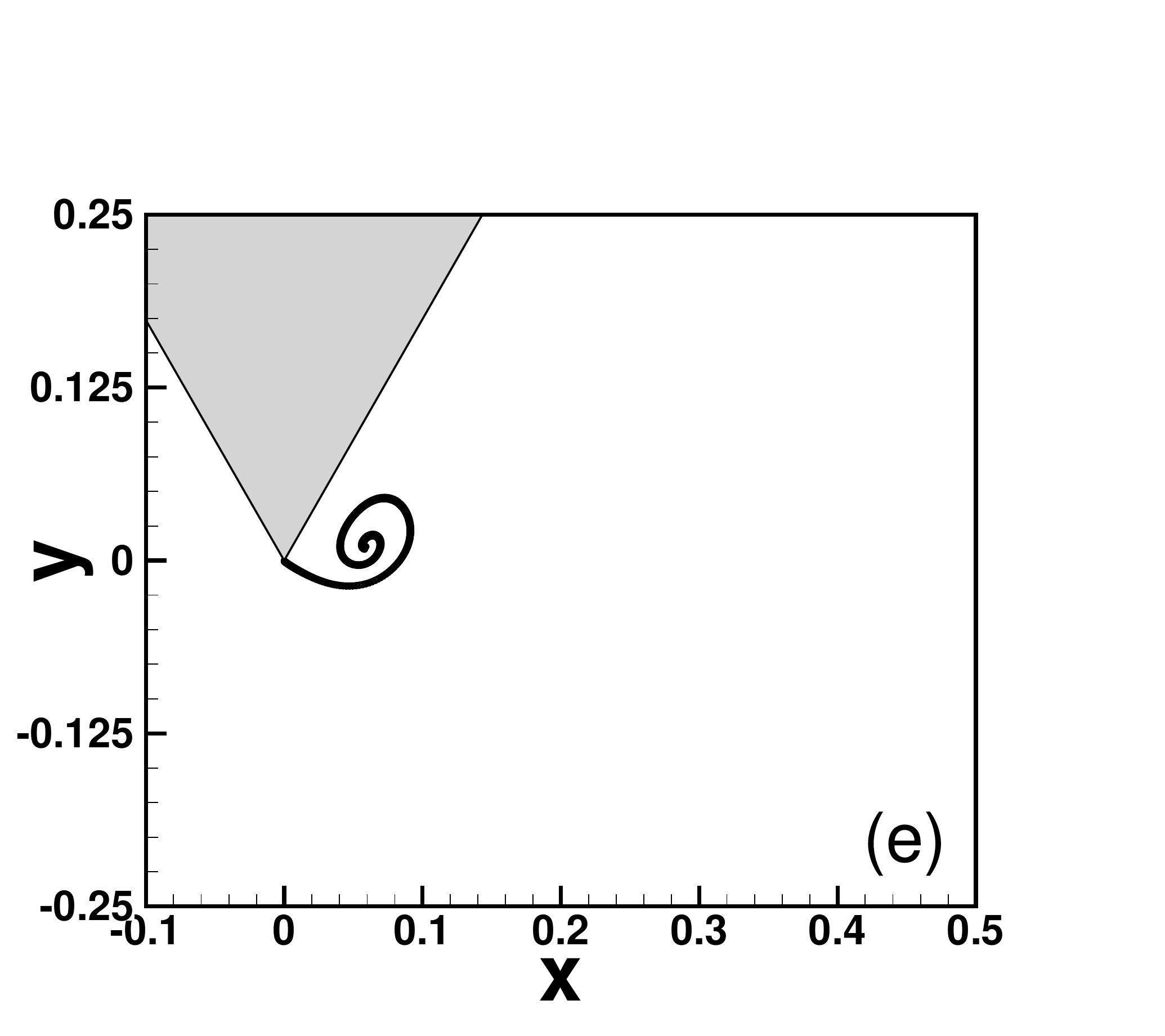}
			&
			\hspace{-0.5cm}\includegraphics[width=0.3\linewidth]{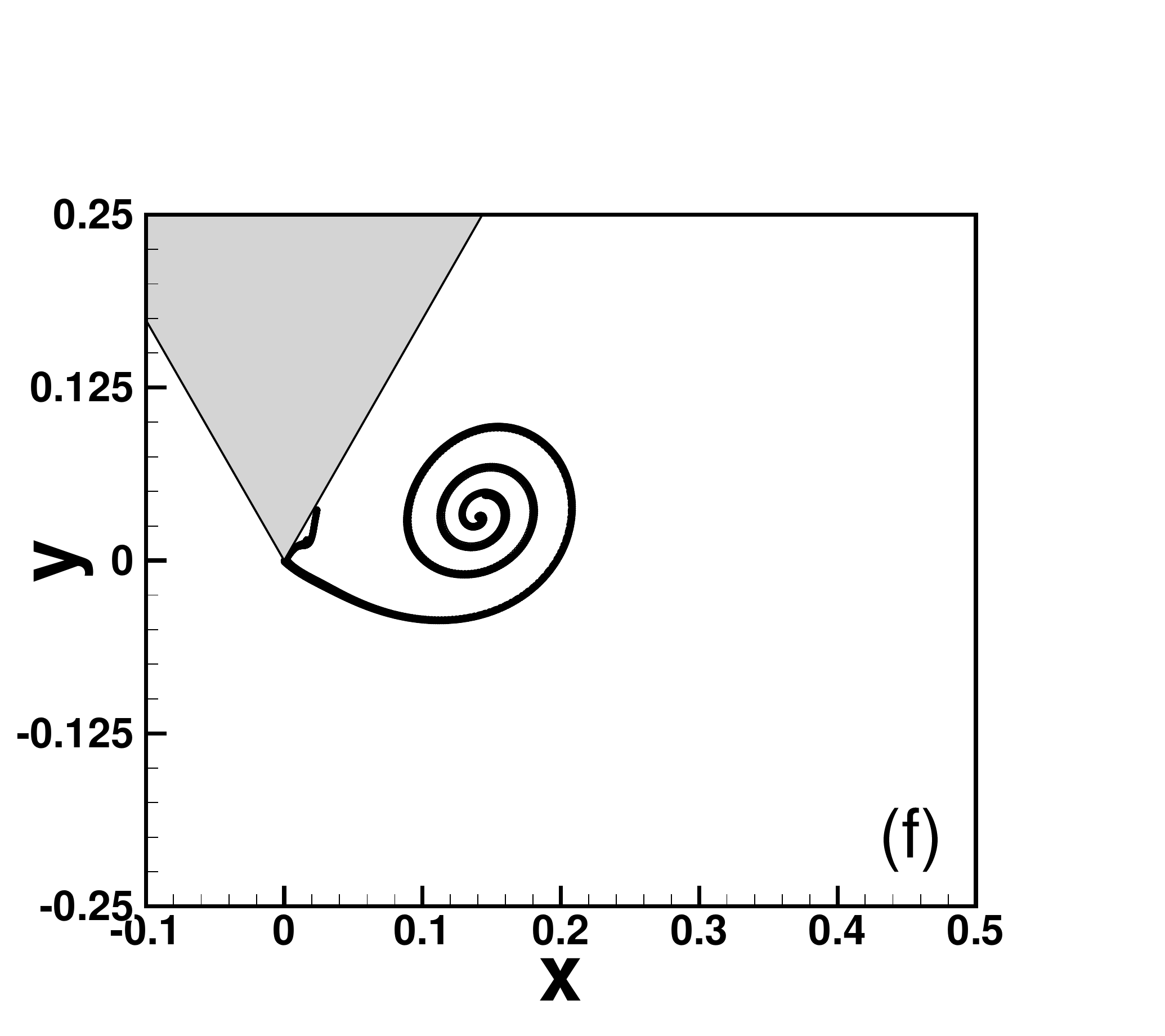}
			\&
			\hspace{-0.5cm}\includegraphics[width=0.3\linewidth]{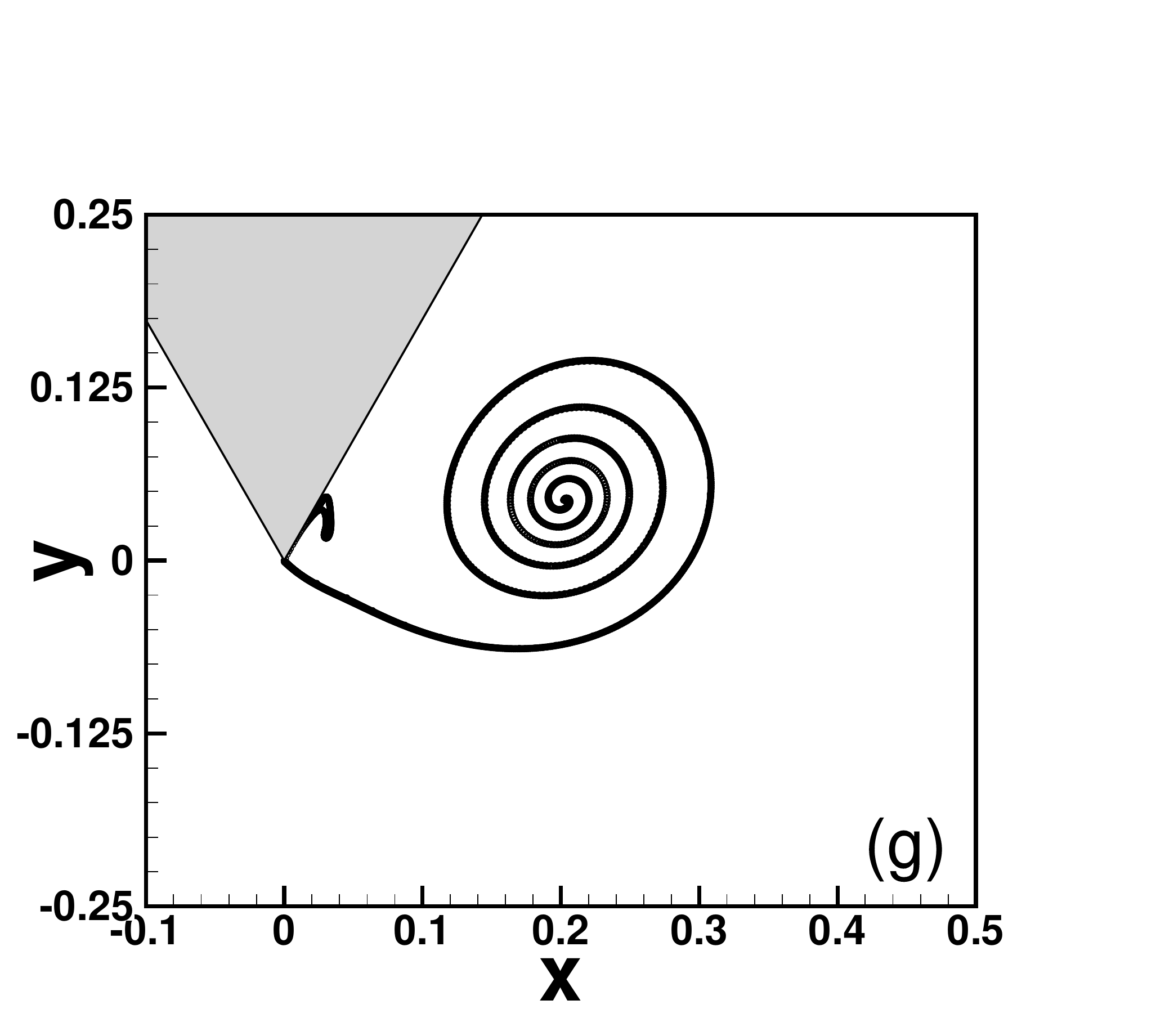}
			&
			\hspace{-0.5cm}\includegraphics[width=0.3\linewidth]{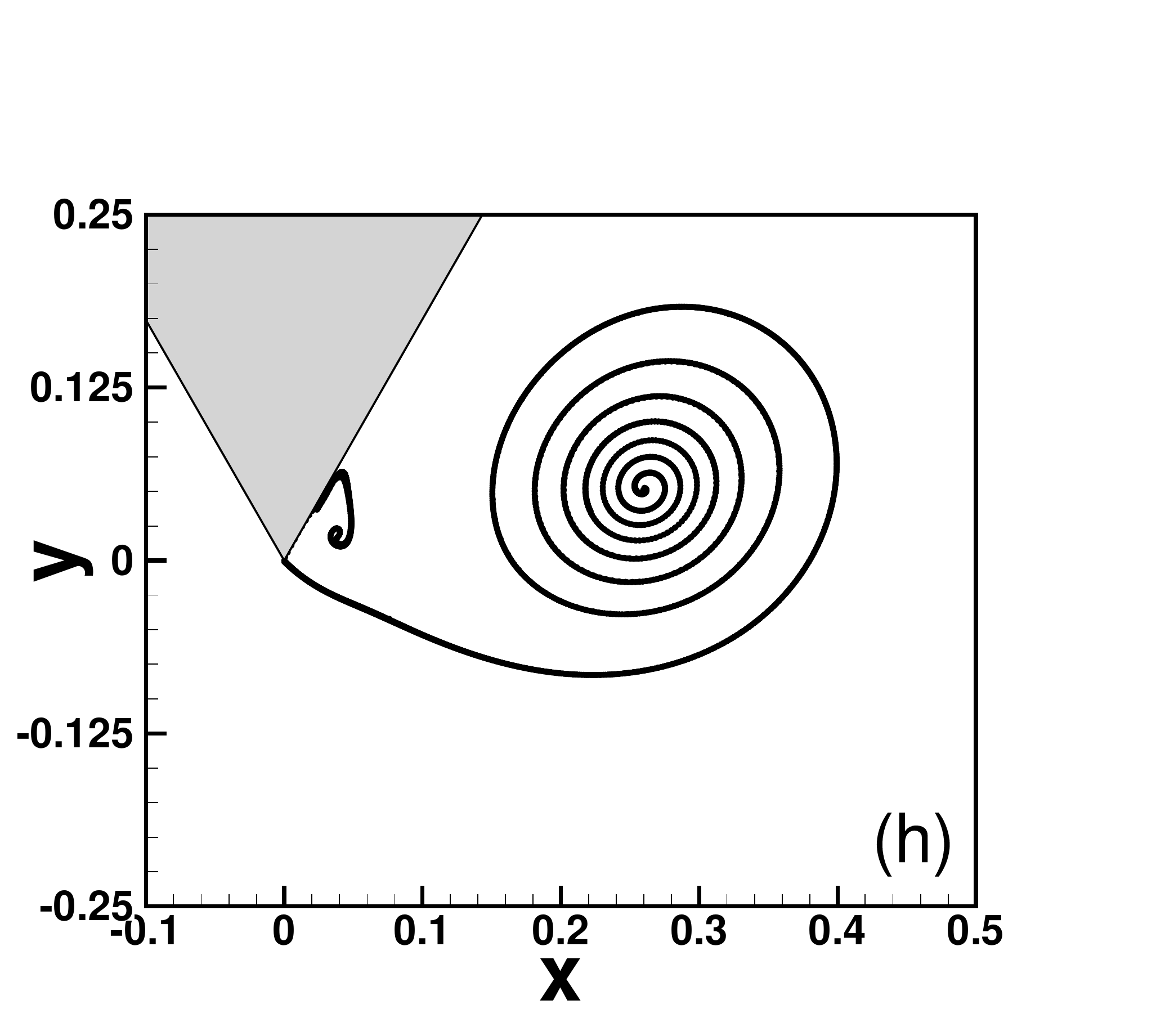}
		\end{tabular}
		\begin{tabular}{cccc}
			\hspace{-1.0cm}\includegraphics[width=0.3\linewidth]{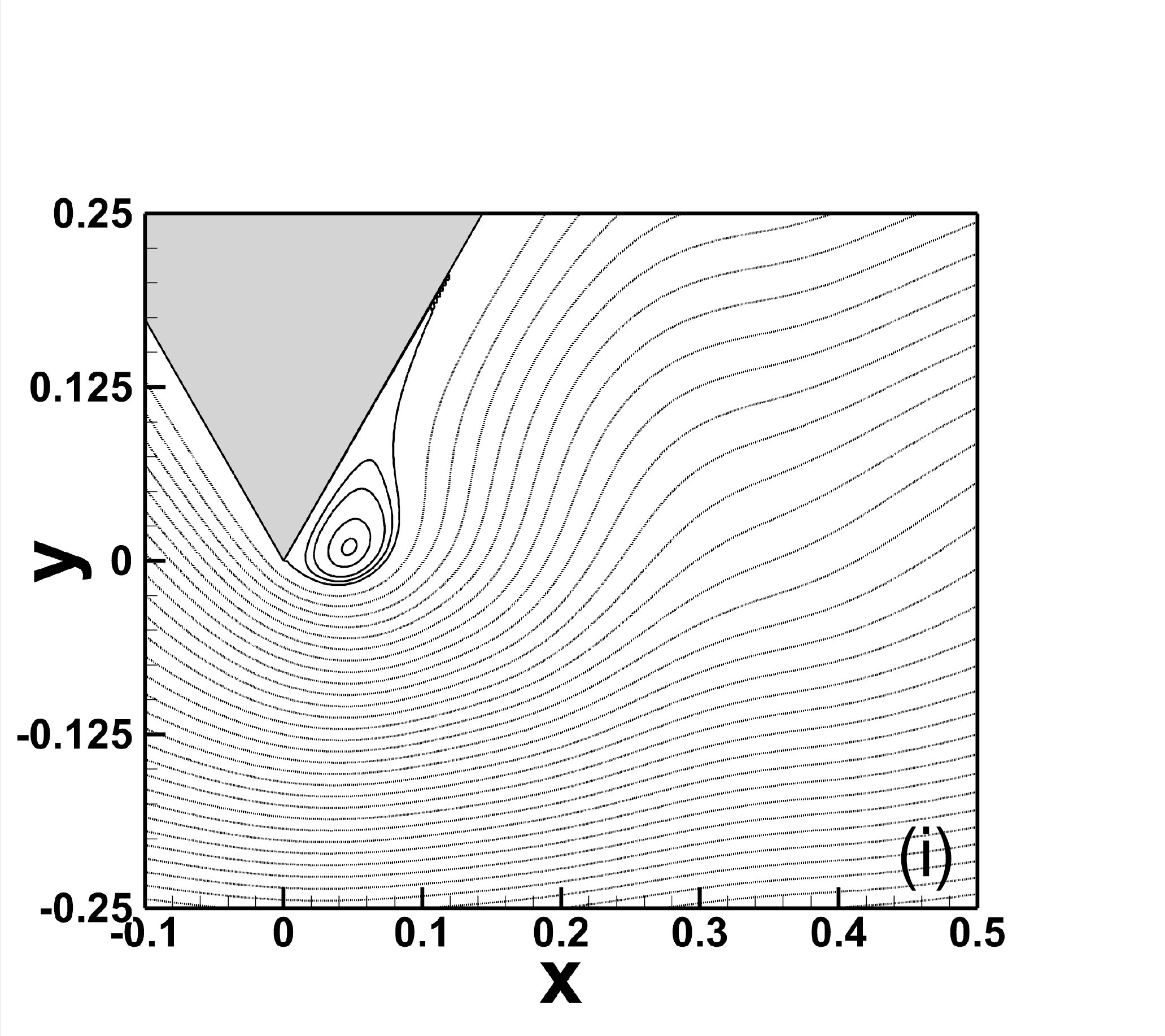}
			&
			\hspace{-0.5cm}\includegraphics[width=0.3\linewidth]{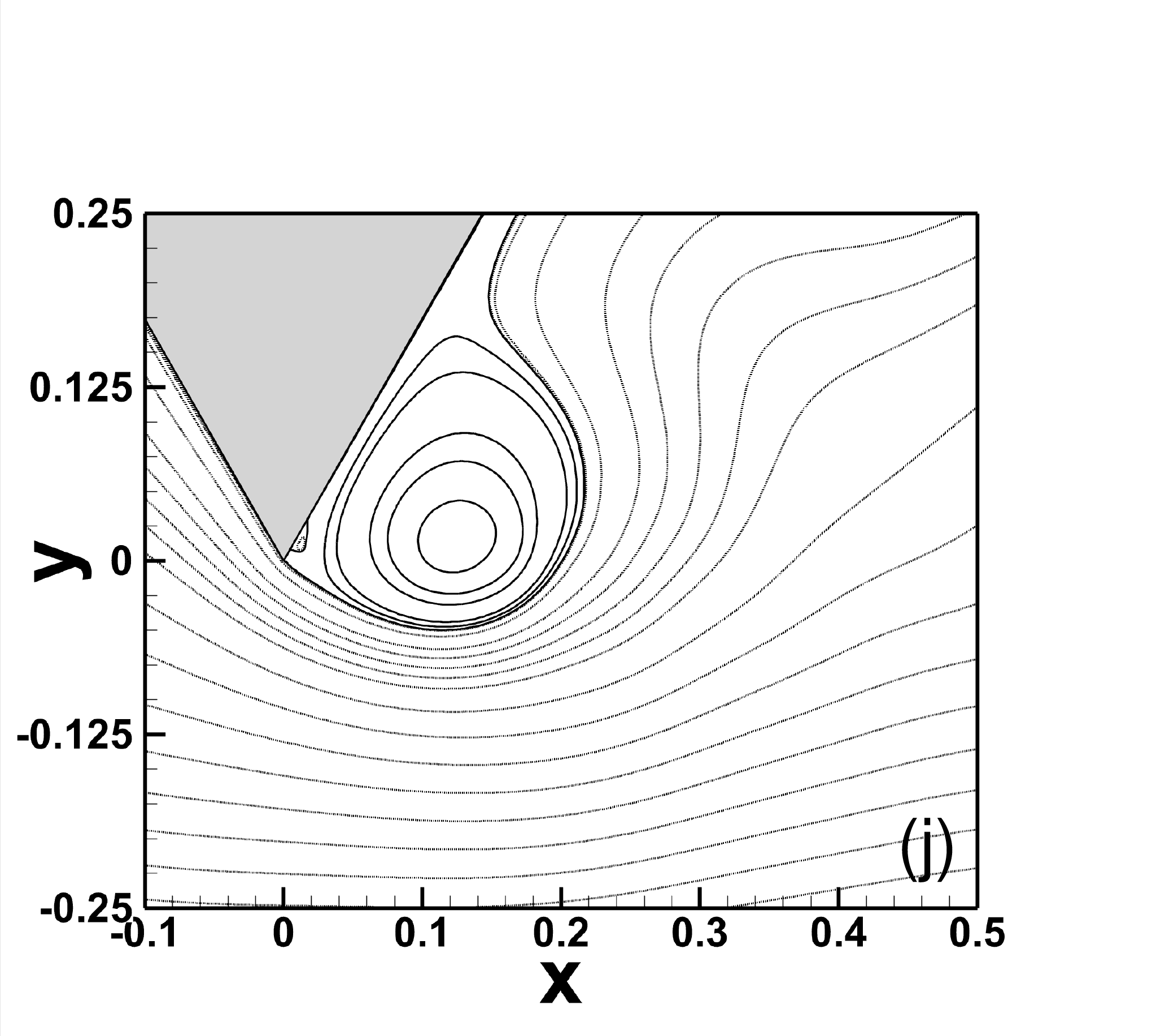}
			\&
			\hspace{-0.5cm}\includegraphics[width=0.3\linewidth]{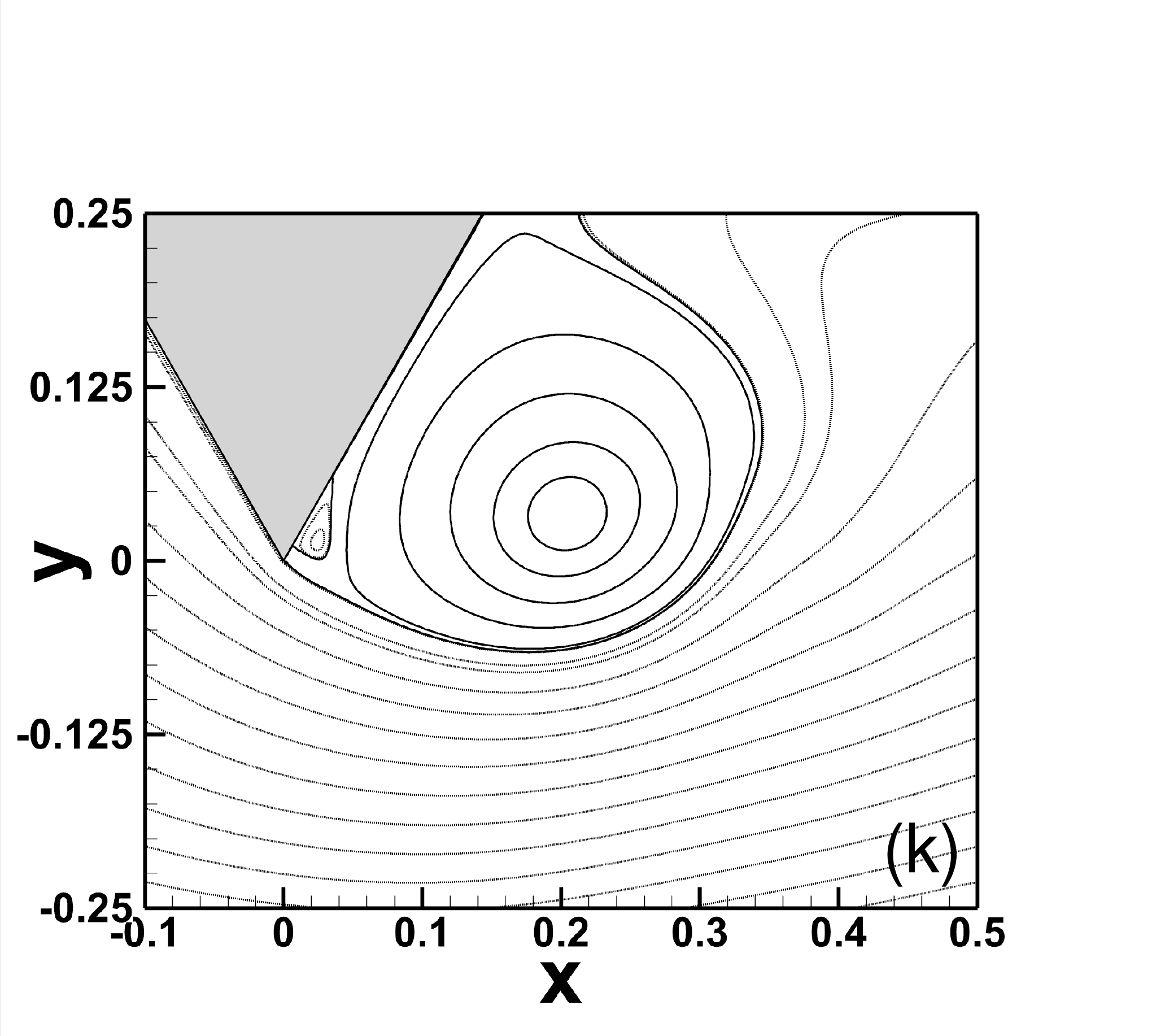}
			&
			\hspace{-0.5cm}\includegraphics[width=0.3\linewidth]{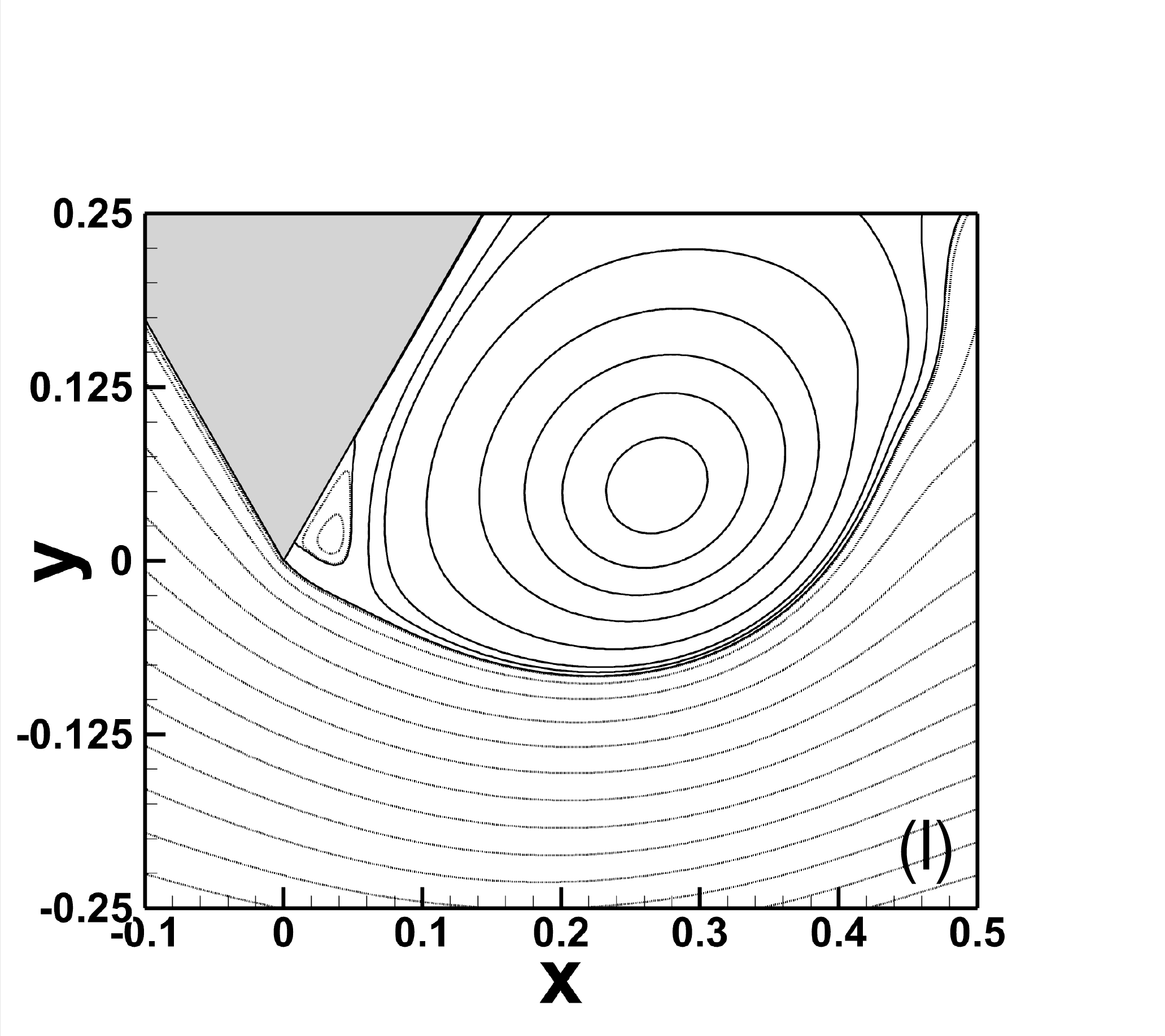}
		\end{tabular}
		\begin{tabular}{cccc}
			\hspace{-1.0cm}\includegraphics[width=0.3\linewidth]{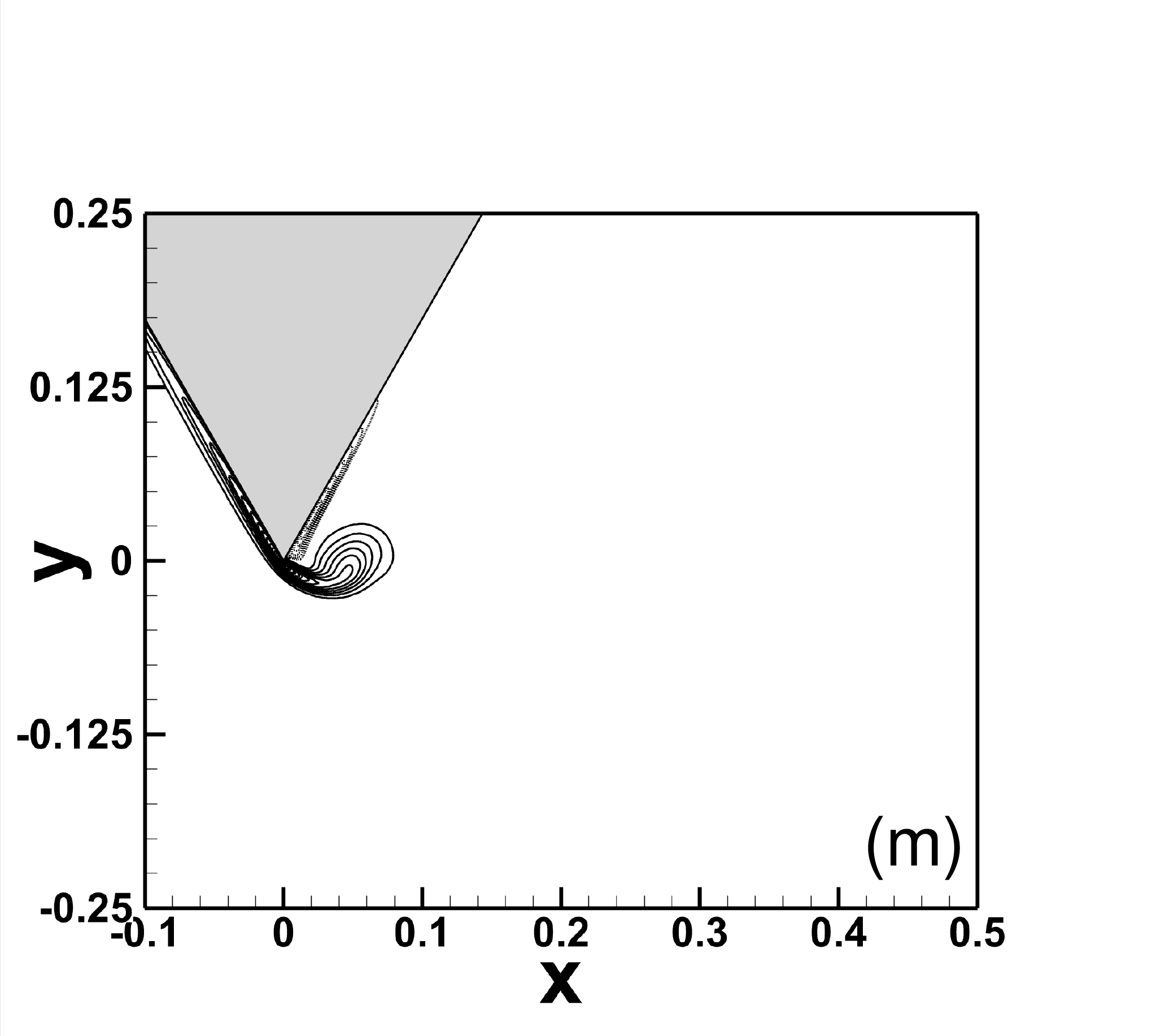}
			&
			\hspace{-0.5cm}\includegraphics[width=0.3\linewidth]{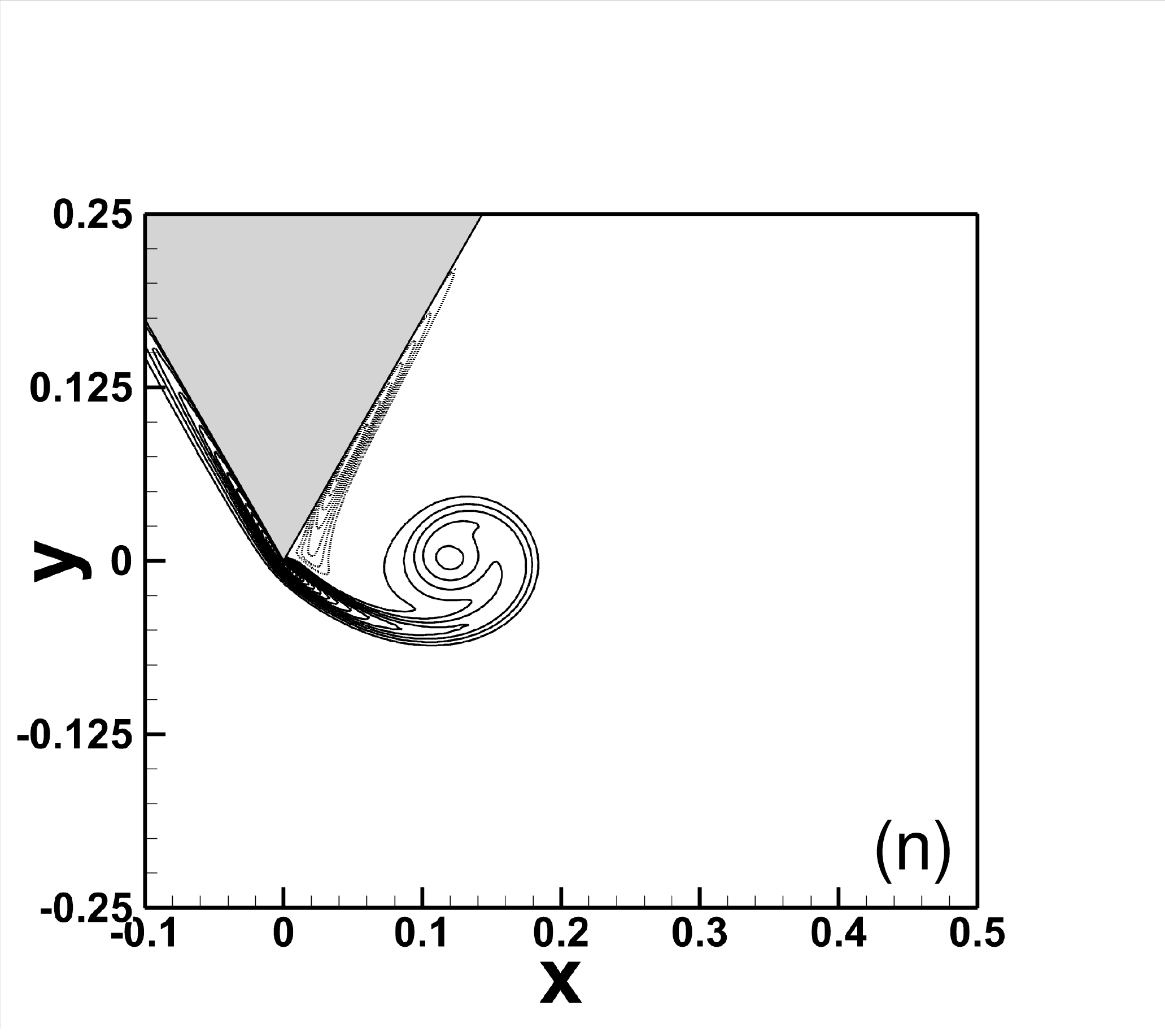}
			\&
			\hspace{-0.5cm}\includegraphics[width=0.3\linewidth]{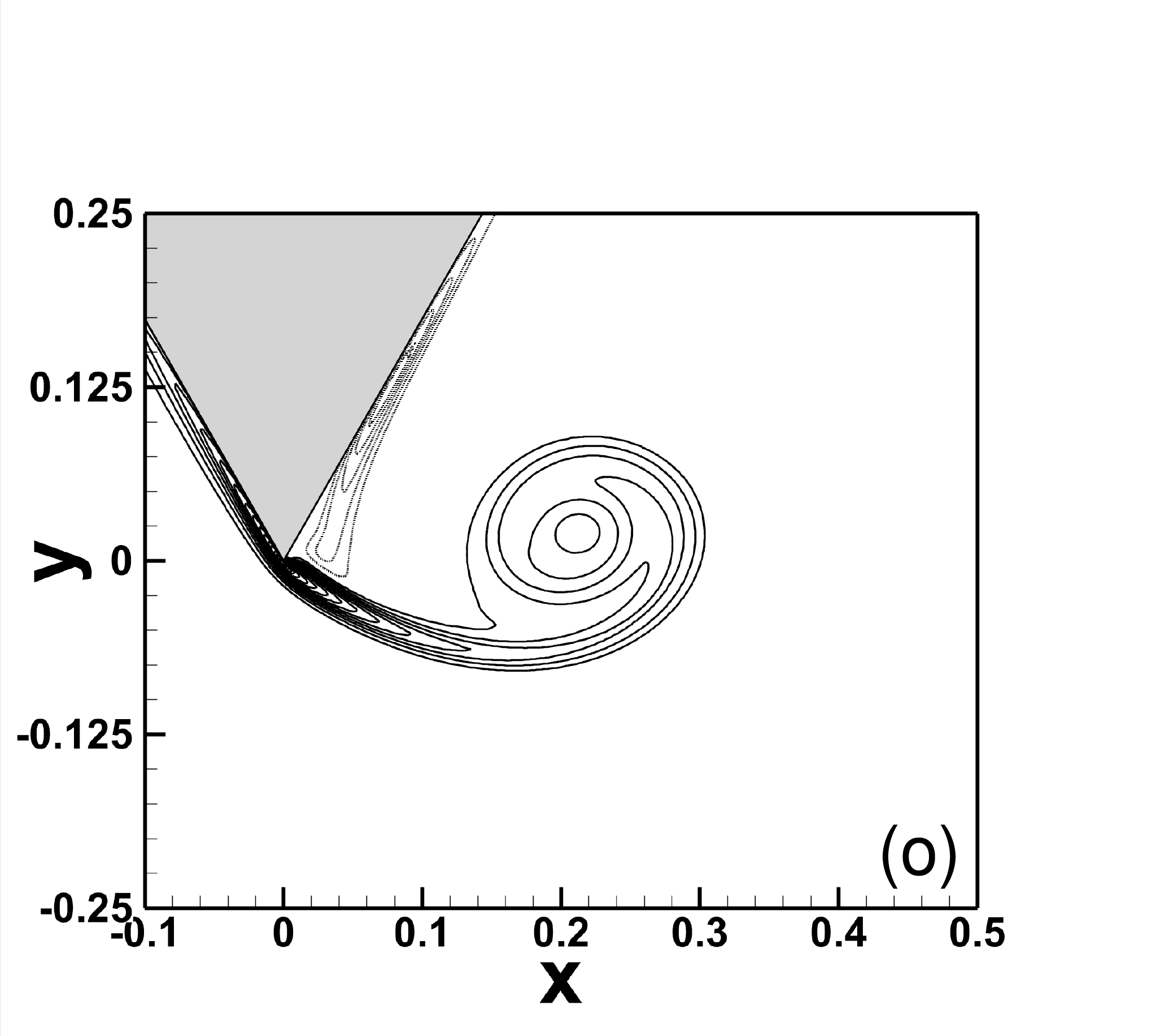}
			&
			\hspace{-0.5cm}\includegraphics[width=0.3\linewidth]{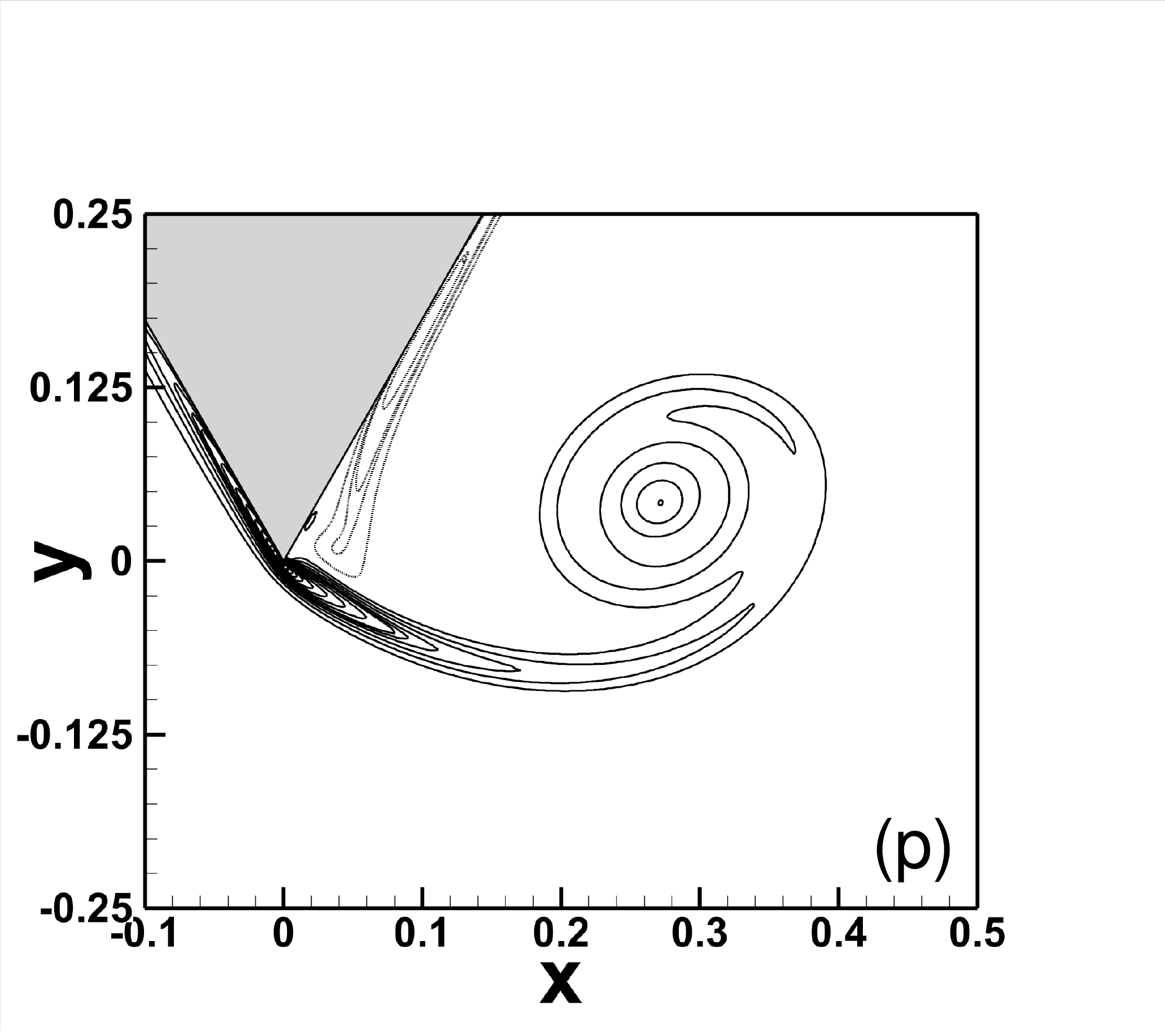}
		\end{tabular}
		\caption{\sl {Streaklines from experimental visualization of \cite{pullin1980} (a-d) and our computation (e-h) and the corresponding streamlines (i-l) and vorticity contours  (m-p) the for flow past a wedge for $Re_c= 1560$ and $m=0$ at instants $\tilde{t}=1\;sec$ ($t=0.024803$),  $\tilde{t}=3\;sec$ ($t=0.074409$), $\tilde{t}=5\;sec$ ($t=0.124016$) and $\tilde{t}=7\;sec$ ($t=0.173622$).}}
		\label{sk_wed_1560}
	\end{figure}
	\begin{figure}
		\begin{tabular}{cccc}
			\hspace{-1.0cm}\includegraphics[width=0.3\linewidth]{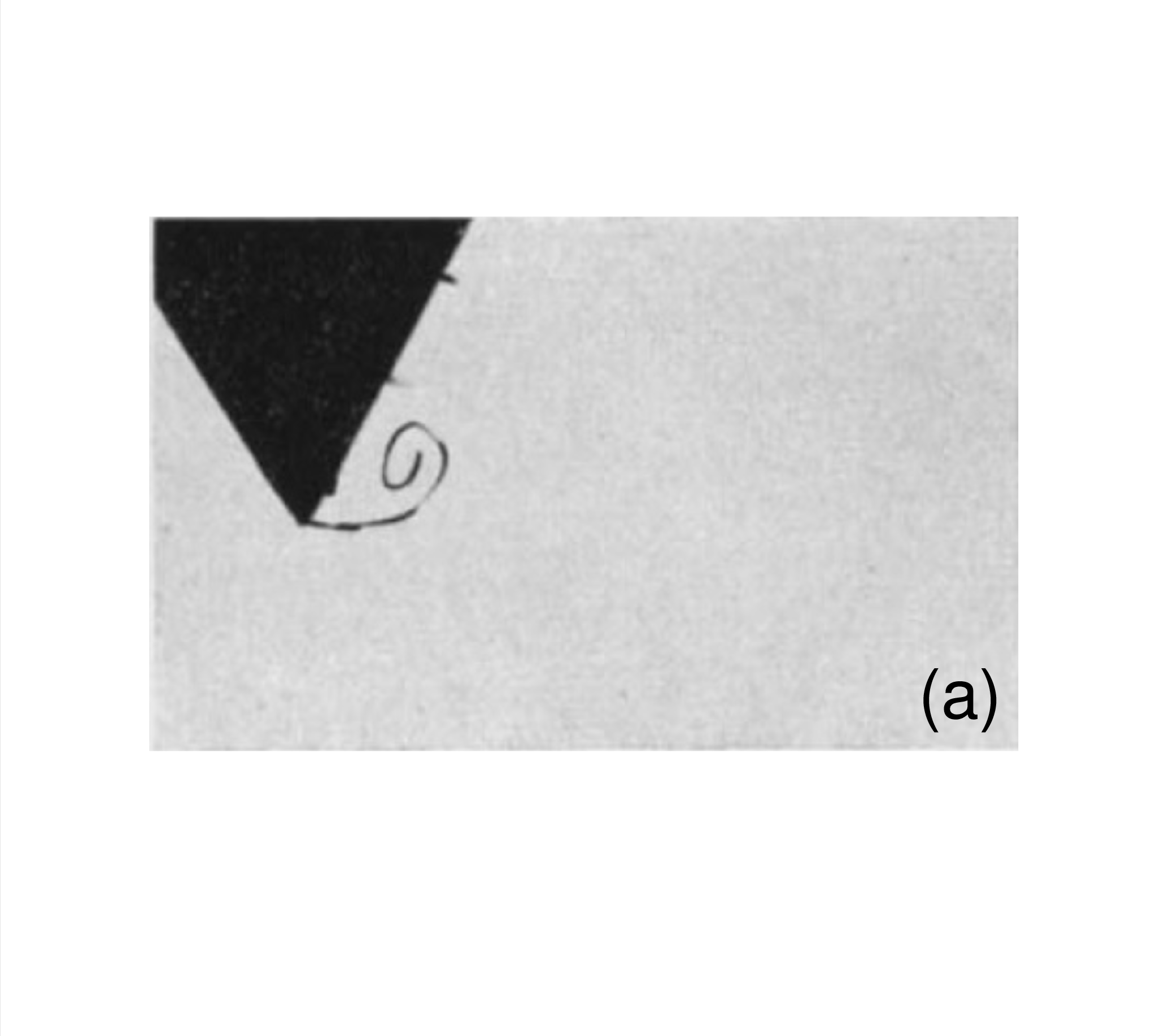}
			&
			\hspace{-0.5cm}\includegraphics[width=0.3\linewidth]{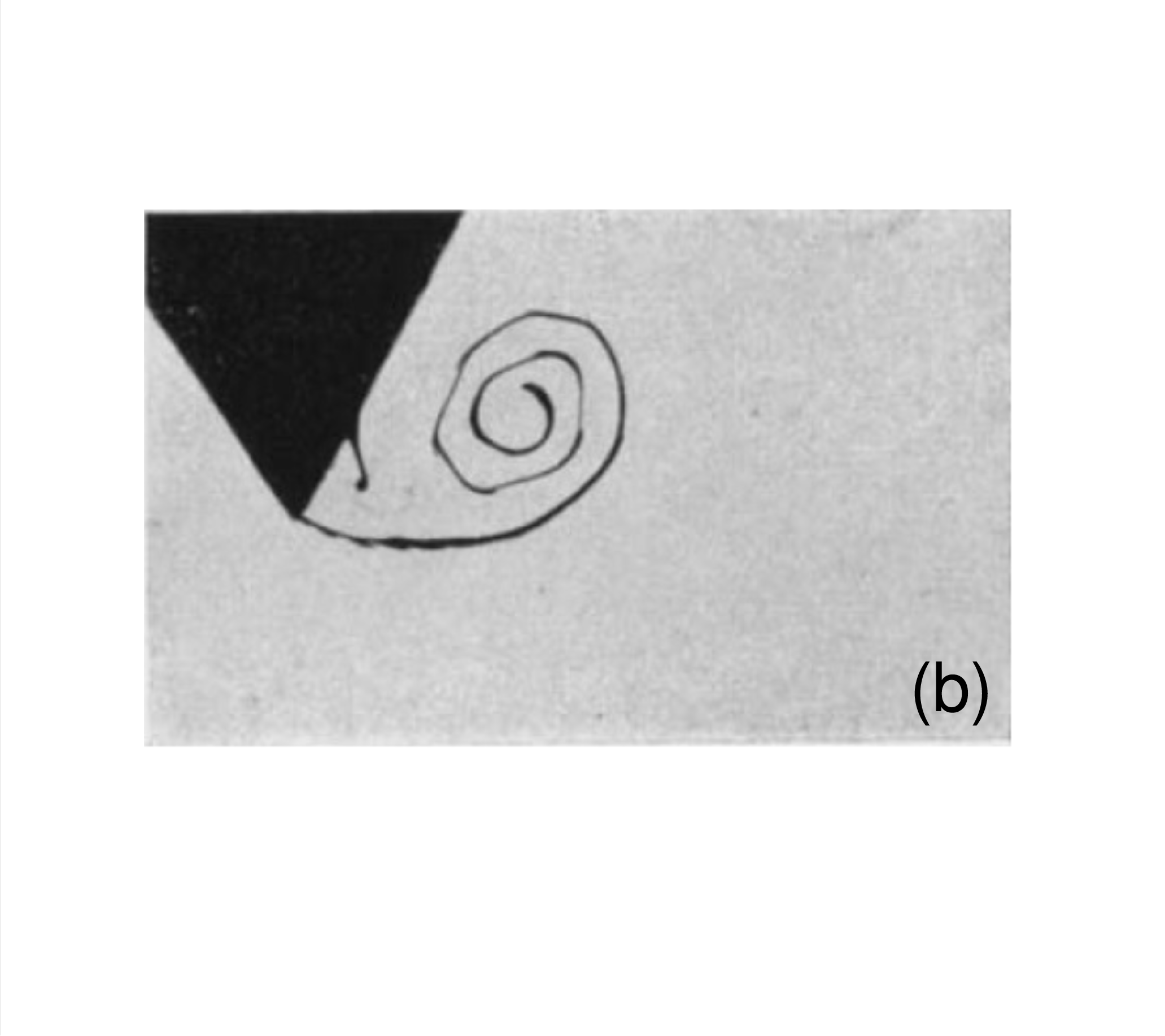}
			\&
			\hspace{-0.5cm}\includegraphics[width=0.3\linewidth]{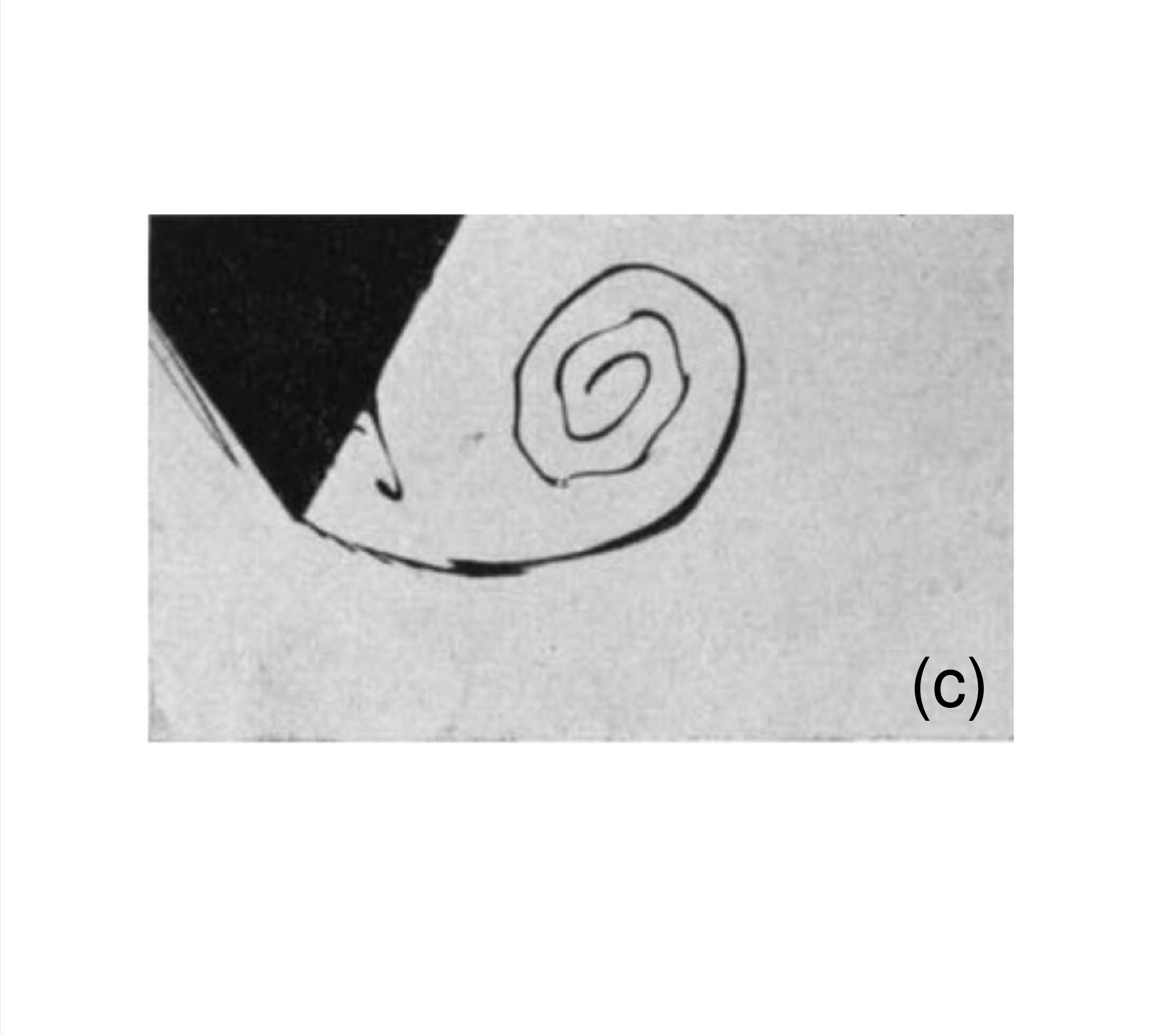}
			&
			\hspace{-0.5cm}\includegraphics[width=0.3\linewidth]{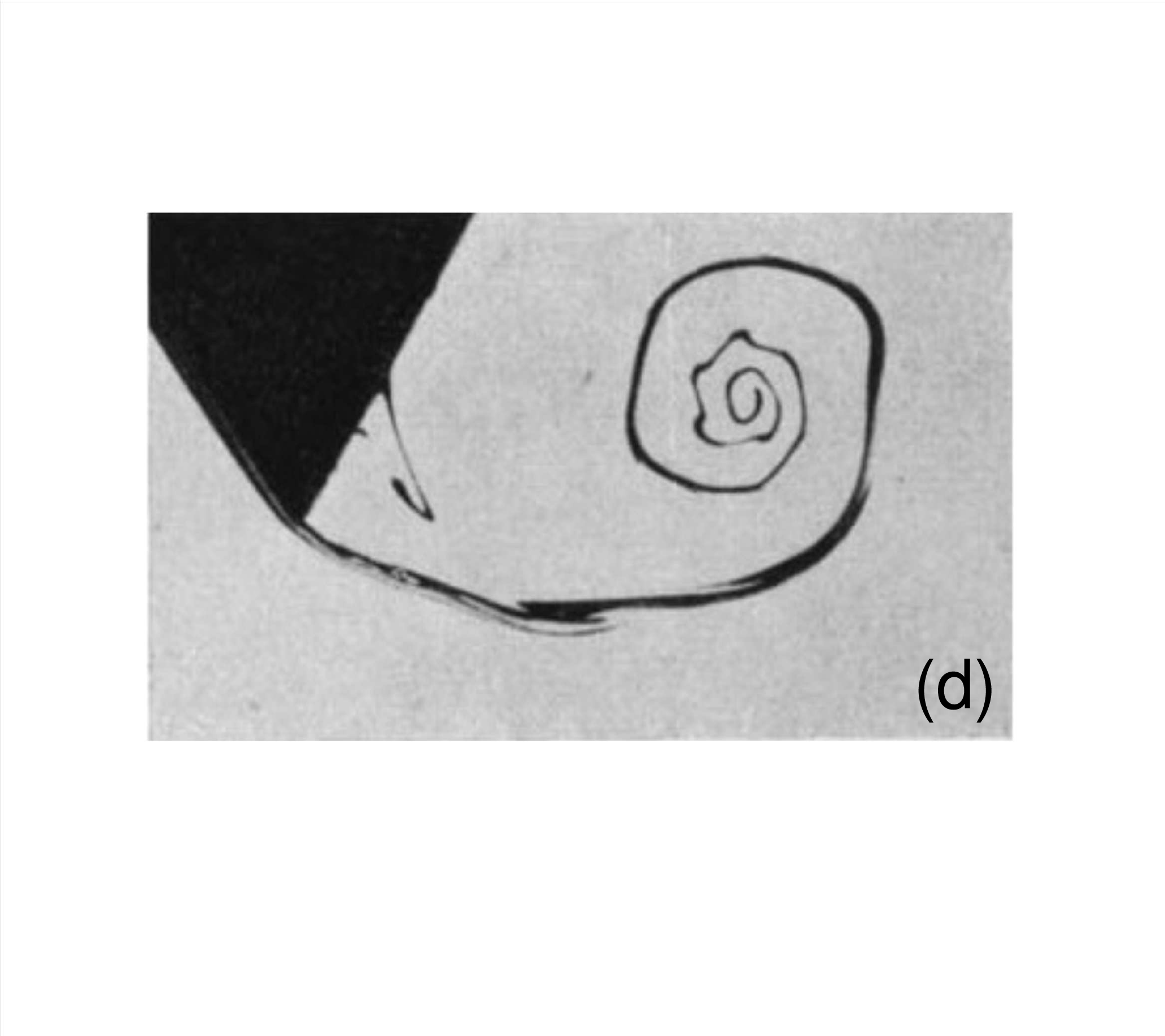}
		\end{tabular}
		\centering
		\begin{tabular}{cccc}
			\hspace{-1.0cm}\includegraphics[width=0.3\linewidth]{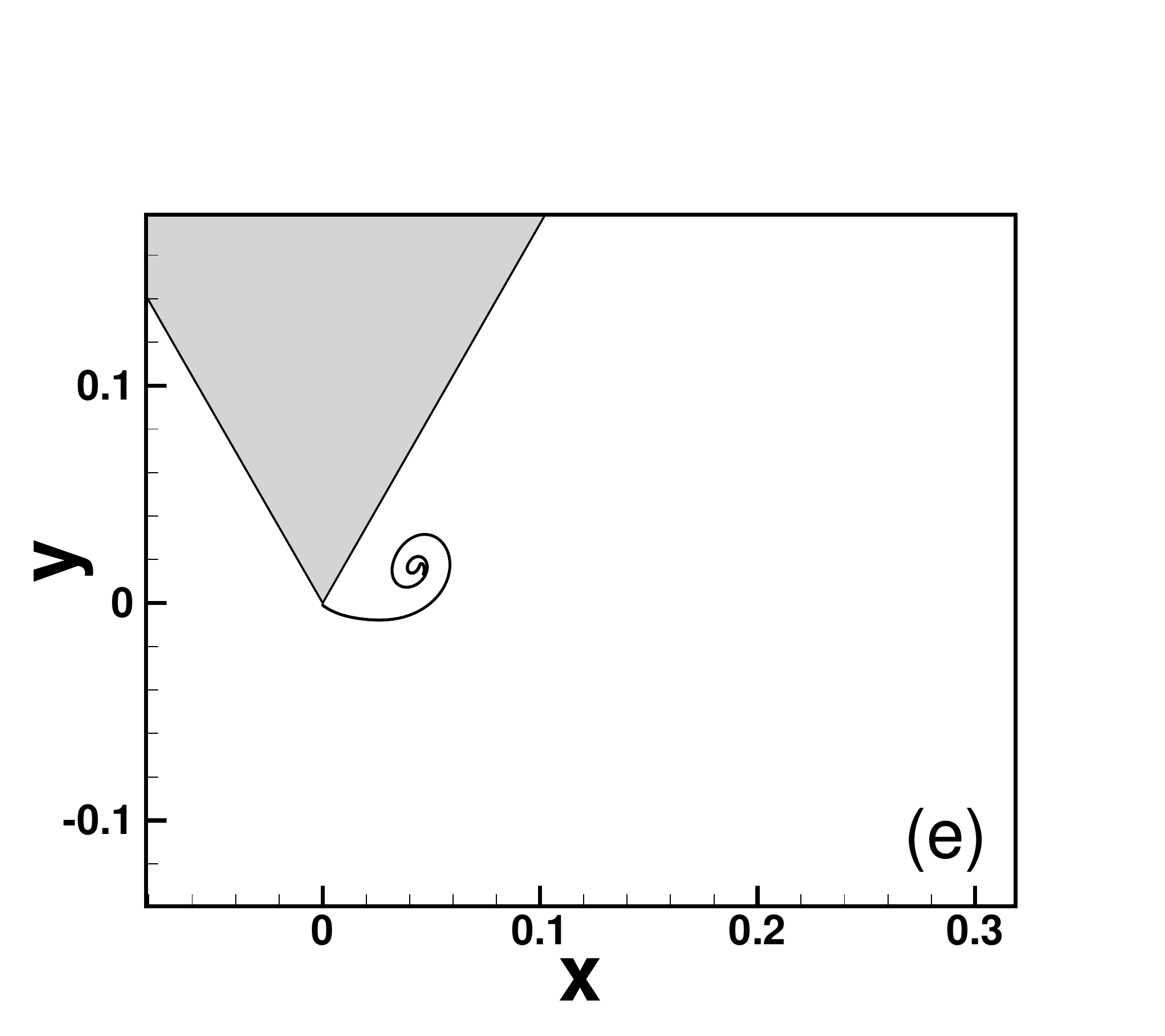}
			&
			\hspace{-0.5cm}\includegraphics[width=0.3\linewidth]{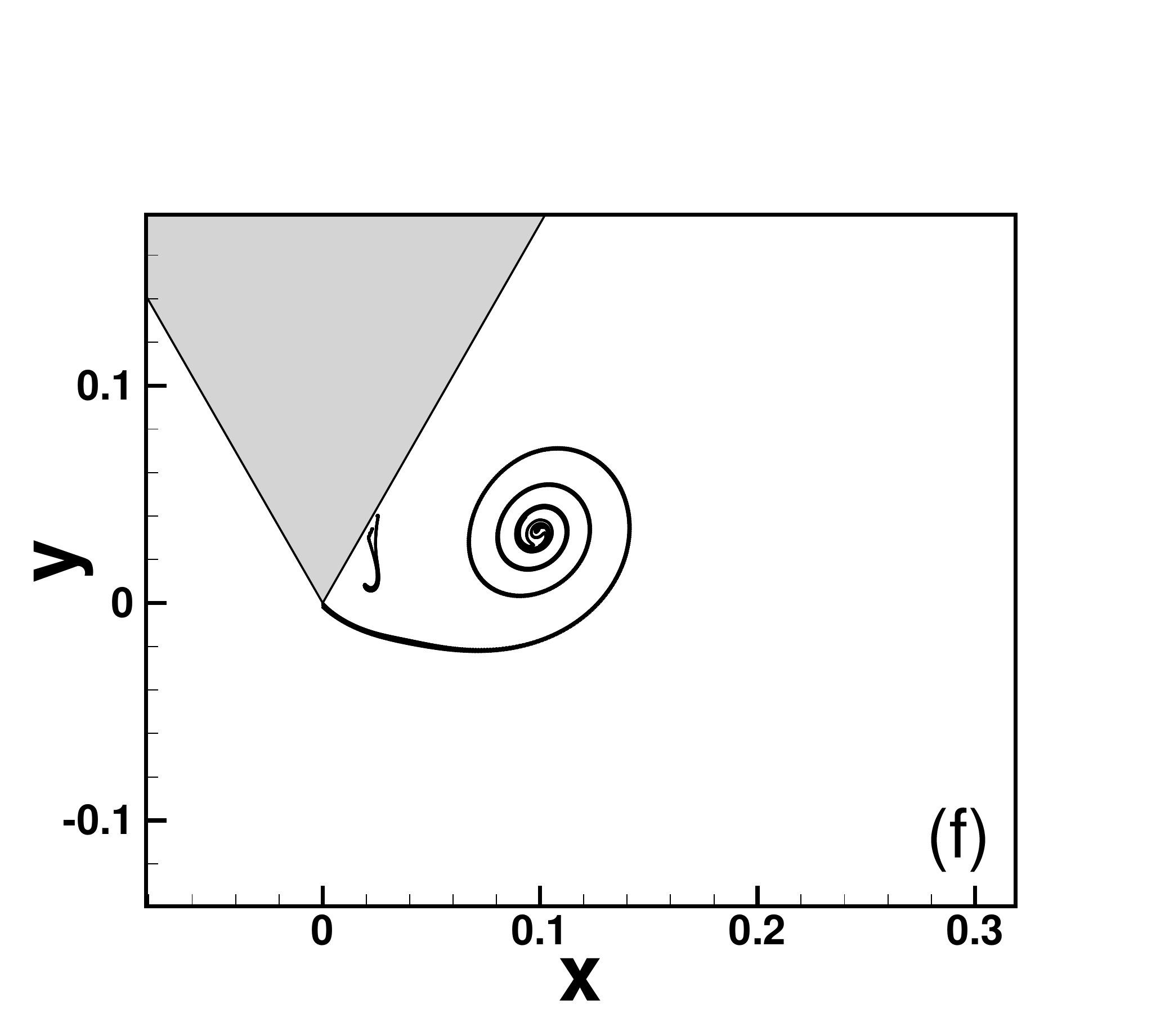}
			\&
			\hspace{-0.5cm}\includegraphics[width=0.3\linewidth]{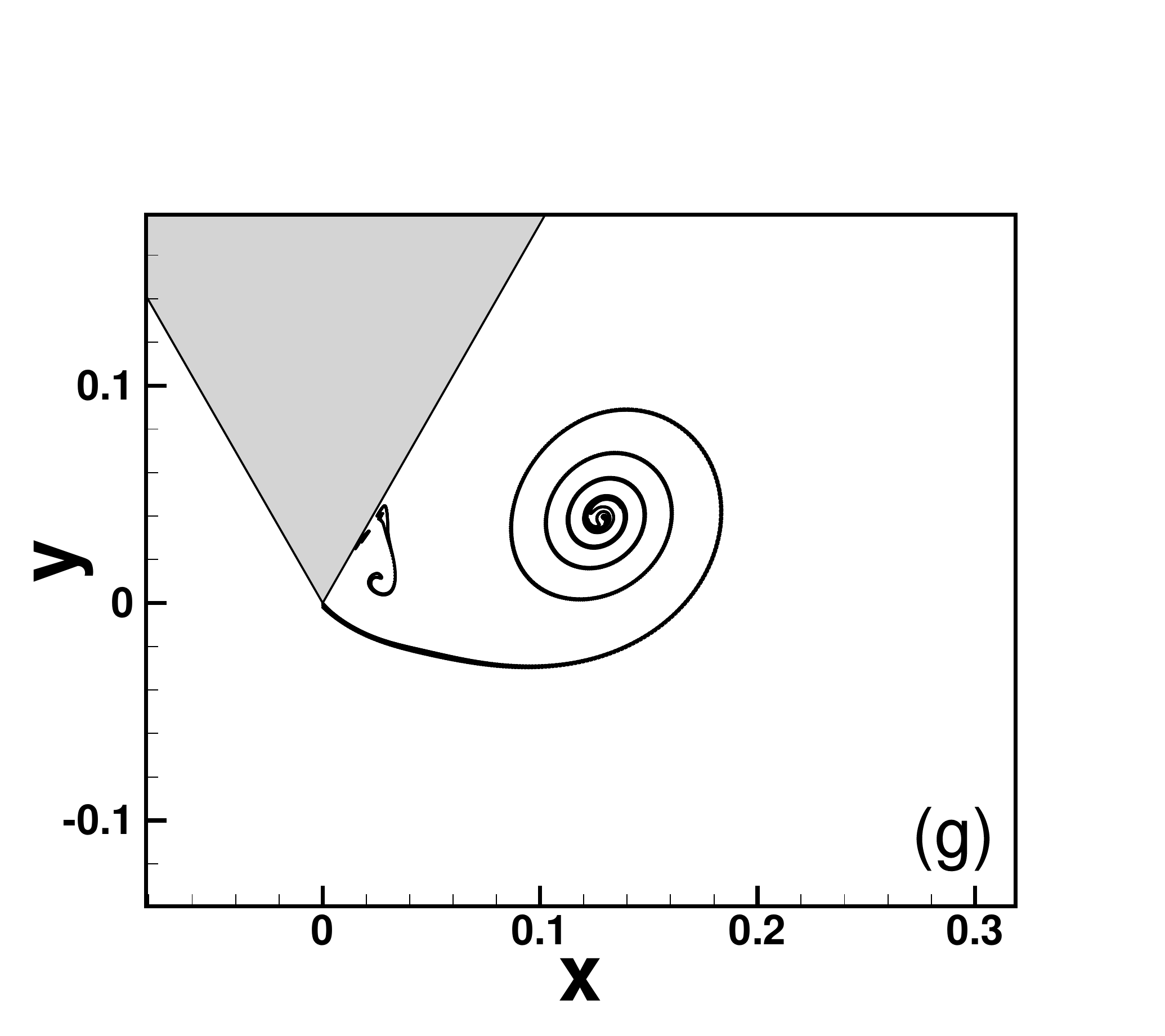}
			&
			\hspace{-0.5cm}\includegraphics[width=0.3\linewidth]{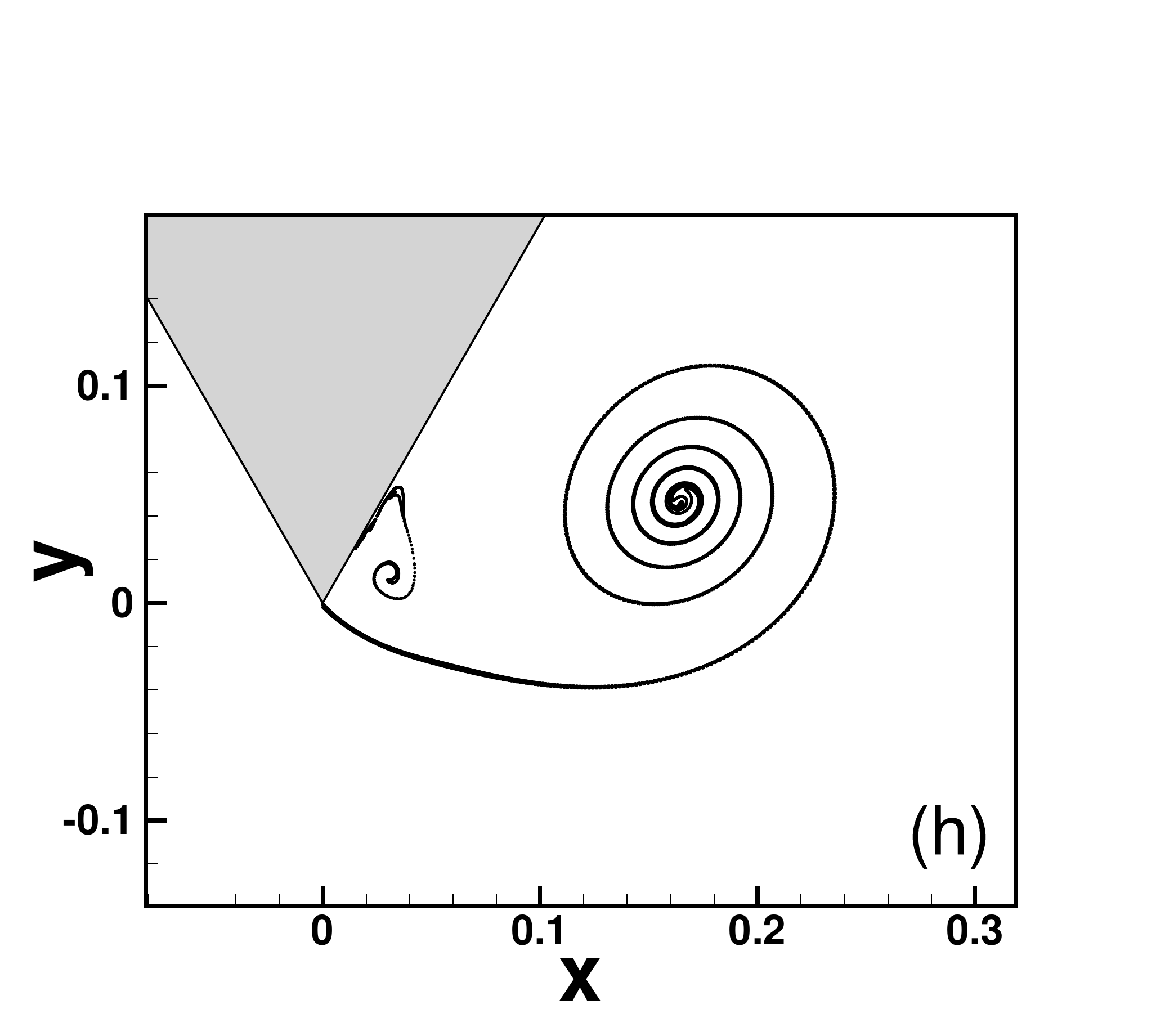}
		\end{tabular}
		\begin{tabular}{cccc}
			\hspace{-1.0cm}\includegraphics[width=0.3\linewidth]{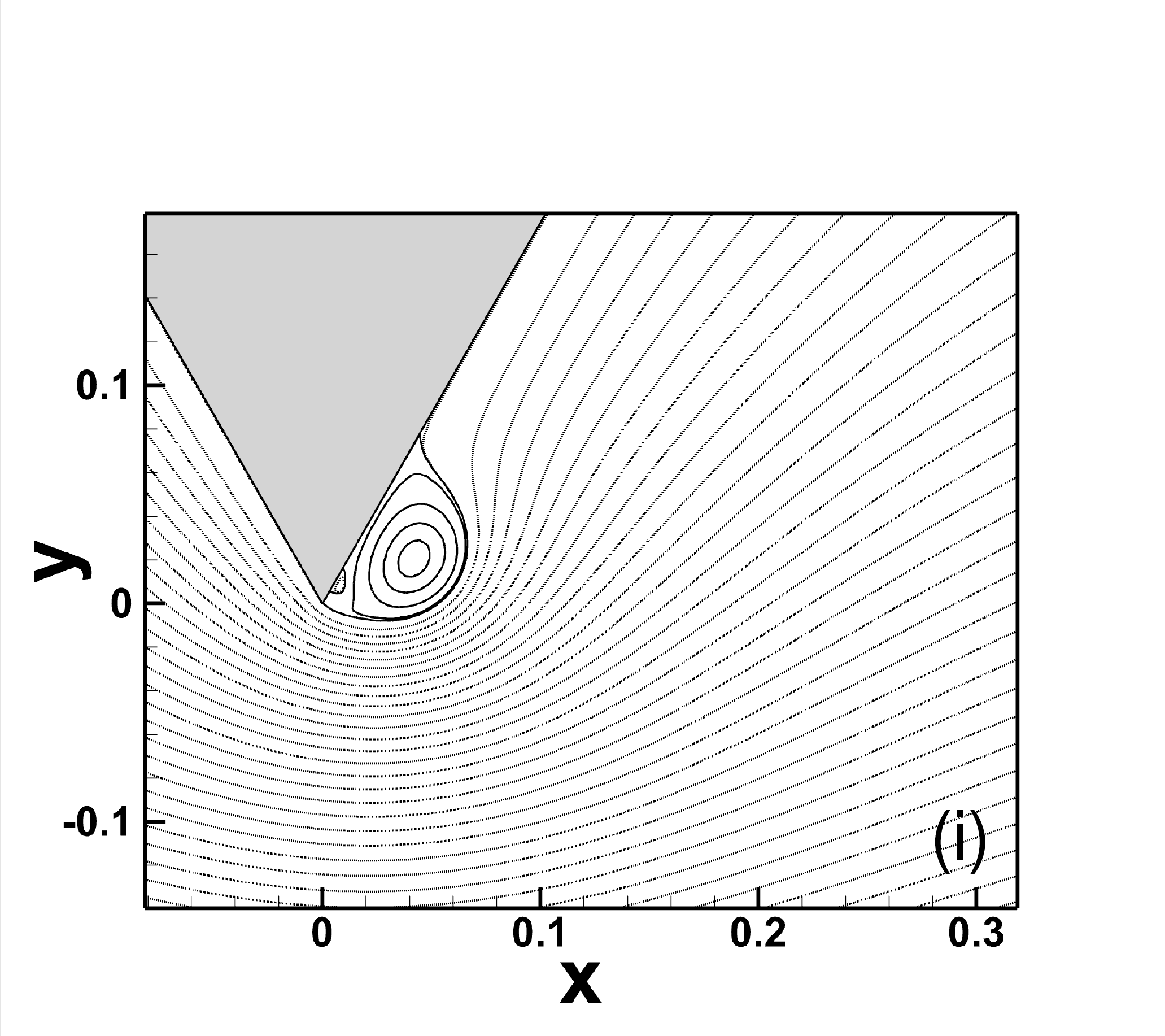}
			&
			\hspace{-0.5cm}\includegraphics[width=0.3\linewidth]{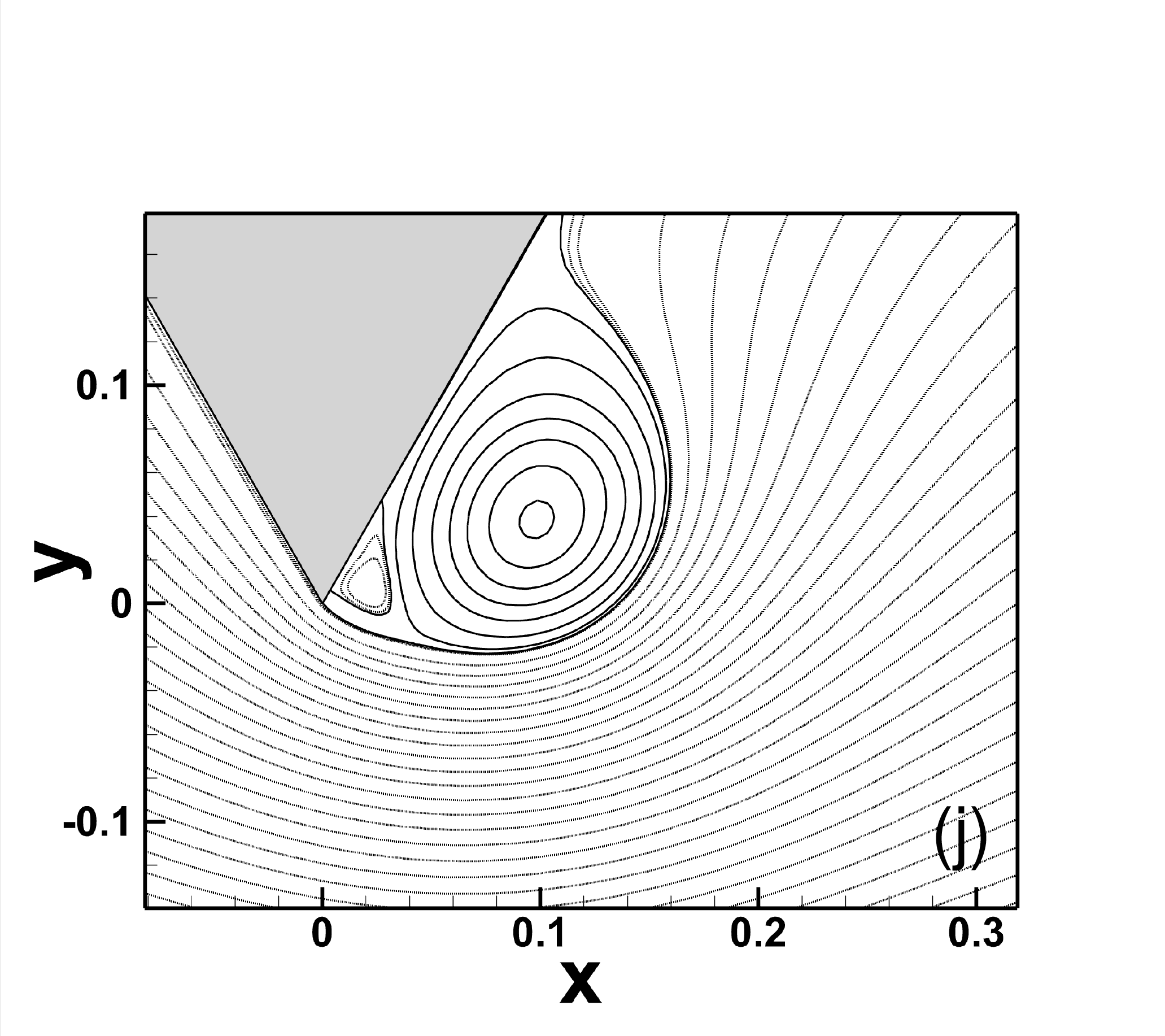}
			\&
			\hspace{-0.5cm}\includegraphics[width=0.3\linewidth]{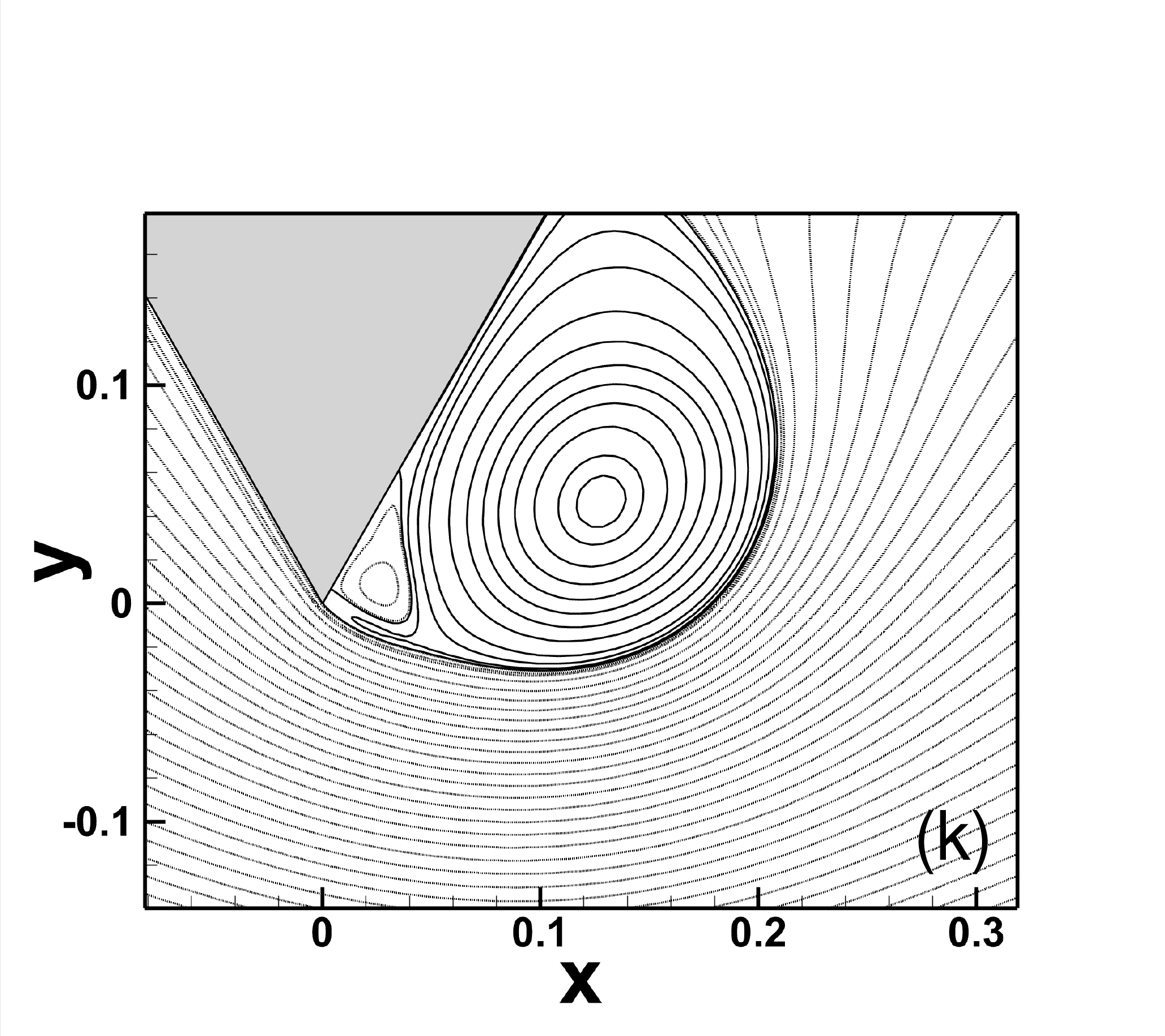}
			&
			\hspace{-0.5cm}\includegraphics[width=0.3\linewidth]{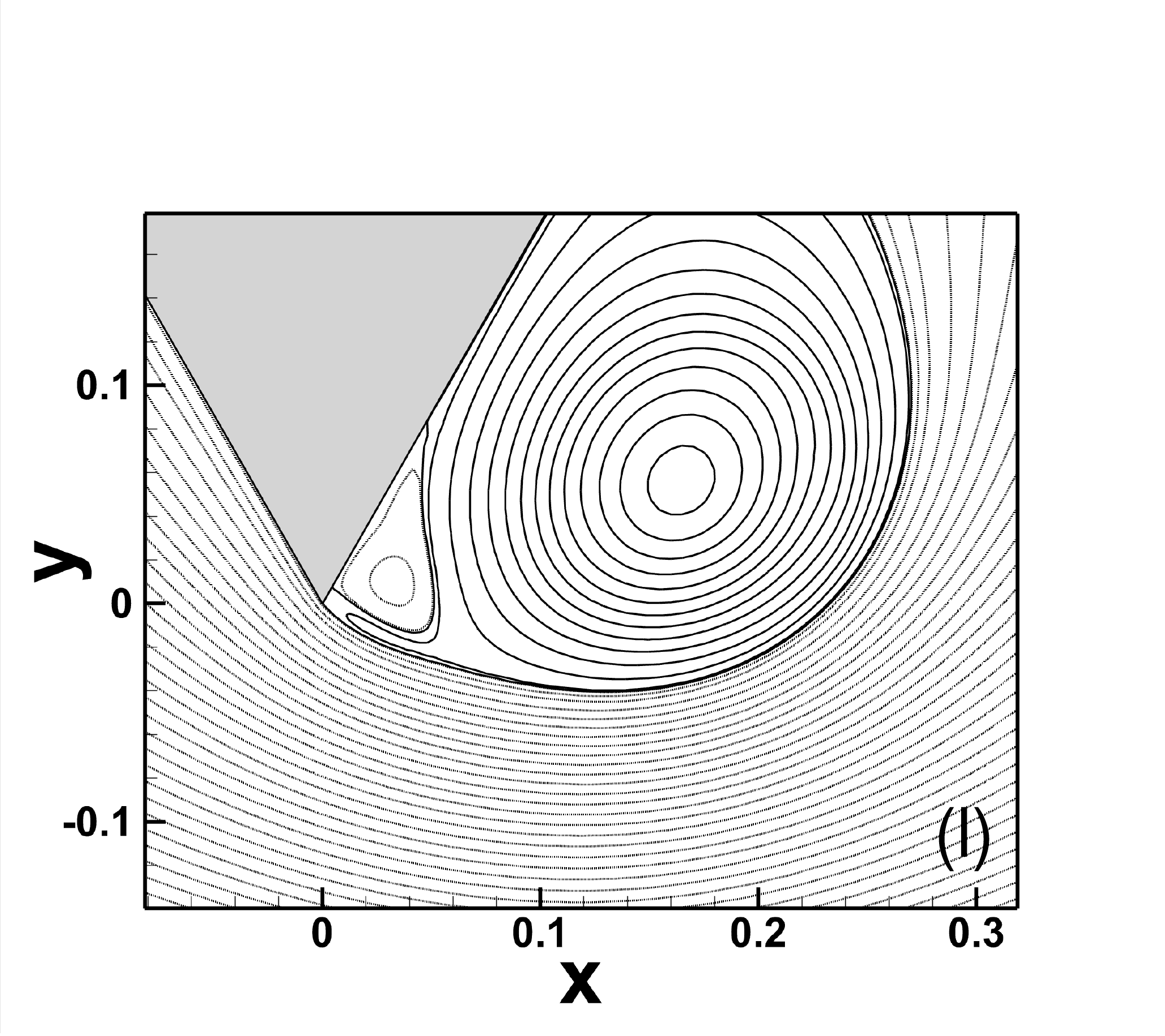}
		\end{tabular}
		\begin{tabular}{cccc}
			\hspace{-1.0cm}\includegraphics[width=0.3\linewidth]{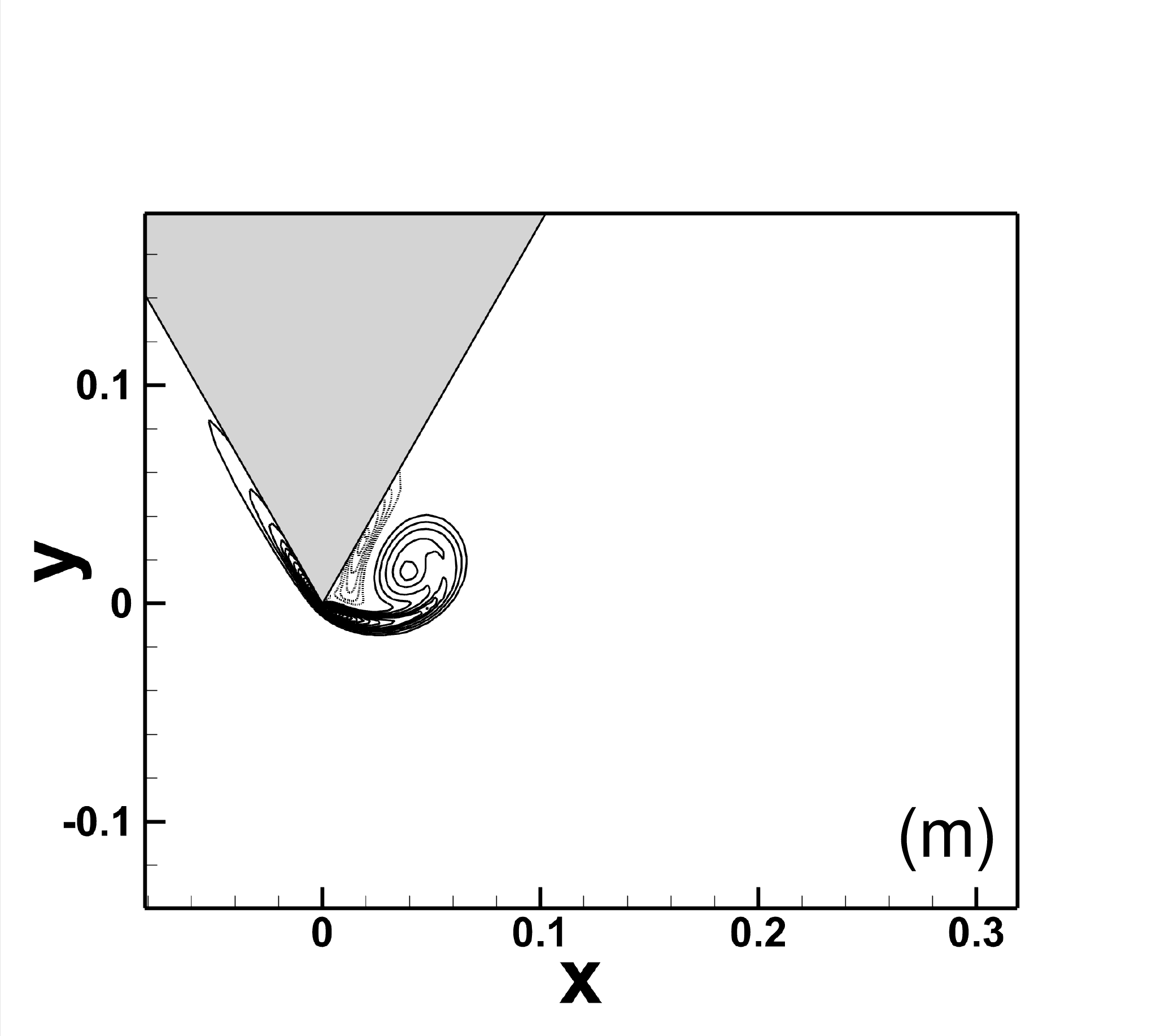}
			&
			\hspace{-0.5cm}\includegraphics[width=0.3\linewidth]{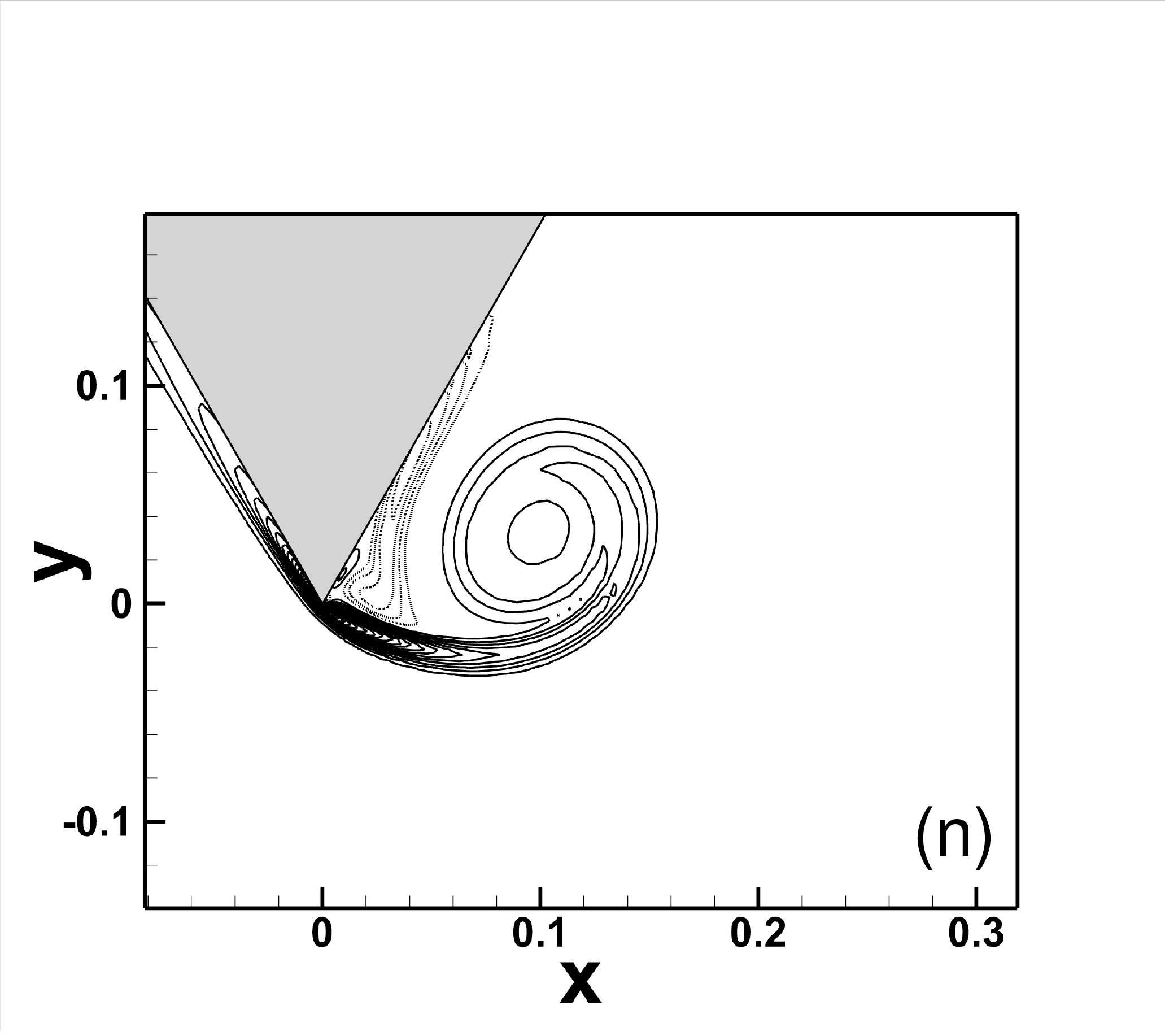}
			\&
			\hspace{-0.5cm}\includegraphics[width=0.3\linewidth]{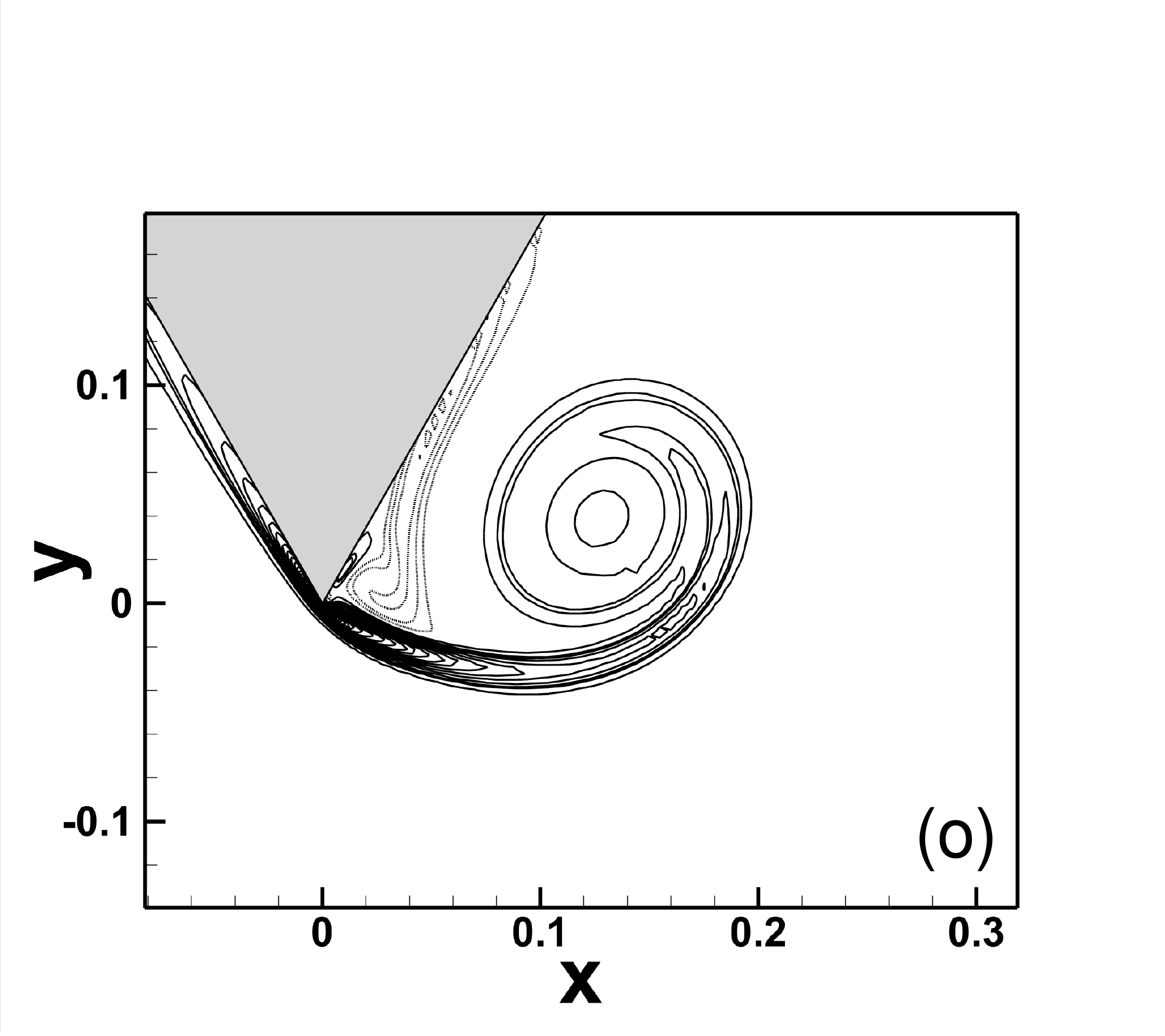}
			&
			\hspace{-0.5cm}\includegraphics[width=0.3\linewidth]{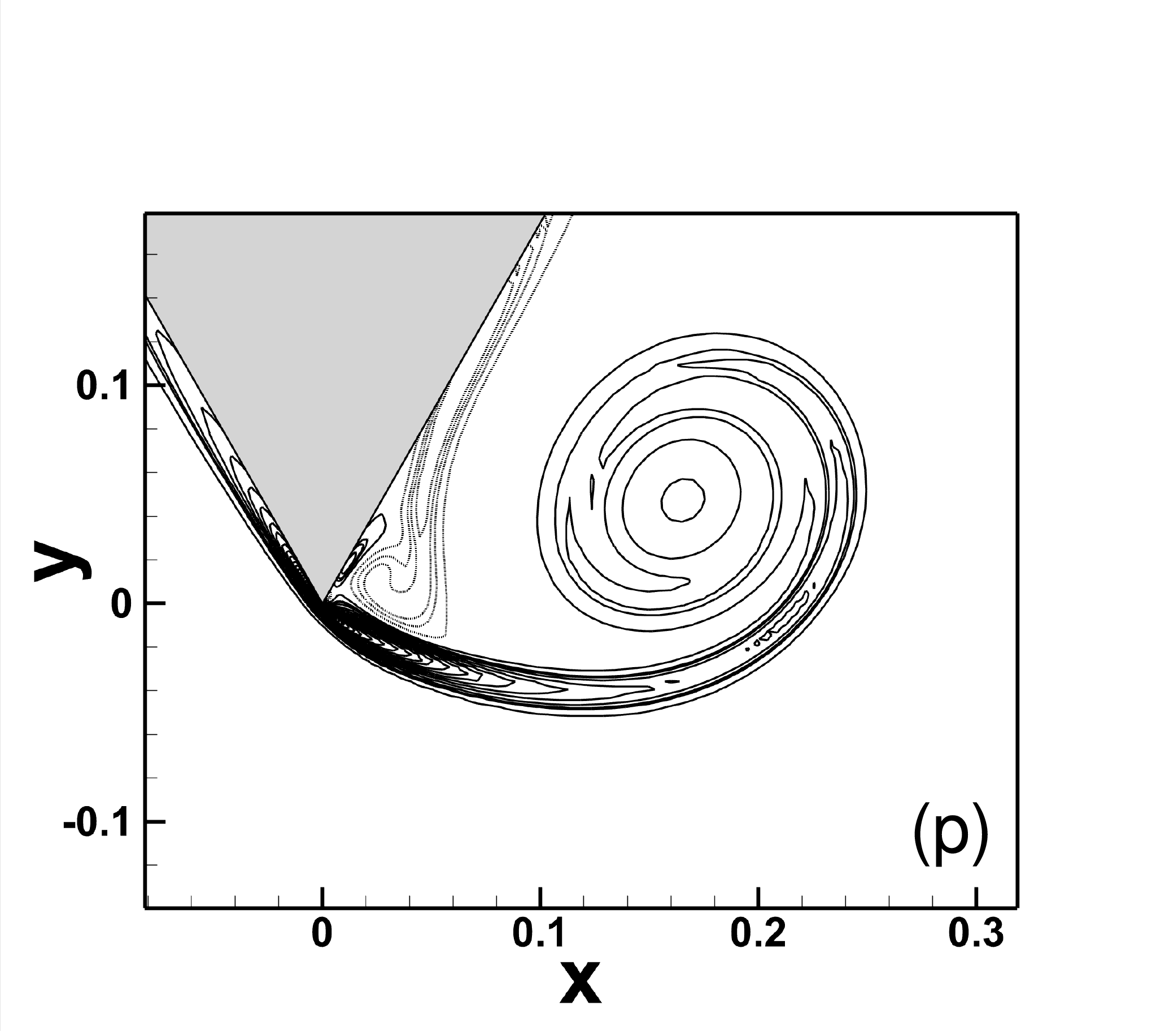}
		\end{tabular}
		\caption{\sl {Streaklines from Pullin and Perry's experimental visualization (a-d) and our computation (e-h) and the corresponding streamlines (i-l) and vorticity contours  (m-p) the for flow past a wedge for $Re_c= 6873$ and $m=0.88$ at instants $\tilde{t}=2.8\;sec$ ($t=0.299493$),  $\tilde{t}=5\;sec$ ($t=0.534809$), $\tilde{t}=6\;sec$ ($t=0.641771$) and $\tilde{t}=7\;sec$ ($t=0.748733$).}}
		\label{sk_wed_6873}
	\end{figure}
	In the next figures \ref{sk_wed_1560}-\ref{sk_wed_6873}, we present the the earliest stage of flow evolution respresnted by streaklines from the experimental visualization of \citet{pullin1980} side by side with from the ones from our simulation along with the corresponding streamlines and vorticity contours for $Re_c=1560$ and $6873$ respectively. The flow characteristics for $Re_c=6621$ is almost similar to that of $Re_c=6873$ and hence not compared here.  For the streamlines and vorticity contours, solid lines represent positive and dotted ones negative values of the contours. It is observed that the start up vortex for $Re_c=1560$ performs more rounds of rotations at time $\tilde{t}=7\;sec$ than $Re_c=6873$ as can seen from the extreme right columns of figures \ref{sk_wed_1560} and \ref{sk_wed_6873}, indicating a stronger circulation for the former in the early stage of the flow. Moreover, the secondary vortex phenomena is more prominent for $Re_c=6873$ than $Re_c=1560$.

	It is heartening to see that our computed results are extremely close to the experimental results of \citet{pullin1980}. The locations of the primary and the secondary vortex centers, their size and shape, and precise instants of their occurrence bear testimonial to the accuracy in capturing the attributes of the fluid flow. Apart from the computational accuracy of the data, this example also aptly demonstrates the capacity of the scheme developed by the authors (\citet{kumar2020}) in handling both Dirichlet and Neumann boundary conditions with equal ease. A close look at the boundary conditions mentioned in section \ref{discr} would reveal that while Dirichlet boundary conditions have been used for $\psi$ at the top and bottom walls, inlet and on the surface of the wedge, Neumann boundary condition has been employed at the outlet. It is worth mentioning that while we have used exact physical boundary conditions matching the ones used in the lab experiment of \citet{pullin1980}, in their flow simulations, \cite{xu2015,xu2016} used potential flow conditions in the far field equivalent to considering an infinite wedge (or flat plate immersed in viscous fluid contained in an unconfined domain).
	
	\subsection{Inviscid scaling}\label{inv_scl}
	\begin{figure}
		\begin{center}
			\includegraphics[width=0.65\textwidth]{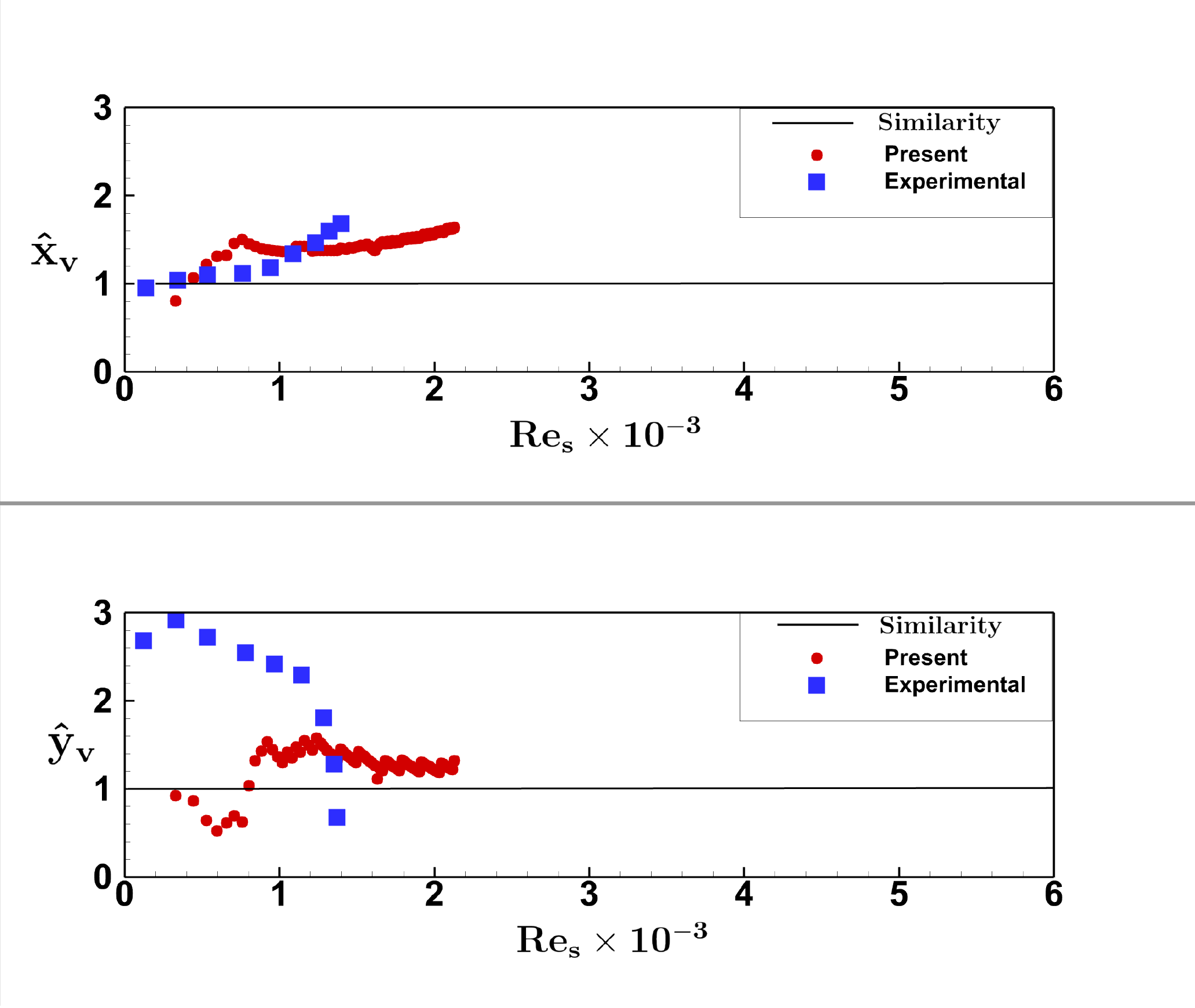}
			{(a)}
			\includegraphics[width=0.65\textwidth]{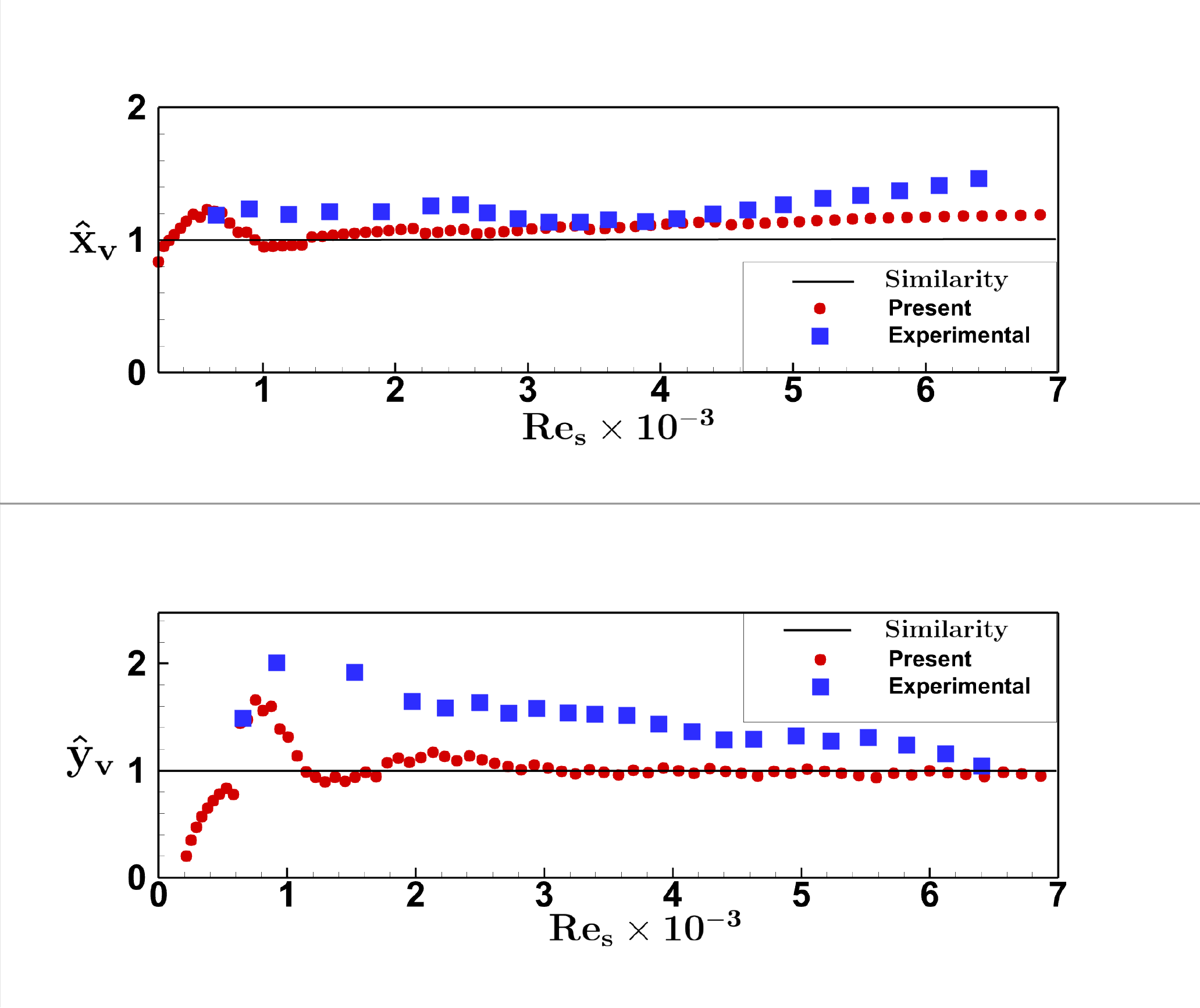}
			{(b)}
		\end{center}
		\caption{{Comparison of experimental (\cite{pullin1980}) and numerical vortex center movements with the similarity solutions for (a) $Re_c=1560$ and (b) $Re_c=6873$.} }
		\label{movement}
	\end{figure}
	\cite{pullin1980} also compared the movement of the vortex centers resulting from their laboratory experiments with that of the rolled up vortex-sheet model found from the scaling laws of Pullin's earlier numerical work on the starting flow past an infinite wedge (\cite{pullin1978}). The initial trajectory of the vortex-centers of these self-similar structures in dimensionless form was given by
	
	\begin{equation}
		\hat{x}_v=\frac{\tilde{x}_M}{a^{1/(2-n)}\tilde{t}^M\mathscr{R}},\;\hat{y}_v=\frac{\tilde{y}_M}{a^{1/(2-n)}\tilde{t}^M\mathscr{I}}
		\label{scale1}
	\end{equation}
	where $a=\alpha_0 A H^{1-n}$, with $\alpha_0$, $n$, $A$ and $H$ are as defined in \eqref{nondim} and \eqref{scale}, and $\displaystyle \mathscr{R}$, $\mathscr{I}$ represent the real and imaginary parts of a complex argument that can be found from table 2 of \cite{pullin1980}. Using the length and time scales defined in equation \eqref{nondim}, \eqref{scale1} reduces to
	\begin{equation}
		\hat{x}_v=\frac{{x}_M}{\alpha_0^{1/(2-n)}{t}^M\mathscr{R}},\;\hat{y}_v=\frac{{y}_M}{\alpha_0^{1/(2-n)}{t}^M\mathscr{I}}
		\label{scale2}
	\end{equation}
	where $(x_M,y_M)$ is the coordinate of the vortex center defined in section \ref{m_effect}.
	
	In figure \ref{movement}, we present the scaled values of the coordinates of the primary vortex center, or in other words, the horizontal and vertical displacements of the vortex centers from the wedge-tip as a function of the scale Reynolds number $Re_s(t)$ defined by \eqref{scale_t} and compare them with the experimental results of \cite{pullin1980} as well as the inviscid similarity theory proposed by \cite{kaden1931}. As $Re_s$ increases, one can clearly see more deviation of the vortex centers from the similarity solutions, more so, for $\hat{x}_v$. \cite{pullin1980} attributed this break-down of similarity behaviour to the influence of the channel boundaries and three-dimensional effect. One can see from figure \ref{movement}(a), for the uniform flow ($Re_c=1560$), our results deviates a bit from the experimental results as had been observed in \cite{xu2016} as well. However, our results are relatively closer to the experimental ones than theirs. On the other hand, for the accelerated case for $Re_c=6873$, our computed solutions are much closer to the experimental observations, and follow the similarity solutions very closely.

	\subsection{The Structure of Vortex shedding beyond the experimental visualization of \citep{pullin1980}}\label{beyond}
	In this section, we provide a detailed description of different stages of evolution for the flow past a wedge mounted on a wall subjected to accelerated flow for $Re_c=6873$ until transition to turbulence. Note that \citet{pullin1980} continued their laboratory experiment only up to a non-dimensional time $t\approx 0.76$ in their study, while we carried out our numerical experiment for the same up to $t=3.10$. In the absence of any experimental visualisation beyond $t=0.76$ for this flow configuration, we relied on the experimental visualisation of \citet{lian1989}, who conducted a series of experiments for the flow past flat plates with sharp edges in the range $2000\leq Re \leq 15000$, to compare our computed results. They used hydrogen bubble technique to visualize the flow and reported three stages of evolution of the starting vortices culminating in a three-fold structure leading to the onset of transition to turbulence. Our investigation also revealed that all these three stages are very much evident during the course of the flow which we detail here.  Note that since \cite{lian1989} provided pictures only for the upper edge of the plate, in some of the figures that follow next, their experimental visualization pictures have been tilted upside down to compare our results with theirs. 
	\subsubsection{Initial Stage}
	\begin{figure}
		\begin{tabular}{ccc}
			\hspace{-1.5cm}\includegraphics[width=0.5\textwidth]{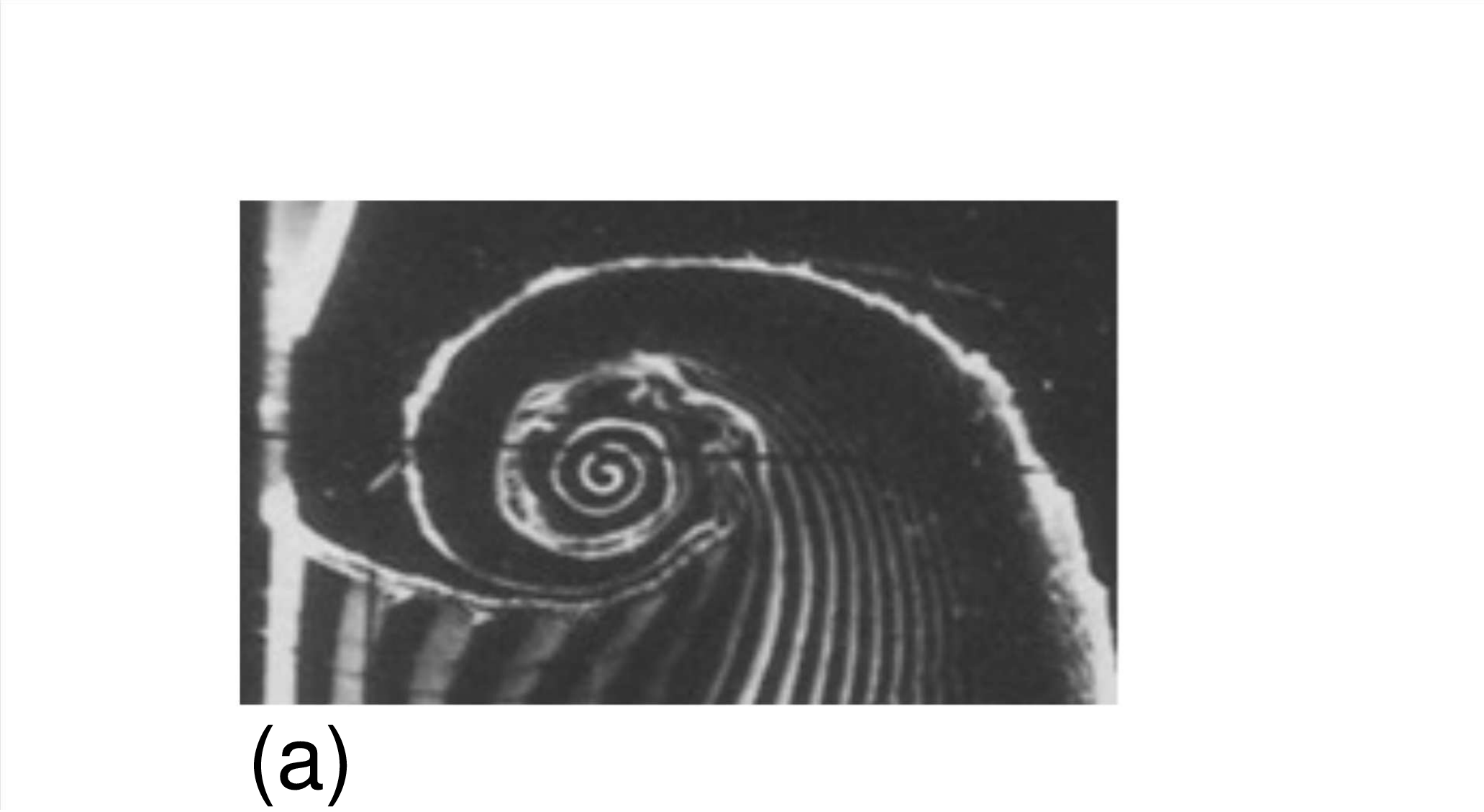}
			&
			\hspace{-1.5cm}\includegraphics[width=0.5\textwidth]{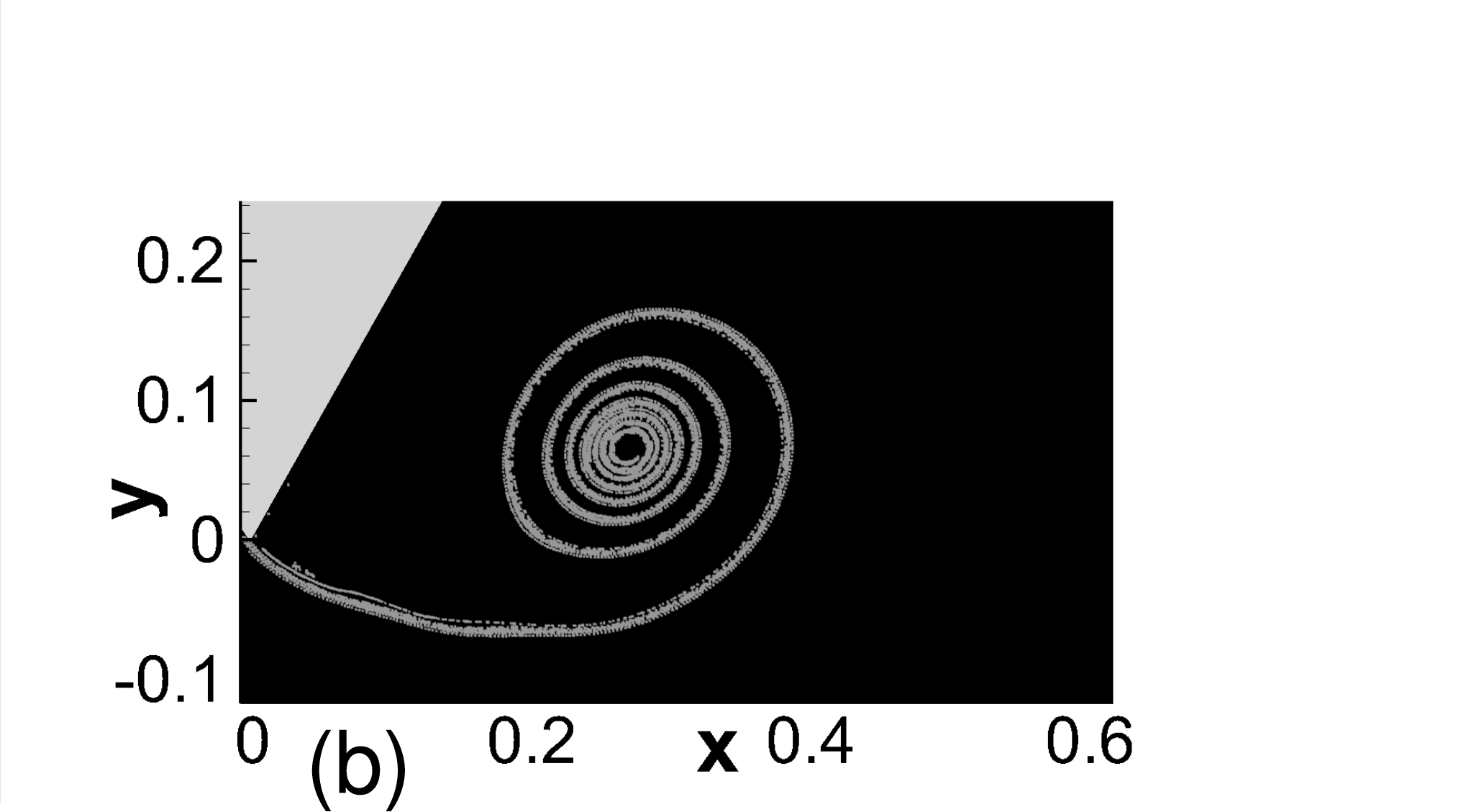}
			&
			\hspace{-1.5cm}\includegraphics[width=0.5\textwidth]{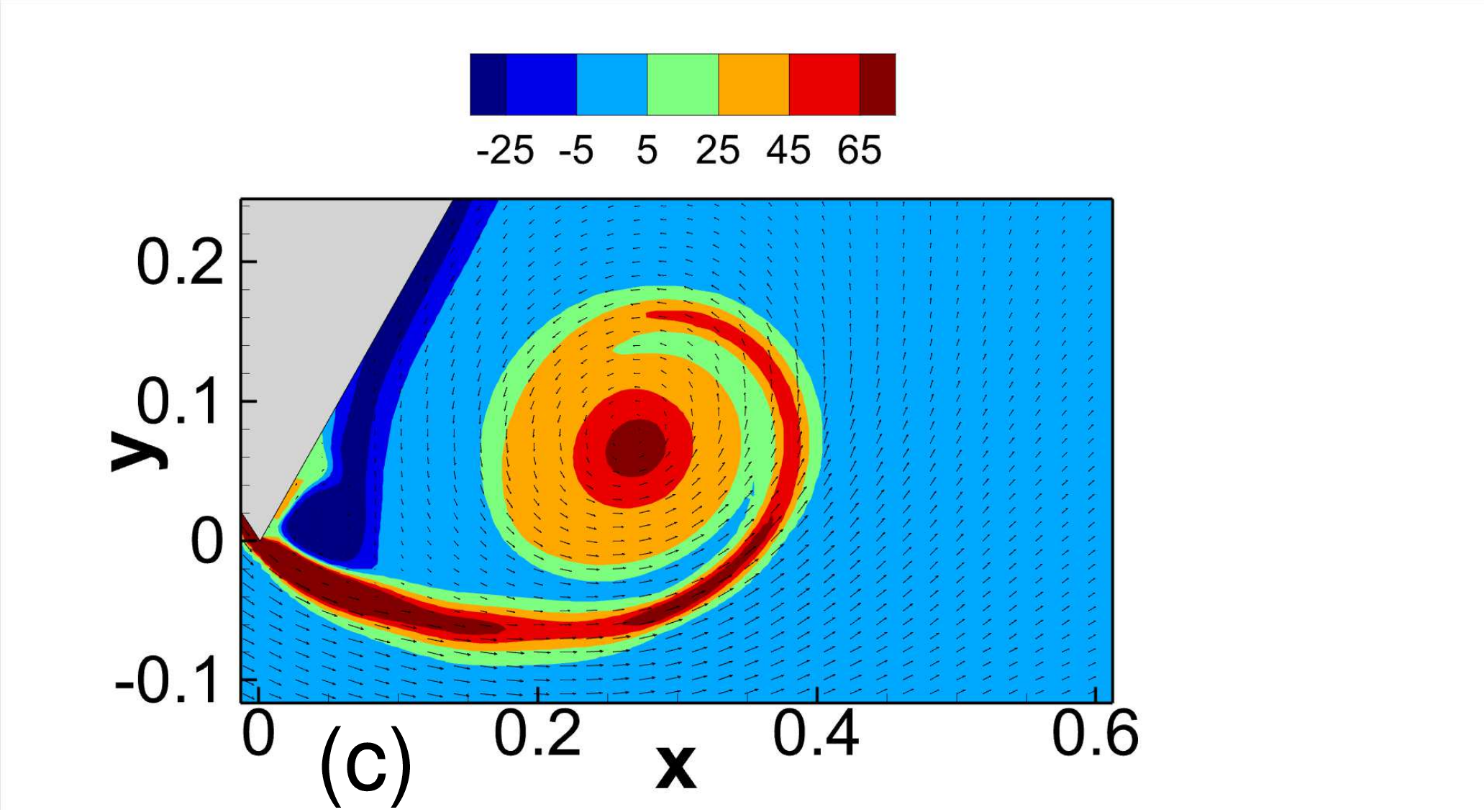}
		\end{tabular}
		\caption{The initial stage: (a) Streaklines from the experimental result of \citet{lian1989} ($t=1.44$), (b) Streaklines from the present numerical computation for $Re_c=6873$ at $\tilde{t}=9.35\;sec$ ($t=1.0$)  and (c) Velocity vectors and vorticity contours from the present computation for $Re_c=6873$ at $\tilde{t}=9.35\;sec$ ($t=1.0$).}
		\label{stage1}
	\end{figure}
	The spiral vortex sheet structure of the starting vortex is depicted in figures \ref{stage1}(a-c), where \citet{lian1989}'s experimental visualization in figure \ref{stage1}(a) is compared to the streaklines resulting from our computation  in figure \ref{stage1}(b). The lead up to the formation of this vortex and its subsequent growth has already been discussed in section \ref{fl_early}. The vortex shedding that originated from the edge has rolled up into a spiral shape with a densely wound-up core layer  (see figures \ref{sk_wed_6873}(a)-(h) also) (\citet{pullin1978}). The viscous effect causes the closely spaced shear layers in real fluids like the ones studied in these studies to combine swiftly into a rigid core, where the concentrated vorticity in the vortex sheet diffuses into uniformly distributed vorticity as can be seen in figure \ref{stage1}(a) and the velocity vectors and the flooded vorticity contours presented in figure \ref{stage1}(c).

	\subsubsection{Second Stage}
	\begin{figure}
		\begin{tabular}{cccc}
			\hspace{-1.0cm}\includegraphics[width=0.3\linewidth]{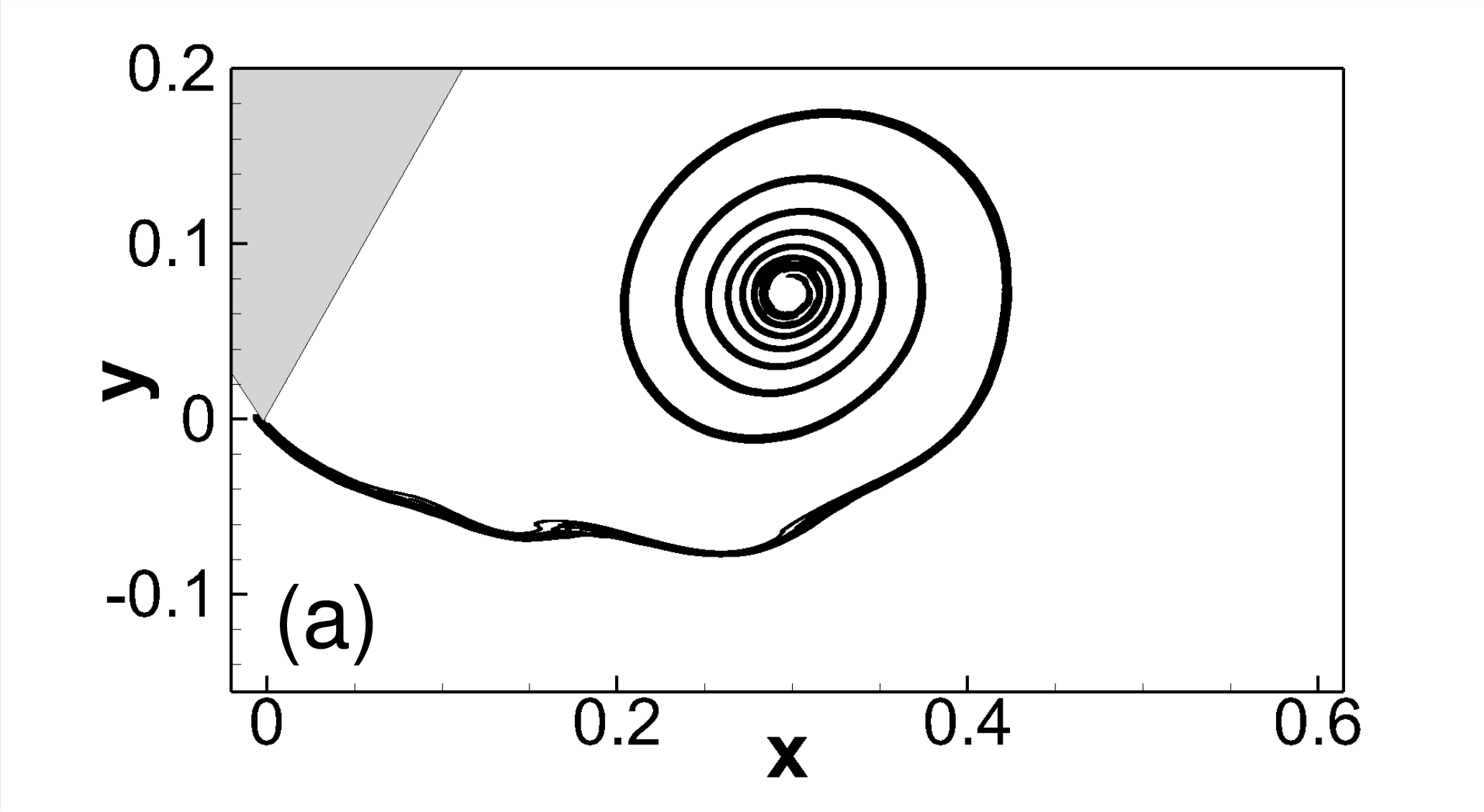}
			&
			\hspace{-0.5cm}\includegraphics[width=0.3\linewidth]{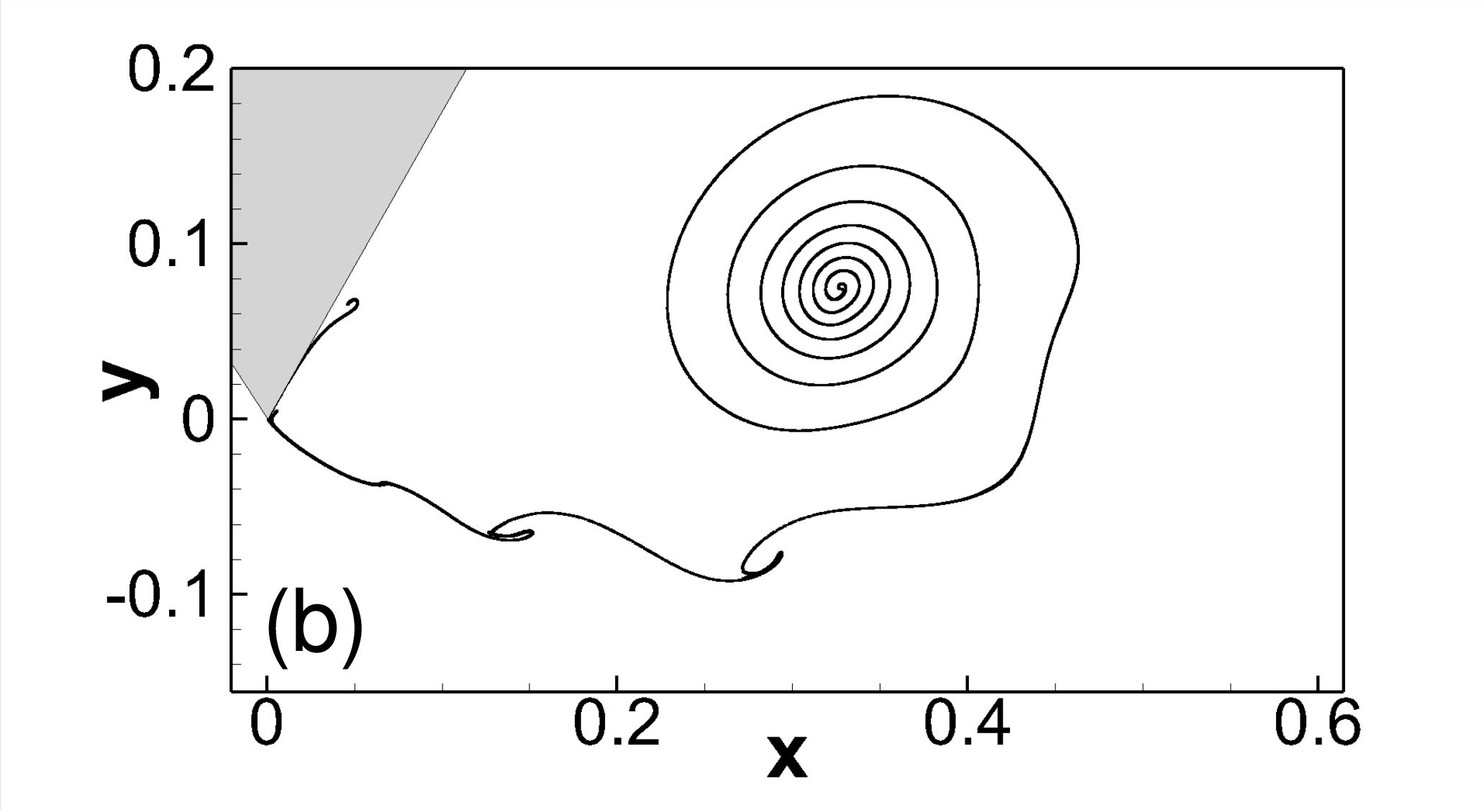}
			\&
			\hspace{-0.5cm}\includegraphics[width=0.3\linewidth]{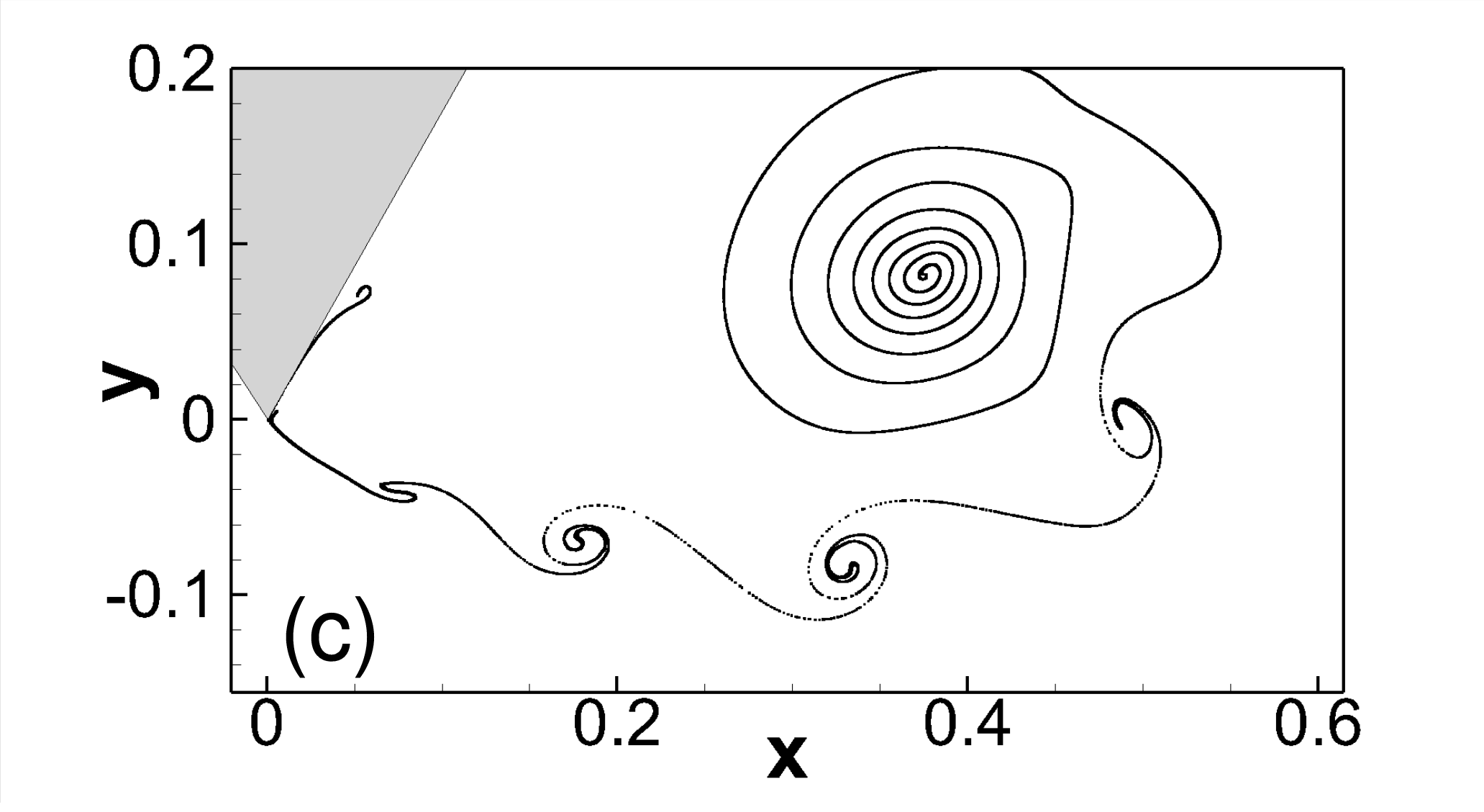}
			&
			\hspace{-0.5cm}\includegraphics[width=0.3\linewidth]{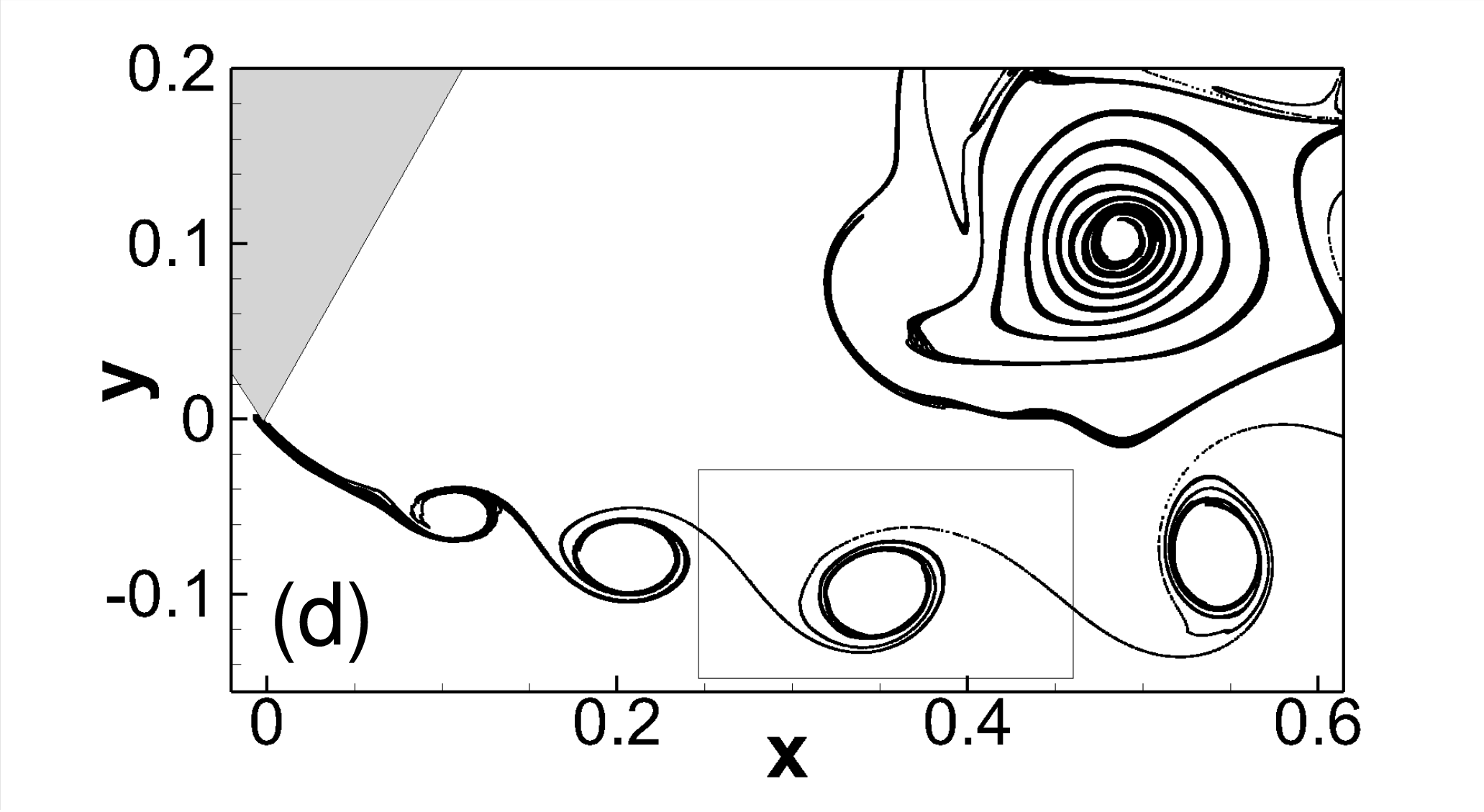}
		\end{tabular}
		\begin{tabular}{cccc}
			\hspace{-1.0cm}\includegraphics[width=0.3\linewidth]{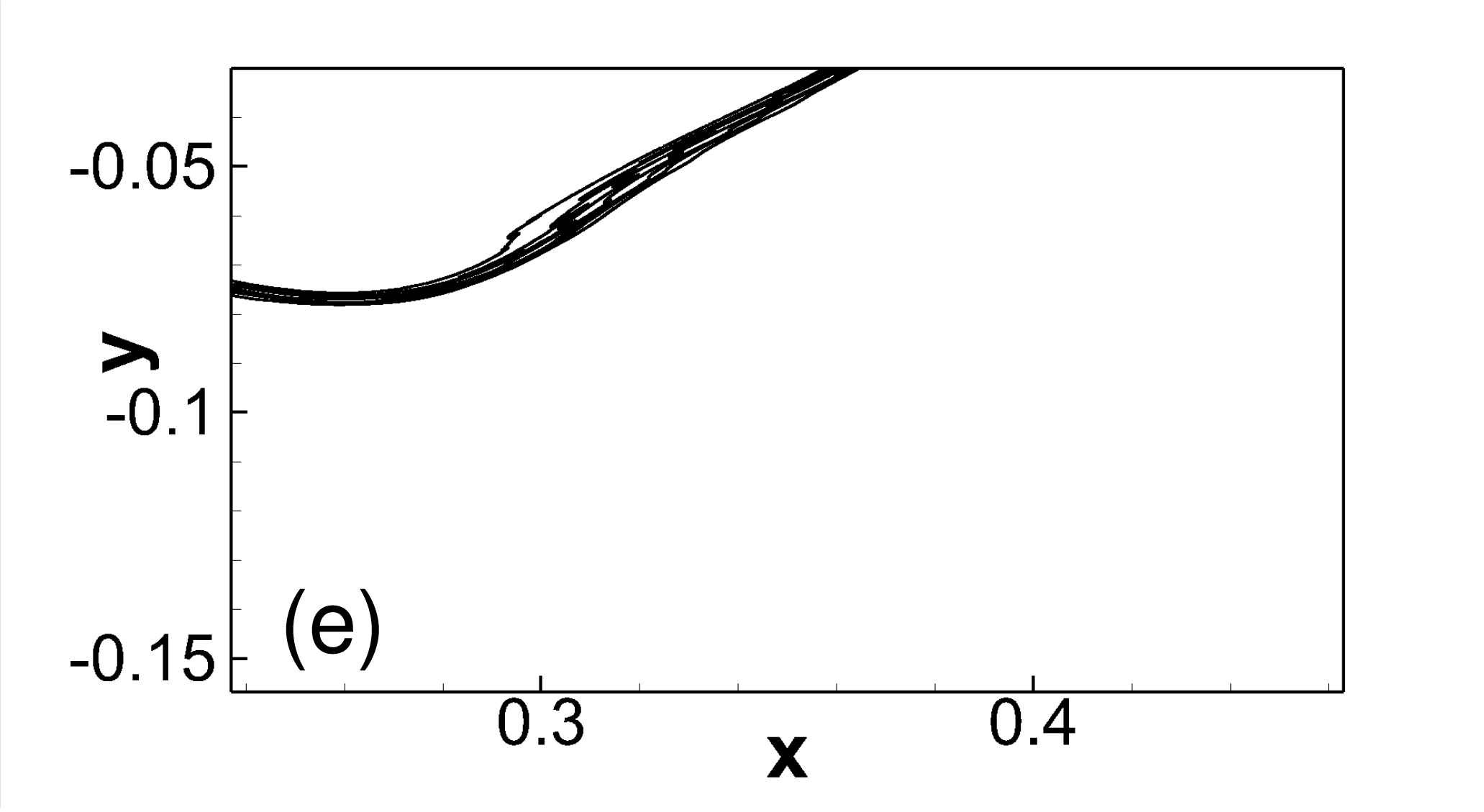}
			&
			\hspace{-0.5cm}\includegraphics[width=0.3\linewidth]{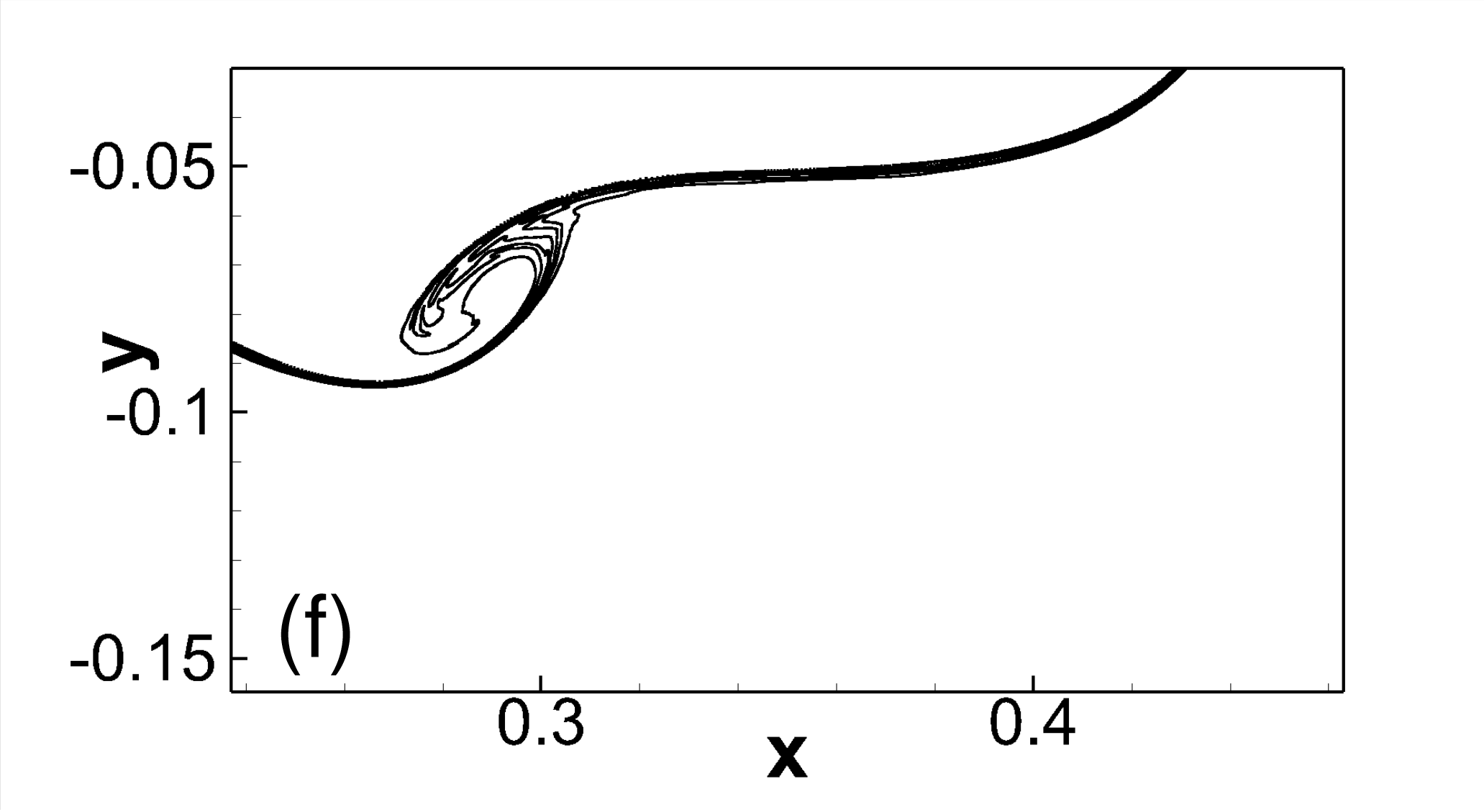}
			\&
			\hspace{-0.5cm}\includegraphics[width=0.3\linewidth]{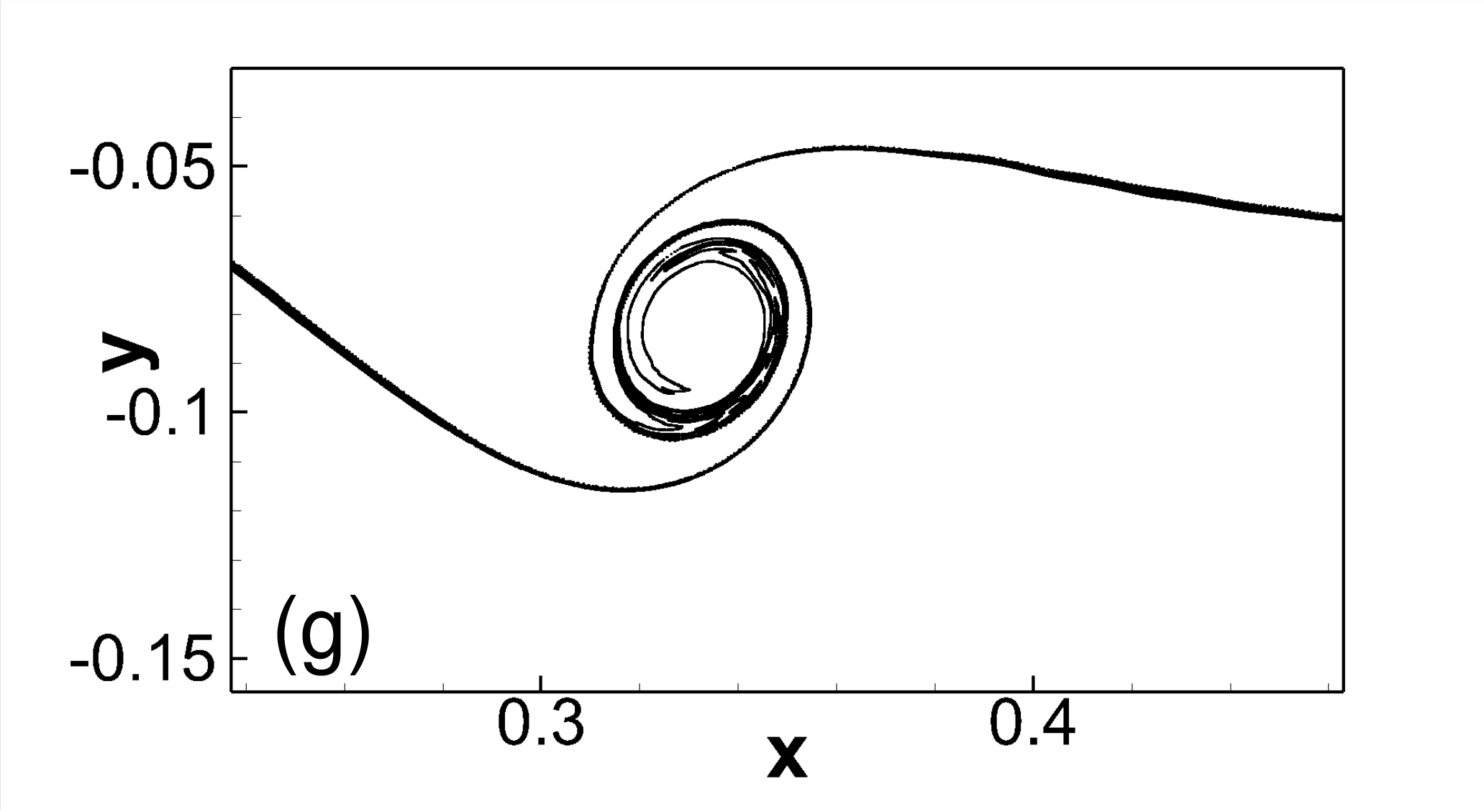}
			&
			\hspace{-0.5cm}\includegraphics[width=0.3\linewidth]{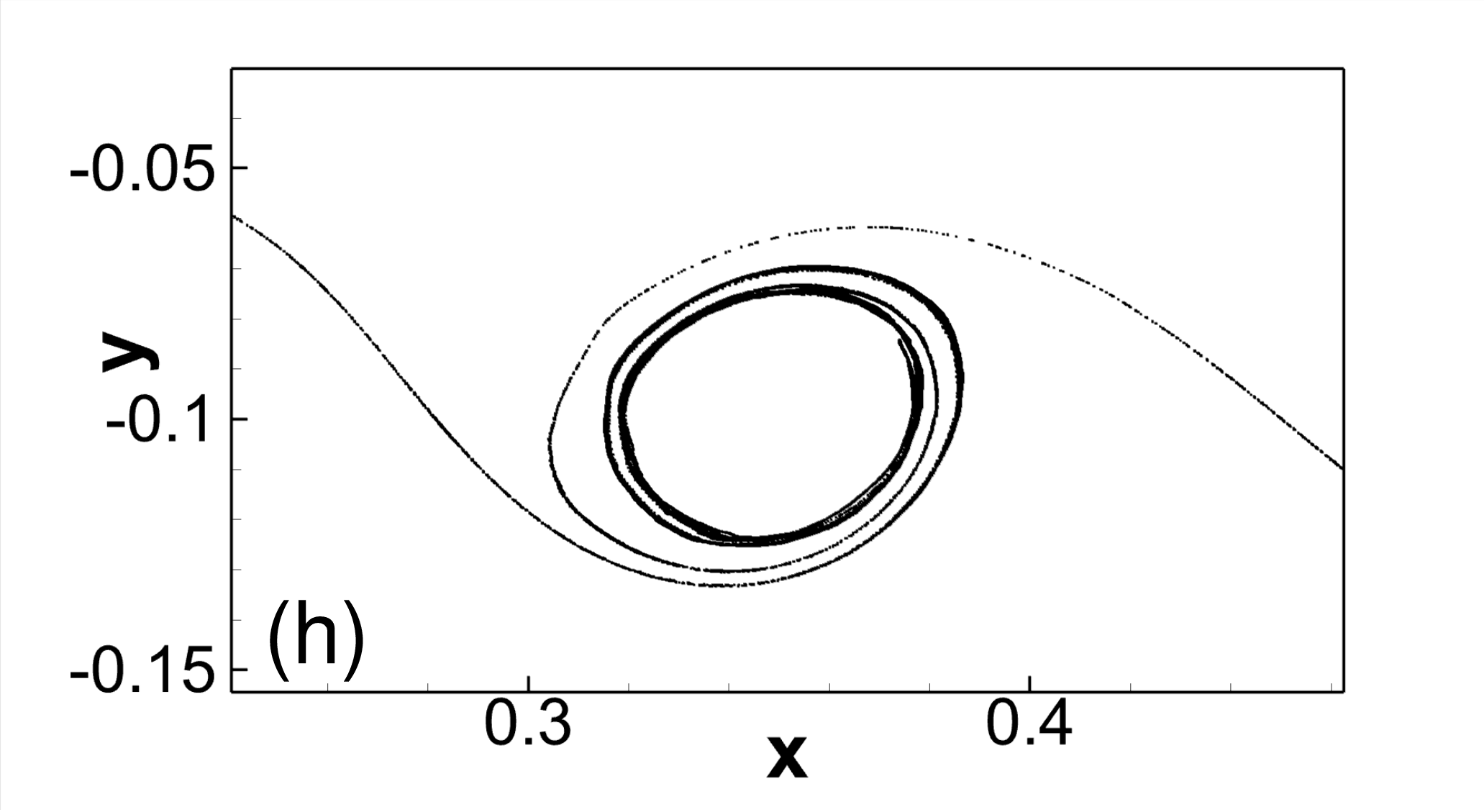}
		\end{tabular}
		\caption{\sl {Onset of shear layer instability and beyond for $Re_c=6873$: Streaklines at times (a) $t=1.06$ ($\tilde{t}=9.91\;sec$), (b) $t=1.12$ ($\tilde{t}=10.47\;sec$), (c) $t=1.20$ ($\tilde{t}=11.22\;sec$) and (d) $t=1.40$ ($\tilde{t}=13.09\;sec$). Close-up view of the evolution of double branching structure corresponding to the rectangular region in \ref{ev_shear}(d) above for the same instants.}}
		\label{ev_shear}
	\end{figure}
\begin{figure}
		\begin{tabular}{cc}
			\hspace{-0.1cm}\includegraphics[width=0.6\linewidth]{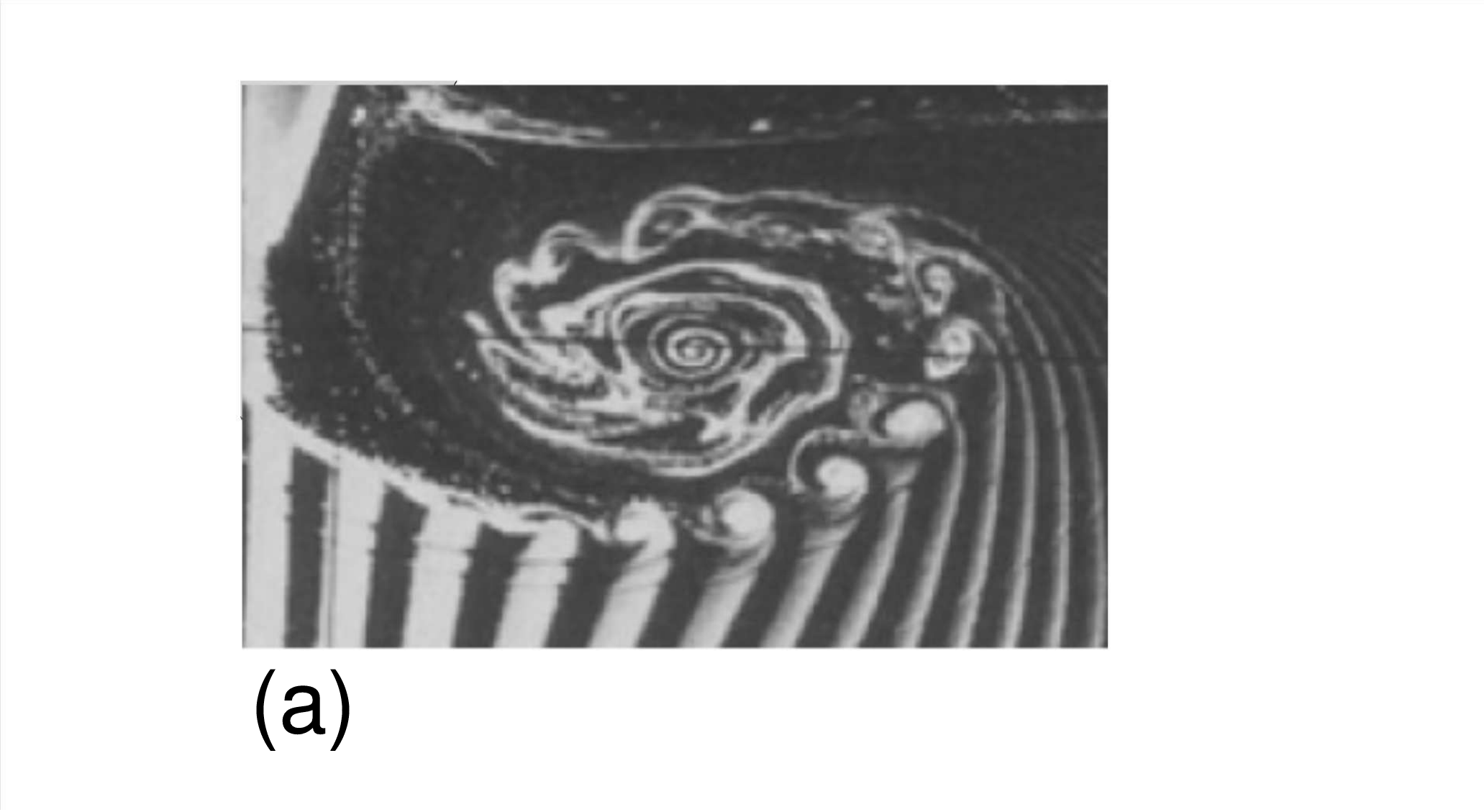}
			&
			\hspace{-2.0cm}\includegraphics[width=0.6\linewidth]{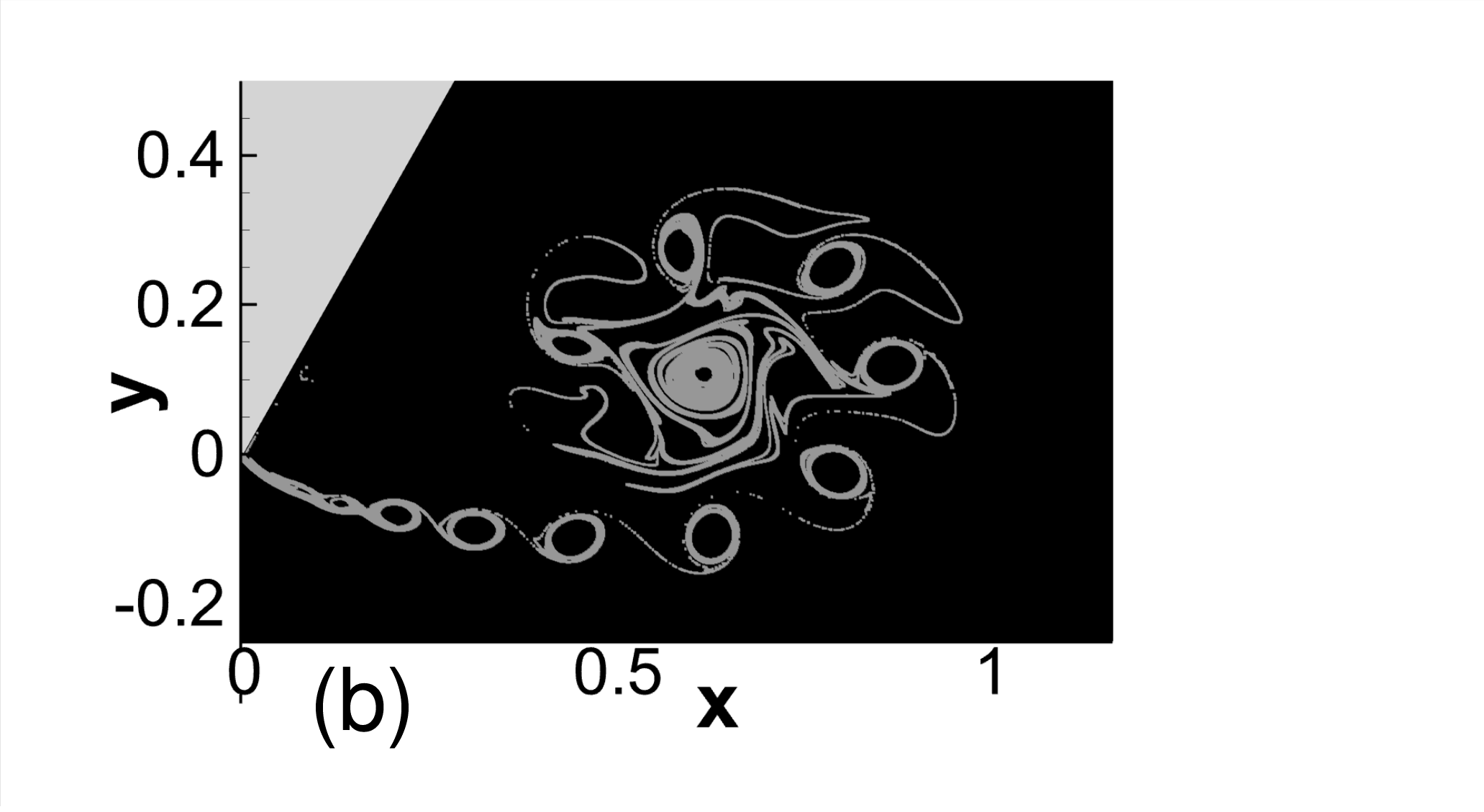}
		\end{tabular}
		\caption{The second stage: Comparison of streaklines between (a) the experimental result of \citet{lian1989}  ($t=2.09$) and (b) the present numerical simulation for $Re_c=6873$ at $\tilde{t}=14.95 \; sec$ ($t=1.6$).}
		\label{stage2}
	\end{figure}

	The second stage is fraught with the phenomenon of shear layer instability described earlier in section \ref{fl_early}, which we depict in figure \ref{ev_shear} where we have plotted streaklines near the wedge-tip at times $t=1.06$, $1.12$, $1.20$ and $1.40$ along with their close-up views.
	Because of the shear layer instability, peripheral vortex layer of the starting vortex becomes wavy as can be seen from figure \ref{ev_shear}(a). Following that, a portion of peripheral  layer of the vortex sheet splits and rolls up into little vortices, as depicted in figures \ref{ev_shear}(b)-(d). The starting vortex breaking into wavy structure is the announcement of shear layer instability and can be seen more clearly from figure \ref{ev_shear}(e)-(h), where we have shown close-up view of the streaklines corresponding to the region in the rectangular box of figure \ref{ev_shear}(d). These small vortices have a double-branched spiral roll structure as shown in figure \ref{ev_shear}(g)-(h), which is typical of shear layer instability (\citet{tsai1993}). In figure \ref{stage2}, we compare our computed streaklines at $t=2.0$ side by side with  the experimental visualization of \citet{lian1989}. The small vortex structures observed in this figure were also observed in the laboratory experiment of flow past an accelerated flat plate by \citet{pierce1961}, and the numerical simulation of \citet{koumoutsakos1996}. One can clearly see from figures \ref{stage2}(a)-(b) that these small vortices are spaced almost uniformly, with their centers lying across the spiral curve of the the large starting vortex. This confirms that they are nothing but parts of the large starting vortex only. For a more vivid visual experience of the evolution of shear layer instability, one may look up the accompanying video "shear.avi".

	\subsubsection{Third Stage: Three fold structure}
	\begin{figure}
		\centering
		\includegraphics[width=3in]{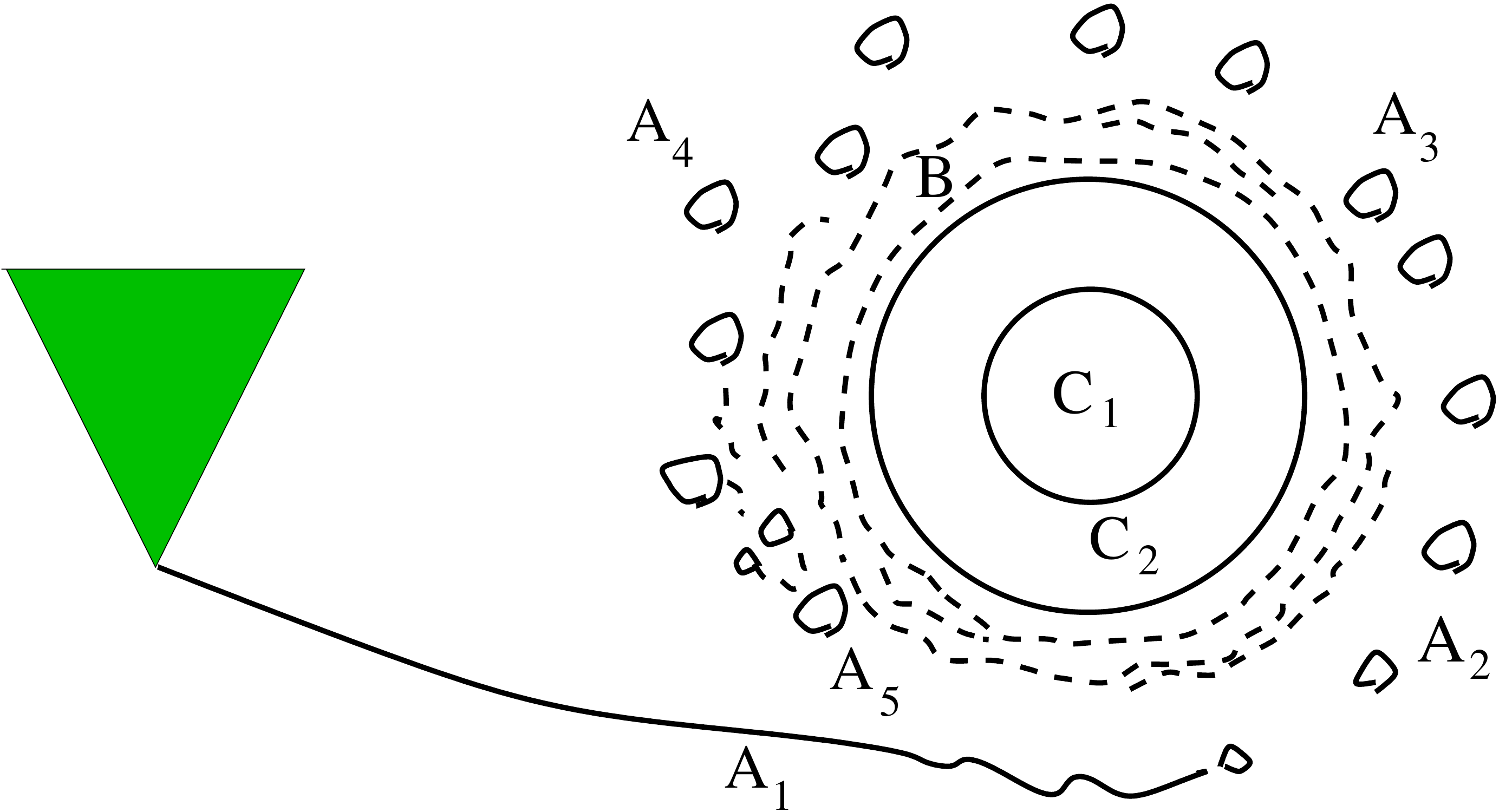}
		\caption{Schematic of the three-fold structure.}
		\label{3f_conf}
	\end{figure}
\begin{figure}
		\begin{tabular}{cc}
			\hspace{-1.5cm}\includegraphics[width=0.6\linewidth]{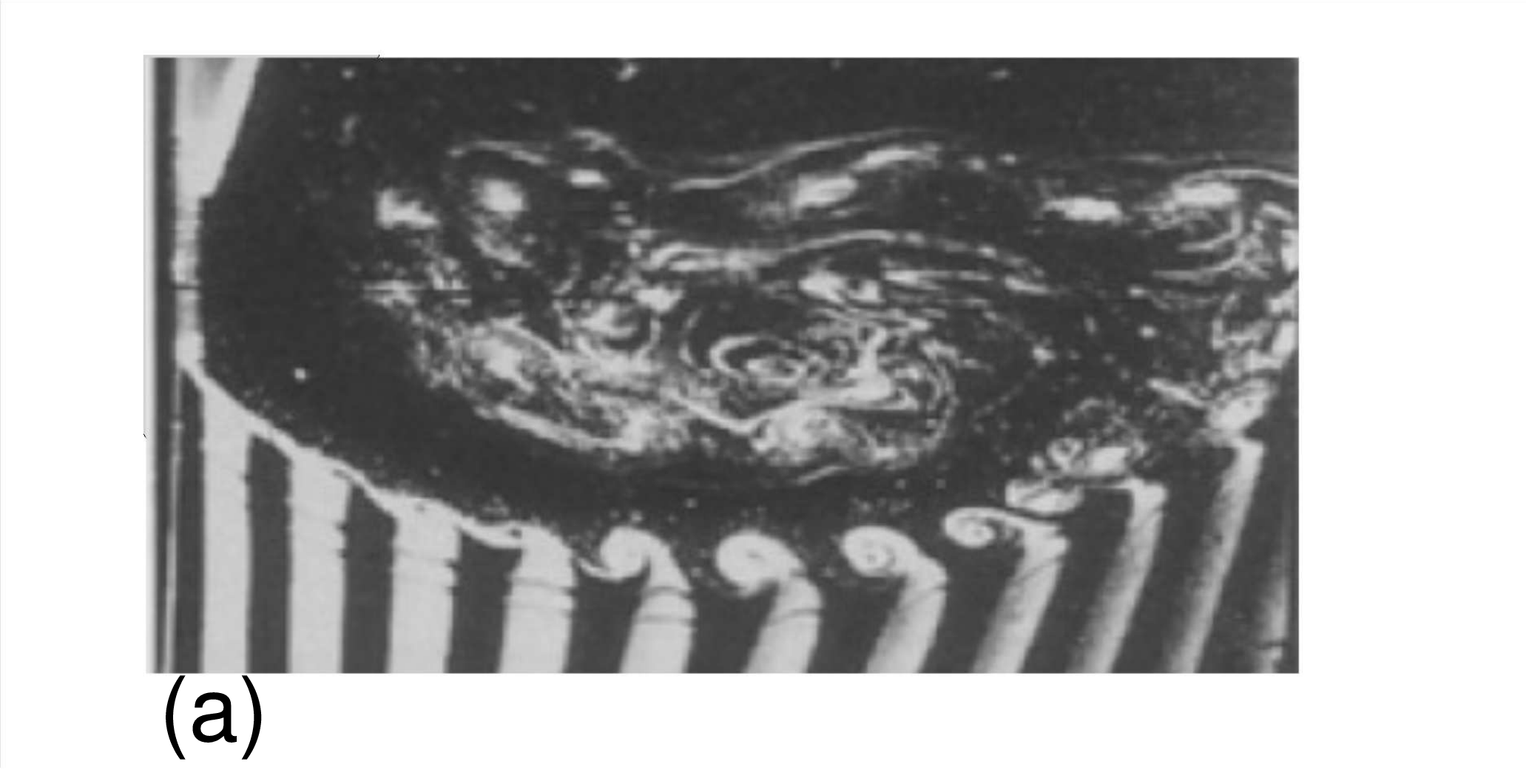}
			&
			\hspace{-2.0cm}\includegraphics[width=0.6\linewidth]{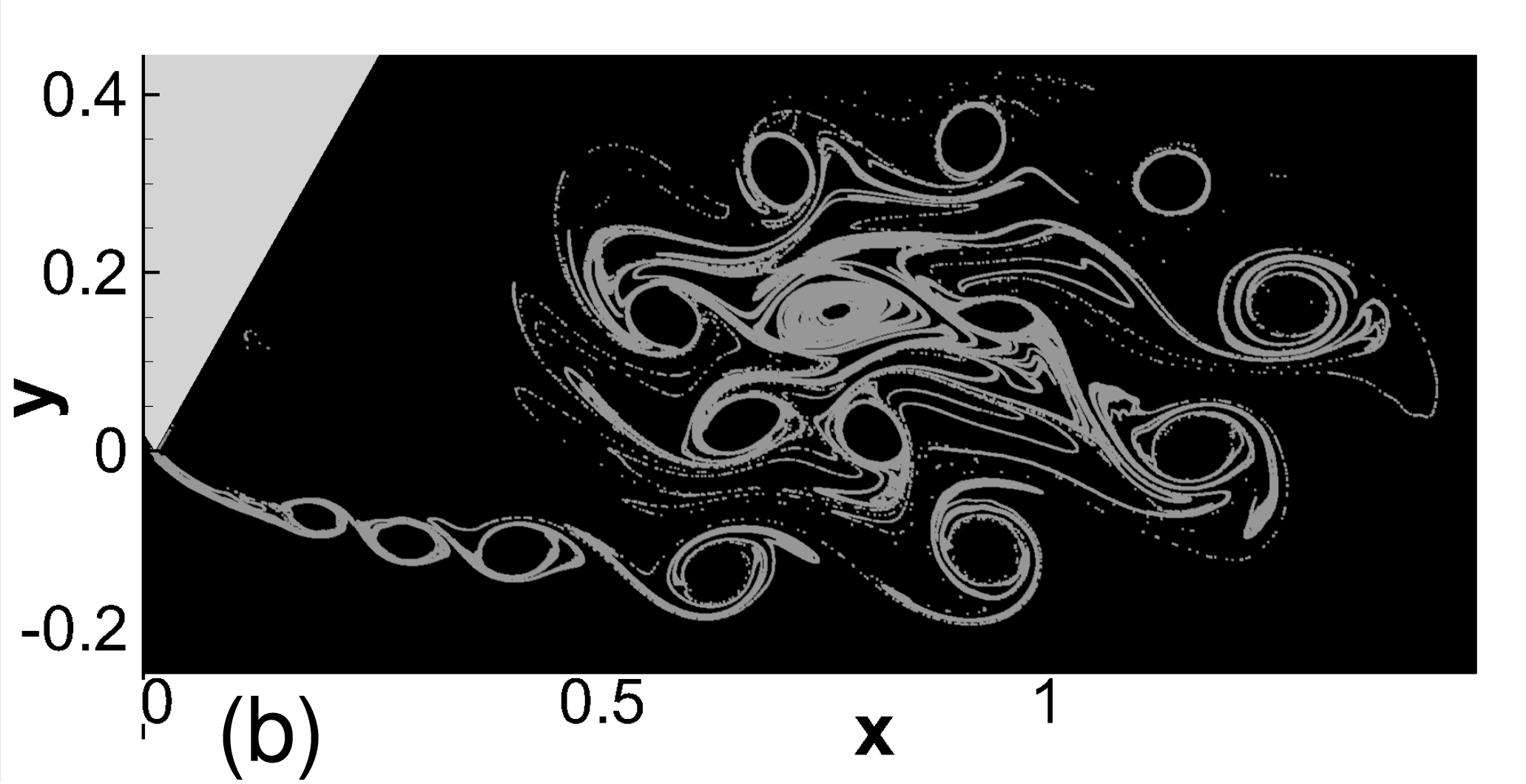}
		\end{tabular}
		\caption{The third stage: Comparison of streaklines between (a) the experimental result of \citet{lian1989} ($t=2.885$) and (b) the present numerical simulation for $Re_c=6873$ at $\tilde{t}=20.57\; sec$ ($t=2.2$).}
		\label{stage3}
	\end{figure}

	\cite{lian1989} described this stage as the one marked by the existence of the three-fold structure. For the benefit of the readers, we re-enact the schematic of this typical structure of the starting vortex behind the wedge in figure \ref{3f_conf}. This stage is  associated with the time frame $t>2$ depicted in figures \ref{center}(b)-(c) beyond the vertical line corresponding to $t=2.0$. In figure \ref{stage3}, we have shown a comparison of streaklines between the experimental results of \citet{lian1989} ($t=2.885$) and the present numerical simulation for $Re_c=6873$ at $\tilde{t}=20.57\; sec$ ($t=2.2$). The three-fold contains the peripheral vortex layer, the core, and the annular zone. The peripheral layer is divided into several sections:
	\begin{itemize}
		\item A1: The part of shear layer, which is just shedded off from the edge;  its starts off smooth but becomes wavy as it moves downstream.
		\item A2: The region where the shear layer breaks into little vortices that are evenly distributed throughout a smooth arc.
		\item A3: This is the region where as a consequence of interaction between the shedded tiny vortices  lose their regular spacing. Because of the instability generated by the interaction, the position of these vortices may become chaotic.
		\item A4: In this portion, the small vortices are scattered even more randomly.
		\item A5: It is located at the very end of the peripheral layer. The flow seems to be quite turbulent here.\\
	\end{itemize}
	
	The core can be divided into two parts: The Central core '$C_1$' and the outer core '$C_2$'.
	\begin{itemize}
		\item $C_1$: This core is formed owing mainly by the act of laminar diffusion. Viscous shear stress causes the concentrated vorticity of the initial spiral shear layer to diffuse. Since no apparent mass transfer is involved in the process, the shape of the spiral line remains undisturbed as shown in the the central part of the starting vortex in figure \ref{stage3}. However, the flow pattern surrounding this part appears very turbulent. This part is a rigid core.
		\item $C_2$: It is the region around the undisturbed core and the streak lines appear very chaotic. However a close look at figure \ref{stage2} would reveal that some instants earlier, in the same region '$C_2$' the streak line used to be smooth, that implies there is no shear layer and no turbulent flow. Through  a series of visualization, \citet{lian1989} had established that but some time before the event of figure \ref{stage2}(a), it was in fact turbulent, the turbulence appearance in the region $C_2$ of figure \ref{stage3} had germinated at that time. Thus the fluid in the outer core $C_2$ has passed three stages; in the initial stage there were spiral shear layer (see the core in figure \ref{stage1}(b)), which then broke into turbulent flow (see the region just outside the core in figure \ref{stage2}(b)), and later, the turbulence has dissipated and the flow has again become laminar with distributed vorticity. This is the reason $C_2$ is sometimes termed as a "relaminarized" region. 
	\end{itemize}
	
	The region marked as $B$ is the annular region which is turbulent in nature, a thin one and sometimes one can see this layer itself breaking into small vortices as in figure \ref{stage2}(a).

	Thus, from the above discussion coupled with streakline visualizations of figure \ref{stage3}(a)-(b), it is clear that our simulation has very aptly captured all three stages of the evolution of the starting vortex leading to the three-fold structure which is exemplified by the comparison of our computed results with those of \cite{lian1989}. To the best of our knowledge, for the flow past a mounted wedge, such structures have not been reported earlier. This further re-establishes the robustness of the scheme developed in \citet{kumar2020}, which despite being primarily developed for laminar flows, has remarkably resolved the early stages of turbulence as well.

	\subsubsection{A short note on the long time flow evolution for $Re_c=1560$}
	\begin{figure}
		\begin{tabular}{cc}
			\hspace{-1.5cm}\includegraphics[width=0.6\linewidth]{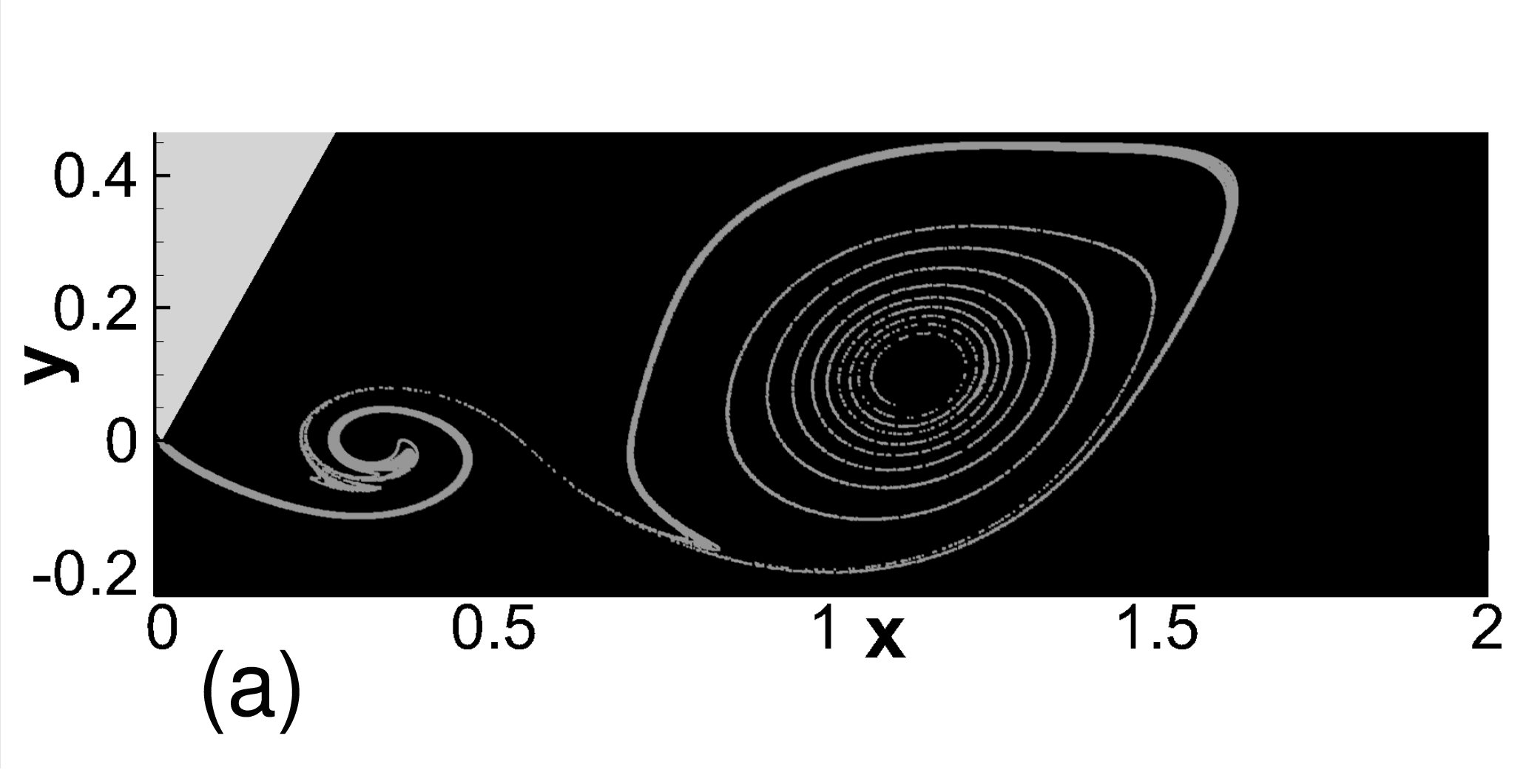}
			&
			\hspace{-2.0cm}\includegraphics[width=0.6\linewidth]{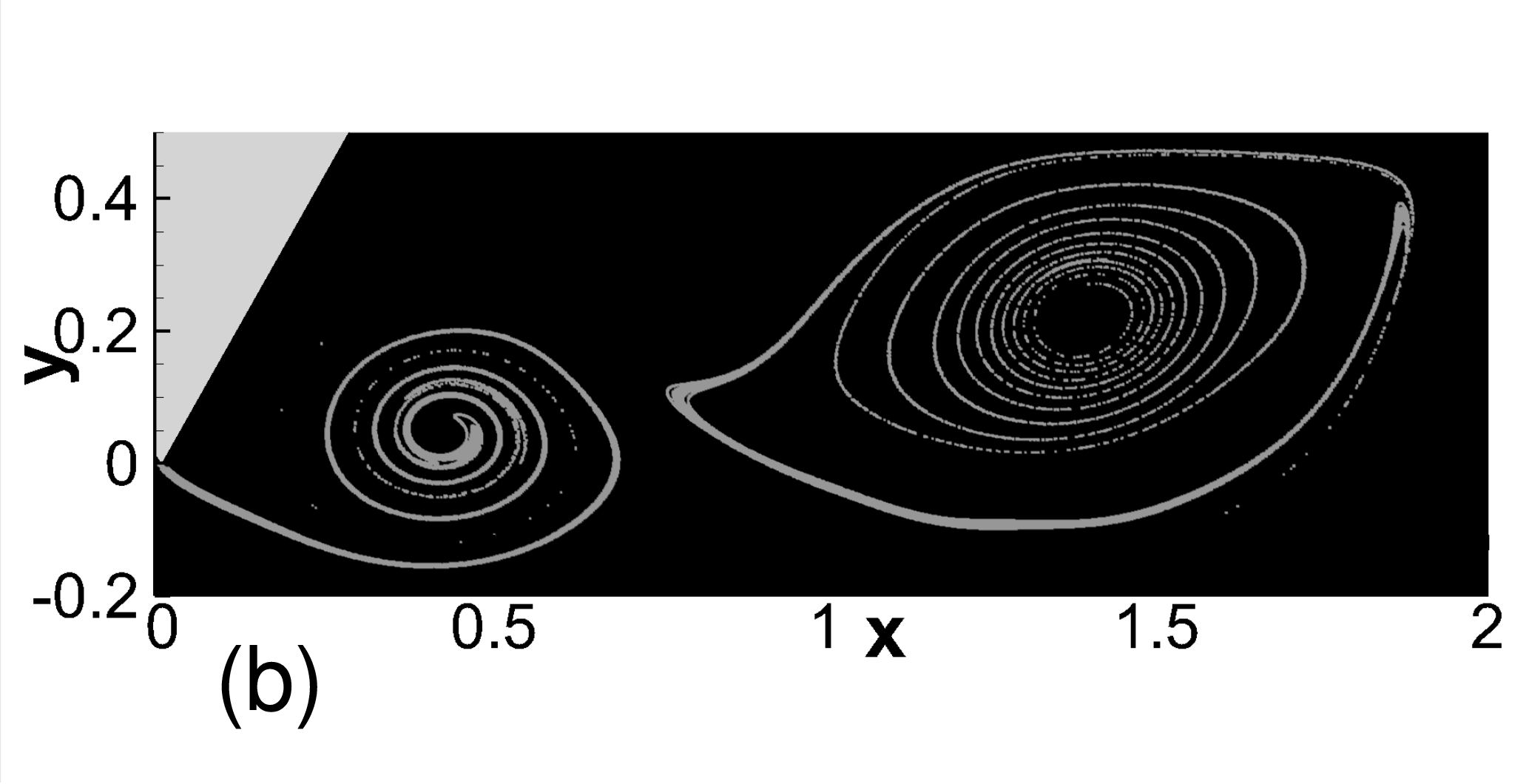}
		\end{tabular}
		\caption{Streaklines for $Re_c=1560$ at instants (a) $\tilde{t}=100.79 \; sec$ ($t=2.5$) and (b) $\tilde{t}=120.95 \; sec$ ($t=3.0$).}
		\label{long_1560}
	\end{figure}
	Our discussions so far in section \ref{beyond} have been confined to flows for $m \neq 0$ only. For $m=0$ and $\beta=1/3$, description for the flow simulation for the initial stage could be found in the work of \cite{xu2016}, albeit for an infinite wedge. Therefore, it would be interesting to see whether the flow patterns observed in the second stage for $m \neq 0$ are also existent for $m=0$ (uniform flow) as well or not. Here, the evidence of a full blown shear layer instability characterised by the double-branching structure was seen around a real physical time $\tilde{t}=100.79 \; sec$ ($t=2.5$) for $Re_c=1560\; (m=0)$  as in figure \ref{long_1560}(a). On the other hand, same was observed for $Re_c=6873$ at a much earlier time $\tilde{t}=19.91\; sec$ ($t=1.06$) as can be seen from the streaklines in figure \ref{ev_shear}(a). However for $Re_c=1560$, the flow remained laminar for the whole duration of time for which simulation was performed as can be seen from figure \ref{ev_shear}(b) at time $\tilde{t}=120.95 \; sec$ ($t=3.0$).
	
	\subsection{Existence of Coherent Structures}
	Coherent structures are structures that remain in the flow for an extended period of time and are not merely a  momentary phenomena (\citet{hussain1986}). Considering any two-dimensional turbulent flow; there are several structures that resemble vortices and remain in the flow for a short period of time. They will have a rotary motion and a circular appearance, and if they don't endure more than a few dynamical times, they will be ephemeral, non-coherent, and classified as Eddies. If, on the other hand, an 'Eddy' tends to stay for an extended period of time and does not appear to vanish, it is no longer an Eddy but a vortex, and it is no longer transient. These structures are coherent and can persist a viscous time scale or longer, depending on the conditions. Multi-scale vorticity fluctuations are a characteristic of turbulent flows. Several methods exist for identifying vortical coherent formations. Closed-loop streamlines and pathlines, minimum pressure areas, and absolute magnitude of vorticity are examples, however since they combine rotational motion and shear, these methods do not truly identify vortex cores. Nevertheless, \citet{jeong1995} demonstrated that the second invariant of the rate of strain-tensor and the negative part of the second largest eigenvalue of the same tensor are reliable indicators of coherent structures in various flow situations. The second invariant in terms of streamfunction is given by  the {\bf Q criterion} (\citet{benzi1987}). 
	\begin{figure}
		\begin{center}
			\includegraphics[width=0.45\textwidth]{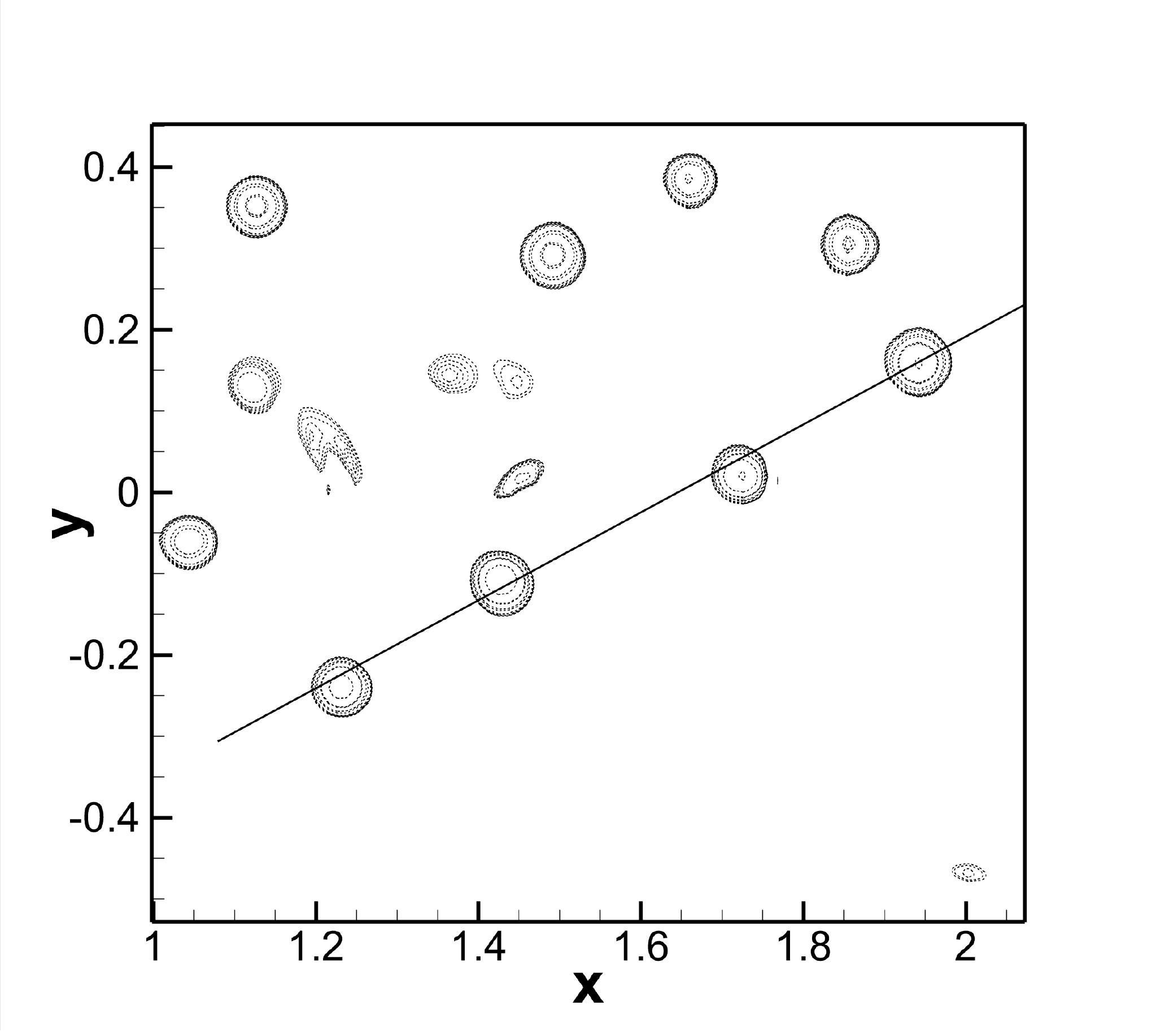}
			{(a)}
			\includegraphics[width=0.45\textwidth]{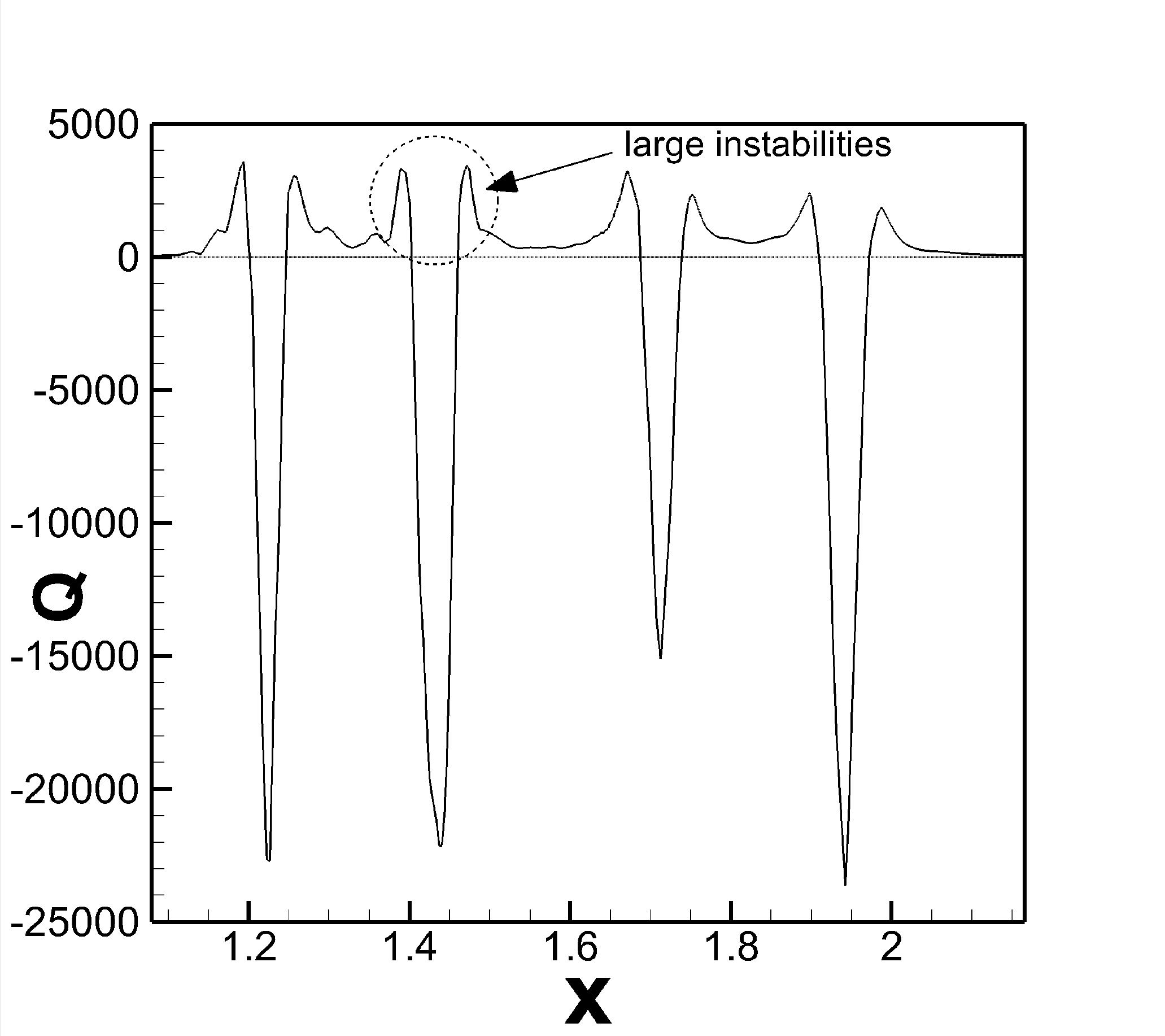}
			{(b)}
			\caption{(a) Contour maps of $Q$ at $\tilde{t}=24.31\;sec$ and (b) Distribution of $Q$ along the line shown in (a) for $Re_c=6873$.}
			\label{q_c}
		\end{center}
	\end{figure}
	
	A quick backdrop of this criterion could be given as follows. The eigenvalue analysis of the perturbed velocity field mentioned above, the eigenvalues are given by the formula $\lambda=\pm \sqrt{Q}$, where
	$$\displaystyle Q=\left (\frac{\partial^2 \psi}{\partial x\partial y}\right )^2-\frac{\partial^2 \psi}{\partial x^2}\frac{\partial^2 \psi}{\partial y^2}.$$	
	It follows that in the regions of the fluid where $Q<0$, the distance between two particles embedded in the original velocity field will not diverge as a function of time. In figure \ref{q_c}(a), we plot the $Q$ contours for the flow past a mounted wedge discussed in the previous section for $Re_c=6873$ at non-dimensional time $\tilde{t}=24.31\; sec$ ($t=2.6$) in a cross-section of the computational domain. Here, the dotted curves represent negative values of $Q$ which corresponds to stable eigenvalues while solid  curves represent positive values corresponding to unstable ones. As such the vortical shapes bounded by closed solid curves are coherent structures. One can see that these structures are always surrounded by dense negative contours indicating that large instabilities occur only at the edge of the vortices. This phenomenon can be understood more clearly by drawing a straight line through the centres of the coherent vortices (see figure \ref{q_c}(a)) away from the wedge in the region $A_3$ of figure \ref{3f_conf}, and plotting the $Q$ distribution along the $x$-coordinates of it as shown in figure \ref{q_c}(b). Here the  segments of the graph below the zero-line represent the coherent structures. One can always see some oscillations on the edge of these structures just above the zero-line indicating the instabilities surrounding them. A similar procedure for the vortical structures resulting from the $Q$-contours for $Re_c=1560$ at $\tilde{t}=120.95\; sec$ ($t=3.0$) revealed no such instabilities. Thus our simulation has also resolved the existence of coherent structures in the flow field indicating the onset of turbulence for $Re_c=6873$.
	
	\section{Conclusion}\label{concl}
	In this study, a recently developed second-order spatially and temporally accurate compact scheme on a non uniform Cartesian grid, developed by the authors, is used to simulate the complex fluid flow past a wedge mounted on a wall placed on a channel with wedge angle $60^{\circ}$ for channel Reynolds number $Re_c=1560$, $6621$ and $6873$. While the first $Re_c$ corresponds to a uniform flow corresponding to a pre-chosen parameter $m=0$, the latter two corresponds to accelerated flow corresponding to $m=0.45$ and $0.88$ respectively.  It is inspired by the famous 1980 laboratory experiment of \cite{pullin1980}. To the best of our knowledge, this is probably the first time that for such flow situation, the boundary conditions used in the actual computation is such that it exactly mimics the physical situation used in the laboratory. Moreover, the authors are not aware of any numerical simulation for the flow past a wedge  exactly of the current configuration for an value of $m$ as high as $0.88$.
	
	Our numerical simulation was able to replicate all the flow features of their laboratory experiment very accurately. A detailed account of the flow separation in the earliest part of the flow along with their time-line has also been provided. This includes tracking the trajectories of the vortex and rotation centers and comparing them with the laboratory experiment as well as inviscid similarity theory. Our computed vortex center was found to be more closer to the scaling of the inviscid similarity theory than the experimental results. We also investigated the effect of the parameter $m$ that determines the intensity of acceleration as well as the intriguing consequence of the non-dimensionalization typical of such flow configuration. Initially a weaker flow field was observed for the accelerated flow cases than the uniform one, marked by a reduction in the size of the starting vortices at non-dimensional time $t<1$ coupled with a weaker circulation. On the other hand, for $t>1$, both the primary vortex center displacements and the size of the vortices are seen to catch each other. For $m \neq 0$, shear layer instability sets in very early in the flow, leading to more complicated vortex dynamics, indicating a transition towards turbulence.
	
	The most significant achievement of the current study is, however, the simulation of the flow for a time range much beyond Pullin and Perry's experimental endeavour. The evolution of shear-layer instability leading to the onset of turbulence, characterised by the existence of coherent structures has been captured very aptly by our simulation. The accuracy of our simulation has been validated not only by the existence of coherent structures in the flow, but also by the remarkable closeness of our simulation to the high Reynolds number experimental results of \cite{lian1989} for the accelerated flow past a normal flat plate. All the three stages of vortex shedding, including the extremely complicated three-fold structure were resolved very efficiently.
	
	\bibliographystyle{unsrtnat}
	\bibliography{references}
\end{document}